\documentclass[12pt,a4paper]{article}

\usepackage{cite}
\usepackage{amsmath}
\usepackage{amssymb}
\usepackage{graphicx}
\usepackage{geometry}

\newcommand{\half}{\frac{1}{2}}
\newcommand{\ep}{\epsilon}
\newcommand{\al}{\alpha_{s}}

\makeatletter
    
    \@addtoreset{equation}{section}
\makeatother

\begin{document}
\setlength{\baselineskip}{18pt}

\begin{titlepage}

\begin{flushright}
HRI-P-14-09-006\\
KOBE-TH-15-07
\end{flushright}
\vspace{1.0cm}
\begin{center}
{\Large\bf A proof algorithm associated with the dipole splitting 
algorithm} 
\end{center}
\vspace{20mm}

\begin{center}
{\large
K. Hasegawa
}\footnote{{\it E-mail address}\,: kouhei@phys.sci.kobe-u.ac.jp\,.}
\end{center}
\vspace{1cm}
\centerline{{\it Department of Physics, Kobe University,
Kobe 657-8501, Japan.}}

%
%
\vspace{2.0cm}
\centerline{\large\bf Abstract}
\vspace{0.5cm}
We present a proof algorithm associated with the dipole 
splitting algorithm (DSA). The proof algorithm (PRA) is 
a straightforward algorithm to prove that the summation 
of all the subtraction terms created by the DSA vanishes. 
The execution of the PRA provides a strong consistency
check including all the subtraction terms --the dipole, 
I, P, and K terms-- in an analytical way. Thus we can obtain
more reliable QCD NLO corrections. We clearly define 
the PRA with all the necessary formulae and demonstrate 
it in the hadron collider processes
$pp \to \mu^{+}\mu^{-},\,2\,jets$, and $n\,jets$.

\end{titlepage}

\tableofcontents
\newpage
\section{Introduction \label{sec_1}}
The present article follows
Ref.\cite{Hasegawa:2014oya}.
We would like to start with a summary of
Ref.\cite{Hasegawa:2014oya}.
At the CERN Large Hadron Collider (LHC), 
the standard-model higgs boson was discovered in 
the year 2012 during 
run\,1 with collision energies of 7 and 8 TeV.
Run\,2, with an energy of 13 TeV, is planned to start 
in 2015.
In order to identify the discovery signals, we need
both precise experimental results
and precise theoretical predictions.
For the precise theoretical predictions, at least
the inclusion of the 
quantum chromodynamics\,(QCD)
next-to-leading order\,(NLO) 
corrections is required.
One of the most successful procedures to obtain 
the QCD NLO corrections for multiparton leg
processes is the Catani--Seymour dipole subtraction
procedure \cite{Catani:1996vz,Catani:2002hc}. 
This procedure has been already applied to the huge
number of processes happening at the LHC.
A partial list of the achievements there is collected
in the bibliography of Ref.\cite{Hasegawa:2014oya}.

Now that the dipole subtraction has been applied to
so many processes, we can see some drawbacks about
the use of the procedure.
Among these drawbacks, we would like to point
out three difficulties. The first difficulty is 
to confirm whether the subtraction terms created
are necessary and sufficient ones, and whether the 
expression of each term includes no mistake. 
The second difficulty is for one person to reproduce
the results in an article written by another person,
in particular, the difficulty of specifying all the 
subtraction terms used. This difficulty is equally 
valid in the inverse case, namely, the difficulty
for one person to tell the other person all 
the subtraction terms used in a reasonably short form
without confusion. The third difficulty is
about the use of computer packages
in which the dipole subtraction procedure is automated. 
Publicly available packages are presented in 
Refs.
\cite{Seymour:2008mu,
Hasegawa:2008ae,
Hasegawa:2009tx,
Hasegawa:2010dz,
Frederix:2008hu,
Frederix:2010cj,
Czakon:2009ss}.
Users sometimes have difficulties
in understanding the algorithms
implemented in the packages
and the outputs of the run.

In order to solve some of these difficulties,
it is required that a practical algorithm to use 
the dipole subtraction is clearly defined, and that
the documentation of the algorithm includes the
clear presentation of all the subjects in the
wishlist as shown in 
Ref.\cite{Hasegawa:2014oya}\,:
\begin{enumerate}
\item Input, output, creation order, and
all formulae in the document,
\vspace{-2mm}
\item Necessary information to specify each subtraction term,
\vspace{-2mm}
\item Summary table of all subtraction terms created,
\vspace{-2mm}
\item Associated proof algorithm.
\end{enumerate}
We succeeded in constructing an algorithm that allows
the clear presentation of all the entries in the
wishlist. It is named the dipole splitting algorithm
(DSA). The DSA has been already presented,
focusing on the entries 1--3 
at the wishlist, in 
Ref.\cite{Hasegawa:2014oya}.
Thus, the purpose of this article is to present
the last entry in the wishlist,
``\,4.\,Associated proof algorithm''.

An associated proof algorithm means a 
straightforward algorithm to prove that the summation
of all the subtraction terms created by the DSA vanishes.
The proof algorithm is abbreviated as PRA hereafter.
What the PRA proves is expressed in a formula, as follow.
The QCD NLO corrections for an arbitrary process in a hadron
collider are written as 
\begin{equation}
\sigma_{\mbox{{\tiny NLO}}} 
= \sigma_{\mbox{{\tiny R}}} 
+ \sigma_{\mbox{{\tiny V}}} 
+ \sigma_{\mbox{{\tiny C}}}\,, 
\label{norm}
\end{equation}
where the symbols
$\sigma_{\mbox{{\tiny R}}}, \sigma_{\mbox{{\tiny V}}}$ 
and 
$\sigma_{\mbox{{\tiny C}}}$
represent the real correction, the virtual correction, 
and the collinear subtraction term, respectively.
In the framework of dipole subtraction, 
the NLO corrections are reconstructed as
\begin{equation}
\sigma_{\mbox{{\tiny NLO}}} 
= (\sigma_{\mbox{{\tiny R}}} 
- \sigma_{\mbox{{\tiny D}}})
+ (\sigma_{\mbox{{\tiny V}}} 
+ \sigma_{\mbox{{\tiny I}}}) 
+ \sigma_{\mbox{{\tiny P}}} 
+ \sigma_{\mbox{{\tiny K}}}\,, 
\label{dipo}
\end{equation}
where the symbols 
$\sigma_{\mbox{{\tiny D}}}, \sigma_{\mbox{{\tiny I}}}$,
$\sigma_{\mbox{{\tiny P}}}$,
and 
$\sigma_{\mbox{{\tiny K}}}$ 
represent the dipole, I, P, and K terms, 
respectively. The quantities 
$(\sigma_{\mbox{{\tiny R}}} 
- \sigma_{\mbox{{\tiny D}}})$,
$(\sigma_{\mbox{{\tiny V}}} 
+ \sigma_{\mbox{{\tiny I}}})$,
$\sigma_{\mbox{{\tiny P}}}$,
and 
$\sigma_{\mbox{{\tiny K}}}$
are separately finite.
When the NLO cross section in
Eq.\,(\ref{norm}) is equated with 
the cross section in
Eq.\,(\ref{dipo}),
we obtain the identity for the arbitrary process as
\begin{equation}
\sigma_{\mbox{{\tiny subt}}} 
= \sigma_{\mbox{{\tiny D}}} 
+ \sigma_{\mbox{{\tiny C}}}
- \sigma_{\mbox{{\tiny I}}} 
- \sigma_{\mbox{{\tiny P}}} - 
\sigma_{\mbox{{\tiny K}}}=0\,. 
\label{subt}
\end{equation}
We call this relation the consistency relation 
of the subtraction terms.
The PRA is a straightforward algorithm to prove the
consistency relation in Eq.\,(\ref{subt}) 
for any given process,
if all the subtraction terms --the dipole, I, P, and 
K terms-- are created by the DSA.

We clarify the main advantages of the PRA.
Since the relation in Eq.\,(\ref{subt}) 
includes all the subtraction terms 
--the dipole, I, P, and K terms-- created,
the proof of the consistency relation gives 
the confirmation
of all the terms. As mentioned above, the subtracted
cross sections
$(\sigma_{\mbox{{\tiny R}}} 
- \sigma_{\mbox{{\tiny D}}})$
and
$(\sigma_{\mbox{{\tiny V}}} 
+ \sigma_{\mbox{{\tiny I}}})$
are separately finite. When we use any wrong
collection or wrong expression for the dipole
or the I term, the subtracted cross sections
would diverge.
In this way, at least the divergent parts of the
dipole and I terms are confirmed by the 
successful cancellation against the real 
and virtual corrections.
Compared to the dipole and I terms,
the P and K terms
$\sigma_{\mbox{{\tiny P}}}$,
and 
$\sigma_{\mbox{{\tiny K}}}$
are separately finite themselves,
and confirmation by the cancellation
is impossible. The PRA can provide
the precious confirmation including the P and K terms.
This is the first advantage of the PRA.
The PRA is executed in an analytical way
and does not rely on any numerical evaluation.
All the dipole terms are integrated over the
soft and collinear regions of the phase space 
in an analytical way in $d$ dimensions.
The expressions for the integrated dipole
terms are available in
Refs.\cite{Catani:1996vz,Catani:2002hc}.
The PRA utilizes the integrated dipole terms and 
all the steps of the PRA are executed in
an analytical way that does not require 
any numerical evaluation.
This is the second advantage of the PRA.
In consequence, we can have a strong consistency check
of all the subtraction terms 
by the execution of the PRA,
and we can obtain more reliable QCD NLO corrections 
as a results.

In the dipole subtraction procedure,
some algorithms to create the subtraction terms
may be constructed.
The construction 
of a straightforward algorithm
to prove the consistency relation
in Eq.\,(\ref{subt}),
is not always possible
for all of them.
The cancellations in the consistency
relation 
are realized between
cross sections with
the same initial states
and the same reduced Born processes.
In the DSA, the subtraction terms created
are classified by real processes
and the kind of the parton splitting.
The classification of the subtraction terms
is converted to the classification by
the initial states and the 
reduced Born processes.
Then we can easily identify the cross
sections that cancel each other
in Eq.\,(\ref{subt}).
The systematical identification
of the cancellations
for an arbitrary process
makes possible for us to construct
a straightforward proof algorithm
of the consistency relation. 
In the original article about
Catani--Seymour dipole subtraction
\cite{Catani:1996vz},
the dipole terms are constructed
to subtract the soft and collinear
divergences from the real corrections,
and then the dipole terms
are analytically integrated over the soft and 
collinear regions in 
$d$-dimensional phase space.
The integrated dipole terms
are transformed to the I, P,
and K terms.
The method of the transformation is
explained for proton--proton collider case
in Sec.\,10 in Ref.\cite{Catani:1996vz}.
The PRA is just the transformation 
in a different order.
The PRA is constructed in such a way that
when the subtraction terms
are created by the DSA,
the consistency
relation of the subtraction 
terms can be proved
just by following well defined 
steps in a straightforward algorithm.

The present paper is organized as follows\,:
The PRA is defined in Sec.\,\ref{sec_2}.
All the necessary formulae are collected
in Appendix \ref{ap_A}.
The PRA is demonstrated in the processes,
$pp \to \mu^{+}\mu^{-}$, 2 jets, and 
$n$ jets in Sec.\,\ref{sec_3}, 
\ref{sec_4}, and \ref{sec_5},
respectively.
The results of the PRA 
in the dijet and $n$ jets
processes
are summarized
in Appendices \ref{ap_B} and \ref{ap_C},
respectively.
Sec.\,\ref{sec_6}
is devoted to the summary.

\clearpage




\section{Proof algorithm \label{sec_2}}
\subsection{Definition \label{s_2_1}}
The prediction of the cross section
including the QCD NLO corrections is
generally written as
\begin{equation}
\sigma_{\mbox{{\tiny prediction}}} = 
\sigma_{\mbox{{\tiny LO}}} + \sigma_{\mbox{{\tiny NLO}}}\,, 
\end{equation}
where the symbol $\sigma_{\mbox{{\tiny LO}}}$ represents
the leading order (LO) cross section or a distribution,
and
the symbol $\sigma_{\mbox{{\tiny NLO}}}$ represents the 
QCD NLO corrections to the LO cross section.
The LO cross section does not appear in the 
present paper hereafter.
For a given collider process
the real emission processes
that contribute to the collider process
are written as $\mbox{R}_{i}$\,.
The set consisting of all the real
emission processes is denoted as
\begin{equation}
\{\mbox{R}_{i}\}=\{
\mbox{R}_{1},\mbox{R}_{2},
...,
\mbox{R}_{\,n_{real}}
\}\,,
\label{setri}
\end{equation}
where $n_{\mbox{{\tiny real}}}$ is the number of all the
real processes. 
When the NLO corrections are treated within the framework of
the DSA\,\cite{Hasegawa:2014oya}, they are expressed
as
\begin{equation}
\sigma_{\mbox{{\tiny NLO}}} = \sum_{i=1}^{n_{\mbox{{\tiny real}}}} 
\sigma(\mbox{R}_{i})\,. \label{masternlo}
\end{equation}
In the DSA, all the corrections are
classified by the real processes $\mbox{R}_{i}$,
and each contribution is denoted as $\sigma(\mbox{R}_{i})$.
The cross section $\sigma(\mbox{R}_{i})$ 
is defined as
\begin{equation}
\sigma(\mbox{R}_{i})=
\bigl(\,
\sigma_{\mbox{{\tiny R}}}(\mbox{R}_{i}) - 
\sigma_{\mbox{{\tiny D}}}(\mbox{R}_{i}) 
\,\bigr) + 
\bigl(\,
\sigma_{\mbox{{\tiny V}}}(\mbox{B}1(\mbox{R}_{i})) + 
\sigma_{\mbox{{\tiny I}}}(\mbox{R}_{i}) 
\,\bigr) + 
\sigma_{\mbox{{\tiny P}}}(\mbox{R}_{i}) + 
\sigma_{\mbox{{\tiny K}}}(\mbox{R}_{i})\,,
\label{master1}
\end{equation}
where the cross sections
$\sigma_{\mbox{{\tiny R}}}(\mbox{R}_{i})$ 
and $\sigma_{\mbox{{\tiny V}}}(\mbox{B}1(\mbox{R}_{i}))$
represent
the real and virtual corrections, respectively.
The process $\mbox{B}1(\mbox{R}_{i})$ is the Born
process reduced from $\mbox{R}_{i}$ by the rule
$\mbox{B}1(\mbox{R}_{i})=
\mbox{R}_{i}\,\mbox{-}(\mbox{a gluon in final state})$,
in Ref.\cite{Hasegawa:2014oya}.
The cross sections 
$\sigma_{\mbox{{\tiny D}}}(\mbox{R}_{i})$,
$\sigma_{\mbox{{\tiny I}}}(\mbox{R}_{i})$,
$\sigma_{\mbox{{\tiny P}}}(\mbox{R}_{i})$,
and
$\sigma_{\mbox{{\tiny K}}}(\mbox{R}_{i})$
represent the contributions of the dipole, I, P, and K terms, 
respectively. 
The cross sections
are factorized into the parton distribution
function (PDF) and the subpartonic cross sections as
\begin{align}
\sigma(\mbox{R}_{i}) 
&= \int dx_{1} \int dx_{2} \ 
f_{\mbox{{\tiny F}}(x_{a})}(x_{1}) 
f_{\mbox{{\tiny F}}(x_{b})}(x_{2}) \ \times  \nonumber \\
& \ \ \biggl[
\bigl(\hat{\sigma}_{\mbox{{\tiny R}}}(\mbox{R}_{i}) - 
\hat{\sigma}_{\mbox{{\tiny D}}}(\mbox{R}_{i}) \bigr) + 
\bigl(\hat{\sigma}_{\mbox{{\tiny V}}}(\mbox{B}1(\mbox{R}_{i})) + 
\hat{\sigma}_{\mbox{{\tiny I}}}(\mbox{R}_{i}) \bigr) + 
\hat{\sigma}_{\mbox{{\tiny P}}}(\mbox{R}_{i}) + 
\hat{\sigma}_{\mbox{{\tiny K}}}(\mbox{R}_{i}) \biggr],  
\label{master}
\end{align}
where $f_{\mbox{{\tiny F}}(x_{a/b})}(x_{1/2})$
represents the PDF and 
the subscript $\mbox{F}(x_{a/b})$ denotes the field 
species of the initial 
state parton in the leg a/b as defined in 
Ref.\cite{Hasegawa:2014oya}.
The symbols $\hat{\sigma}_{\mbox{{\tiny R}}}(\mbox{R}_{i})$ and 
$\hat{\sigma}_{\mbox{{\tiny V}}}(\mbox{B}1(\mbox{R}_{i}))$
represent the partonic real and virtual corrections,
respectively. The quantities $\hat{\sigma}(\mbox{R}_{i})$,
with the subscripts, D, I, P, and K, represent 
the contributions of the dipole, I, P, and K terms 
to the partonic cross sections, respectively.
The definitions of all the partonic cross sections
are collected in Appendix \ref{ap_A_1}.
In order to specify the jet observables
the corresponding jet functions,
$F_J^{(n/n+1)}$,
must be multiplied by all the cross sections.
The use of the jet functions
in the dipole subtraction is explained
in Ref.\cite{Catani:1996vz}.
For compact expression, we do not show 
the jet functions explicitly in the present article.

In the original calculation method
for the QCD NLO corrections, 
the corrections can be constructed as
\begin{equation}
\sigma_{\mbox{{\tiny NLO}}} = \sum_{i=1}^{n_{\mbox{{\tiny real}}}} 
\sigma_{orig}(\mbox{R}_{i}), \label{nlotrad}
\end{equation}
where the symbol $\sigma_{orig}(\mbox{R}_{i})$ denotes
each correction belonging to the real process $\mbox{R}_{i}$.
$n_{\mbox{{\tiny real}}}$ is the same number 
as in Eq.\,(\ref{masternlo}). 
The cross section $\sigma_{orig}(\mbox{R}_{i})$
consists of three terms\,:
\begin{equation}
\sigma_{orig}(\mbox{R}_{i})=
\sigma_{\mbox{{\tiny R}}}(\mbox{R}_{i})
+
\sigma_{\mbox{{\tiny V}}}(\mbox{B}1(\mbox{R}_{i})) 
+ 
\sigma_{\mbox{{\tiny C}}}(\mbox{R}_{i})\,,
\label{trad}
\end{equation}
where the symbols $\sigma_{\mbox{{\tiny R}}}(\mbox{R}_{i})$ 
and $\sigma_{\mbox{{\tiny V}}}(\mbox{B}1(\mbox{R}_{i}))$
are the same real and virtual corrections
as appear in Eq.\,(\ref{master1}).
$\sigma_{\mbox{{\tiny C}}}(\mbox{R}_{i})$
represents the collinear subtraction term.
When the NLO cross section in
Eq.\,(\ref{nlotrad}) is equated with
the cross section in
Eq.\,(\ref{masternlo}), we obtain the identity
for an arbitrary process as
\begin{equation}
\sum_{i=1}^{n_{\mbox{{\tiny real}}}} 
\sigma_{subt}(\mbox{R}_{i})=0\,,
\label{purp}
\end{equation}
where the cross section 
$\sigma_{subt}(\mbox{R}_{i})$
is defined as
\begin{align}
\sigma_{subt}(\mbox{R}_{i})
&=
\sigma_{orig}(\mbox{R}_{i})
-
\sigma(\mbox{R}_{i})
\nonumber \\
&=
\sigma_{\mbox{{\tiny D}}}(\mbox{R}_{i})
+
\sigma_{\mbox{{\tiny C}}}(\mbox{R}_{i})
-
\sigma_{\mbox{{\tiny I}}}(\mbox{R}_{i})
-
\sigma_{\mbox{{\tiny P}}}(\mbox{R}_{i})
-
\sigma_{\mbox{{\tiny K}}}(\mbox{R}_{i})
\,.
\label{defsub}
\end{align}
The cross section 
$\sigma_{subt}(\mbox{R}_{i})$
includes all the subtraction terms.
We call the relation in
Eq.\,(\ref{purp})
the consistency relation of the
subtraction terms.
The aim of the PRA is to prove
the consistency relation
for an arbitrary collider process.
In order to construct the proof algorithm in
separate steps, we reconstruct the dipole term
$\sigma_{\mbox{{\tiny D}}}(\mbox{R}_{i})$ into 
four terms as
\begin{equation}
\sigma_{\mbox{{\tiny D}}}(\mbox{R}_{i})=
\sigma_{\mbox{{\tiny D}}}(\mbox{R}_{i},\,\mbox{I})
+
\sigma_{\mbox{{\tiny D}}}(\mbox{R}_{i},\,\mbox{P})
+
\sigma_{\mbox{{\tiny D}}}(\mbox{R}_{i},\,\mbox{K})
+
\sigma_{\mbox{{\tiny D}}}(\mbox{R}_{i},\,{\tt dip}2)\,.
\label{intdipf}
\end{equation}
The definitions of the four terms will be given in 
Sec.\,\ref{s_2_3}--\ref{s_2_6},
respectively.
Using the four terms, we can rewrite
the $\sigma_{subt}(\mbox{R}_{i})$ in
Eq.\,(\ref{defsub}) as
\begin{align}
\sigma_{subt}(\mbox{R}_{i})
&=
[\,\sigma_{\mbox{{\tiny D}}}(\mbox{R}_{i},\,\mbox{I})
-
\sigma_{\mbox{{\tiny I}}}(\mbox{R}_{i})
\,]
\,+\,
[\,
\sigma_{\mbox{{\tiny D}}}(\mbox{R}_{i},\,\mbox{P})
+
\sigma_{\mbox{{\tiny C}}}(\mbox{R}_{i})
-
\sigma_{\mbox{{\tiny P}}}(\mbox{R}_{i})
\,]
\nonumber \\
&\hspace{5mm}+
[\,
\sigma_{\mbox{{\tiny D}}}(\mbox{R}_{i},\,\mbox{K})
-
\sigma_{\mbox{{\tiny K}}}(\mbox{R}_{i})
\,]
\,+\,
\sigma_{\mbox{{\tiny D}}}(\mbox{R}_{i},\,{\tt dip}2)\,.
\label{sigdfour}
\end{align}
The execution of
the proof algorithm proceeds according to the steps 
in such a way
that the first three terms in square brackets
are calculated in turn. 
At this stage, we can define 
all the six steps of the PRA as follows\,:
\begin{align}
\mbox{{\bf Step 1.}}  & \ \ 
\mbox{Convert the dipole terms}
\ \sigma_{\mbox{{\tiny D}}}(\mbox{R}_{i})
\ \, \mbox{to the integrated form}, 
\nonumber\\[7pt] 
\mbox{{\bf 2.}} & \ \ 
\sigma_{\mbox{{\tiny D}}}(\mbox{R}_{i},\,\mbox{I})
-
\sigma_{\mbox{{\tiny I}}}(\mbox{R}_{i})
=
-\sigma_{\mbox{{\tiny I}}}(\mbox{R}_{i},(2)\,
\mbox{-}1/2,N_{f}{\cal V}_{f\bar{f}})
\,, \nonumber\\[7pt] 
\mbox{{\bf 3.}} & \ \
\sigma_{\mbox{{\tiny D}}}(\mbox{R}_{i},\,\mbox{P})
+
\sigma_{\mbox{{\tiny C}}}(\mbox{R}_{i})
-
\sigma_{\mbox{{\tiny P}}}(\mbox{R}_{i})=0
\,, \nonumber\\[7pt] 
\mbox{{\bf 4.}} & \ \
\sigma_{\mbox{{\tiny D}}}(\mbox{R}_{i},\,\mbox{K})
-
\sigma_{\mbox{{\tiny K}}}(\mbox{R}_{i})
=
-\sigma_{\mbox{{\tiny K}}}(\mbox{R}_{i}, {\tt dip}1,
(3)/(4)\,\mbox{-}1,N_{f}h)
\,, \nonumber\\[7pt] 
\mbox{{\bf 5.}} & \ \
\sigma_{subt}(\mbox{R}_{i})=
-\sigma_{\mbox{{\tiny I}}}(\mbox{R}_{i},(2)\,\mbox{-}1/2,
N_{f}{\cal V}_{f\bar{f}})
-\sigma_{\mbox{{\tiny K}}}(\mbox{R}_{i},{\tt dip}1,
(3)/(4)\,\mbox{-}1,N_{f}h)
\nonumber\\[3pt] 
&\hspace{24mm}+
\sigma_{\mbox{{\tiny D}}}(\mbox{R}_{i},\,{\tt dip}2)
\,, \nonumber\\[7pt] 
\mbox{{\bf 6.}} & \ \ 
\sum_{i=1}^{n_{\mbox{{\tiny real}}}} 
\sigma_{subt}(\mbox{R}_{i})
=0\,.
\label{PRAstep}
\end{align}
All the six steps will be separately explained in 
Sec.\,\ref{s_2_2}--\ref{s_2_7},
respectively.
The premise for the execution of the PRA for a given 
process is that all the dipole, I, P, and K 
terms are created by the DSA in Ref.\cite{Hasegawa:2014oya}.
%
%
%
\subsection{Step 1 : Integrated dipole terms
$\sigma_{\mbox{{\tiny D}}}$ 
\label{s_2_2}}
{\bfseries Step\,1}
of the PRA is to convert all the dipole terms,
which are created by the DSA,
into the integrated form.
The contribution of each dipole term to the 
partonic cross section is generally written 
in $d$ dimensions,
\begin{equation}
\hat{\sigma}_{\mbox{{\tiny D}}}(\mbox{R}_{i}, {\tt dip}j) 
= \frac{1}{S_{\mbox{{\tiny R}}_{i}}} \ \Phi(\mbox{R}_{i})_{d} 
\cdot \frac{1}{n_{s}(a) n_{s}(b)} \, 
\mbox{D}(\mbox{R}_{i}, {\tt dip}j\,)_{IJ,K}\,,
\label{dipori}
\end{equation}
where $\mbox{R}_{i}$ is a real correction process
and ${\tt dip}j$ is the category to which the dipole 
term belongs.
$S_{\mbox{{\tiny R}}_{i}}$ is the symmetric factor 
of the process $\mbox{R}_{i}$.
The spin degree of freedom, $n_{s}(a/b)$, 
is determined as
$n_{s}(\mbox{quark})=2$ and $n_{s}(\mbox{gluon})=d-2=2(1-\ep)$. 
Each dipole term is specified with three legs $I,J,$ and 
$K$ of the real process $\mbox{R}_{i}$.
The dipole term is generally written as
\begin{equation}
\mbox{D}(\mbox{R}_{i}, {\tt dip}j\,)_{IJ,K} =
 -\frac{1}{s_{{\scriptscriptstyle IJ}}} 
\frac{1}{x_{{\scriptscriptstyle IJK}}}
\frac{1}{\mbox{T}_{\mbox{{\tiny F}}(y_{emi})}^{2}}
\,\langle \mbox{B}j \ | 
\mbox{T}_{y_{emi}} \cdot 
\mbox{T}_{y_{spe}}
\ \mbox{V}_{{\scriptscriptstyle IJ,K}}^{y_{emi}}
 | \ \mbox{B}j \rangle\,,
\label{dipgen}
\end{equation}
where the details of the notation expressing the dipole terms
are explained in Ref.\cite{Hasegawa:2014oya}.
The category Dipole\,$j$
\,(in short, ${\tt dip}j$)
and the subcategory of the splittings are shown in
Fig.\,\ref{fig_A2_int_D} 
in Appendix \ref{ap_A_2}.
It is noted that in the DSA we introduce 
the {\em field mapping}, 
$y=f(\tilde{x})$, and in Eq.\,(\ref{dipgen})
the legs of the reduced Born process
$\mbox{B}j$, on which the color and helicity operators act,
are specified with the elements $(y_{emi},y_{spe})$
of set $\{y\}$.

The partonic cross section of the dipole term 
in Eq.\,(\ref{dipori})
is converted to the integrated form as 
\begin{equation}
\hat{\sigma}_{\mbox{{\tiny D}}}(\mbox{R}_{i}, {\tt dip}j, x_{a/b}) 
=
-\frac{A_{d}}{S_{\mbox{{\tiny R}}_{i}}} 
\cdot
\int_{0}^{1}dx \,
\frac{1}{\mbox{T}_{\mbox{{\tiny F}}(y_{emi})}^{2}}\, 
{\cal V}(x;\ep) 
\cdot
\Phi_{a/b}(\mbox{B}_{j},x)_{d}\,
[\, y_{emi}, \,y_{spe}\,]\,,
\label{intmast}
\end{equation}
where the overall factor $A_{d}$ is defined as
\begin{equation}
A_{d}=\frac{\al}{2\pi} \frac{(4\pi \mu^{2})^{\ep}}{\Gamma(1-\ep)}\,.
\label{ad}
\end{equation}
The phase space $\Phi_{a/b}(\mbox{B}_{j},x)_{d}$
is defined in $d$ dimensions in
Eq.\,(\ref{pkps2a})/(\ref{pkps2b}).
The explicit expressions of the integrated dipole 
terms depend on
four types of dipole terms, final--final (FF),
final--initial (FI), initial--final (IF), and
initial--initial (II).
The four types are defined in
Ref.\cite{Catani:1996vz} and 
\cite{Hasegawa:2014oya}.
The types of dipole terms are denoted by 
subcategories\,:
\begin{equation}
\hat{\sigma}_{\mbox{{\tiny D}}}(\mbox{R}_{i},\,{\tt dip}j\,)
\
\supset 
\
\hat{\sigma}_{\mbox{{\tiny D}}}(\mbox{R}_{i},\,{\tt dip}j,\,
\mbox{FF} / \mbox{FI} / \mbox{IF} / \mbox{II} \ )\,.
\end{equation}
The expressions are separately shown in
Eqs.\,(\ref{ffmast}), (\ref{fimast}), (\ref{ifmast}),
and (\ref{iimast}), respectively in 
Appendix \ref{ap_A_2}.
The expression of the the factor ${\cal V}(x;\ep)$
is determined in each type as
\begin{equation}
{\cal V}(x;\ep)
=
\left\{
\begin{array}{ll}
{\cal V}_{\mbox{{\tiny F}}(x_{i}) \mbox{{\tiny F}}(x_{j})}(\ep)
\,\delta(1-x)
& : \mbox{FF} \ (ij,k) \,,
\\
{\cal V}_{\mbox{{\tiny F}}(x_{i}) \mbox{{\tiny F}}(x_{j})} (x\,;\ep)
& : \mbox{FI} \ \ (ij,a) \,,
\\
{\cal V}^{\mbox{{\tiny F}}(x_{a/b}), \, \mbox{{\tiny F}}(y_{emi})}
(x\,;\ep)
& : \mbox{IF} \ \ (ai,k)\,,
\\
{\widetilde {\cal V}}^{\mbox{{\tiny F}}(x_{a/b}),\, 
\mbox{{\tiny F}}(y_{emi})}(x\,;\ep)
& : \mbox{II} \ \ \ (ai,b)\,.
\end{array}
\right.
\end{equation}
The concrete expressions of 
the universal singular functions, 
${\cal V}_{\mbox{{\tiny F}}(x_{i}) \mbox{{\tiny F}}(x_{j})}(\ep)$,
${\cal V}_{\mbox{{\tiny F}}(x_{i}) \mbox{{\tiny F}}(x_{j})}
(x\,;\ep)$,
${\cal V}^{\mbox{{\tiny F}}(x_{a/b}), \,  \mbox{{\tiny F}}(y_{emi})}
(x\,;\ep)$,
and 
${\widetilde {\cal V}}^{\mbox{{\tiny F}}(x_{a/b}),\, 
\mbox{{\tiny F}}(y_{emi})}(x\,;\ep)$,
further depend on the kinds of splittings,
which are written 
with the factor
$1/\mbox{T}_{\mbox{{\tiny F}}(y_{emi})}^{2}$
in Eqs.\,(\ref{singff}), (\ref{singfi}), (\ref{singif}),
and (\ref{singii}), respectively.
The symbol, $[\, y_{emi}, \,y_{spe}\,]$, is defined as
\begin{equation}
[ \, y_{emi}, \,y_{spe} ]
=s^{-\ep} 
\cdot
\langle \mbox{B}j \ | 
\mbox{T}_{y_{emi}} \cdot 
\mbox{T}_{y_{spe}}
 | \ \mbox{B}j \rangle_{d}\,, 
\end{equation}
where the Lorentz scalar, $s$, also depends on the types,
FF, FI, IF, and II, which are defined in 
Eqs.\,(\ref{lsff}), (\ref{lsfi}), (\ref{lsif}),
and (\ref{lsii}), respectively.
The quantity $\langle \mbox{B}j \ | 
\mbox{T}_{y_{emi}} \cdot 
\mbox{T}_{y_{spe}}
 | \ \mbox{B}j \rangle$
is the color-correlated Born squared amplitude
after the spin--color summed and averaged in
$d$ dimensions.

Thanks to the notation employed in the DSA
\cite{Hasegawa:2014oya}, we have a simple and
universal expression for the integrated dipole
term in Eq.\,(\ref{intmast}).
The expression can be further abbreviated in 
a fixed form as
\begin{equation}
\hat{\sigma}_{\mbox{{\tiny D}}}(\mbox{R}_{i}, {\tt dip}j, 
x_{a/b}) 
=
-\frac{A_{d}}{S_{\mbox{{\tiny R}}_{i}}} 
\cdot
(\mbox{Factor 1})
\cdot
(\mbox{Factor 2})\,,
\label{fixf}
\end{equation}
where Factors 1 and 2 are denoted as
\begin{align}
(\mbox{Factor 1})&=\int_{0}^{1}dx \,
\frac{1}{\mbox{T}_{\mbox{{\tiny F}}(y_{emi})}^{2}}\, 
{\cal V}(x;\ep)\,,
\\
(\mbox{Factor 2})&=\Phi_{a/b}(\mbox{B}_{j},x)_{d}\,
[\, y_{emi}, \,y_{spe}\,]\,.
\end{align}
Once the concrete expressions of Factors 1 and 2
are determined, the integrated dipole term is also 
uniquely determined.
Factor 1 is universal in the category
with the same splitting and has a spectator
in the same state among the initial and final states.
In order to summarize all the integrated dipole terms
in a short form, it is sufficient that
the information on the reduced Born process, the
kind of splitting, and Factors 1 and 2
is supplied in a table format.
The summary tables for the dijet process are shown 
in Tables \ref{ap_B_1_tab1}--\ref{ap_B_1_tab11} 
in Appendix \ref{ap_B}.
These tables can be a template of the format
for the summary tables for an arbitrary process.

We would like to show here two examples
in the real process
$\mbox{R}_{1}=u\bar{u} \to u\bar{u}g$,
which contributes to the dijet process
$pp \to 2\,jets + X$.
The creation of the dipole terms by the DSA 
is explained in Ref.\cite{Hasegawa:2014oya}.
\begin{itemize}
%
%
\item Example 1: Table 4 
in Appendix B.1 in Ref.\cite{Hasegawa:2014oya},
\,1.\,(13,\,2)\,. \\
The reduced Born process is taken as
\begin{equation}
\mbox{B}1=u(y_{a})\bar{u}(y_{b}) \to u(y_{1})\bar{u}(y_{2})\,,
\label{ex1b1}
\end{equation}
and the field mapping for this dipole term
is made as
\begin{equation}
(y_{a},y_{b};y_{1}, y_{2})=(a,b \ ;\widetilde{13},\widetilde{2})\,.
\end{equation}
The contribution to the cross section is written as
\begin{equation}
\hat{\sigma}_{\mbox{{\tiny D}}}(\mbox{R}_{1}, {\tt dip}1) 
= \frac{1}{S_{\mbox{{\tiny R}}_{1}}} \ \Phi(\mbox{R}_{1})_{d} 
\cdot \frac{1}{n_{s}(u) n_{s}(\bar{u})} \, 
\mbox{D}(\mbox{R}_{1}, {\tt dip}1)_{13,2}\,,
\end{equation}
and the dipole term is written as
\begin{equation}
\mbox{D}(\mbox{R}_{1}, {\tt dip}1\ )_{13,2} =
-\frac{1}{s_{13}} 
\frac{1}{\mbox{C}_{\mbox{{\tiny F}}}} 
\ \mbox{V}_{13,2}
\,\langle \mbox{B}1 \ | 
\mbox{T}_{1} \cdot 
\mbox{T}_{2}
\,| \ \mbox{B}1 \rangle\,.
\end{equation}
The dipole term is of final--final type
term and is converted to the integrated form in 
Eq.\,(\ref{ffmast}) as
\begin{equation}
\hat{\sigma}_{\mbox{{\tiny D}}}(\mbox{R}_{1},\,{\tt dip}1,\, 
\mbox{FF}\,) 
=
-\frac{A_{d}}{S_{\mbox{{\tiny R}}_{1}}}
\cdot
\frac{1}{\mbox{C}_{\mbox{{\tiny F}}}}
{\cal V}_{fg} (\ep) 
\cdot 
\Phi(\mbox{B}1)_{d} \,
[\,1,2\,]\,,
\end{equation}
where the color--correlated Born squared amplitude
is denoted as
\begin{equation}
[\,1,2\,]=(s_{{\scriptscriptstyle y_{1}, y_{2}}})^{-\ep} 
\cdot
\langle \mbox{B}1 \ | 
\mbox{T}_{y_{1}} \cdot 
\mbox{T}_{y_{2}}
 | \ \mbox{B}1 \rangle_{d}\,, 
\end{equation}
with the Lorentz scalar
$s_{{\scriptscriptstyle y_{1}, y_{2}}}=
2 \, \mbox{P}(y_{1}) \cdot \mbox{P}(y_{2})$.
The expression of the dipole term is further abbreviated
in the fixed form in Eq.\,(\ref{fixf})
with Factors 1 and 2 as
\begin{align}
(\mbox{Factor 1})&=
\frac{1}{\mbox{C}_{\mbox{{\tiny F}}}}
{\cal V}_{fg} (\ep)\,,
\\
(\mbox{Factor 2})&=
\Phi(\mbox{B}1)_{d} \,
[\,1,2\,]\,.
\end{align}
In the summary table, in addition to Factors 1 and 2,
we can explicitly specify the reduced Born process
and the kind of splitting, which are 
$\mbox{B}1$ and (1)-1 in the present case.
Actually, the integrated dipole term is shown with
the above information in the first one,
$1.\,(a,b\,;\widetilde{12},\widetilde{3})$,
in Table\,\ref{ap_B_1_tab1}
in Appendix \ref{ap_B_1}.
%
%
\item Example 2: Table 4 in Ref.\cite{Hasegawa:2014oya},
\,18.\,(a1,\,b)\,. 
\\
The reduced Born process and the field mapping are
determined as 
\begin{equation}
\mbox{B}3u=
g(y_{a})\bar{u}(y_{b}) \to \bar{u}(y_{1})g(y_{2})
\ \ \mbox{and} \ \
(y_{a},y_{b};y_{1}, y_{2}) =
(\widetilde{a1},\widetilde{b}\,;2,3)\,.
\end{equation}
The cross section and the dipole term are written as
\begin{align}
\hat{\sigma}_{\mbox{{\tiny D}}}(\mbox{R}_{1}, {\tt dip}3u) 
&= \frac{1}{S_{\mbox{{\tiny R}}_{1}}} \ \Phi(\mbox{R}_{1})_{d} 
\cdot \frac{1}{n_{s}(u) n_{s}(\bar{u})} \, 
\mbox{D}(\mbox{R}_{1}, {\tt dip}3u)_{a1,b}\,,
\\
\mbox{D}(\mbox{R}_{1}, {\tt dip}3u)_{a1,b}
&= -\frac{1}{s_{a1}} \frac{1}{x_{1,ab}}
\frac{1}{\mbox{C}_{\mbox{{\tiny A}}}} \ 
\langle \, \mbox{B}3u \,|
\mbox{T}_{y_{a}} 
\cdot \mbox{T}_{y_{b}} \ 
\mbox{V}_{{\scriptscriptstyle a1,b}}^{y_{a}} | 
\ \mbox{B}3u \rangle\,.
\end{align}
The dipole term is of initial-initial type
and is converted to the integrated form in 
Eq.\,(\ref{iimast}) as
\begin{equation}
\hat{\sigma}_{\mbox{{\tiny D}}}(\mbox{R}_{1},\,{\tt dip}3u,\, 
\mbox{II},\,x_{a}) 
=
-\frac{A_{d}}{S_{\mbox{{\tiny R}}_{1}}} 
\cdot
\int_{0}^{1}dx \,
\frac{1}{\mbox{C}_{\mbox{{\tiny A}}}} 
{\widetilde {\cal V}}^{f,\,g}(x;\ep)
\cdot
\Phi_{a}(\mbox{B}3u,x)_{d} 
\,[ \,a,b\,]\,,
\end{equation}
where the symbol $[\,a,b\,]$ is denoted as
\begin{equation}
[\,a,b\,]=
(s_{x_{a},\,y_{b}})^{-\ep} 
\cdot
\langle \mbox{B}3u \ | 
\mbox{T}_{y_{a}} \cdot 
\mbox{T}_{y_{b}}
 | \ \mbox{B}3u \rangle_{d}\,, 
\end{equation}
with the Lorentz scalar
$s_{x_{a},\,y_{b}}=2\,p_{a} \cdot p_{b}$.
Factors 1 and 2 
in Eq.\,(\ref{fixf})
are determined as
\begin{align}
(\mbox{Factor 1})&=
\int_{0}^{1}dx \,
\frac{1}{\mbox{C}_{\mbox{{\tiny A}}}} 
{\widetilde {\cal V}}^{f,\,g}(x;\ep)\,,
\\
(\mbox{Factor 2})&=
\Phi_{a}(\mbox{B}3u,x)_{d} 
\,[ \,a,b\,]\,.
\end{align}
The integrated dipole term is shown in the entry
$18.\,(\widetilde{a1},\widetilde{b}\,;2,3)$
in Table\,\ref{ap_B_1_tab1}.
\end{itemize}
%
%
In this way, all the dipole terms can be converted 
into the integrated form.
Then the summary tables of all the dipole terms
created by the DSA 
are converted to summary tables of the integrated
dipole terms with the necessary information.
One original dipole term is converted to
one integrated dipole term, and the total number
of dipole terms is conserved through the
conversion.
As an example, for the dijet process,
all the summary tables of the dipole terms
created by the DSA are shown at Tables 4--14
in Appendix B.1 in \cite{Hasegawa:2014oya}.
All the summary tables are converted into
the summary tables of the integrated dipole
terms as Tables 
\ref{ap_B_1_tab1}--\ref{ap_B_1_tab11} 
in Appendix \ref{ap_B}
in the present article.
%
%
%
\subsection{Step 2 : 
$\sigma_{\mbox{{\tiny D}}}(\mbox{I}) - \sigma_{\mbox{{\tiny I}}}$
\label{s_2_3}}
{\bfseries Step 2} of the PRA is to prove the relation
\begin{equation}
\sigma_{\mbox{{\tiny D}}}(\mbox{R}_{i},\,\mbox{I})
-
\sigma_{\mbox{{\tiny I}}}(\mbox{R}_{i})
=
-\sigma_{\mbox{{\tiny I}}}(\mbox{R}_{i},(2)\,
\mbox{-}1/2,N_{f}{\cal V}_{f\bar{f}})\,.
\label{prast2}
\end{equation}
The relation in Eq.\,(\ref{prast2}) stands for an arbitrary
process and is regarded as an identity.
The left-hand side of Eq.\,(\ref{prast2})
is the first term in square brackets in 
Eq.\,(\ref{sigdfour}).
We define the three cross sections
$\sigma_{\mbox{{\tiny D}}}(\mbox{R}_{i},\,\mbox{I})$,
$\sigma_{\mbox{{\tiny I}}}(\mbox{R}_{i})$,
and
$\sigma_{\mbox{{\tiny I}}}(\mbox{R}_{i},(2)\,
\mbox{-}1/2,N_{f}{\cal V}_{f\bar{f}})$
in Eq.\,(\ref{prast2})
as follows.

We first define the cross section
$\sigma_{\mbox{{\tiny D}}}(\mbox{R}_{i},\,\mbox{I})$.
In {\bfseries Step\,1} 
all the dipole terms are converted into the integrated
form. Among them, we take only the dipole terms
in the Dipole\,1 category.
The Dipole\,1 includes splittings
(1)--(4) with the cases of the spectator
in the final/initial state denoted as 
subcategory -1/2, which are shown 
in Fig.\,\ref{fig_A3_int_I} in 
Appendix \ref{ap_A_3}.
We extract the following parts 
from all the expressions,
depending on splittings
(1)--(4) and on the subcategory -1/2.
For the dipole terms with splitting (1)-1,
we take all of the singular function
${\cal V}_{fg}(\ep)$ in Eq.\,(\ref{isfg}).
For the dipole terms with splitting (1)-2,
we extract the part $\delta(1-x) \,
{\cal V}_{fg}(\ep)$ from the singular
function ${\cal V}_{fg}(x\,;\ep)$
in Eq.\,(\ref{nuxepfg}).
Then, for both of the splittings (1)-1 and -2,
the partonic cross sections
$\hat{\sigma}_{\mbox{{\tiny D}}}(
\mbox{R}_{i},\mbox{I},(1)\mbox{-}1/2)$
are extracted as the same expression\,:
\begin{equation}
\hat{\sigma}_{\mbox{{\tiny D}}}(
\mbox{R}_{i},\,\mbox{I},\,(1)\mbox{-}1/2)
=
-\frac{A_{d}}{S_{\mbox{{\tiny R}}_{i}}} 
\cdot
\frac{1}{\mbox{C}_{\mbox{{\tiny F}}}}
{\cal V}_{fg} (\ep) 
\cdot
\Phi(\mbox{B}1)_{d}\,
[\, y_{emi}, \,y_{spe}\,]\,.
\label{di11}
\end{equation}
The cross section is obtained
by multiplying the PDFs by the partonic
cross section in Eq.\,(\ref{di11}),
as shown in Eq.\,(\ref{master}).
In the cases of the other splittings,
the partonic cross sections are defined as follows,
and the cross sections are similarly obtained
by the multiplication of the PDFs.
For splitting (2)-1, we take all
of ${\cal V}_{gg}(\ep)$ in Eq.\,(\ref{isgg}),
and, for the splitting (2)-2, we extract the part
$\delta(1-x) \,{\cal V}_{gg}(\ep)$
from ${\cal V}_{gg}(x\,;\ep)$
in Eq.\,(\ref{nuxepgg}).
Then, for both splittings (2)-1/2,
the cross section
$\hat{\sigma}_{\mbox{{\tiny D}}}(
\mbox{R}_{i},\mbox{I},
(2)\mbox{-}1/2)$
is defined as
\begin{equation}
\hat{\sigma}_{\mbox{{\tiny D}}}(
\mbox{R}_{i},\,\mbox{I},\,(2)\mbox{-}1/2)
=
-\frac{A_{d}}{S_{\mbox{{\tiny R}}_{i}}} 
\cdot
\frac{1}{\mbox{C}_{\mbox{{\tiny A}}}}
{\cal V}_{gg} (\ep) 
\cdot
\Phi(\mbox{B}1)_{d}\,
[\, y_{emi}, \,y_{spe}\,]\,.
\label{di12}
\end{equation}
For splitting (3)-1, we extract the part
$\delta(1-x) \,{\cal V}_{fg}(\ep)$
from ${\cal V}^{f,f}(x\,;\ep)$ in 
Eq.\,(\ref{sinifff}),
and, for splitting (3)-2, we extract the 
same part $\delta(1-x) \,{\cal V}_{fg}(\ep)$
from ${\cal V}^{f,f}(x\,;\ep)$ in
${\widetilde {\cal V}}^{f,f}(x;\ep)$
in Eq.\,(\ref{siniiff}).
Then, for both splittings (3)-1/2,
the partonic cross sections
$\hat{\sigma}_{\mbox{{\tiny D}}}(
\mbox{R}_{i},\mbox{I},(3)\mbox{-}1/2)$
are extracted
as the identical expression in Eq.\,(\ref{di11}).
For splitting (4)-1, we extract the part
$\delta(1-x) \,{\cal V}_{g}(\ep)$
from ${\cal V}^{g,g}(x\,;\ep)$ in 
Eq.\,(\ref{sinifgg}),
and, for the splitting (4)-2,
we extract the same quantity 
$\delta(1-x) \,{\cal V}_{g}(\ep)$
from ${\cal V}^{g,g}(x\,;\ep)$ in
${\widetilde {\cal V}}^{g,g}(x;\ep)$
in Eq.\,(\ref{siniigg}).
Then, for both splittings (4)-1/2,
the cross sections 
$\hat{\sigma}_{\mbox{{\tiny D}}}(
\mbox{R}_{i},\mbox{I},(4)\mbox{-}1/2)$
are defined as
\begin{equation}
\hat{\sigma}_{\mbox{{\tiny D}}}(
\mbox{R}_{i},\,\mbox{I},\,(4)\mbox{-}1/2)
=
-\frac{A_{d}}{S_{\mbox{{\tiny R}}_{i}}} 
\cdot
\frac{1}{\mbox{C}_{\mbox{{\tiny A}}}}
{\cal V}_{g} (\ep) 
\cdot
\Phi(\mbox{B}1)_{d}\,
[\, y_{emi}, \,y_{spe}\,]\,.
\label{di14}
\end{equation}
In Eqs.\,(\ref{di11}), (\ref{di12}), and
(\ref{di14}), 
the color-correlated Born squared amplitude, 
$[\, y_{emi}, \,y_{spe}\,]$,
is denoted as
\begin{equation}
[ \, y_{emi}, \,y_{spe} ]
=(s_{y_{emi}, \,y_{spe}})^{-\ep} 
\cdot
\langle \mbox{B}1 \ | 
\mbox{T}_{y_{emi}} \cdot 
\mbox{T}_{y_{spe}}
 | \ \mbox{B}1 \rangle_{d}\,,
\end{equation}
with the Lorentz scalar
$s_{y_{emi}, y_{spe}}=2 \, \mbox{P}(y_{emi}) 
\cdot \mbox{P}(y_{spe})$.
The partonic cross section 
$\hat{\sigma}_{\mbox{{\tiny D}}}(
\mbox{R}_{i},\,\mbox{I})$
is the summation of all the existing 
partonic cross sections
$\hat{\sigma}_{\mbox{{\tiny D}}}(
\mbox{R}_{i},\mbox{I},(1)\sim(4)\mbox{-}1/2)$
defined above.
The formulae for
$\hat{\sigma}_{\mbox{{\tiny D}}}(
\mbox{R}_{i},\,\mbox{I})$
are collected in Appendix \ref{ap_A_3}.

Next we define the cross section
$\sigma_{\mbox{{\tiny I}}}(\mbox{R}_{i})$. 
The partonic cross section
$\hat{\sigma}_{\mbox{{\tiny I}}}(\mbox{R}_{i})$
is the contributions of the I terms that
are created by the DSA \cite{Hasegawa:2014oya}.
The contribution of each I term is written as
\begin{equation}
\hat{\sigma}_{\mbox{{\tiny I}}}(\mbox{R}_{i})_{IK} 
= -\frac{A_{d}}{S_{\mbox{{\tiny B1}}}} \cdot
\frac{1}{\mbox{T}_{\mbox{{\tiny F}}(I)}^{2}}
{\cal V}_{\mbox{{\tiny F}}(I)}
\cdot
\ \Phi(\mbox{B}1)_{d} \,
[\,I,K \,] \,,
\end{equation}
where again the notation is defined in
Ref.\cite{Hasegawa:2014oya}.
The $S_{\mbox{{\tiny B1}}}$ is the symmetric factor
of the reduced Born process $\mbox{B}1(\mbox{R}_{i})$.
In the DSA, the creation of the I terms is
ordered by the species of the first leg $I$,
(1)--(4),
which are shown
in Fig.\,\ref{fig_A8_I} in Appendix \ref{ap_A_8}.
There are four cases for the choices for
the first leg $I$, and each case has
further choices for the second leg $K$ in the 
final/initial state denoted as -1/2.
The factor 
${\cal V}_{\mbox{{\tiny F}}(I)}/
\mbox{T}_{\mbox{{\tiny F}}(I)}^{2}$
is determined in each case as
\begin{equation}
\frac{1}{\mbox{T}_{\mbox{{\tiny F}}(I)}^{2}}
{\cal V}_{\mbox{{\tiny F}}(I)}
=
\left\{
\begin{array}{ll}
\frac{1}{\mbox{C}_{\mbox{{\tiny F}}}}
{\cal V}_{f}(\ep)
& : (1),(3)-1/2\,,
\\
\frac{1}{\mbox{C}_{\mbox{{\tiny A}}}} 
{\cal V}_{g}(\ep)
& : (2),(4)-1/2\,,
\end{array}
\right.
\end{equation}
where the universal singular functions
${\cal V}_{f}(\ep)$ and
${\cal V}_{g}(\ep)$
are defined in 
Eqs.\,(\ref{nufeq}) and (\ref{singgg}), 
respectively.
The symbol $[\,I,K \,]$ is denoted as
\begin{equation}
[ \, I, \, K \,]
=s_{IK}^{-\ep} 
\,.
\langle \mbox{B}1 \ | 
\mbox{T}_{I} \cdot 
\mbox{T}_{K}
 | \ \mbox{B}1 \rangle_{d}\,,
\end{equation}
with the Lorentz scalar
$s_{IK}=2\,p_{I} \cdot p_{K}$.
The partonic cross section 
$\hat{\sigma}_{\mbox{{\tiny I}}}(\mbox{R}_{i})$
is the summation of all the created I terms as
\begin{equation}
\hat{\sigma}_{\mbox{{\tiny I}}}(\mbox{R}_{i})
=
\sum_{I,K} 
\hat{\sigma}_{\mbox{{\tiny I}}}(\mbox{R}_{i})_{IK}.
\label{sigisum}
\end{equation}
Similar to the case 
$\sigma_{\mbox{{\tiny D}}}(\mbox{R}_{i},\,\mbox{I})$,
the cross section $\sigma_{\mbox{{\tiny I}}}(\mbox{R}_{i})$
is obtained by multiplying the PDFs by the partonic
cross section 
$\hat{\sigma}_{\mbox{{\tiny I}}}(\mbox{R}_{i})$.
The formulae for the I term
$\sigma_{\mbox{{\tiny I}}}(\mbox{R}_{i})$
are collected in Appendix 
\ref{ap_A_8}.

The third cross section
$\sigma_{\mbox{{\tiny I}}}(\mbox{R}_{i},(2)\,
\mbox{-}1/2,N_{f}{\cal V}_{f\bar{f}})$
is defined as follows.
We take only the I terms that belong to
the splitting (2)-1/2
in Fig.\,\ref{fig_A8_I}.
Among them we extract
the part $N_{f}{\cal V}_{f\bar{f}}(\ep)$
in the function
${\cal V}_{g}(\ep)$ in Eq.\,(\ref{singgg}).
The extracted part is defined as the partonic
cross section, 
\begin{equation}
\hat{\sigma}_{\mbox{{\tiny I}}}(
\mbox{R}_{i},(2)\,
\mbox{-}1/2,N_{f}{\cal V}_{f\bar{f}}
)_{IK} 
= -\frac{A_{d}}{S_{\mbox{{\tiny B1}}}} \cdot
\frac{N_{f}}{\mbox{C}_{\mbox{{\tiny A}}}}
{\cal V}_{f\bar{f}}(\ep)
\cdot
\ \Phi(\mbox{B}1)_{d} \,
[\,I,K \,] \,,
\label{siginf}
\end{equation}
where the leg $I$ is 
in the final state and 
the field species is a gluon.
By definition, the term exists only if 
the reduced Born process $\mbox{B}1(\mbox{R}_{i})$
includes any gluon in the final state.
This statement is equivalent to saying that 
the term exists only if the real process 
$\mbox{R}_{i}$ includes two or more gluons
in the final state.
The partonic cross section 
$\hat{\sigma}_{\mbox{{\tiny I}}}(\mbox{R}_{i},(2)\,
\mbox{-}1/2,N_{f}{\cal V}_{f\bar{f}})$
is the summation of all the existing 
cross sections in Eq.\,(\ref{siginf}).
The formula is also added in 
Appendix \ref{ap_A_8}.
Now that 
all three terms in 
Eq.\,(\ref{prast2})
are defined, 
we can interpret
the equation in such a way that
the extracted part from the 
integrated dipole term,
$\hat{\sigma}_{\mbox{{\tiny D}}}(
\mbox{R}_{i},\,\mbox{I})$,
cancels the I term,  
$\sigma_{\mbox{{\tiny I}}}(\mbox{R}_{i})$,
except for the part
$\sigma_{\mbox{{\tiny I}}}(\mbox{R}_{i},(2)\,
\mbox{-}1/2,N_{f}{\cal V}_{f\bar{f}})$.
The remaining part
$\sigma_{\mbox{{\tiny I}}}(\mbox{R}_{i},(2)\,
\mbox{-}1/2,N_{f}{\cal V}_{f\bar{f}})$
is canceled by a term
created in a different real process.
The mechanism of the cancellation will
be clarified in {\bfseries Step 6} 
in Sec.\,\ref{s_2_7}.

Finally, we see one example.
We take the same real emission process
as used in
{\bfseries Step 1},
$\mbox{R}_{1}=u\bar{u} \to u\bar{u}g$.
The reduced Born process,
$\mbox{B}1(\mbox{R}_{1})=u\bar{u} \to u\bar{u}$,
does not include any gluon in the final state,
and the right-hand side of Eq.\,(\ref{prast2})
does not exist. Then the relation to be proved 
is written as
\begin{equation}
\sigma_{\mbox{{\tiny D}}}(\mbox{R}_{1},\,\mbox{I})
-
\sigma_{\mbox{{\tiny I}}}(\mbox{R}_{1})
=0\,.
\label{ex1tar}
\end{equation}
First we construct the partonic cross section
$\hat{\sigma}_{\mbox{{\tiny D}}}(\mbox{R}_{1},\,\mbox{I})$.
All the integrated dipole terms of
$\hat{\sigma}_{\mbox{{\tiny D}}}(\mbox{R}_{1})$
are summarized in Table
\ref{ap_B_1_tab1} in Appendix \ref{ap_B_1}.
Among all the 21 dipole terms, we take only
the first twelve terms in the category Dipole\,1.
Following the definition of
$\sigma_{\mbox{{\tiny D}}}(\mbox{R}_{i},\,\mbox{I})$
given above,
we extract the cross section as
\begin{align}
\hat{\sigma}_{\mbox{{\tiny D}}}(
\mbox{R}_{1},\,\mbox{I})
&=
-\frac{A_{d}}{S_{\mbox{{\tiny R}}_{1}}} 
\cdot
\frac{1}{\mbox{C}_{\mbox{{\tiny F}}}}
{\cal V}_{fg}(\ep) 
\cdot
\Phi(\mbox{B}1)_{d} \
\Bigl(\,
[1,2]+[2,1]+[1,a]+[1,b]+[2,a]+[2,b]
\nonumber\\
&\hspace{30mm}+[a,1]+[a,2]+[b,1]+[b,2]+[a,b]+[b,a]
\Bigr)\,,
\end{align}
with the symmetric factor 
$S_{\mbox{{\tiny R}}_{1}}=1$.
Second, we construct the contribution of the I terms,
$\sigma_{\mbox{{\tiny I}}}(\mbox{R}_{1})$.
The I terms that are created by the DSA are
summarized in Table 15 in Appendix B.2 in 
ref.\cite{Hasegawa:2014oya}. 
The twelve I terms are written as
\begin{align}
\hat{\sigma}_{\mbox{{\tiny I}}}(\mbox{R}_{1})
&=
-\frac{A_{d}}{S_{\mbox{{\tiny B1}}}} \cdot
\frac{1}{\mbox{C}_{\mbox{{\tiny F}}}}
{\cal V}_{f}(\ep)
\cdot
\ \Phi(\mbox{B}1)_{d} \
\Bigl(\,
[1,2]+[2,1]+[1,a]+[1,b]+[2,a]+[2,b]
\nonumber\\
&\hspace{30mm}+[a,1]+[a,2]+[b,1]+[b,2]+[a,b]+[b,a]
\Bigr)\,,
\end{align}
with the symmetric factor 
$S_{\mbox{{\tiny B1}}}=1$.
Using the relations
$S_{\mbox{{\tiny R}}_{1}}=S_{\mbox{{\tiny B1}}}=1$
and  
${\cal V}_{f}(\ep)={\cal V}_{fg}(\ep)$,
we prove the relation
in Eq.\,(\ref{ex1tar}).
{\bfseries Step 2}
for the process $\mbox{R}_{1}$ is
completed.
The result is shown in
Eq.\,(\ref{r1st2})
in Appendix \ref{ap_B_1}.
As mentioned above, in this example,
the right-hand side of Eq.\,(\ref{prast2})
does not exist.
Cases where the right-hand side exists
will be seen in Sec.\,\ref{sec_4}
and Sec.\,\ref{sec_5}.
%
%
%
\subsection{Step 3 : 
$\sigma_{\mbox{{\tiny D}}}(\mbox{P}) + \sigma_{c} - 
\sigma_{\mbox{{\tiny P}}}$
\label{s_2_4}}
{\bfseries Step 3} of the PRA is to prove the relation
\begin{equation}
\sigma_{\mbox{{\tiny D}}}(\mbox{R}_{i},\,\mbox{P})
+
\sigma_{\mbox{{\tiny C}}}(\mbox{R}_{i})
-
\sigma_{\mbox{{\tiny P}}}(\mbox{R}_{i})=0\,.
\label{prast3}
\end{equation}
The left-hand side of Eq.\,(\ref{prast3})
is the second term in the square brackets in
Eq.\,(\ref{sigdfour}).
We define the three cross sections
$\sigma_{\mbox{{\tiny D}}}(\mbox{R}_{i},\,\mbox{P})$,
$\sigma_{\mbox{{\tiny C}}}(\mbox{R}_{i})$,
and
$\sigma_{\mbox{{\tiny P}}}(\mbox{R}_{i})$
in Eq.\,(\ref{prast3}) as follows.

First we define the cross section 
$\sigma_{\mbox{{\tiny D}}}(\mbox{R}_{i},\,\mbox{P})$.
The quantity is extracted from the integrated dipole terms
converted in {\bfseries Step 1}
in the following way.
We choose only the dipole terms with splittings
(3), (4), (6), and (7), 
as shown in Fig.\,\ref{fig_A4_int_P}
in Appendix \ref{ap_A_4}.
For the initial--final dipole terms,
namely, splittings (3)-,\,(4)-,\,(6)-, and (7)-1,
we extract the factors
$(-1/\ep + \ln x) \cdot P^{ff,\,gg,\,fg,\,gf}(x)$
in the functions 
${\cal V}^{f,f}(x\,;\ep)$, 
${\cal V}^{g,g}(x\,;\ep)$,
${\cal V}^{f,g}(x\,;\ep),$
and ${\cal V}^{g,f}(x\,;\ep)$
in Eqs.\,(\ref{sinifff})--(\ref{sinifgf}), respectively. 
For the initial--initial dipole terms,
namely, splittings, (3)-,\,(4)-,\,(6)-, and (7)-2,
we extract the same factors 
in the functions 
${\widetilde {\cal V}}^{f,f}(x;\ep)$,
${\widetilde {\cal V}}^{g,g}(x;\ep)$,
${\widetilde {\cal V}}^{f,g}(x;\ep)$,
and ${\widetilde {\cal V}}^{g,f}(x;\ep)$
in Eqs.\,(\ref{siniiff})--(\ref{siniigf}).
For all the dipole terms with splittings
(3)-,\,(4)-,\,(6)-, and (7)-1/2,
the partonic cross sections
$\hat{\sigma}_{\mbox{{\tiny D}}}(\mbox{R}_{i},\,\mbox{P})$
are defined as the universal expression
\begin{align}
\hat{\sigma}_{\mbox{{\tiny D}}}
(\,
\mbox{R}_{i},\,
\mbox{P},\,{\tt dip}j,
\,
x_{a/b}
) 
&=
\frac{A_{d}}{S_{\mbox{{\tiny R}}_{i}}} 
\int_{0}^{1}dx \,
\biggl(\frac{1}{\ep} - \ln x \biggr)
\frac{1}{\mbox{T}_{\mbox{{\tiny F}}(y_{emi})}^{2}} 
P^{\mbox{{\tiny F}}(x_{a/b}) \, \mbox{{\tiny F}}(y_{emi}) }(x)
\,\times
\nonumber \\
& \hspace{50mm}
\Phi_{a/b}(\mbox{B}_{j},x)_{d} 
\, [ \, y_{emi}, \,y_{spe} ]\,, 
\end{align}
where the factor
$P^{\mbox{{\tiny F}}(x_{a/b}) \, \mbox{{\tiny F}}(y_{emi}) }(x)/
\mbox{T}_{\mbox{{\tiny F}}(y_{emi})}^{2}$
depends on the splittings
as written in Eq.\,(\ref{sigdpp}).
The symbol $[ \, y_{emi}, \,y_{spe} ]$
is defined in 
Eqs.\,(\ref{clbsaif})
and
(\ref{clbsaii})
for the initial--final 
and 
initial-initial
dipole terms, respectively.
The cross section 
$\sigma_{\mbox{{\tiny D}}}(\mbox{R}_{i},\,\mbox{P})$
is obtained by multiplying the PDFs with the
partonic cross section
$\hat{\sigma}_{\mbox{{\tiny D}}}
(\,\mbox{R}_{i},\,\mbox{P}\,)$.
The formulae for
$\sigma_{\mbox{{\tiny D}}}(\mbox{R}_{i},\,\mbox{P})$
are collected in Appendix \ref{ap_A_4}.

Second, we define the cross section
$\sigma_{\mbox{{\tiny C}}}(\mbox{R}_{i})$.
This quantity is the collinear subtraction term
that is introduced in the QCD NLO corrections
as in Eq.\,(\ref{trad}).
Some algorithms to create collinear
subtraction terms for an arbitrary process
may be available. We introduce here an
algorithm that is analogous with 
the algorithm to create the P term in
the DSA \cite{Hasegawa:2014oya}.
The input is taken in a real process
$\mbox{R}_{i}$\,.
As with the DSA, $\mbox{R}_{i}$ defines
the set
$\{x\}=\{x_{a},x_{b}\,;x_{1},...,x_{n+1}\}$.
The field species and the momenta are 
denoted as
$\mbox{F}(\{x\})=\{
\mbox{F}(x_{a}),\mbox{F}(x_{b})\,;
\mbox{F}(x_{1}), ...,\mbox{F}(x_{n+1})
\}$
and $\{p_{a},p_{b}\,; p_{1},...,p_{n+1}\}$.
Then we check whether the process 
$\mbox{R}_{i}$ can have splittings
(3), (4), (6), and (7), shown
in Fig.\,\ref{fig_A7_coll}
in Appendix \ref{ap_A_7}.
We start with splitting (3), including the 
leg-a\,($x_{a}$). 
When the process $\mbox{R}_{i}$ can have
splitting (3),
a pair $(x_{a},x_{i})$ is
chosen and the new element $x_{\widetilde{ai}}$
is created with the field species
$\mbox{F}(x_{\widetilde{ai}})$,
which is the species of the root of the
splitting\,: in the present splitting (3), 
a quark.
The reduced Born process $\mbox{B}1$
is taken in the same one as determined 
for the dipole terms in the DSA.
The $\mbox{B}1$ associates 
the set $\{y\}=\{y_{a},y_{b}\,;y_{1},...,y_{n}\}$,
the field species
$\mbox{F}(\{y\})=\{
\mbox{F}(y_{a}),\mbox{F}(y_{b})\,;
\mbox{F}(y_{1}), ...,\mbox{F}(y_{n})
\}$,
and the momenta,
$\mbox{P}(\{y\})=\{
\mbox{P}(y_{a}),\mbox{P}(y_{b})\,;
\mbox{P}(y_{1}), ...,\mbox{P}(y_{n})
\}$.
There are two possible cases,
$\mbox{F}(x_{\widetilde{ai}})=\mbox{F}(y_{a})$
or
$\mbox{F}(x_{\widetilde{ai}})=\mbox{F}(y_{b})$,
which are denoted as $y_{emi}=y_{a}$
or $y_{emi}=y_{b}$, respectively.
For both cases, the collinear subtraction terms
with leg-a\,($x_{a}$) are created as
\begin{equation}
\hat{\sigma}_{\mbox{{\tiny C}}}(\mbox{R}_{i},
{\tt dip}1,\,x_{a}) 
= 
\frac{A_{d}}{S_{\mbox{{\tiny B1}}}}
\int_{0}^{1}dx\,
\Bigl(\,
\frac{1}{\ep}
-
\ln \mu_{F}^{2}
\Bigr)
P^{ff}(x) 
\cdot
\Phi_{a}(\mbox{B}_{1},x)_{d}\,
\langle \mbox{B}1 \rangle\,,
\end{equation}
where $S_{\mbox{{\tiny B1}}}$ is the symmetric
factor of the Born process $\mbox{B}1$
and 
the Altarelli--Parisi splitting function $P^{ff}(x)$
is defined in Eq.\,(\ref{alpff}).
The symbol $\langle \mbox{B}1 \rangle$
represents the square of the matrix elements
of the process $\mbox{B}1$
after the spin-color is summed and averaged
in $d$ dimensions, which is written with
the input momenta as
\begin{align}
\langle \mbox{B}1 \rangle 
=|\,\mbox{M}_{\mbox{{\tiny B1}}}
(\,\mbox{P}(y_{a}),\mbox{P}(y_{b})\,\to
\mbox{P}(y_{1}), ...,\mbox{P}(y_{n})
\,)\,|^{2}\,.
\end{align}
In the case in which $y_{emi}=y_{a}$ or $y_{emi}=y_{b}$\,,
the input momenta in the initial state
are determined as
$(\mbox{P}(y_{a}),\mbox{P}(y_{b}))
=(xp_{a},p_{b})$ or $(p_{b},xp_{a})$,
respectively.
It is noted that, when the final state of
$\mbox{R}_{i}$ includes identical fields,
one kind of splitting has as many possible pairs
$(x_{a},x_{i})$ as
the number of identical fields.
In this case, only one pair is taken and the other
pairs must be discarded.
For instance, the process 
$\mbox{R}_{i}=u(x_{a})\bar{u}(x_{b}) 
\to g(x_{1})g(x_{2})g(x_{3})$
has three pairs $(x_{a},x_{1})$, $(x_{a},x_{2})$,
and $(x_{a},x_{3})$ for splitting (3).
Among them, only one pair, for instance, 
$(x_{a},x_{1})$, is taken and
the others, $(x_{a},x_{2})$
and $(x_{a},x_{3})$,
must be discarded.
The discard rule is the same as the creation algorithm
of the P term in the DSA.
The creation algorithm shown above is similarly
applied for leg-b\,($x_{b}$) and 
for the other
splittings (4), (6), and (7).
We summarize the general formulae 
for the collinear subtraction term
in an arbitrary process
as follows. The collinear subtraction terms consist
of terms with the different splittings,
and, equivalently, different reduced Born processes as
\begin{equation}
\hat{\sigma}_{\mbox{{\tiny C}}}(\mbox{R}_{i}) 
= \sum_{\mbox{{\tiny dip}}j} 
\hat{\sigma}_{\mbox{{\tiny C}}}(\mbox{R}_{i},\,{\tt dip}j),
\end{equation}
where each term 
$\hat{\sigma}_{\mbox{{\tiny C}}}(\mbox{R}_{i},{\tt dip}j)$
can have the contributions with
leg-a\,($x_{a}$) and leg-b\,($x_{b}$) as
\begin{equation}
\hat{\sigma}_{\mbox{{\tiny C}}}(\mbox{R}_{i}, {\tt dip}j) 
=\hat{\sigma}_{\mbox{{\tiny C}}}(\mbox{R}_{i}, {\tt dip}j,\,x_{a}) 
+
\hat{\sigma}_{\mbox{{\tiny C}}}(\mbox{R}_{i}, {\tt dip}j,\,x_{b})\,.
\end{equation}
The collinear subtraction term with the reduced 
Born process $\mbox{B}j$ and leg-a/b is 
universally written as
\begin{equation}
\hat{\sigma}_{\mbox{{\tiny C}}}(\mbox{R}_{i}, {\tt dip}j,\,x_{a/b}) 
= 
\frac{A_{d}}{S_{\mbox{{\tiny B}}j}}
\int_{0}^{1}dx\,
\Bigl(\,
\frac{1}{\ep}
-
\ln \mu_{F}^{2}
\Bigr)
P^{
\mbox{{\tiny F}}(x_{a/b})\mbox{{\tiny F}}(y_{emi})
}(x) 
\cdot
\Phi_{a/b}(\mbox{B}_{j},x)_{d}\,
\langle \mbox{B}j \rangle\,,
\end{equation}
where the splitting function
$P^{\mbox{{\tiny F}}(x_{a/b})\mbox{{\tiny F}}(y_{emi})}(x)$
is determined as shown in 
Eq.\,(\ref{collap}).
The formulae for the 
collinear subtraction term
are collected in Appendix \ref{ap_A_7}.

The third cross section
$\sigma_{\mbox{{\tiny P}}}(\mbox{R}_{i})$
is the contribution of the P terms created
by the DSA \cite{Hasegawa:2014oya}.
The input for the creation is taken
in a real process $\mbox{R}_{i}$ 
and the creation of the P terms 
is ordered by splittings,
(3),\,(4),\,(6), and (7), and also by
the spectators in final/initial states
denoted as -1/2, which are shown in
Fig.\,\ref{fig_A9_P}
in Appendix \ref{ap_A_9}.
The partonic cross section
$\hat{\sigma}_{\mbox{{\tiny P}}}(\mbox{R}_{i})$
is written in 4 dimensions as the universal form
\begin{align}
\hat{\sigma}_{\mbox{{\tiny P}}}(\mbox{R}_{i},\,  {\tt dip}j,
x_{a/b}) 
&= 
\frac{A_{4}}{S_{\mbox{{\tiny B}}_{j} }}
\int_{0}^{1}dx \,
\frac{1}{\mbox{T}_{\mbox{{\tiny F}}(y_{emi})}^{2}} \,
P^{
\mbox{{\tiny F}}(x_{a/b})\mbox{{\tiny F}}(y_{emi})
}(x) \cdot
\ln \frac{\mu_{F}^{2}}{x\,s_{x_{a/b},y_{spe}}}
\,\cdot
\nonumber \\
& \hspace{35mm}
\Phi_{a/b}(\mbox{B}_{j},x)_{4} \,
\langle \, y_{emi}, \,y_{spe} \rangle\,,
\end{align}
where the factor $A_{4}$ is denoted as
$A_{4}=\al/2\pi$ and $S_{\mbox{{\tiny B}}_{j}}$
is the symmetric factor of the process 
$\mbox{B}_{j}$.
The definition of the factor
$P^{
\mbox{{\tiny F}}(x_{a/b})\mbox{{\tiny F}}(y_{emi})
}(x)/
\mbox{T}_{\mbox{{\tiny F}}(y_{emi})}^{2}
$
is the same as the factor for the cross section
$\hat{\sigma}_{\mbox{{\tiny D}}}
(\mbox{R}_{i},\mbox{P})$
in Eq.\,(\ref{sigdpp}).
The definition of the Lorentz scalar 
$s_{x_{a/b},y_{spe}}$ is also the same as
in Eq.\,(\ref{sigdplo}).
The symbol 
$\langle \, y_{emi}, \,y_{spe} \rangle$
represents the color-correlated Born squared
amplitude in 4 dimensions as
$\langle \, y_{emi}, \,y_{spe} \rangle\,
=
\langle \mbox{B}j \ | 
\mbox{T}_{y_{emi}} \cdot 
\mbox{T}_{y_{spe}}
 | \ \mbox{B}j \rangle_{4}$\,.
The formulae for the P term
$\sigma_{\mbox{{\tiny P}}}(\mbox{R}_{i})$
are collected in Appendix \ref{ap_A_9}.

We see one example in the same process
as used in {\bfseries Step 1} and 
{\bfseries 2},
$\mbox{R}_{1}=u\bar{u} \to u\bar{u}g$.
The relation to be proved for the process
is written as
\begin{equation}
\sigma_{\mbox{{\tiny D}}}(\mbox{R}_{1},\,\mbox{P})
+
\sigma_{\mbox{{\tiny C}}}(\mbox{R}_{1})
-
\sigma_{\mbox{{\tiny P}}}(\mbox{R}_{1})=0\,.
\label{prs3ex}
\end{equation}
As shown in Table\,\ref{ap_B_1_tab1}
in Appendix \ref{ap_B_1}, the dipole terms
include splittings, (3), (6)$u$, 
and (6)$\bar{u}$,
which have the following reduced Born processes
respectively\,:
\begin{align}
\mbox{B}1&=u\bar{u} \to u\bar{u}\,,
\\
\mbox{B}3u&=g\bar{u} \to \bar{u}g\,,
\\
\mbox{B}3\bar{u}&=ug \to ug\,.
\end{align}
Then the relation in
Eq.\,(\ref{prs3ex}) is divided into 
three independent relations\,:
\begin{equation}
\sigma_{\mbox{{\tiny D}}}(\mbox{R}_{1},\,\mbox{P}, {\tt dip}j)
+
\sigma_{\mbox{{\tiny C}}}(\mbox{R}_{1}, {\tt dip}j)
-
\sigma_{\mbox{{\tiny P}}}(\mbox{R}_{1}, {\tt dip}j)=0
\label{prs3exj}
\end{equation}
for ${\tt dip}j= {\tt dip}\,1,3u$, and $3\bar{u}$.
First, we construct 
$\sigma_{\mbox{{\tiny D}}}(\mbox{R}_{1},\,\mbox{P})$.
We only show the expressions for the terms
$\sigma_{\mbox{{\tiny D}}}(\mbox{R}_{1},\,\mbox{P},{\tt dip}1,x_{a})$
and
$\sigma_{\mbox{{\tiny D}}}(\mbox{R}_{1},\,\mbox{P}, {\tt dip}3u)$
for convenience as
\begin{align}
\sigma_{\mbox{{\tiny D}}}(\mbox{R}_{1},\,\mbox{P}, {\tt dip}1,x_{a})
&=
\frac{A_{d}}{S_{\mbox{{\tiny R}}_{1}}} 
\int_{0}^{1}dx \,
\biggl(\frac{1}{\ep} - \ln x \biggr)
\,\frac{1}{\mbox{C}_{\mbox{{\tiny F}}}} 
P^{ff}(x)\,\Phi_{a}(\mbox{B}1,x)_{d}
\,\times
\nonumber \\
& \hspace{50mm}
\, \bigl(\,[a,1]+ [a,2]+[a,b]\,\bigr)\,,
\label{ex1d1xa}
\\
\sigma_{\mbox{{\tiny D}}}(\mbox{R}_{1},\,\mbox{P}, {\tt dip}3u)
&=\frac{A_{d}}{S_{\mbox{{\tiny R}}_{1}}} 
\int_{0}^{1}dx \,
\biggl(\frac{1}{\ep} - \ln x \biggr)
\,\frac{1}{\mbox{C}_{\mbox{{\tiny A}}}} 
P^{fg} (x)\,\Phi_{a}(\mbox{B}3u,x)_{d}
\,\times
\nonumber \\
& \hspace{50mm}
\, \bigl(\,[a,1]+ [a,2]+[a,b]\,\bigr)\,,
\label{ex1d3u}
\end{align}
with the symmetric factor
$S_{\mbox{{\tiny R}}_{1}}=1$.
In Eq.\,(\ref{ex1d1xa}), the color-correlated
Born squared amplitudes are denoted as
\begin{equation}
[\,a,1/2/b\,]=
(s_{x_{a},\,y_{1/2/b}})^{-\ep} 
\cdot
\langle \mbox{B}1 \ | 
\mbox{T}_{y_{a}} \cdot 
\mbox{T}_{y_{1/2/b}}
 | \ \mbox{B}1 \rangle_{d}\,,
\label{ex1lf1}
\end{equation}
where the Lorentz factors
are written as
$s_{x_{a},\,y_{1/2}}=2\,p_{a} \cdot \mbox{P}(y_{1/2})$ 
and $s_{x_{a},\,y_{b}}=2\,p_{a} \cdot p_{b}$.
The corresponding quantities in 
Eq.\,(\ref{ex1d3u})
are written in such a way that $\mbox{B}1$ is
replaced with $\mbox{B}3u$
in Eq.\,(\ref{ex1lf1}).
Next we construct the collinear subtraction term
$\sigma_{\mbox{{\tiny C}}}(\mbox{R}_{1})$.
The algorithm to create the collinear subtraction
term is executed as follows.
The process $\mbox{R}_{1}$ associates set
$\{x\}$ as
$\mbox{R}_{1}=u(x_{a})\bar{u}(x_{b}) \to 
u(x_{1})\bar{u}(x_{2})g(x_{3})$.
The process $\mbox{R}_{1}$
can have splitting (3) with leg-a
and the pair $(x_{a},x_{3})$
is chosen.
The reduced Born process $\mbox{B}1$
associates set $\{y\}$ as
$\mbox{B}1=u(y_{a})\bar{u}(y_{b}) \to 
u(y_{1})\bar{u}(y_{2})$,
and the relation 
$\mbox{F}(x_{\widetilde{a3}})=\mbox{F}(y_{a})=u$
stands.
The process $\mbox{R}_{1}$
can also have splitting (6)$u$ with leg-a
and the pair $(x_{a},x_{1})$ is chosen.
The reduced Born process $\mbox{B}3u$
associates set $\{y\}$ as
$\mbox{B}3u=g(y_{a})\bar{u}(y_{b}) \to 
\bar{u}(y_{1})g(y_{2})$
and the relation 
$\mbox{F}(x_{\widetilde{a1}})=\mbox{F}(y_{a})=g$
stands.
Then we write down
the collinear subtraction terms as
\begin{align}
\hat{\sigma}_{\mbox{{\tiny C}}}(\mbox{R}_{1}, {\tt dip}1,
\,x_{a}) 
&= 
\frac{A_{d}}{S_{\mbox{{\tiny B1}}}}
\int_{0}^{1}dx\,
\Bigl(\,
\frac{1}{\ep}
-
\ln \mu_{F}^{2}
\Bigr)
P^{ff}(x) 
\cdot
\Phi_{a}(\mbox{B}1,x)_{d}\,
\langle \mbox{B}1 \rangle\,,
\\
\hat{\sigma}_{\mbox{{\tiny C}}}(\mbox{R}_{1}, {\tt dip}3u) 
&= 
\frac{A_{d}}{S_{\mbox{{\tiny B3}}}}
\int_{0}^{1}dx\,
\Bigl(\,
\frac{1}{\ep}
-
\ln \mu_{F}^{2}
\Bigr)
P^{fg}(x) 
\cdot
\Phi_{a}(\mbox{B}3u,x)_{d}\,
\langle \mbox{B}3u \rangle\,,
\end{align}
with the symmetric factors 
$S_{\mbox{{\tiny B1}}}=S_{\mbox{{\tiny B3}}}=1$.
Then we construct the third term  
$\sigma_{\mbox{{\tiny P}}}(\mbox{R}_{1})$.
The P terms created by the DSA are summarized
in Table 20 in Appendix B.3 
in Ref.\cite{Hasegawa:2014oya}.
We write down the terms 
$\sigma_{\mbox{{\tiny P}}}(\mbox{R}_{1}, {\tt dip}1,x_{a})$
and
$\sigma_{\mbox{{\tiny P}}}(\mbox{R}_{1}, {\tt dip}3u)$
as
\begin{align}
\hat{\sigma}_{\mbox{{\tiny P}}}(\mbox{R}_{i},\,  {\tt dip}1,
x_{a}) 
&= 
\frac{A_{4}}{S_{\mbox{{\tiny B1}} }}
\int_{0}^{1}dx \,
\frac{1}{\mbox{C}_{\mbox{{\tiny F}}}} 
P^{ff}(x)
\cdot
\Phi_{a}(\mbox{B}1,x)_{4}
\, \times
\nonumber\\
&\hspace{13mm}\biggl[
\ln \frac{\mu_{F}^{2}}{x\,s_{x_{a},y_{1}}}
\langle a,1 \rangle
+
\ln \frac{\mu_{F}^{2}}{x\,s_{x_{a},y_{2}}}
\langle a,2 \rangle
+
\ln \frac{\mu_{F}^{2}}{x\,s_{x_{a},y_{b}}}
\langle a,b \rangle
\biggr]\,,
\label{ex1pd1xa}
\\
\hat{\sigma}_{\mbox{{\tiny P}}}(\mbox{R}_{1},\,  {\tt dip}3u) 
&= 
\frac{A_{4}}{S_{\mbox{{\tiny B3}} }}
\int_{0}^{1}dx \,
\frac{1}{\mbox{C}_{\mbox{{\tiny A}}}} 
P^{fg}(x)
\cdot
\Phi_{a}(\mbox{B}3u,x)_{4}
\, \times
\nonumber\\
&\hspace{13mm}\biggl[
\ln \frac{\mu_{F}^{2}}{x\,s_{x_{a},y_{1}}}
\langle a,1 \rangle
+
\ln \frac{\mu_{F}^{2}}{x\,s_{x_{a},y_{2}}}
\langle a,2 \rangle
+
\ln \frac{\mu_{F}^{2}}{x\,s_{x_{a},y_{b}}}
\langle a,b \rangle
\biggr]\,.
\end{align}

Now that all the terms are explicitly written down,
we start the calculation from the summation,
$\hat{\sigma}_{\mbox{{\tiny D}}}(\mbox{R}_{1},\,\mbox{P}, 
{\tt dip}1,x_{a})
+
\hat{\sigma}_{\mbox{{\tiny C}}}(\mbox{R}_{1}, {\tt dip}1,
\,x_{a})$,
as
\begin{align}
\hat{\sigma}_{\mbox{{\tiny D}}}
(\mbox{R}_{1},\,\mbox{P}, {\tt dip}1,x_{a})
+
\hat{\sigma}_{\mbox{{\tiny C}}}(\mbox{R}_{1}, {\tt dip}1,
\,x_{a})
&=
\frac{A_{d}}{S_{\mbox{{\tiny B1}}}} 
\int_{0}^{1}dx \,
P^{ff}(x)\,\Phi_{a}(\mbox{B}1,x)_{d}
\,\biggl[
\nonumber\\
\biggl(\frac{1}{\ep} - \ln x \biggr)
\,\frac{1}{\mbox{C}_{\mbox{{\tiny F}}}}&
\Bigl(
s_{x_{a},\,y_{1}}^{-\ep}\langle a,1 \rangle
+
s_{x_{a},\,y_{2}}^{-\ep}\langle a,2 \rangle
+
s_{x_{a},\,y_{b}}^{-\ep}\langle a,b \rangle
\Bigr)
\nonumber \\
&\hspace{25mm}
+ \Bigl(\,
\frac{1}{\ep}
-
\ln \mu_{F}^{2}
\Bigr)
\langle \mbox{B}1 \rangle\,
\biggr]\,,
\label{ex1sum1}
\end{align}
where the relation
$S_{\mbox{{\tiny R}}_{1}}=S_{\mbox{{\tiny B1}}}=1$
is used.
We expand the factor with the Lorentz scalar as
$s^{-\ep} = 1 -\ep \ln s$,
and also expand the squared amplitude
$\langle \mbox{B}1 \rangle$
by using the color conservation law as
\begin{equation}
\langle \mbox{B}1 \rangle
=
\frac{1}{\mbox{C}_{\mbox{{\tiny F}}}}
\langle \mbox{B}1 |
\,\mbox{T}_{y_{a}} \cdot 
\mbox{T}_{y_{a}}
 | \ \mbox{B}1 \rangle
=
-\frac{1}{\mbox{C}_{\mbox{{\tiny F}}}}
\bigl(
\langle a,1 \rangle + \langle a,2 \rangle
+ \langle a,b \rangle
\bigr).
\end{equation}
The use of the color conservation law
in the dipole subtraction
is explained in
Ref.\cite{Catani:1996vz}.
Then we obtain the expression as
\begin{align}
\hat{\sigma}_{\mbox{{\tiny D}}}
(\mbox{R}_{1},\,\mbox{P}, {\tt dip}1,x_{a})
+
\hat{\sigma}_{\mbox{{\tiny C}}}(\mbox{R}_{1}, {\tt dip}1,
\,x_{a})
&=
\frac{A_{d}}{S_{\mbox{{\tiny B1}} }}
\int_{0}^{1}dx \,
\frac{1}{\mbox{C}_{\mbox{{\tiny F}}}} 
P^{ff}(x)
\cdot
\Phi_{a}(\mbox{B}1,x)_{d}
\, \times
\nonumber\\
\biggl[
\ln \frac{\mu_{F}^{2}}{x\,s_{x_{a},y_{1}}}
&\langle a,1 \rangle
+
\ln \frac{\mu_{F}^{2}}{x\,s_{x_{a},y_{2}}}
\langle a,2 \rangle
+
\ln \frac{\mu_{F}^{2}}{x\,s_{x_{a},y_{b}}}
\langle a,b \rangle
\biggr]\,.
\end{align}
Since the expression is now finite in 4 dimensions,
it is safely reduced back to 4 dimensions,
which becomes nothing but the term
$\hat{\sigma}_{\mbox{{\tiny P}}}(\mbox{R}_{i},\,  {\tt dip}1,
x_{a})$ in Eq.\,(\ref{ex1pd1xa}).
In this way, the relation
\begin{equation}
\sigma_{\mbox{{\tiny D}}}(\mbox{R}_{1},\,\mbox{P}, {\tt dip}1
,x_{a})
+
\sigma_{\mbox{{\tiny C}}}(\mbox{R}_{1}, {\tt dip}1,x_{a})
-
\sigma_{\mbox{{\tiny P}}}(\mbox{R}_{1}, {\tt dip}1,x_{a})
=0
\end{equation}
is proved. Finally, we calculate the summation
$\hat{\sigma}_{\mbox{{\tiny D}}}(\mbox{R}_{1},\,\mbox{P}, {\tt dip}3u)
+
\hat{\sigma}_{\mbox{{\tiny C}}}(\mbox{R}_{1}, {\tt dip}3u)$
as
\begin{align}
\hat{\sigma}_{\mbox{{\tiny D}}}
(\mbox{R}_{1},\,\mbox{P}, {\tt dip}3u)
+
\hat{\sigma}_{\mbox{{\tiny C}}}(\mbox{R}_{1}, {\tt dip}3u)
&=
\frac{A_{d}}{S_{\mbox{{\tiny B3}}}} 
\int_{0}^{1}dx \,
P^{fg}(x)\,\Phi_{a}(\mbox{B}3u,x)_{d}
\,\biggl[
\nonumber\\
\biggl(\frac{1}{\ep} - \ln x \biggr)
\,\frac{1}{\mbox{C}_{\mbox{{\tiny A}}}}&
\Bigl(
s_{x_{a},\,y_{1}}^{-\ep}\langle a,1 \rangle
+
s_{x_{a},\,y_{2}}^{-\ep}\langle a,2 \rangle
+
s_{x_{a},\,y_{b}}^{-\ep}\langle a,b \rangle
\Bigr)
\nonumber \\
&\hspace{25mm}
+ \Bigl(\,
\frac{1}{\ep}
-
\ln \mu_{F}^{2}
\Bigr)
\langle \mbox{B}3u \rangle\,
\biggr]\,.
\label{ex1sum2}
\end{align}
Similar to the calculation in Eq.\,(\ref{ex1sum1}),
we expand the factor with the Lorentz scalar 
and use the color conservation law as
$\langle \mbox{B}3u \rangle=
-\bigl(
\langle a,1 \rangle + \langle a,2 \rangle
+ \langle a,b \rangle
\bigr)/\mbox{C}_{\mbox{{\tiny A}}}$.
Then the quantity in Eq.\,(\ref{ex1sum2})
becomes finite in 4 dimensions, which is equal
to the term 
$\hat{\sigma}_{\mbox{{\tiny P}}}(\mbox{R}_{1},\,  {\tt dip}3u)$.
The relation in Eq.\,(\ref{prs3exj}) 
for ${\tt dip}3u$ is proved.
The relations in Eq.\,(\ref{prs3exj}) 
for the remaining cases,
Dipole\,1 with leg-b and 
Dipole\,3$\bar{u}$, are similarly proved.
Thus the relation in Eq.\,(\ref{prs3ex})
is proved and {\bfseries Step 3} for
the process $\mbox{R}_{1}$ is completed.
The results are shown in
Eqs.\,(\ref{r1st3}) and (\ref{r1st3ind})
in Appendix \ref{ap_B_1}.
%
%
%
\subsection{Step 4 : 
$\sigma_{\mbox{{\tiny D}}}(\mbox{K}) - \sigma_{\mbox{{\tiny K}}}$
\label{s_2_5}}
{\bfseries Step 4} of the PRA is to prove the relation,
\begin{equation}
\sigma_{\mbox{{\tiny D}}}(\mbox{R}_{i},\,\mbox{K})
-
\sigma_{\mbox{{\tiny K}}}(\mbox{R}_{i})
=
-\sigma_{\mbox{{\tiny K}}}(\mbox{R}_{i}, {\tt dip}1,
(3)/(4)\,\mbox{-}1,N_{f}h)\,.
\label{prast4}
\end{equation}
The left-hand side of Eq.\,(\ref{prast4})
is the third term in the square brackets
in Eq.\,(\ref{sigdfour}).
We define the three cross sections in 
Eq.\,(\ref{prast4})
as follows.

We first define the cross section
$\sigma_{\mbox{{\tiny D}}}(\mbox{R}_{i},\,\mbox{K})$.
The integrated dipole terms are separated
into the four terms shown in 
Eq.\,(\ref{intdipf}).
Among them, we have already defined the terms,
$\sigma_{\mbox{{\tiny D}}}(\mbox{R}_{i},\,\mbox{I})$
and
$\sigma_{\mbox{{\tiny D}}}(\mbox{R}_{i},\,\mbox{P})$
in Sec.\,\ref{s_2_3} and \ref{s_2_4}, respectively.
The term
$\sigma_{\mbox{{\tiny D}}}(\mbox{R}_{i},\,{\tt dip}2)$
is defined as the dipole term with
splitting (5) in the category
Dipole\,2, which is concretely shown
in Sec.\,\ref{s_2_6}.
Then the term
$\sigma_{\mbox{{\tiny D}}}(\mbox{R}_{i},\,\mbox{K})$
is defined as the remaining term in
Eq.\,(\ref{intdipf}) as
\begin{equation}
\sigma_{\mbox{{\tiny D}}}(\mbox{R}_{i},\,\mbox{K})
=
\sigma_{\mbox{{\tiny D}}}(\mbox{R}_{i})
-
\sigma_{\mbox{{\tiny D}}}(\mbox{R}_{i},\,\mbox{I})
-
\sigma_{\mbox{{\tiny D}}}(\mbox{R}_{i},\,\mbox{P})
-
\sigma_{\mbox{{\tiny D}}}(\mbox{R}_{i},\,{\tt dip}2)\,.
\end{equation}
All the poles $1/\ep^{2}$ and $1/\ep$ in
the integrated dipole term
$\sigma_{\mbox{{\tiny D}}}(\mbox{R}_{i})$
are extracted by the three terms
$\sigma_{\mbox{{\tiny D}}}(\mbox{R}_{i},
\mbox{I/P}/{\tt dip}2)$.
Then the cross section
$\sigma_{\mbox{{\tiny D}}}(\mbox{R}_{i},\,\mbox{K})$
is finite and defined in $4$ dimensions.
The term
$\sigma_{\mbox{{\tiny D}}}(\mbox{R}_{i},\,\mbox{K})$
is classified by the splittings
shown in Fig.\,\ref{fig_A5_int_K} in 
Appendix \ref{ap_A_5}.
We start with the category Dipole\,1,
which includes
the splittings (1),\,(2),\,(3), and (4).
For splittings (1)/(2)-2,
the remaining factors
$\mbox{C}_{\mbox{{\tiny F}}} (g(x)
- 3 h(x)/2)$
and
$\mbox{C}_{\mbox{{\tiny A}}} 
(2 g(x) - 11 h(x)/3)$
in the functions
${\cal V}_{fg}(x\,;\ep)$
and
${\cal V}_{gg}(x\,;\ep)$
in Eqs.\,(\ref{nuxepfg})
and (\ref{nuxepgg}),
respectively, are taken.
We only show here the full expression for the splitting
(1)-2 with leg-a as
\begin{equation}
\hat{\sigma}_{\mbox{{\tiny D}}}(\,
\mbox{R}_{i},\,
\mbox{K}, {\tt dip}1,
(1)\mbox{-}2,
x_{a}
\,) 
=
-\frac{A_{4}}{S_{\mbox{{\tiny R}}_{i}}} 
\int_{0}^{1}dx \,
\biggl(
g(x)-\frac{3}{2}h(x)
\biggr)
\cdot 
\Phi_{a}(\mbox{B}1,x)_{4} 
\, \langle  \, y_{emi}, \,y_{spe} \,\rangle \,.
\label{intdip1k12}
\end{equation}
For splittings (3)/(4)-1, the remaining terms
${\cal V}_{other}^{f,f/g,g}(x\,;\ep)$
in the functions
${\cal V}^{f,f/g,g}(x\,;\ep)$
in Eqs.\,(\ref{sinifff})/(\ref{sinifgg}),
respectively, are taken.
The cross section for splitting (3)-1
with leg-a is written as
\begin{equation}
\hat{\sigma}_{\mbox{{\tiny D}}}(\,
\mbox{R}_{i},\,
\mbox{K}, {\tt dip}1,
(3)\mbox{-}1,
x_{a}
\,) 
=
-\frac{A_{4}}{S_{\mbox{{\tiny R}}_{i}}} 
\int_{0}^{1}dx \,
\frac{1}{\mbox{C}_{\mbox{{\tiny F}}}}
{\cal V}_{other}^{f,f}(x\,;\ep)
\cdot 
\Phi_{a}(\mbox{B}1,x)_{4} 
\, \langle  \, y_{emi}, \,y_{spe} \,\rangle \,.
\end{equation}
For splittings (3)/(4)-2, the terms
$({\cal V}_{other}^{f,f}(x;\ep) +
\mbox{C}_{\mbox{{\tiny F}}} \, g(x)
+ {\widetilde K}^{ff}(x))$
and
$({\cal V}_{other}^{g,g}(x;\ep) +
\mbox{C}_{\mbox{{\tiny A}}} \, g(x)
+ {\widetilde K}^{gg}(x))$,
in ${\widetilde {\cal V}}^{f,f}(x;\ep)$
and
${\widetilde {\cal V}}^{g,g}(x;\ep)$
in Eqs.\,(\ref{siniiff})
and
(\ref{siniigg}),
respectively, are extracted.
The expression for splitting (3)-2 is written as
\begin{align}
\hat{\sigma}_{\mbox{{\tiny D}}}(\,
\mbox{R}_{i},\,
\mbox{K}, {\tt dip}1,
(3)\mbox{-}2,
x_{a}
\,) 
&=
-\frac{A_{4}}{S_{\mbox{{\tiny R}}_{i}}} 
\int_{0}^{1}dx \,
\frac{1}{\mbox{C}_{\mbox{{\tiny F}}}}
\bigl(\,
{\cal V}_{other}^{f,f}(x\,;\ep)
+
\mbox{C}_{\mbox{{\tiny F}}} \, g(x)
+ {\widetilde K}^{ff}(x)\,
\bigr)\,\times
\nonumber \\
& 
\hspace{40mm}
\Phi_{a}(\mbox{B}1,x)_{4} 
\, \langle  \, y_{emi}, \,y_{spe} \,\rangle \,.
\label{intdip1k32}
\end{align}
As mentioned above, the cross section
$\sigma_{\mbox{{\tiny D}}}(\mbox{R}_{i},\,\mbox{K})$
is defined in 4 dimensions, and so in
Eqs.\,(\ref{intdip1k12})--(\ref{intdip1k32})
the overall factor $A_{4}$, the phase space 
$\Phi_{a}(\mbox{B}1,x)_{4}$,
and the reduced Born squared amplitude
$\langle  \, y_{emi}, \,y_{spe} \,\rangle$,
are all defined in 4 dimensions.
The formulae for
$\hat{\sigma}_{\mbox{{\tiny D}}}(\mbox{R}_{i},
\mbox{K}, {\tt dip}1)$
are collected in 
Eqs.\,(\ref{a5eq1})--(\ref{a5eq3})
in Appendix \ref{ap_A_5}.
Then we proceed to the Dipole 3 and 4 category, i.e.,
splittings (6) and (7), respectively.
For splittings (6)/(7)-1, the terms
${\cal V}_{other}^{f,g/g,f}$
$(x\,;\ep)$
in
Eqs.\,(\ref{siniffg})/
(\ref{sinifgf})
are taken.
For splittings (6)/(7)-2, the terms
$({\cal V}^{f,g/g,f}(x;\ep)_{other}$
$+{\widetilde K}^{fg/gf}(x))$
in
Eqs.\,(\ref{siniifg})/
(\ref{siniigf})
are taken.
The formulae for splittings (6)/(7)-1/2
are collected in
Eqs.\,(\ref{a5eq4}) and (\ref{a5eq5}).

Next we define the cross section
$\sigma_{\mbox{{\tiny K}}}(\mbox{R}_{i})$.
This quantity is the contribution of the K terms
created by the DSA \cite{Hasegawa:2014oya}.
The K terms are classified into categories
Dipole\,1 and 3/4 with subcategories
(3)/(4)-0/1/2 and (6)/(7)-0/2,
respectively,
as shown in Fig.\,\ref{fig_A10_K} in 
Appendix \ref{ap_A_10}.
The contributions to the partonic cross sections
are denoted as
$\hat{\sigma}_{\mbox{{\tiny K}}}(\mbox{R}_{i},{\tt dip}1,\,
(3)/(4)\mbox{-}0/1/2)$
and
$\hat{\sigma}_{\mbox{{\tiny K}}}(\mbox{R}_{i},{\tt dip}3/4,\,
(6)/(7)\mbox{-}0/2)$,
respectively.
The explicit expressions for the cross sections
are collected in 
Eqs.\,(\ref{a10eq1})--(\ref{a10eq3})
and
Eqs.\,(\ref{a10eq5}) and (\ref{a10eq7}),
respectively.
We show here the expression for
the cross section in Dipole\,1 with
splitting (3)/(4)-1 and with leg-a as
\begin{equation}
\hat{\sigma}_{\mbox{{\tiny K}}}(\mbox{R}_{i},{\tt dip}1,\,
(3)/(4)\mbox{-}1,x_{a}) 
= 
\frac{A_{4}}{S_{\mbox{{\tiny B}}_{1}}}
\int_{0}^{1}dx \
\frac{\gamma_{\mbox{{\tiny F}}(y_{spe})}}
{\mbox{T}_{\mbox{{\tiny F}}(y_{spe})}^{2}} \, h(x)
\cdot
\Phi_{a}(\mbox{B}_{1},x)_{4} \,
\langle \, y_{emi}, \,y_{spe} \rangle\,\,.
\label{dipk341}
\end{equation}
This term has a special feature that the factor
$\gamma_{\mbox{{\tiny F}}(y_{spe})}/
\mbox{T}_{\mbox{{\tiny F}}(y_{spe})}^{2}$
is determined by the field species of the
spectator, $\mbox{F}(y_{spe})$,
unlike the other terms.

The third cross section, 
$\sigma_{\mbox{{\tiny K}}}(\mbox{R}_{i}, {\tt dip}1,
(3)/(4)\,\mbox{-}1,N_{f}h)$,
is defined as follows.
When the cross section
$\hat{\sigma}_{\mbox{{\tiny K}}}(\mbox{R}_{i},{\tt dip}1,\,
(3)/(4)\mbox{-}1)$
in Eq.\,(\ref{dipk341}),
which is in the category Dipole\,1 with 
splitting (3)/(4)-1,
has the gluon in the final state
of the reduced Born process
as the spectator, i.e.,
$\mbox{F}(y_{spe})=\mbox{gluon}$,
the factor
$\gamma_{\mbox{{\tiny F}}(y_{spe})}/
\mbox{T}_{\mbox{{\tiny F}}(y_{spe})}^{2}$
is determined as
$\gamma_{g}/\mbox{T}_{g}^{2}
=11/6 - 
2\,\mbox{T}_{\mbox{{\tiny R}}} \mbox{$N$}_{f}/
3\mbox{C}_{\mbox{{\tiny A}}}$
in Eq.\,(\ref{gamts}).
From the factor we extract the second term,
$-2\,\mbox{T}_{\mbox{{\tiny R}}} \mbox{$N$}_{f}/
3\mbox{C}_{\mbox{{\tiny A}}}$,
and define
$\sigma_{\mbox{{\tiny K}}}(\mbox{R}_{i}, {\tt dip}1,
(3)/(4)\,\mbox{-}1,N_{f}h)$
as
\begin{equation}
\sigma_{\mbox{{\tiny K}}}(\mbox{R}_{i}, {\tt dip}1,
(3)/(4)\,\mbox{-}1,N_{f}h)
= 
\frac{A_{4}}{S_{\mbox{{\tiny B}}_{1}}}
\int_{0}^{1}dx \
\biggl(-\frac{2}{3}
\frac{\mbox{T}_{\mbox{{\tiny R}}} \mbox{$N$}_{f}}
{\mbox{C}_{\mbox{{\tiny A}}}}
\biggr)
\, h(x)
\cdot
\Phi_{a/b}(\mbox{B}1,x)_{4} \,
\langle \, y_{emi}, \,y_{spe} \rangle\,.
\label{dipk341nf}
\end{equation}
Now all three cross sections 
in Eq.\,(\ref{prast4})
are defined.

The relation in
Eq.\,(\ref{prast4})
is separated into three independent ones,
\begin{align}
\sigma_{\mbox{{\tiny D}}}(\mbox{R}_{i},\,\mbox{K},{\tt dip}1)
-
\sigma_{\mbox{{\tiny K}}}(\mbox{R}_{i},{\tt dip}1)
&=
-\sigma_{\mbox{{\tiny K}}}(\mbox{R}_{i}, {\tt dip}1,
(3)/(4)\,\mbox{-}1,N_{f}h)\,,
\label{s4d1}
\\
\sigma_{\mbox{{\tiny D}}}(\mbox{R}_{i},\,\mbox{K},{\tt dip}3)
-
\sigma_{\mbox{{\tiny K}}}(\mbox{R}_{i},{\tt dip}3)
&=0\,,
\label{s4d2}
\\
\sigma_{\mbox{{\tiny D}}}(\mbox{R}_{i},\,\mbox{K},{\tt dip}4)
-
\sigma_{\mbox{{\tiny K}}}(\mbox{R}_{i},{\tt dip}4)
&=0\,,
\label{s4d3}
\end{align}
because the cancellations are realized between
the cross sections with the same reduced 
Born processes.
In order to prove the relations in
Eqs.\,(\ref{s4d1})--(\ref{s4d3})
in a systematic way,
we also divide {\bfseries Step 4}
into substeps defined as follows.

For the relation with Dipole\,1 
in Eq.\,(\ref{s4d1}),
{\bfseries Step 4} is divided into three substeps as
\begin{align}
&{\bf 4\mbox{-}1.} \ \ 
\hat{\sigma}_{\mbox{{\tiny D}}}(\mbox{K},{\tt dip}1,
(1)\mbox{--}(4)\mbox{-}2,g
)^{\circlearrowright}
+
\hat{\sigma}_{\mbox{{\tiny D}}}(\mbox{K},{\tt dip}1,
(3)/(4)\mbox{-}1/2,
{\cal V}_{other}^{a,a}
)^{\circlearrowright}
\nonumber \\
&\hspace{90mm}-
\hat{\sigma}_{\mbox{{\tiny K}}}({\tt dip}1,(3)/(4)\mbox{-}0)
=0\,,
\label{step41}
\\
&{\bf 4\mbox{-}2.} \ \ 
\hat{\sigma}_{\mbox{{\tiny D}}}(\mbox{K},{\tt dip}1,
(1)/(2)\mbox{-}2,h)
-
\hat{\sigma}_{\mbox{{\tiny K}}}({\tt dip}1,
(3)/(4)\mbox{-}1)
=-\hat{\sigma}_{\mbox{{\tiny K}}}({\tt dip}1,
(3)/(4)\,\mbox{-}1,N_{f}h)\,,
\label{step42}
\\
&{\bf 4\mbox{-}3.} \ \ 
\hat{\sigma}_{\mbox{{\tiny D}}}(\mbox{K},{\tt dip}1,
(3)/(4)\mbox{-}2,\widetilde{\mbox{K}}^{aa})
-
\hat{\sigma}_{\mbox{{\tiny K}}}({\tt dip}1,
(3)/(4)\mbox{-}2)
=0\,,
\label{step43}
\end{align}
where the argument $\mbox{R}_{i}$
in the cross sections is omitted as
$\hat{\sigma}(\mbox{R}_{i},\mbox{X}) \to
\hat{\sigma}(\mbox{X})$
for the compact notation.
The cross section
$\sigma_{\mbox{{\tiny D}}}(\mbox{R}_{i},\,\mbox{K},
{\tt dip}1)$
is reconstructed into four terms as
\begin{align}
\sigma_{\mbox{{\tiny D}}}(\mbox{R}_{i},\,\mbox{K},
{\tt dip}1)
&=
\hat{\sigma}_{\mbox{{\tiny D}}}(\mbox{K},{\tt dip}1,
(1)\mbox{--}(4)\mbox{-}2,g
)^{\circlearrowright}
+
\hat{\sigma}_{\mbox{{\tiny D}}}(\mbox{K},{\tt dip}1,
(3)/(4)\mbox{-}1/2,
{\cal V}_{other}^{a,a}
)^{\circlearrowright}
\nonumber\\
& \ \
+
\hat{\sigma}_{\mbox{{\tiny D}}}(\mbox{K},{\tt dip}1,
(1)/(2)\mbox{-}2,h)
+
\hat{\sigma}_{\mbox{{\tiny D}}}(\mbox{K},{\tt dip}1,
(3)/(4)\mbox{-}2,\widetilde{\mbox{K}}^{aa}(x))\,,
\label{sec25d1kf}
\end{align}
which appear in 
Eqs.\,(\ref{step41})--(\ref{step43}).
We define the four terms as follows.
The term 
$\hat{\sigma}_{\mbox{{\tiny D}}}(\mbox{K},{\tt dip}1,$
$(1)\mbox{--}(4)\mbox{-}2,g)$
is defined in such a way that,
in the case with splittings
(1)--(4)-2 in
Eqs.\,(\ref{a5eq1}) and (\ref{a5eq3}),
the terms including the function $g(x)$ 
are extracted. The expressions
are shown in Eq.\,(\ref{dkd1g}).
The symbol 
\textquoteleft$\circlearrowright$\textquoteright\,
in the term
$\hat{\sigma}_{\mbox{{\tiny D}}}(\mbox{K},{\tt dip}1,
(1)\sim(4)\mbox{-}2,g
)^{\circlearrowright}$
means that, when all the existing terms of the
$\hat{\sigma}_{\mbox{{\tiny D}}}(\mbox{K},{\tt dip}1,
(1)\sim(4)\mbox{-}2,g)$
are summed, the expressions include the summation
of the color-correlated Born squared amplitudes
such as
\begin{equation}
\langle a,1 \rangle + \langle a,2 \rangle
+ ... +
\langle a,n \rangle
+
\langle a,b \rangle\,.
\end{equation}
The color conservation law can be
applied to the summation
and the color correlations are 
factorized as
\begin{equation}
\sum_{i=1}^{n}\,
\langle a,i \rangle
+
\langle a,b \rangle
=-\langle a,a \rangle\
=
-\mbox{T}_{\mbox{{\tiny F}}(y_{a})}^{2}
\cdot 
\langle \mbox{B}1 \rangle\,.
\end{equation}
The next term
$\hat{\sigma}_{\mbox{{\tiny D}}}(\mbox{K},{\tt dip}1,
(3)/(4)\mbox{-}1/2,
{\cal V}_{other}^{a,a})$
is defined in such a way that,
in the case with splittings 
(3)/(4)-1/2 in
Eqs.\,(\ref{a5eq2}) and (\ref{a5eq3}),
the terms ${\cal V}_{other}^{a,a}(x\,;\ep)$ 
are extracted. The expressions
are shown in 
Eqs.\,(\ref{dkd1oth1}) and (\ref{dkd1oth2}).
The meaning of the symbol 
\textquoteleft$\circlearrowright$\textquoteright\,
in the term
$\hat{\sigma}_{\mbox{{\tiny D}}}(\mbox{K},{\tt dip}1,
(3)/(4)\mbox{-}1/2,{\cal V}_{other}^{a,a}
)^{\circlearrowright}$
is the same as in the previous term
$\hat{\sigma}_{\mbox{{\tiny D}}}(\mbox{K},{\tt dip}1,
(1)\mbox{--}(4)\mbox{-}2,g
)^{\circlearrowright}$.
The third term
$\hat{\sigma}_{\mbox{{\tiny D}}}(\mbox{K},{\tt dip}1,
(1)/(2)\mbox{-}2,h)$
is defined in such a way that
the terms including the function $h(x)$ 
in splittings (1)/(2)-2 in 
Eq.\,(\ref{a5eq1})
are extracted.
The expressions are written in
Eq.\,(\ref{dkd1h}).
The last term
$\hat{\sigma}_{\mbox{{\tiny D}}}(\mbox{K},{\tt dip}1,
(3)/(4)\mbox{-}2,\widetilde{\mbox{K}}^{aa})$
is defined in such a way that the terms 
$\widetilde{\mbox{K}}^{ff/gg}(x)$
in splittings (3)/(4)-2 in 
Eq.\,(\ref{a5eq3})
are extracted, which are written in
Eq.\,(\ref{dkd1ktil}).
The way of executing substeps {\tt 4-1}--{\tt 4-3} 
is demonstrated in the example later in the present 
subsection.

For the relations with Dipole\,3 and 4
in Eqs.\,(\ref{s4d2}) and (\ref{s4d3}),
{\bfseries Step 4} is divided into
two substeps as
\begin{align}
&{\bf 4\mbox{-}1.} \ \
\hat{\sigma}_{\mbox{{\tiny D}}}(\mbox{K},{\tt dip}3/4,
(6)/(7)\mbox{-}1/2,
{\cal V}_{other}^{a,b}
)^{\circlearrowright}
-
\hat{\sigma}_{\mbox{{\tiny K}}}({\tt dip}3/4,
(6)/(7)\mbox{-}0)
=0\,,
\label{dk34step41}
\\
&{\bf 4\mbox{-}2.} \ \ 
\hat{\sigma}_{\mbox{{\tiny D}}}(\mbox{K},{\tt dip}3/4,
(6)/(7)\mbox{-}2,\widetilde{\mbox{K}}^{ab})
-
\hat{\sigma}_{\mbox{{\tiny K}}}({\tt dip}3/4,
(6)/(7)\mbox{-}2)
=0\,,
\label{dk34step42}
\end{align}
where again the argument $\mbox{R}_{i}$ is
abbreviated. The cross section
$\sigma_{\mbox{{\tiny D}}}(\mbox{R}_{i},\,\mbox{K},
{\tt dip}3/4)$
is reconstructed into two terms as
\begin{equation}
\sigma_{\mbox{{\tiny D}}}(\mbox{R}_{i},\,\mbox{K},
{\tt dip}3/4)
=
\hat{\sigma}_{\mbox{{\tiny D}}}(\mbox{K},{\tt dip}3/4,
(6)/(7)\mbox{-}1/2,
{\cal V}_{other}^{a,b}
)
+
\hat{\sigma}_{\mbox{{\tiny D}}}(\mbox{K},{\tt dip}3/4,
(6)/(7)\mbox{-}2,\widetilde{\mbox{K}}^{ab})\,.
\end{equation}
The terms ${\cal V}_{other}^{a,b}(x\,;\ep)$ 
in Eqs.\,(\ref{a5eq4}) and (\ref{a5eq5})
are extracted and 
the cross sections
$\hat{\sigma}_{\mbox{{\tiny D}}}(\mbox{K},$
${\tt dip}3/4, (6)/(7)\mbox{-}1/2,$
${\cal V}_{other}^{a,b}
)$
are defined in 
Eqs.\,(\ref{dkd34oth1}) and (\ref{dkd34oth2}),
respectively.
The meaning of the symbol 
\textquoteleft$\circlearrowright$\textquoteright\
for this term
is the same as in the case of Dipole\,1.
The terms $\widetilde{\mbox{K}}^{ab}(x)$
in Eq.\,(\ref{a5eq5})
are extracted and
the cross sections
$\hat{\sigma}_{\mbox{{\tiny D}}}(
\mbox{K},
{\tt dip}3/4,
(6)/(7)\mbox{-}2,
\widetilde{\mbox{K}}^{ab})$
are defined in 
Eq.\,(\ref{dkd34ktil}).
The method of execution is demonstrated
in the following example.

We demonstrate the execution of
{\bfseries Step 4} in the same process
as used in the previous sections,
$\mbox{R}_{1}=u\bar{u} \to u\bar{u}g$.
The relation in
Eq.\,(\ref{prast4})
is written for the process $\mbox{R}_{1}$ as
\begin{equation}
\sigma_{\mbox{{\tiny D}}}(\mbox{R}_{1},\,\mbox{K})
-
\sigma_{\mbox{{\tiny K}}}(\mbox{R}_{1})
=0\,,
\label{step4ex1}
\end{equation}
where the right-hand side of Eq.\,(\ref{prast4})
does not exist, because the reduced Born process
$\mbox{B}1(\mbox{R}_{1})$
$=u\bar{u} \to u\bar{u}$
does not include any gluon in the final state.
The integrated dipole terms
$\sigma_{\mbox{{\tiny D}}}(\mbox{R}_{1})$
are summarized in Table\,\ref{ap_B_1_tab1}
in Appendix \ref{ap_B_1}.
The dipole terms include
three categories,
Dipole\,1, 3$u$, and 3$\bar{u}$, and
the relation in 
Eq.\,(\ref{step4ex1})
is separated into the three as
\begin{align}
\sigma_{\mbox{{\tiny D}}}(\mbox{R}_{1},\,\mbox{K},{\tt dip}1)
-
\sigma_{\mbox{{\tiny K}}}(\mbox{R}_{1},{\tt dip}1)
&=0\,,
\label{ex1s4d1}
\\
\sigma_{\mbox{{\tiny D}}}(\mbox{R}_{1},\,\mbox{K},{\tt dip}3u)
-
\sigma_{\mbox{{\tiny K}}}(\mbox{R}_{1},{\tt dip}3u)
&=0\,,
\label{ex1s4d2}
\\
\sigma_{\mbox{{\tiny D}}}(\mbox{R}_{1},\,\mbox{K},{\tt dip}3\bar{u})
-
\sigma_{\mbox{{\tiny K}}}(\mbox{R}_{1},{\tt dip}3\bar{u})
&=0\,.
\label{ex1s4d3}
\end{align}

First, we prove the relation for Dipole\,1
in Eq.\,(\ref{ex1s4d1}).
In order to prove it, we have the three steps
as shown in
Eqs.\,(\ref{step41})--(\ref{step43})\,:
\begin{align}
&{\bf 4\mbox{-}1.} \ \ 
\hat{\sigma}_{\mbox{{\tiny D}}}(\mbox{K},{\tt dip}1,
(1)/(3)\mbox{-}2,g
)^{\circlearrowright}
+
\hat{\sigma}_{\mbox{{\tiny D}}}(\mbox{K},{\tt dip}1,
(3)\mbox{-}1/2,
{\cal V}_{other}^{f,f}
)^{\circlearrowright}
\nonumber \\
&\hspace{90mm}
-\hat{\sigma}_{\mbox{{\tiny K}}}({\tt dip}1,(3)\mbox{-}0)
=0\,,
\label{exst41}
\\
&{\bf 4\mbox{-}2.} \ \ 
\hat{\sigma}_{\mbox{{\tiny D}}}(\mbox{K},{\tt dip}1,
(1)\mbox{-}2,h)
-
\hat{\sigma}_{\mbox{{\tiny K}}}({\tt dip}1,
(3)\mbox{-}1)
=0\,,
\label{exst42}
\\
&{\bf 4\mbox{-}3.} \ \ 
\hat{\sigma}_{\mbox{{\tiny D}}}(\mbox{K},{\tt dip}1,
(3)\mbox{-}2,\widetilde{\mbox{K}}^{ff})
-
\hat{\sigma}_{\mbox{{\tiny K}}}({\tt dip}1,
(3)\mbox{-}2)
=0\,.
\label{exst43}
\end{align}
$\hat{\sigma}_{\mbox{{\tiny D}}}(\mbox{R}_{1},\,\mbox{K},
{\tt dip}1)$ with leg-a
is written as a summation of the possible
splittings\,:
\begin{align}
\hat{\sigma}_{\mbox{{\tiny D}}}(\mbox{R}_{1},\,\mbox{K},
{\tt dip}1,x_{a})
&=
\hat{\sigma}_{\mbox{{\tiny D}}}(\mbox{R}_{1},\,\mbox{K},
{\tt dip}1,(1)\mbox{-}2,x_{a})
+
\hat{\sigma}_{\mbox{{\tiny D}}}(\mbox{R}_{1},\,\mbox{K},
{\tt dip}1,(3)\mbox{-}1,x_{a})
\nonumber\\
&\hspace{40mm}+
\hat{\sigma}_{\mbox{{\tiny D}}}(\mbox{R}_{1},\,\mbox{K},
{\tt dip}1,(3)\mbox{-}2,x_{a})\,.
\end{align}
We explicitly prove the relations in 
Eqs.\,(\ref{exst41})--(\ref{exst43})
with leg-a as follows.
The relations with leg-b are also 
similarly proved.
$\hat{\sigma}_{\mbox{{\tiny D}}}(\mbox{R}_{1},\,\mbox{K},
{\tt dip}1,x_{a})$
is reconstructed into the four terms in
Eq.\,(\ref{sec25d1kf})
as
\begin{align}
\sigma_{\mbox{{\tiny D}}}(\mbox{R}_{1},\,\mbox{K},
{\tt dip}1)
&=
\hat{\sigma}_{\mbox{{\tiny D}}}(\mbox{K},{\tt dip}1,
(1)/(3)\mbox{-}2,g
)^{\circlearrowright}
+
\hat{\sigma}_{\mbox{{\tiny D}}}(\mbox{K},{\tt dip}1,
(3)\mbox{-}1/2,
{\cal V}_{other}^{f,f}
)^{\circlearrowright}
\nonumber\\
& \ \
+
\hat{\sigma}_{\mbox{{\tiny D}}}(\mbox{K},{\tt dip}1,
(1)\mbox{-}2,h)
+
\hat{\sigma}_{\mbox{{\tiny D}}}(\mbox{K},{\tt dip}1,
(3)\mbox{-}2,\widetilde{\mbox{K}}^{ff})\,,
\end{align}
where each term is written as
\begin{align}
\hat{\sigma}_{\mbox{{\tiny D}}}(\mbox{K},{\tt dip}1,
(1)/(3)\mbox{-}2,g)^{\circlearrowright}
&=
-\frac{A_{4}}{S_{\mbox{{\tiny R}}_{1}}} 
\int_{0}^{1}dx \,
g(x)\,
\Phi_{a}(\mbox{B}1)_{4} 
\, 
(\,\langle 1,a \rangle + \langle 2,a \rangle +
\langle a,b \rangle
\,)\,,
\\
\hat{\sigma}_{\mbox{{\tiny D}}}(\mbox{K},{\tt dip}1,
(3)\mbox{-}1/2,
{\cal V}_{other}^{f,f}
)^{\circlearrowright}
&=
-\frac{A_{4}}{S_{\mbox{{\tiny R}}_{1}}} 
\int_{0}^{1}dx \,
\frac{1}{\mbox{C}_{\mbox{{\tiny F}}}}
{\cal V}_{other}^{f,f}(x\,;\ep)
\cdot 
\Phi_{a}(\mbox{B}1)_{4} 
\nonumber\\
&\hspace{40mm}(\,\langle a,1 \rangle + \langle a,1 \rangle +
\langle a,b \rangle
\,)\,,
\\
\hat{\sigma}_{\mbox{{\tiny D}}}(\mbox{K},{\tt dip}1,
(1)\mbox{-}2,h)
&=
-\frac{A_{4}}{S_{\mbox{{\tiny R}}_{1}}} 
\int_{0}^{1}dx \,
\biggl(
-\frac{2}{3}h(x)
\biggr)
\,\Phi_{a}(\mbox{B}1)_{4} 
\, 
(\,\langle 1,a \rangle + \langle 2,a \rangle \,)\,,
\\
\hat{\sigma}_{\mbox{{\tiny D}}}(\mbox{K},{\tt dip}1,
(3)\mbox{-}2,\widetilde{\mbox{K}}^{ff})
&=
-\frac{A_{4}}{S_{\mbox{{\tiny R}}_{1}}} 
\int_{0}^{1}dx \,
\frac{1}{\mbox{C}_{\mbox{{\tiny F}}}}
{\widetilde K}^{ff}(x)
\cdot
\Phi_{a}(\mbox{B}1)_{4} 
\, \langle  \, a,b \,\rangle \,.
\label{ex1d1ktil}
\end{align}
Here the argument $x_{a}$ is suppressed.
Next we write down the cross sections
of the K terms,
$\hat{\sigma}_{\mbox{{\tiny K}}}(\mbox{R}_{1},
{\tt dip}1,(3)\mbox{-}0/1/2)$.
The K terms created by the DSA
are summarized in Table.\,20 in Appendix B.3
in Ref.\cite{Hasegawa:2014oya}.
They are written down as
\begin{align}
\hat{\sigma}_{\mbox{{\tiny K}}}(\mbox{R}_{1},
{\tt dip}1,(3)\mbox{-}0,x_{a})
&=
\frac{A_{4}}{S_{\mbox{{\tiny B}}_{1}}}
\int_{0}^{1}dx \
\overline{\mbox{K}}^{ff}(x)
\cdot
\Phi_{a}(\mbox{B}1)_{4} \,
\langle \mbox{B}1 \rangle\,, 
\\
\hat{\sigma}_{\mbox{{\tiny K}}}(\mbox{R}_{1},
{\tt dip}1,(3)\mbox{-}1,x_{a})
&=
\frac{A_{4}}{S_{\mbox{{\tiny B}}_{1}}}
\int_{0}^{1}dx \
\frac{3}{2}h(x)
\cdot
\Phi_{a}(\mbox{B}1)_{4} \,
(\,\langle a,1 \rangle + \langle a,2 \rangle \,)\,,
\\
\hat{\sigma}_{\mbox{{\tiny K}}}(\mbox{R}_{1},
{\tt dip}1,(3)\mbox{-}2,x_{a})
&=\frac{A_{4}}{S_{\mbox{{\tiny B}}_{1}}}
\int_{0}^{1}dx \
\frac{-1}{\mbox{C}_{\mbox{{\tiny F}}}}
\widetilde{\mbox{K}}^{ff}(x)
\cdot
\Phi_{a}(\mbox{B}1)_{4} \,
\langle \, a,b \,\rangle\,.
\label{ex1kktil}
\end{align}
Then we execute from step {\bf 4-1}.
We calculate the summation
$\hat{\sigma}_{\mbox{{\tiny D}}}(\mbox{K},{\tt dip}1,
(1)/(3)\mbox{-}2,g
)^{\circlearrowright}$
$+$
$\hat{\sigma}_{\mbox{{\tiny D}}}(\mbox{K},{\tt dip}1,
(3)\mbox{-}1/2,
{\cal V}_{other}^{f,f}
)^{\circlearrowright}$
as
\begin{align}
\hat{\sigma}_{\mbox{{\tiny D}}}(\mbox{K},{\tt dip}1,
(1)/(3)\mbox{-}2,g
)^{\circlearrowright}
&+
\hat{\sigma}_{\mbox{{\tiny D}}}(\mbox{K},{\tt dip}1,
(3)\mbox{-}1/2,
{\cal V}_{other}^{f,f}
)^{\circlearrowright}
=
\frac{A_{4}}{S_{\mbox{{\tiny R}}_{1}}} 
\int_{0}^{1}dx \
\Phi_{a}(\mbox{B}1)_{4}\, \times
\nonumber\\
&\hspace{55mm}\langle \mbox{B}1 \rangle\,
(\,
\mbox{C}_{\mbox{{\tiny F}}} g(x)
+
{\cal V}_{other}^{f,f}(x\,;\ep)
\,)\,,
\end{align}
where we use the color conservation law as
$\langle a,1 \rangle + \langle a,1 \rangle +
\langle a,b \rangle
=
-\mbox{C}_{\mbox{{\tiny F}}}
\langle \mbox{B}1 \rangle$.
Noting the relations
$\overline{\mbox{K}}^{ff}(x)
={\cal V}_{other}^{f,f}(x\,;\ep) 
+ \mbox{C}_{\mbox{{\tiny F}}} \, g(x)$,
and
$S_{\mbox{{\tiny R}}_{1}}=S_{\mbox{{\tiny B}}_{1}}=1$,
the relation in Eq.\,(\ref{exst41})
is proved.
Then we proceed to step {\bf 4-2}.
The left-hand side of the relation
in Eq.\,(\ref{exst42})
is calculated as
\begin{align}
\hat{\sigma}_{\mbox{{\tiny D}}}(\mbox{K},{\tt dip}1,
(1)\mbox{-}2,h)
-
\hat{\sigma}_{\mbox{{\tiny K}}}({\tt dip}1,
(3)\mbox{-}1)
&=
\frac{A_{4}}{S_{\mbox{{\tiny B}}_{1}}}
\int_{0}^{1}dx \
\Phi_{a}(\mbox{B}1)_{4} \,
\frac{3}{2}h(x)\, \times \bigl[
\nonumber\\
& \ \ \ \ 
(\,
\langle 1,a \rangle + \langle 2,a \rangle \,)
-
(\,
\langle a,1 \rangle + \langle a,2 \rangle \,)
\,\bigr]
\nonumber\\
&=0\,,
\end{align}
where the relation 
$\langle a,1/2 \rangle=\langle 1/2,a \rangle$
is used. In step {\bf 4-3}
the relation in Eq.\,(\ref{exst43})
is trivial with 
Eqs.\,(\ref{ex1d1ktil}) and (\ref{ex1kktil}).
In this way the relation with leg-a in 
Eq.\,(\ref{ex1s4d1})
is proved.

Finally, we prove the relation for Dipole\,3$u$
in Eq.\,(\ref{ex1s4d2}).
To prove it, we have two substeps as
\begin{align}
&{\bf 4\mbox{-}1.} \ \
\hat{\sigma}_{\mbox{{\tiny D}}}(\mbox{K},{\tt dip}3u,
(6)\mbox{-}1/2,
{\cal V}_{other}^{f,g}
)^{\circlearrowright}
-
\hat{\sigma}_{\mbox{{\tiny K}}}({\tt dip}3u,
(6)\mbox{-}0)
=0\,,
\label{ex134step41}
\\
&{\bf 4\mbox{-}2.} \ \ 
\hat{\sigma}_{\mbox{{\tiny D}}}(\mbox{K},{\tt dip}3u,
(6)\mbox{-}2,\widetilde{\mbox{K}}^{fg})
-
\hat{\sigma}_{\mbox{{\tiny K}}}({\tt dip}3u,
(6)\mbox{-}2)
=0\,.
\label{ex134step42}
\end{align}
$\hat{\sigma}_{\mbox{{\tiny D}}}(\mbox{R}_{1},\,\mbox{K},
{\tt dip}3u)$
is written as the summation
\begin{equation}
\hat{\sigma}_{\mbox{{\tiny D}}}(\mbox{R}_{1},\,\mbox{K},
{\tt dip}3u)
=
\hat{\sigma}_{\mbox{{\tiny D}}}(\mbox{R}_{1},\,\mbox{K},
{\tt dip}3u,(6)\mbox{-}1)
+
\hat{\sigma}_{\mbox{{\tiny D}}}(\mbox{R}_{1},\,\mbox{K},
{\tt dip}3u,(6)\mbox{-}2),
\end{equation}
which is reconstructed as
\begin{equation}
\hat{\sigma}_{\mbox{{\tiny D}}}(\mbox{R}_{1},\,\mbox{K},
{\tt dip}3u)
=
\hat{\sigma}_{\mbox{{\tiny D}}}(\mbox{K},{\tt dip}3u,
(6)\mbox{-}1/2,
{\cal V}_{other}^{f,g}
)^{\circlearrowright}
+
\hat{\sigma}_{\mbox{{\tiny D}}}(\mbox{K},{\tt dip}3u,
(6)\mbox{-}2,\widetilde{\mbox{K}}^{fg})\,.
\end{equation}
The two terms are written as
\begin{align}
\hat{\sigma}_{\mbox{{\tiny D}}}(\mbox{K},{\tt dip}3u,
(6)\mbox{-}1/2,
{\cal V}_{other}^{f,g}
)^{\circlearrowright}
&=
-\frac{A_{4}}{S_{\mbox{{\tiny R}}_{1}}} 
\int_{0}^{1}dx \,
\frac{1}{\mbox{C}_{\mbox{{\tiny A}}}}
{\cal V}_{other}^{f,g}(x\,;\ep)
\cdot 
\Phi_{a}(\mbox{B}3u)_{4} 
\nonumber\\
&\hspace{40mm}(\,\langle a,1 \rangle + \langle a,2 \rangle 
+ \langle a,b \rangle
\,)\,,
\\
\hat{\sigma}_{\mbox{{\tiny D}}}(\mbox{K},{\tt dip}3u,
(6)\mbox{-}2,\widetilde{\mbox{K}}^{fg})
&=
-\frac{A_{4}}{S_{\mbox{{\tiny R}}_{1}}} 
\int_{0}^{1}dx \,
\frac{1}{\mbox{C}_{\mbox{{\tiny A}}}}
{\widetilde K}^{fg}(x)
\cdot
\Phi_{a}(\mbox{B}3u)_{4} 
\, \langle  \, a,b \,\rangle \,.
\label{ex1d3ktil}
\end{align}
Referring to the K terms with splittings
(6)$u$-0/2 in Table.20 in Ref.\cite{Hasegawa:2014oya},
the contributions of the K terms
are written down as
\begin{align}
\hat{\sigma}_{\mbox{{\tiny K}}}({\tt dip}3u,
(6)\mbox{-}0)
&=
\frac{A_{4}}{S_{\mbox{{\tiny B}}_{3u}}}
\int_{0}^{1}dx \
\overline{\mbox{K}}^{fg}(x)
\cdot
\Phi_{a}(\mbox{B}3u)_{4} \,
\langle \mbox{B}3u \rangle\,, 
\\
\hat{\sigma}_{\mbox{{\tiny K}}}({\tt dip}3u,
(6)\mbox{-}2)
&=
\frac{A_{4}}{S_{\mbox{{\tiny B}}_{3u}}}
\int_{0}^{1}dx \
\frac{-1}{\mbox{C}_{\mbox{{\tiny A}}}}
\widetilde{\mbox{K}}^{fg}(x)
\cdot
\Phi_{a}(\mbox{B}3u)_{4} \,
\langle \, a,b \,\rangle\,.
\label{ex1kd3ktil}
\end{align}
Then we start from step\,{\bf 4-1}.
The left-hand side of Eq.\,(\ref{ex134step41}) 
is calculated as
\begin{align}
\hat{\sigma}_{\mbox{{\tiny D}}}(\mbox{K},{\tt dip}3u,
(6)\mbox{-}1/2,
{\cal V}_{other}^{f,g}
)^{\circlearrowright}
-
\hat{\sigma}_{\mbox{{\tiny K}}}({\tt dip}3u,
(6)\mbox{-}0)
&=
\frac{A_{4}}{S_{\mbox{{\tiny B}}_{3u}}} 
\int_{0}^{1}dx \
\Phi_{a}(\mbox{B}3u)_{4}\, \times
\nonumber\\
&\hspace{0mm}\langle \mbox{B}3u \rangle\,
(\,
{\cal V}_{other}^{f,g}(x\,;\ep)
-
\overline{\mbox{K}}^{fg}(x)
\,)
\nonumber\\
&=0\,,
\end{align}
where we use the relations
$S_{\mbox{{\tiny R}}_{1}}=
S_{\mbox{{\tiny B}}_{3u}}=1$,
the color conservation
$\langle a,1 \rangle + \langle a,1 \rangle +
\langle a,b \rangle
=
-\mbox{C}_{\mbox{{\tiny A}}}
\langle \mbox{B}3u \rangle$,
and 
$\overline{\mbox{K}}^{fg/gf}(x) 
={\cal V}_{other}^{f,g/g,f}(x\,;\ep)$.
In step {\bf 4-2},
the relation in Eq.\,(\ref{ex134step42})
trivially stands with 
Eqs.\,(\ref{ex1d3ktil}) and (\ref{ex1kd3ktil}).
Then the relation in 
Eq.\,(\ref{ex1s4d2})
is proved.

In a similar way, the relations 
for Dipole\,1 with leg-b and 
for Dipole\,3$\bar{u}$ can be proved. 
Then the relation in
Eq.\,(\ref{step4ex1})
is proved and {\bfseries Step 4} for
the process $\mbox{R}_{1}$ is completed.
The results are shown in
Eqs.\,(\ref{r1st4}) and (\ref{r1st41})
in Appendix\,\ref{ap_B_1}.
In this example, the right-hand side of
Eq.\,(\ref{prast4}),
$\sigma_{\mbox{{\tiny K}}}(\mbox{R}_{i}, 
{\tt dip}1,(3)/(4)\,\mbox{-}1,N_{f}h)$,
does not exist. The presence or
absence of the right-hand side
is the same as the right-hand side
$\sigma_{\mbox{{\tiny I}}}(\mbox{R}_{i},(2)\,
\mbox{-}1/2,N_{f}{\cal V}_{f\bar{f}})$
in Eq.\,(\ref{prast2})
in {\bfseries Step 2}.
The cases where the right-hand side
exists will be seen in 
Sec.\,\ref{sec_4} and \ref{sec_5}.
%
%
%
\subsection{Step 5 : $\sigma_{subt}$ \label{s_2_6}}
{\bfseries Step 5} of the PRA is to write down
the cross section $\sigma_{subt}(\mbox{R}_{i})$
in Eq.\,(\ref{sigdfour}).
We substitute the first three terms 
in square brackets by the proved relations
in Eqs.\,(\ref{prast2}), (\ref{prast3}), 
and (\ref{prast4})
in {\bfseries Steps 2}, {\bfseries 3}, and 
{\bfseries 4},
respectively.
Then we obtain
$\sigma_{subt}(\mbox{R}_{i})$
in the expression
\begin{equation}
\sigma_{subt}(\mbox{R}_{i})=
-\sigma_{\mbox{{\tiny I}}}(\mbox{R}_{i},(2)\,\mbox{-}1/2,
N_{f}{\cal V}_{f\bar{f}})
-\sigma_{\mbox{{\tiny K}}}(\mbox{R}_{i},{\tt dip}1,
(3)/(4)\,\mbox{-}1,N_{f}h)
+
\sigma_{\mbox{{\tiny D}}}(\mbox{R}_{i},\,{\tt dip}2)\,.
\label{prast5}
\end{equation}
The cross sections
$\sigma_{\mbox{{\tiny I}}}(\mbox{R}_{i},(2)\,\mbox{-}1/2,
N_{f}{\cal V}_{f\bar{f}})$ and 
$\sigma_{\mbox{{\tiny K}}}(\mbox{R}_{i},{\tt dip}1,
(3)/(4)\,\mbox{-}1,N_{f}h)$
are defined in {\bfseries Step\,2} and {\bfseries 4}
in Sec.\,\ref{s_2_3} and \ref{s_2_5},
respectively.
We here define the cross section
$\sigma_{\mbox{{\tiny D}}}(\mbox{R}_{i},\,{\tt dip}2)$
as follows.

The term
$\sigma_{\mbox{{\tiny D}}}(\mbox{R}_{i},\,{\tt dip}2)$
is nothing but the integrated dipole terms
in the category Dipole\,2 with splitting
(5)-1/2 in Fig.\,\ref{fig_A2_int_D}
in Appendix \ref{ap_A_2}.
The term is written as the summation of 
the two terms with splittings (5)-1 and -2\,:
\begin{equation}
\hat{\sigma}_{\mbox{{\tiny D}}}(\mbox{R}_{i},\,{\tt dip}2)
=
\hat{\sigma}_{\mbox{{\tiny D}}}
(\,
\mbox{R}_{i},\,
{\tt dip}2,(5)\mbox{-}1\,) 
+
\hat{\sigma}_{\mbox{{\tiny D}}}
(\,
\mbox{R}_{i},\,
{\tt dip}2,(5)\mbox{-}2\,)\,.
\label{dip212}
\end{equation}
The expressions for the two terms are shown 
in 
Eqs.\,(\ref{d251}) and (\ref{d252}),
respectively,
in Appendix \ref{ap_A_6}.
For use in the last step,
{\bfseries Step 6},
we reconstruct the two terms 
into two different ones as
\begin{equation}
\hat{\sigma}_{\mbox{{\tiny D}}}(\mbox{R}_{i},\,{\tt dip}2)
=
\hat{\sigma}_{\mbox{{\tiny D}}}
(\,
\mbox{R}_{i},\,
{\tt dip}2,(5)\mbox{-}1/2,
\,
{\cal V}_{f\bar{f}}
\,) 
+
\hat{\sigma}_{\mbox{{\tiny D}}}
(\,
\mbox{R}_{i},\,
{\tt dip}2,(5)\mbox{-}2,
\,h)\,.
\label{dip2recon}
\end{equation}
The term
$\hat{\sigma}_{\mbox{{\tiny D}}}
(\,
\mbox{R}_{i},\,
{\tt dip}2,(5)\mbox{-}1/2,
\,
{\cal V}_{f\bar{f}}
\,)$
is defined in such a way that,
for the splitting (5)-1,
all the terms are taken, and,
for the splitting (5)-2,
the term including
the function
${\cal V}_{f\bar{f}}(\ep)$
in
Eq.\,(\ref{nuxepffb})
is extracted.
The term
$\hat{\sigma}_{\mbox{{\tiny D}}}
(\,
\mbox{R}_{i},\,
{\tt dip}2,(5)\mbox{-}2,
\,h)$
is defined in such a way that,
for the splitting (5)-2,
the term with
the function
$h(x)$ in
Eq.\,(\ref{nuxepffb})
is extracted.
The expressions are written as
\begin{align}
\hat{\sigma}_{\mbox{{\tiny D}}}
(\,
\mbox{R}_{i},\,
{\tt dip}2,(5)\mbox{-}1/2,
\,
{\cal V}_{f\bar{f}}
\,) 
&=
-\frac{A_{d}}{S_{\mbox{{\tiny R}}_{i}}} 
\cdot
\frac{1}{\mbox{C}_{\mbox{{\tiny A}}}} 
{\cal V}_{f\bar{f}} (\ep) 
\cdot 
\Phi(\mbox{B}2)_{d} \,
[ \, y_{emi}, \,y_{spe} ]\,,
\label{sigd2nuff}
\\
\hat{\sigma}_{\mbox{{\tiny D}}}
(\,
\mbox{R}_{i},\,
{\tt dip}2,(5)\mbox{-}2,
\,h) 
&=
-\frac{A_{4}}{S_{\mbox{{\tiny R}}_{i}}} 
\int_{0}^{1}dx \,
\frac{\mbox{T}_{\mbox{{\tiny R}}}}
{\mbox{C}_{\mbox{{\tiny A}}}} 
\,\frac{2}{3}h(x)
\cdot 
\Phi_{a/b}(\mbox{B}2,x)_{4}\,
\langle \, y_{emi}, \,y_{spe} \rangle\,.
\label{sigd2h}
\end{align}
The first term
$\hat{\sigma}_{\mbox{{\tiny D}}}
(\mbox{R}_{i},
{\tt dip}2,(5)\mbox{-}1/2,
{\cal V}_{f\bar{f}})$
is defined in $d$ dimensions
with the color-correlated Born 
squared amplitude,
$[ \, y_{emi}, \,y_{spe} ]$
$=$
$(s_{y_{emi}, \,y_{spe}})^{-\ep}
\cdot$
$\langle \, y_{emi}, \,y_{spe} \rangle_{d}$\,.
The second term
$\hat{\sigma}_{\mbox{{\tiny D}}}
(\mbox{R}_{i},$
${\tt dip}2,(5)\mbox{-}2,
\,h)$
is finite in 4 dimensions,
and so the squared amplitude 
is defined in 4 dimensions as
$\langle \, y_{emi}, \,y_{spe} \rangle_{4}$.
The formulae for the term
$\sigma_{\mbox{{\tiny D}}}(\mbox{R}_{i},{\tt dip}2)$
are collected in 
Appendix \ref{ap_A_6}.
As mentioned in {\bfseries Steps 2} and 
{\bfseries 4},
the terms
$\sigma_{\mbox{{\tiny I}}}(\mbox{R}_{i},(2)\,
\mbox{-}1/2,N_{f}{\cal V}_{f\bar{f}})$
and
$\sigma_{\mbox{{\tiny K}}}(\mbox{R}_{i},$
${\tt dip}1,
(3)/(4)\,\mbox{-}1,N_{f}h)$
on the right-hand side of Eq.\,(\ref{prast5})
exist only if the process $\mbox{R}_{i}$
includes two or more gluons in the final state.
The term 
$\sigma_{\mbox{{\tiny D}}}(\mbox{R}_{i},\,{\tt dip}2)$
exists only if $\mbox{R}_{i}$
includes a quark--antiquark pair 
($q\bar{q}$ pair)
in the final state.
For instance, if the final state of
$\mbox{R}_{i}$ includes neither
a gluon nor a $q\bar{q}$ pair,
the cross section $\sigma_{subt}(\mbox{R}_{i})$
vanishes itself as
$\sigma_{subt}(\mbox{R}_{i})=0$\,.

Finally, we show one example in the process
used in the previous sections,
$\mbox{R}_{1}=u\bar{u} \to u\bar{u}g$.
We have seen the results for $\mbox{R}_{1}$
in {\bfseries Steps 2}, {\bfseries 3},
and {\bfseries 4}
in Eqs.\,(\ref{ex1tar}), (\ref{prs3ex}),
and (\ref{step4ex1}), respectively.
In Eq.\,(\ref{sigdfour}),
we substitute the three terms in
the square brackets by the three relations
and obtain the expression for
$\sigma_{subt}(\mbox{R}_{1})$ as
\begin{equation}
\sigma_{subt}(\mbox{R}_{1})=
\sigma_{\mbox{{\tiny D}}}(\mbox{R}_{1},\,{\tt dip}2)\,.
\end{equation}
Since the process $\mbox{R}_{1}$ includes
only one gluon in the final state,
the first two terms on the right-hand side
of Eq.\,(\ref{prast5})
do not exist.
The cross section
$\sigma_{\mbox{{\tiny D}}}(\mbox{R}_{1},{\tt dip}2)$
is written as the summation of the two terms
as in Eq.\,(\ref{dip212}).
Referring to Table\,\ref{ap_B_1_tab1}
in Appendix \ref{ap_B_1},
the two terms 
in the category Dipole\,2
are written as
\begin{align}
\hat{\sigma}_{\mbox{{\tiny D}}}
(\,
\mbox{R}_{1},\,
{\tt dip}2,(5)\mbox{-}1) 
&=
-\frac{A_{d}}{S_{\mbox{{\tiny R}}_{1}}} 
\cdot
\frac{1}{\mbox{C}_{\mbox{{\tiny A}}}} 
{\cal V}_{f\bar{f}} (\ep) 
\cdot 
\Phi(\mbox{B}2)_{d} \,
[1,2]\,,
\\
\hat{\sigma}_{\mbox{{\tiny D}}}
(\,
\mbox{R}_{1},\,
{\tt dip}2,(5)\mbox{-}2) 
&=
-\frac{A_{d}}{S_{\mbox{{\tiny R}}_{1}}} 
\int_{0}^{1}dx \,
\frac{1}{\mbox{C}_{\mbox{{\tiny A}}}}
{\cal V}_{f\bar{f}}(x\,;\ep)
\,
\biggl[\,
\Phi_{a}(\mbox{B}2,x)_{d}\,
[ 1,a ]
+
\Phi_{b}(\mbox{B}2,x)_{d}\,
[ 1,b ]
\,\biggr]\,,
\end{align}
where the reduced Born process is
determined as
$\mbox{B}2=u(y_{a})\bar{u}(y_{b}) \to 
g(y_{1})g(y_{2})$.
$\sigma_{\mbox{{\tiny D}}}(\mbox{R}_{1},
{\tt dip}2)$
is reconstructed into the two terms 
in Eq.\,(\ref{dip2recon}) as
\begin{align}
\hat{\sigma}_{\mbox{{\tiny D}}}
(\,
\mbox{R}_{1},\,
{\tt dip}2,(5)\mbox{-}1/2,
\,
{\cal V}_{f\bar{f}}
\,) 
&=
-\frac{A_{d}}{S_{\mbox{{\tiny R}}_{1}}} 
\cdot
\frac{1}{\mbox{C}_{\mbox{{\tiny A}}}} 
{\cal V}_{f\bar{f}} (\ep) 
\cdot 
\Phi(\mbox{B}2)_{d} \,
\bigl(\,
[1,2]+[1,a]+[1,b]
\,\bigr)\,,
\label{exst5d21}
\\
\hat{\sigma}_{\mbox{{\tiny D}}}
(\,
\mbox{R}_{1},\,
{\tt dip}2,(5)\mbox{-}2,
\,h) 
&=
-\frac{A_{4}}{S_{\mbox{{\tiny R}}_{1}}} 
\int_{0}^{1}dx \,
\frac{\mbox{T}_{\mbox{{\tiny R}}}}
{\mbox{C}_{\mbox{{\tiny A}}}} 
\,\frac{2}{3}h(x)
\, \times
\nonumber\\
&\hspace{20mm}\Bigl(\,
\Phi_{a}(\mbox{B}2,x)_{4}\,
\langle 1,a \rangle
+
\Phi_{b}(\mbox{B}2,x)_{4}\,
\langle 1,b \rangle
\,\Bigr)\,.
\label{exst5d22}
\end{align}
The results are summarized in
Eqs.\,(\ref{r1st51})--(\ref{r1st5last})
in Appendix \ref{ap_B_1}.
In this way,
we write down the term
$\sigma_{subt}(\mbox{R}_{1})$
and
{\bfseries Step 5} for $\mbox{R}_{1}$
is completed.
In Sec.\,\ref{sec_4} and \ref{sec_5},
we will see
the cases where the two terms
$\sigma_{\mbox{{\tiny I}}}(\mbox{R}_{i},(2)\,\mbox{-}1/2,
N_{f}{\cal V}_{f\bar{f}})$
and
$\sigma_{\mbox{{\tiny K}}}(\mbox{R}_{i},{\tt dip}1,
(3)/(4)\,\mbox{-}1,N_{f}h)$
in Eq.\,(\ref{prast5})
exist.
%
%
%
\subsection{Step 6 : 
$\sum_{i} \sigma_{subt}(\mbox{R}_{i})=0$
\label{s_2_7}}
For a given collider process,
the set of all the real emission
processes is denoted as
$\{\mbox{R}_{i}\}=\{
\mbox{R}_{1},\mbox{R}_{2},
...,
\mbox{R}_{\,n_{real}}
\}$
in Eq.\,(\ref{setri}).
{\bfseries Steps 1}--{\bfseries 5}
are repeated over all the real processes,
$\mbox{R}_{1},\mbox{R}_{2},
...,$
$\mbox{R}_{\,n_{real}}$,
and we obtain the corresponding
cross sections as
$\sigma_{subt}(\mbox{R}_{i})$
\begin{equation}
\{
\sigma_{subt}(\mbox{R}_{i})
\}=\{
\sigma(\mbox{R}_{1}),
\sigma(\mbox{R}_{2}),
...,
\sigma(\mbox{R}_{n_{real}})
\}\,.
\end{equation}
Each cross section is obtained in the expression
in Eq.\,(\ref{prast5}).
{\bfseries Step\,6} of the PRA
is to prove that
the summation of all the cross sections
$\sigma_{subt}(\mbox{R}_{i})$
vanishes as
\begin{equation}
\sum_{i=1}^{n_{\mbox{{\tiny real}}}} 
\sigma_{subt}(\mbox{R}_{i})
=0\,.
\label{prast6}
\end{equation}

We can prove the relation 
in Eq.\,(\ref{prast6})
in a systematic way as follows.
For the process $\mbox{R}_{i}$,
which has two or more gluons
in the final state,
we introduce a set,
$\mbox{Con}(\mbox{R}_{i})$,
defined as
$\mbox{Con}(\mbox{R}_{i})
=\{
\mbox{R}_{i},\{\mbox{R}_{i}^{\,{\tiny C}}\}
\}$,
where the subset 
$\{\mbox{R}_{i}^{\,{\tiny C}}\}$
is defined 
with the massless quark flavors
as
$\{\mbox{R}_{i}^{\,{\tiny C}}\}
=$
$\{
\mbox{R}_{i}^{\,{\tiny C,u\bar{u}}},$
$\mbox{R}_{i}^{\,{\tiny C,d\bar{d}}},$
$\mbox{R}_{i}^{\,{\tiny C,s\bar{s}}},
...
\}$.
The element of the process 
$\mbox{R}_{i}^{\,{\tiny C,q\bar{q}}}$
is defined in such a way that
two gluons in the final state of
$\mbox{R}_{i}$ are replaced with
the $q\bar{q}$ pair as
\begin{equation}
\mbox{R}_{i}^{\,{\tiny C,q\bar{q}}}=
\mbox{R}_{i} - (gg)_{final} + 
(q\bar{q})_{final}.
\end{equation}
We call the process
$\mbox{R}_{i}^{\,{\tiny C,q\bar{q}}}$
the {\em $gg\,\mbox{--}\,q\bar{q}$ conjugation} of 
the process $\mbox{R}_{i}$.
The process 
$\mbox{R}_{i}^{\,{\tiny C,q\bar{q}}}$
exists for all the massless quark flavors,
for instance, $q=u,d,s,c,$ and $b$,
at the energy scale of the LHC.
Then we introduce the cross section 
$\sigma(\mbox{Con}(\mbox{R}_{i}))$
as
\begin{align}
\sigma(\mbox{Con}(\mbox{R}_{i}))
&=
-\sigma_{\mbox{{\tiny I}}}(\mbox{R}_{i},(2)\,\mbox{-}1/2,
N_{f}{\cal V}_{f\bar{f}})
-\sigma_{\mbox{{\tiny K}}}(\mbox{R}_{i},{\tt dip}1,
(3)/(4)\,\mbox{-}1,N_{f}h)
\nonumber\\
&\hspace{5mm}+
\sigma_{\mbox{{\tiny D}}}(\mbox{R}_{i}^{\,{\tiny C,q\bar{q}}},
\,{\tt dip}2)\cdot N_{f}\,.
\label{sigcon}
\end{align}
Here the cross sections
with different quark flavors
$\sigma_{\mbox{{\tiny D}}}(\mbox{R}_{i}^{\,{\tiny C,q\bar{q}}},
\,{\tt dip}2)$
have the same contributions 
and
the summation of the cross sections with all the 
massless quark
flavors is equal to the multiplication of
the number of massless quark flavors,
$N_{f}$,
by the cross section with the single flavor.
The cross section
$\sigma_{\mbox{{\tiny D}}}(\mbox{R}_{i}^{\,{\tiny C,q\bar{q}}},
\,{\tt dip}2)$
is separated into two terms,
$\hat{\sigma}_{\mbox{{\tiny D}}}
(\,
\mbox{R}_{i}^{\,{\tiny C,q\bar{q}}},
{\tt dip}2,(5)\mbox{-}1/2,
\,
{\cal V}_{f\bar{f}}
\,)$
and
$\hat{\sigma}_{\mbox{{\tiny D}}}
(\,
\mbox{R}_{i}^{\,{\tiny C,q\bar{q}}},
{\tt dip}2,(5)\mbox{-}2,
\,h)$,
as shown in Eq.\,(\ref{dip2recon}).
On the right-hand side of
Eq.\,(\ref{sigcon}),
we prove the two relations of the cancellation as
\begin{align}
-\sigma_{\mbox{{\tiny I}}}(\mbox{R}_{i},(2)\,\mbox{-}1/2,
N_{f}{\cal V}_{f\bar{f}})
+
\hat{\sigma}_{\mbox{{\tiny D}}}
(\,
\mbox{R}_{i}^{\,{\tiny C,q\bar{q}}},\,
{\tt dip}2,(5)\mbox{-}1/2,
\,
{\cal V}_{f\bar{f}}
\,)\cdot N_{f}
&=0\,,
\label{can1}
\\
-\sigma_{\mbox{{\tiny K}}}(\mbox{R}_{i},{\tt dip}1,
(3)/(4)\,\mbox{-}1,N_{f}h)
+
\hat{\sigma}_{\mbox{{\tiny D}}}
(\,
\mbox{R}_{i}^{\,{\tiny C,q\bar{q}}},\,
{\tt dip}2,(5)\mbox{-}2,
\,h)\cdot N_{f}
&=0\,,
\label{can2}
\end{align}
which leads to the cancellation of
$\sigma(\mbox{Con}(\mbox{R}_{i}))$ as
\begin{equation}
\sigma(\mbox{Con}(\mbox{R}_{i}))=0\,.
\label{s6sigconv}
\end{equation}
We introduce the set
$\{\mbox{Con}(\mbox{R}_{i})\}$,
which consists of all the possible sets
$\mbox{Con}(\mbox{R}_{i})$
for the processes $\{\mbox{R}_{i}\}$.
We further introduce the set
$\mbox{Self}$.
The set $\mbox{Self}$ contains 
the process
$\mbox{R}_{i}$,
the final state of which
includes one or no gluon,
and the final state includes
no $q\bar{q}$ pair.
As mentioned in the previous
sections, for any element
$\mbox{R}_{i}$ of the set $\mbox{Self}$,
the cross section 
$\sigma_{subt}(\mbox{R}_{i})$
itself vanishes as
\begin{equation}
\sigma_{subt}(\mbox{R}_{i})=0\,.
\end{equation}
Using the sets
$\{\mbox{Con}(\mbox{R}_{i})\}$
and
$\mbox{Self}$,
the left-hand side of Eq.\,(\ref{prast6})
is always reconstructed as
\begin{equation}
\sum_{i=1}^{n_{\mbox{{\tiny real}}}} 
\sigma_{subt}(\mbox{R}_{i})
=
\sum_{{\tiny \{\mbox{Con}(\mbox{R}_{i})\}}}
\sigma(\mbox{Con}(\mbox{R}_{i}))
+
\sum_{{\tiny \mbox{Self}\,\supset\,\mbox{R}_{j}}}
\sigma_{subt}(\mbox{R}_{j})\,,
\label{prast6rec}
\end{equation}
where the symbol
$\sum_{{\tiny \{\mbox{Con}(\mbox{R}_{i})\}}}$
represents the summation over all the
sets
$\mbox{Con}(\mbox{R}_{i})$
of the set
$\{\mbox{Con}(\mbox{R}_{i})\}$,
and
the symbol
$\sum_{{\tiny \mbox{Self}\,\supset\,\mbox{R}_{j}}}$
represents the summation over all the
processes $\mbox{R}_{j}$ of the set 
$\mbox{Self}$.
Due to the cancellations,
$\sigma(\mbox{Con}(\mbox{R}_{i}))=0$
and 
$\sigma_{subt}(\mbox{R}_{j})=0$,
the relation in Eq.\,(\ref{prast6})
is proved.

We briefly explain why the terms
$\sigma_{\mbox{{\tiny I}}}(\mbox{R}_{i},(2)\,\mbox{-}1/2,
N_{f}{\cal V}_{f\bar{f}})$
and
$\sigma_{\mbox{{\tiny K}}}(\mbox{R}_{i},{\tt dip}1,
(3)/(4)\,\mbox{-}1,N_{f}h)$,
and the term
$\sigma_{\mbox{{\tiny D}}}(\mbox{R}_{i}^{\,{\tiny C,q\bar{q}}},
\,{\tt dip}2)$
cancel each other,
as shown in Eq.\,(\ref{s6sigconv}).
The cancellations in Eq.\,(\ref{prast6})
are always realized between the cross sections
with the same initial states
and the same reduced Born processes.
Among them,
the cancellation inside
$\sigma_{subt}(\mbox{R}_{i})$ itself
is calculated in 
{\bfseries Steps\,1}--{\bfseries 5},
and the results are generally written
in Eq.\,(\ref{prast5}).
When the process $\mbox{R}_{i}$
includes two or more gluons
in the final state,
the reduced Born process
$\mbox{B}1(\mbox{R}_{i})$
includes one or more gluon
in the final state.
The 1-loop virtual correction of the process
$\mbox{B}1(\mbox{R}_{i})$
is denoted as
$(\mbox{M}_{\mbox{\tiny LO}}(\mbox{B}1(\mbox{R}_{i}))
\cdot
\mbox{M}_{\mbox{\tiny 1-loop}}
(\mbox{B}1(\mbox{R}_{i}))^{\star}
+ c.c.)$\,.
In the original calculation 
of the QCD NLO correction without
the dipole subtraction,
the soft and collinear divergences
of the 1-loop corrections
are canceled by the
real corrections of the processes
$\mbox{R}_{i}$
and
$\mbox{R}_{i}^{\,{\tiny C,\,q\bar{q}}}$.
The 1-loop corrections 
to the gluon legs in the final state
include the contribution of the quark loops
to the gluon propagators.
The collinear divergence
from the quark loop
is canceled by 
the collinear divergence
from the collinear limit 
of the $q\bar{q}$ splitting
in the real correction
$\mbox{R}_{i}^{\,{\tiny C,\,q\bar{q}}}$.
In the dipole subtraction
the collinear divergence
from the quark loop is
subtracted by the I term
$\hat{\sigma}_{\mbox{{\tiny I}}}
(\mbox{R}_{i},(2)\mbox{-}1/2,N_{f}{\cal V}_{f\bar{f}})$,
and the collinear divergence
from the collinear limit 
of the $q\bar{q}$ splitting
is subtracted by the dipole term
$\hat{\sigma}_{\mbox{{\tiny D}}}
(\,
\mbox{R}_{i}^{\,{\tiny C,q\bar{q}}},\,
{\tt dip}2,(5)\mbox{-}1/2,
\,
{\cal V}_{f\bar{f}}
\,)$.
Then the cancellations of the collinear
divergences from the quark loop 
and the $q\bar{q}$ splitting in the 
original calculation
are represented in the dipole subtraction
as in Eq.\,(\ref{can1}).
It is supposed that
the remaining finite term,
$\hat{\sigma}_{\mbox{{\tiny D}}}
(\mbox{R}_{i}^{\,{\tiny C,q\bar{q}}},\,
{\tt dip}2,(5)\mbox{-}2, h)$,
in
$\hat{\sigma}_{\mbox{{\tiny D}}}
(\mbox{R}_{i}^{\,{\tiny C,q\bar{q}}},\,
{\tt dip}2)$
is transferred into the K term
$\sigma_{\mbox{{\tiny K}}}(\mbox{R}_{i})$,
and identified as the term
$\sigma_{\mbox{{\tiny K}}}(\mbox{R}_{i},{\tt dip}1,
(3)/(4)\,\mbox{-}1,N_{f}h)$
in the construction of
the dipole subtraction
by the authors in Ref.\cite{Catani:1996vz}.
The relation is represented in
Eq.\,(\ref{can2}).
In this way, the cancellation
$\sigma(\mbox{Con}(\mbox{R}_{i}))=0$
is understood.
The above explanation
simultaneously also becomes
the explanation of why
$\sigma_{subt}(\mbox{R}_{j})$
for any $\mbox{R}_{j} \subset$ Self
completely cancels
inside itself as
$\sigma_{subt}(\mbox{R}_{j})=0$.

We see one example of the cancellation
$\sigma(\mbox{Con}(\mbox{R}_{i}))=0$.
We take the process
$\mbox{R}_{8u}=u\bar{u} \to ggg$
in Table\,\ref{ap_B_1_tab8}
in Appendix \ref{ap_B_8}
in the dijet process.
After the execution of 
{\bfseries Steps\,1}--{\bfseries 5},
we obtain the cross section
$\sigma_{subt}(\mbox{R}_{8u})$
in Eq.\,(\ref{r8st5}) as
\begin{equation}
\hat{\sigma}_{subt}(\mbox{R}_{8u})=
-\,
\hat{\sigma}_{\mbox{{\tiny I}}}
(\mbox{R}_{8u},(2)\mbox{-}1/2,N_{f}{\cal V}_{f\bar{f}})
-\,
\hat{\sigma}_{\mbox{{\tiny K}}}
(\mbox{R}_{8u},\,{\tt dip}1,
(3)\mbox{-}1,N_{f}h)\,,
\end{equation}
where the two terms are written 
in Eqs.\,(\ref{r8st2nf}) 
and (\ref{r8st4nf})
as
\begin{align}
\hat{\sigma}_{\mbox{{\tiny I}}}
(\mbox{R}_{8u},(2)\mbox{-}1/2,N_{f}{\cal V}_{f\bar{f}})
&=
-\frac{A_{d}}{S_{\mbox{{\tiny B}}_{1}}} 
\cdot
\frac{N_{f}}{\mbox{C}_{\mbox{{\tiny A}}}} 
{\cal V}_{f\bar{f}} (\ep) 
\Phi(\mbox{B}1)_{d} \,
\cdot
\nonumber\\
&\ \ \ 
\bigl(
\,
[1,2]+
[2,1]+
[1,a]+
[1,b]+
[2,a]+
[2,b]
\,
\bigr)\,,
\\[7pt]
\hat{\sigma}_{\mbox{{\tiny K}}}
(\mbox{R}_{8u},\,{\tt dip}1,
(3)\mbox{-}1,N_{f}h)
&=
-\frac{A_{4}}{S_{\mbox{{\tiny B}}_{1}}} 
\int_{0}^{1}dx \,
\frac{\mbox{T}_{\mbox{{\tiny R}}}N_{f}}
{\mbox{C}_{\mbox{{\tiny A}}}} 
\,\frac{2}{3}h(x)
\ \times
\nonumber\\
&
\Bigl[
\Phi_{a}(\mbox{B}1,x)_{4}\,
\bigl(
\langle a,1 \rangle
+
\langle a,2 \rangle
\bigr)
+
\Phi_{b}(\mbox{B}1,x)_{4}\,
\bigl(
\langle b,1 \rangle
+
\langle b,2 \rangle
\bigr)
\Bigr]\,.
\end{align}
For the process $\mbox{R}_{8u}$,
the set $\mbox{Con}(\mbox{R}_{8u})$
is determined as
$\mbox{Con}(\mbox{R}_{8u})
=\{
\mbox{R}_{8u},\{\mbox{R}_{8u}^{\,{\tiny C}}\}
\}$
with the subset
$\{\mbox{R}_{8u}^{\,{\tiny C}}\}
=
\{
\mbox{R}_{8u}^{\,{\tiny C,u\bar{u}}},
\mbox{R}_{8u}^{\,{\tiny C,d\bar{d}}},
...
\}$.
The $gg\,\mbox{--}\,u\bar{u}$ conjugation of 
the process $\mbox{R}_{8u}$
is written as
$\mbox{R}_{8u}^{\,{\tiny C,u\bar{u}}}
=u\bar{u} \to u\bar{u}g$,
which is nothing but the process $\mbox{R}_{1}$
used in the examples in the previous sections.
Then the cross section
$\sigma(\mbox{Con}(\mbox{R}_{8u}))$
is written down as
\begin{align}
\sigma(\mbox{Con}(\mbox{R}_{8u}))
&=
-\sigma_{\mbox{{\tiny I}}}(\mbox{R}_{8u},(2)\,\mbox{-}1/2,
N_{f}{\cal V}_{f\bar{f}})
-\sigma_{\mbox{{\tiny K}}}(\mbox{R}_{8u},{\tt dip}1,
(3)/(4)\,\mbox{-}1,N_{f}h)
\nonumber\\
&+
\hat{\sigma}_{\mbox{{\tiny D}}}
(\,
\mbox{R}_{1},\,
{\tt dip}2,(5)\mbox{-}1/2,
\,
{\cal V}_{f\bar{f}}
\,) \cdot N_{f}
+
\hat{\sigma}_{\mbox{{\tiny D}}}
(\,
\mbox{R}_{1},\,
{\tt dip}2,(5)\mbox{-}2,
\,h) \cdot N_{f}
\,,
\end{align}
where the last two terms are obtained 
in Eqs.\,(\ref{exst5d21}) 
and (\ref{exst5d22}).
The color-correlated Born squared
amplitudes in
Eqs.\,(\ref{exst5d21}) 
and (\ref{exst5d22})
are extracted as
$([1,2]+[1,a]+[1,b])$
and
$\langle 1,a/b \rangle$.
The reduced Born process of $\mbox{R}_{1}$ 
in the category Dipole\,2
is fixed as
$\mbox{B}2(\mbox{R}_{1})=u(y_{a})\bar{u}(y_{a})
\to g(y_{1})g(y_{2})$.
The reduced Born process
$\mbox{B}2(\mbox{R}_{1})$
has two gluons in the final state
and, in that sense, the legs of two gluons,
$y_{1}$ and $y_{2}$,
are symmetric.
When we require 
in Eq.\,(\ref{sigcon})
that 
the contributions of the I and K terms,
$\sigma_{\mbox{{\tiny I}}}(\mbox{R}_{i},(2)\,\mbox{-}1/2,
N_{f}{\cal V}_{f\bar{f}})$
and
$\sigma_{\mbox{{\tiny K}}}(\mbox{R}_{i},{\tt dip}1,
(3)/(4)\,\mbox{-}1,N_{f}h)$,
are canceled by
the integrated dipole term
$\hat{\sigma}_{\mbox{{\tiny D}}}
(\,\mbox{R}_{i}^{\,{\tiny C,u\bar{u}}},\,
{\tt dip}2)$
at the level of the square of the matrix
elements, $|\mbox{M}|^{2}$, 
on each phase-space point,
the summation of the color-correlated Born squared
amplitudes in $\hat{\sigma}_{\mbox{{\tiny D}}}
(\,\mbox{R}_{i}^{\,{\tiny C,u\bar{u}}},\,
{\tt dip}2)$
must be {\em symmetrized} 
about the color factor insertion operators
on the identical fields.
In the present case, the summation in the
term
$\hat{\sigma}_{\mbox{{\tiny D}}}
(\mbox{R}_{1},
{\tt dip}2)$
must be symmetrized about
the color factor operators,
$\mbox{T}_{y_{1}}$ and $\mbox{T}_{y_{2}}$,
on the two gluon legs,
$y_{1}$ and $y_{2}$, as
\begin{align}
[1,2]+[1,a]+[1,b]
&=\half
\bigl(
\,
[1,2]+
[2,1]+
[1,a]+
[2,a]+
[1,b]+
[2,b]
\,
\bigr)\,,
\label{maxsym1}
\\
\langle 1,a/b \rangle
&=\half
\bigl(
\langle 1,a/b \rangle
+
\langle 2,a/b \rangle
\bigr)\,.
\label{maxsym2}
\end{align}
Using the symmetrized expressions in 
Eqs.\,(\ref{maxsym1}) and (\ref{maxsym2}),
and the relation of the symmetric factors,
$S_{
\mbox{{\tiny B}}_{1}
(\mbox{{\tiny R}}_{8u})
}
=
2\,S_{
\mbox{{\tiny R}}_{1}
}
=2
$,
we prove the two relations of the cancellation
in
Eqs.\,(\ref{can1}) and (\ref{can2})
for $\mbox{R}_{8u}$.
Then we prove the cancellation as
$\sigma(\mbox{Con}(\mbox{R}_{8u}))=0$.
We will see the full treatment
of {\bfseries Step 6.}
$\sum_{i} \sigma_{subt}(\mbox{R}_{i})=0$,
in the dijet process in Sec.\,\ref{sec_4}.






\section{Drell-Yan\,: $pp \to \mu^{+}\mu^{-} + X$ \label{sec_3}}
\subsection{Results of the DSA \label{s_3_1}}
The subtraction terms
--the dipole, I, P and K terms--
for the Drell--Yan process have been created by the
DSA in Sec.\,3 in Ref.\cite{Hasegawa:2014oya}.
Here we summarize the results.
There are three real emission processes as follows\,:
\begin{align}
\mbox{R}_{1} &= u\bar{u} \to \mu^{-}\mu^{+}g\,, \nonumber\\
\mbox{R}_{2} &= ug \to \mu^{-}\mu^{+}u\,,  \nonumber\\
\mbox{R}_{3} &= \bar{u}g \to \mu^{-}\mu^{+}\bar{u}\,,
\end{align}
and the set $\{\mbox{R}_{i}\}$ is denoted as
$\{\mbox{R}_{i}\}=
\{\mbox{R}_{1},\mbox{R}_{2},\mbox{R}_{3}\}$.
The dipole, I, and P/K terms
that belong to the process 
$\mbox{R}_{1}$
are summarized in Tables 1, 2, and 3,
respectively, in Ref.\cite{Hasegawa:2014oya}.
The dipole and P/K terms that
belong to the process 
$\mbox{R}_{2}$
are summarized in Tables 1 and 3,
respectively, in ref.\cite{Hasegawa:2014oya}.
Those for the $\mbox{R}_{3}$
are summarized in Tables 1 and 3
in Ref.\cite{Hasegawa:2014oya}.
\subsection{Execution of the PRA \label{s_3_2}}
For the subtraction terms created
by the DSA we execute the PRA as follows.
We start with the execution for the process
$\mbox{R}_{1}$.
Since the final state of
$\mbox{R}_{1}$ does not include
two or more gluons,
nor any $q\bar{q}$ pair,
the relation to be proved is written as
\begin{equation}
\sigma_{subt}(\mbox{R}_{1})=0\,.
\label{s3r1}
\end{equation}
To prove the relation,
we start from {\bfseries Step\,1}.
The dipole terms created by the DSA 
in Table 1 in Ref.\cite{Hasegawa:2014oya}
are converted to the integrated dipole
terms shown in Table\,\ref{sec3_tab1}
in the present article.
\begin{table}[tb]
  \centering
\begin{align}
&\hat{\sigma}_{\mbox{{\tiny D}}}
(\mbox{R}_{1u} = u\bar{u} \to \mu^{-}\mu^{+}g): 
\ \ S_{\mbox{{\tiny R}}_{1}}=1,
\nonumber\\
&
\begin{array}{|c|c|c|c|c|c|} \hline
{\tt Dip}\,j
& \mbox{B}j 
& \mbox{\small Splitting} 
& (y_{a},y_{b};y_{1},y_{2}) 
& \mbox{Factor 1}
& \Phi(\mbox{B}_{j})
\,[y_{emi}, y_{spe}]
\\[4pt] \hline
& & & & &  \\[-12pt]
{\tt Dip\,1}
& u\bar{u} \to \mu^{-}\mu^{+}
& (3)-2   
& 1.\,(\widetilde{a3},\widetilde{b} \ ;1,2) 
& \int dx
{\widetilde {\cal V}}^{f,f} (x)
/\mbox{C}_{\mbox{{\tiny F}}}
& \Phi_{a}(\mbox{B}1)  \, [a,b]
\\[4pt]  
&  
&   
& 2.\,(\widetilde{a},\widetilde{b3} \ ;1,2)
&
& \Phi_{b}(\mbox{B}1)  \, [b,a]
\\[4pt] \hline
\end{array} \nonumber
\end{align}
\caption{ Integrated dipole terms:
$\sigma_{\mbox{{\tiny D}}}(\mbox{R}_{1})$\,.
\label{sec3_tab1}}
\end{table}
Then we proceed to {\bfseries Step\,2},
where we prove the relation
\begin{equation}
\sigma_{\mbox{{\tiny D}}}(\mbox{R}_{1},\,\mbox{I})
-
\sigma_{\mbox{{\tiny I}}}(\mbox{R}_{1})
=0\,.
\label{s3r1st2}
\end{equation}
Referring to Table\,\ref{sec3_tab1}
and Appendix \ref{ap_A_3},
the integrated dipole term
$\hat{\sigma}_{\mbox{{\tiny D}}}(\mbox{R}_{i},
\mbox{I})$
is written as
\begin{align}
\hat{\sigma}_{\mbox{{\tiny D}}}(
\mbox{R}_{1},\,\mbox{I})
&=
-\frac{A_{d}}{S_{\mbox{{\tiny R}}_{1}}} 
\cdot
\frac{1}{\mbox{C}_{\mbox{{\tiny F}}}}
{\cal V}_{f}(\ep) 
\cdot
\Phi(\mbox{B}1)_{d} \
\bigl(\,
[a,b]+[b,a]
\,\bigr)\,.
\label{s3di}
\end{align}
Referring to Table 1
in Ref.\cite{Hasegawa:2014oya}
and Appendix \ref{ap_A_8},
the contribution of the I terms
$\sigma_{\mbox{{\tiny I}}}(\mbox{R}_{1})$
is written as the same expression
in Eq.\,(\ref{s3di})
with the relation 
$S_{\mbox{{\tiny R}}_{1}}=S_{\mbox{{\tiny B1}}}=1$.
Then the relation in
Eq.\,(\ref{s3r1st2}) is proved.
In {\bfseries Step\,3}, we prove the relation
\begin{equation}
\sigma_{\mbox{{\tiny D}}}(\mbox{R}_{1},\,\mbox{P})
+
\sigma_{\mbox{{\tiny C}}}(\mbox{R}_{1})
-
\sigma_{\mbox{{\tiny P}}}(\mbox{R}_{1})=0\,.
\label{s3r1st3}
\end{equation}
Referring to Table\,\ref{sec3_tab1}
and Appendix \ref{ap_A_4},
the integrated dipole term
$\sigma_{\mbox{{\tiny D}}}(\mbox{R}_{1},\,\mbox{P})$
is written as
\begin{equation}
\hat{\sigma}_{\mbox{{\tiny D}}}(\mbox{R}_{1},\,\mbox{P})
=
\frac{A_{d}}{S_{\mbox{{\tiny R}}_{1}}} 
\int_{0}^{1}dx \,
\biggl(\frac{1}{\ep} - \ln x \biggr)
\,\frac{1}{\mbox{C}_{\mbox{{\tiny F}}}} 
P^{ff}(x)\,
\biggl[
\Phi_{a}(\mbox{B}1,x)_{d}\,
[a,b]
+
\Phi_{b}(\mbox{B}1,x)_{d}\,
[b,a]
\biggr]\,.
\end{equation}
With Appendix \ref{ap_A_7},
the collinear subtraction term
is created as
\begin{equation}
\hat{\sigma}_{\mbox{{\tiny C}}}(\mbox{R}_{1})
=
\frac{A_{d}}{S_{\mbox{{\tiny B1}}}}
\int_{0}^{1}dx\,
\Bigl(\,
\frac{1}{\ep}
-
\ln \mu_{F}^{2}
\Bigr)
P^{ff}(x) 
\biggl[
\Phi_{a}(\mbox{B}1,x)_{d}\,
\langle \mbox{B}1 \rangle
+
\Phi_{b}(\mbox{B}1,x)_{d}\,
\langle \mbox{B}1 \rangle
\biggr]\,.
\end{equation}
Then we calculate the summation as
\begin{equation}
\hat{\sigma}_{\mbox{{\tiny D}}}(\mbox{R}_{1},\,\mbox{P})
+
\hat{\sigma}_{\mbox{{\tiny C}}}(\mbox{R}_{1})
=
\frac{A_{d}}{S_{\mbox{{\tiny B1}} }}
\int_{0}^{1}dx \,
\frac{1}{\mbox{C}_{\mbox{{\tiny F}}}} 
P^{ff}(x)
\ln \frac{\mu_{F}^{2}}{x\,s_{ab}}
\biggl[
\Phi_{a}(\mbox{B}1,x)_{d}
\,\langle a,b \rangle
+
\Phi_{b}(\mbox{B}1,x)_{d}
\,\langle b,a \rangle
\biggr]\,,
\label{s3dydpsum}
\end{equation}
where we use the color conservation as 
$\langle \mbox{B}1 \rangle
=-\langle a,b \rangle/\mbox{C}_{\mbox{{\tiny F}}}
=-\langle b,a \rangle/\mbox{C}_{\mbox{{\tiny F}}}$.
The summation 
in Eq.\,(\ref{s3dydpsum})
is finite and reduced to 4 dimensions.
The summation in 4 dimensions is shown to
be equal to the P term
$\sigma_{\mbox{{\tiny P}}}(\mbox{R}_{1})$,
which is created in Table 3
in Ref.\cite{Hasegawa:2014oya}
and is written down in Appendix \ref{ap_A_9}.
In this way, the relation in
Eq.\,(\ref{s3r1st3}) is proved.
In {\bfseries Step\,4}, we prove the relation
\begin{equation}
\sigma_{\mbox{{\tiny D}}}(\mbox{R}_{1},\,\mbox{K})
-
\sigma_{\mbox{{\tiny K}}}(\mbox{R}_{1})
=0\,.
\label{s3r1st4}
\end{equation}
With Table\,\ref{sec3_tab1}
and Appendix \ref{ap_A_5},
the term
$\hat{\sigma}_{\mbox{{\tiny D}}}(\mbox{R}_{1},\,\mbox{K})$
is written as
\begin{align}
\hat{\sigma}_{\mbox{{\tiny D}}}(\,
\mbox{R}_{1},\,
\mbox{K}\,) 
&=
-\frac{A_{4}}{S_{\mbox{{\tiny R}}_{1}}} 
\int_{0}^{1}dx \,
\frac{1}{\mbox{C}_{\mbox{{\tiny F}}}}
\bigl(\,
{\cal V}_{other}^{f,f}(x\,;\ep)
+
\mbox{C}_{\mbox{{\tiny F}}} \, g(x)
+ {\widetilde K}^{ff}(x)\,
\bigr)\,\times
\nonumber \\
& 
\hspace{40mm}
\bigl[
\Phi_{a}(\mbox{B}1,x)_{4} 
\, \langle a,b \rangle
+
\Phi_{b}(\mbox{B}1,x)_{4} 
\, \langle b,a \rangle
\bigr]\,.
\end{align}
With Table 3
in Ref.\cite{Hasegawa:2014oya}
and Appendix \ref{ap_A_10},
the term
$\hat{\sigma}_{\mbox{{\tiny K}}}(\mbox{R}_{1})$
is written as
\begin{equation}
\hat{\sigma}_{\mbox{{\tiny K}}}(\mbox{R}_{1})
=
\frac{A_{4}}{S_{\mbox{{\tiny B}}_{1}}}
\int_{0}^{1}dx \
\biggl[
\Phi_{a}(\mbox{B}1,x)_{4}\,
\biggl(
\overline{\mbox{K}}^{ff}(x)
\,
\langle \mbox{B}1 \rangle
-
\frac{1}{\mbox{C}_{\mbox{{\tiny F}}}}
\widetilde{\mbox{K}}^{ff}(x)\,
\langle \, a,b \,\rangle
\biggr)
+
\,(a \leftrightarrow b)\,
\biggl]
\end{equation}
The present process,
$\mbox{R}_{1}=u\bar{u} \to \mu^{-}\mu^{+}g$,
is so simple that
we do not have to divide
{\bfseries Step\,4} into
substeps
in Eqs.\,(\ref{step41})--(\ref{step43}).
Using the relations
$\overline{\mbox{K}}^{ff/gg}(x)
={\cal V}_{other}^{f,f/g,g}(x\,;\ep) 
+ \mbox{C}_{\mbox{{\tiny F}}} \, g(x)$,
$\langle a,b \rangle
=-
\mbox{C}_{\mbox{{\tiny F}}}
\langle \mbox{B}1 \rangle$,
and
$S_{\mbox{{\tiny R}}_{1}}=S_{\mbox{{\tiny B}}_{1}}=1$,
the relation
in Eq.\,(\ref{s3r1st4})
is proved.
Then we obtain the relation in
Eq.\,(\ref{s3r1})
in {\bfseries Step\,5}.

Next we apply the PRA to the process 
$\mbox{R}_{2}=ug \to \mu^{-}\mu^{+}u$.
The relation to be proved is written as
\begin{equation}
\sigma_{subt}(\mbox{R}_{2})=0\,.
\label{s3r2}
\end{equation}
In {\bfseries Step\,1},
the dipole term
in Table 1 in Ref.\cite{Hasegawa:2014oya}
is converted to the integrated dipole
term shown in Table\,\ref{sec3_tab2}.
\begin{table}[tb]
  \centering
\begin{align}
&\hat{\sigma}_{\mbox{{\tiny D}}}
(\mbox{R}_{2} = ug \to \mu^{-}\mu^{+}u): 
\ \ S_{\mbox{{\tiny R}}_{2}}=1,
\nonumber\\
&
\begin{array}{|c|c|c|c|c|c|} \hline
{\tt Dip}\,j
& \mbox{B}j 
& \mbox{\small Splitting} 
& (y_{a},y_{b};y_{1},y_{2}) 
& \mbox{Factor 1}
& \Phi(\mbox{B}_{j})
\,[y_{emi}, y_{spe}]
\\[4pt] \hline
& & & & &  \\[-12pt]
{\tt Dip\,4}u
& u\bar{u} \to \mu^{-}\mu^{+}
& (7)u-2   
& 1.\,(\widetilde{a},\widetilde{b3} \ ;1,2)
& \int dx
{\widetilde {\cal V}}^{g,f} (x)
/\mbox{C}_{\mbox{{\tiny F}}}
& \Phi_{b}(\mbox{B}4u)  \, [b,a]
\\[4pt] \hline
\end{array} \nonumber
\end{align}
\caption{ Integrated dipole terms:
$\sigma_{\mbox{{\tiny D}}}(\mbox{R}_{2})$\,.
\label{sec3_tab2}}
\end{table}
For the process $\mbox{R}_{2}$,
{\bfseries Step\,2}
does not exist because
the final state does not include 
any gluon, and the reduced
Born process $\mbox{B}1(\mbox{R}_{2})$
does not exist.
Then we proceed to
{\bfseries Step\,3},
where we prove the relation
\begin{equation}
\sigma_{\mbox{{\tiny D}}}(\mbox{R}_{2},\,\mbox{P})
+
\sigma_{\mbox{{\tiny C}}}(\mbox{R}_{2})
-
\sigma_{\mbox{{\tiny P}}}(\mbox{R}_{2})=0\,.
\label{s3r2st3}
\end{equation}
Referring to Table\,\ref{sec3_tab1}
and Appendix \ref{ap_A_4},
the term
$\hat{\sigma}_{\mbox{{\tiny D}}}(\mbox{R}_{2},\,\mbox{P})$
is written as
\begin{equation}
\hat{\sigma}_{\mbox{{\tiny D}}}(\mbox{R}_{2},\,\mbox{P})
=
\frac{A_{d}}{S_{\mbox{{\tiny R}}_{2}}} 
\int_{0}^{1}dx \,
\biggl(\frac{1}{\ep} - \ln x \biggr)
\,\frac{1}{\mbox{C}_{\mbox{{\tiny F}}}} 
P^{gf}(x)\,
\Phi_{b}(\mbox{B}4u,x)_{d}\,
[b,a]\,.
\end{equation}
With Appendix \ref{ap_A_7},
the collinear subtraction term
$\sigma_{\mbox{{\tiny C}}}(\mbox{R}_{2})$
is created,
and, with Table 3
in Ref.\cite{Hasegawa:2014oya}
and
Appendix \ref{ap_A_9},
the P term 
$\sigma_{\mbox{{\tiny P}}}(\mbox{R}_{2})$
is written down.
It is shown that the three terms
$\hat{\sigma}_{\mbox{{\tiny D}}}(\mbox{R}_{2},
\mbox{P})$,
$\hat{\sigma}_{\mbox{{\tiny C}}}(\mbox{R}_{2})$,
and
$\hat{\sigma}_{\mbox{{\tiny P}}}(\mbox{R}_{2})$
satisfy
the relation in
Eq.\,(\ref{s3r2st3}).
In {\bfseries Step\,4}
we prove the relation
\begin{equation}
\sigma_{\mbox{{\tiny D}}}(\mbox{R}_{2},\,\mbox{K})
-
\sigma_{\mbox{{\tiny K}}}(\mbox{R}_{2})
=0\,.
\label{s3r2st4}
\end{equation}
With Table\,\ref{sec3_tab1}
and Appendix \ref{ap_A_5},
the term
$\hat{\sigma}_{\mbox{{\tiny D}}}(\mbox{R}_{2},\,\mbox{K})$
is written as
\begin{align}
\hat{\sigma}_{\mbox{{\tiny D}}}(\,
\mbox{R}_{2},\,
\mbox{K}\,) 
&=
-\frac{A_{4}}{S_{\mbox{{\tiny R}}_{1}}} 
\int_{0}^{1}dx \,
\frac{1}{\mbox{C}_{\mbox{{\tiny F}}}}
\bigl(\,
{\cal V}_{other}^{g,f}(x\,;\ep)
+ {\widetilde K}^{gf}(x)\,
\bigr)\,
\Phi_{b}(\mbox{B}4u,x)_{4} 
\, \langle b,a \rangle\,.
\label{s3dyr2dk}
\end{align}
With Table 3
in Ref.\cite{Hasegawa:2014oya}
and Appendix \ref{ap_A_10},
the term
$\hat{\sigma}_{\mbox{{\tiny K}}}(\mbox{R}_{2})$
is written down,
and is shown to be equal to
$\hat{\sigma}_{\mbox{{\tiny D}}}(\mbox{R}_{2},
\mbox{K})$
in Eq.\,(\ref{s3dyr2dk}).
In {\bfseries Step\,5},
we obtain the relation in Eq.\,(\ref{s3r2}).
In a similar way, we can prove the
relation for the process $\mbox{R}_{3}$ as
\begin{equation}
\sigma_{subt}(\mbox{R}_{3})=0\,.
\label{s3r3}
\end{equation}

Finally, we come to the last step,
{\bfseries Step\,6}.
All the real processes in the set
$\{\mbox{R}_{i}\}$ belong to
the set Self as
$\mbox{Self}=\{\mbox{R}_{1},\mbox{R}_{2},
\mbox{R}_{3}\}$,
and any set $\mbox{Con}(\mbox{R}_{i})$
does not exist.
Then we calculate the summation of the
cross sections
$\sigma_{subt}(\mbox{R}_{i})$ as
\begin{equation}
\sum_{i=1}^{3} 
\sigma_{subt}(\mbox{R}_{i})
=
\sum_{{\tiny \mbox{Self}\,\supset\,\mbox{R}_{j}}}
\sigma_{subt}(\mbox{R}_{j})
=0\,,
\end{equation}
where we use the cancellations
in Eqs.\,(\ref{s3r1}), (\ref{s3r2}),
and (\ref{s3r3})
for $\mbox{R}_{1}$, $\mbox{R}_{2}$,
and
$\mbox{R}_{3}$, respectively.
The contributions of the different
flavors, $d,s,c$, and $b$,
are identical to the process
with the up quark, $u$,
and 
the executions of 
{\bfseries Step\,1}--{\bfseries 6}
are also identical.
Thus the execution of the PRA
for the Drell--Yan process
is completed.







\section{Dijet\,: $pp \to 2\,jets + X$ \label{sec_4}}
\subsection{Results of the DSA \label{s_4_1}}
The DSA for the dijet process has been executed
in Sec.\,4 in
Ref.\cite{Hasegawa:2014oya}.
We here summarize the results.
The real processes are denoted as
\begin{align}
\mbox{R}_{1u} 
&= u\bar{u} \to u\bar{u}g\,,
\ \ \ (\mbox{R}_{1d})
\nonumber\\
\mbox{R}_{2u} 
&= uu \to uug\,,
\ \ \ (\mbox{R}_{2\bar{u}},\mbox{R}_{2d},
\mbox{R}_{2\bar{d}})  
\nonumber\\
\mbox{R}_{3u} 
&= ug \to uu\bar{u}\,,
\ \ \ (\mbox{R}_{3\bar{u}},\mbox{R}_{3d},
\mbox{R}_{3\bar{d}})  
\nonumber\\
\mbox{R}_{4u} 
&= u\bar{u} \to d\bar{d}g\,,
\ \ \ (\mbox{R}_{4d}) 
\nonumber\\
\mbox{R}_{5ud} 
&= ud \to udg\,,
\ \ \ (\mbox{R}_{5\bar{u}\bar{d}}) 
\nonumber\\
\mbox{R}_{6u\bar{d}} 
&= u\bar{d} \to u\bar{d}g\,,
\ \ \ (\mbox{R}_{6\bar{u}d}) 
\nonumber\\
\mbox{R}_{7u} 
&= ug \to ud\bar{d}\,,
\ \ \  (\mbox{R}_{7\bar{u}},\mbox{R}_{7d},
\mbox{R}_{7\bar{d}}) 
\nonumber\\
\mbox{R}_{8u} 
&= u\bar{u} \to ggg\,,
\ \ \ (\mbox{R}_{8d}) 
\nonumber\\
\mbox{R}_{9u} 
&= ug \to ugg\,,
\ \ \  (\mbox{R}_{9\bar{u}},\mbox{R}_{9d},
\mbox{R}_{9\bar{d}}) 
\nonumber\\
\mbox{R}_{10u} 
&= gg \to u\bar{u}g\,,
\ \ \ (\mbox{R}_{10d}) 
\nonumber\\
\mbox{R}_{11} 
&= gg \to ggg\,,
\label{s4listri}
\end{align}
where the processes that have identical
expressions for the cross sections
with different quark flavors
are written in round brackets,
e.g., $(\mbox{R}_{1d})$ for the process
$\mbox{R}_{1u}$.
The dipole terms are summarized in
Tables 4--14 in Appendix B.1
in Ref.\cite{Hasegawa:2014oya},
the I terms in Tables 15--19 in Appendix B.2
in Ref.\cite{Hasegawa:2014oya},
and the P and K terms in
Tables 20--30 in Appendix B.3
in Ref.\cite{Hasegawa:2014oya}.
The details of the creation are explained 
in Ref.\cite{Hasegawa:2014oya}.
%
%
%
\subsection{Execution of the PRA \label{s_4_2}}
We start from {\bfseries Step\,1}.
The dipole terms
$\sigma_{\mbox{{\tiny D}}}(\mbox{R}_{i})$
for the processes,
$\mbox{R}_{1},..., \mbox{R}_{11}$,
at Tables 4--14 in Appendix B.1
in Ref.\cite{Hasegawa:2014oya},
are converted into integrated ones
in Tables 
\ref{ap_B_1_tab1}--\ref{ap_B_1_tab11}
in Appendices 
\ref{ap_B_1}--\ref{ap_B_11}
in the present article,
respectively.
The results of
{\bfseries Steps\,2}--{\bfseries 5}
for $\mbox{R}_{1},..., \mbox{R}_{11}$
are shown after
Tables 
\ref{ap_B_1_tab1}--\ref{ap_B_1_tab11}
in Appendices 
\ref{ap_B_1}--\ref{ap_B_11},
respectively.
Here we write down the results of
{\bfseries Step\,5} as follow\,:
\begin{align}
&\hat{\sigma}_{subt}(\mbox{R}_{1u})
=
\hat{\sigma}_{\mbox{{\tiny D}}}(\mbox{R}_{1u},\,{\tt dip}2)\,,
\nonumber\\
&\hat{\sigma}_{subt}(\mbox{R}_{2u})
=0\,,
\nonumber\\
&\hat{\sigma}_{subt}(\mbox{R}_{3u})
=
\hat{\sigma}_{\mbox{{\tiny D}}}(\mbox{R}_{3u},\,{\tt dip}2)\,,
\nonumber\\
&\hat{\sigma}_{subt}(\mbox{R}_{4u})
=
\hat{\sigma}_{\mbox{{\tiny D}}}(\mbox{R}_{4u},\,{\tt dip}2)\,,
\nonumber\\
&\hat{\sigma}_{subt}(\mbox{R}_{5ud})
=0\,,
\nonumber\\
&\hat{\sigma}_{subt}(\mbox{R}_{6u\bar{d}})
=0\,,
\nonumber\\
&\hat{\sigma}_{subt}(\mbox{R}_{7u})
=
\hat{\sigma}_{\mbox{{\tiny D}}}(\mbox{R}_{7u},\,{\tt dip}2)\,,
\nonumber\\
&\hat{\sigma}_{subt}(\mbox{R}_{8u})
=
-\,
\hat{\sigma}_{\mbox{{\tiny I}}}
(\mbox{R}_{8u},(2)\mbox{-}1/2,N_{f}{\cal V}_{f\bar{f}})
-\,
\hat{\sigma}_{\mbox{{\tiny K}}}
(\mbox{R}_{8u},\,{\tt dip}1,
(3)\mbox{-}1,N_{f}h)\,,
\nonumber\\
&\hat{\sigma}_{subt}(\mbox{R}_{9u})
=
-\,
\hat{\sigma}_{\mbox{{\tiny I}}}
(\mbox{R}_{9u},(2)\mbox{-}1/2,N_{f}{\cal V}_{f\bar{f}})
-\,
\hat{\sigma}_{\mbox{{\tiny K}}}
(\mbox{R}_{9u},\,{\tt dip}1,
(3)/(4)\mbox{-}1,N_{f}h)\,,
\nonumber\\
&\hat{\sigma}_{subt}(\mbox{R}_{10u})
=
\hat{\sigma}_{\mbox{{\tiny D}}}(\mbox{R}_{10u},\,{\tt dip}2)\,,
\nonumber\\
&\hat{\sigma}_{subt}(\mbox{R}_{11})
=
-\,
\hat{\sigma}_{\mbox{{\tiny I}}}
(\mbox{R}_{11},(2)\mbox{-}1/2,N_{f}{\cal V}_{f\bar{f}})
-\,
\hat{\sigma}_{\mbox{{\tiny K}}}
(\mbox{R}_{11},\,{\tt dip}1,
(4)\mbox{-}1,N_{f}h)\,,
\end{align}
where the expressions for the cross sections
on the right-hand sides are all written
in Appendix \ref{ap_B}.
Then we proceed to {\bfseries Step\,6}.
We construct the sets
$\mbox{Con}(\mbox{R}_{i})$
and $\mbox{Self}$ as
\begin{align}
\mbox{Con}(\mbox{R}_{8u})
&=\{
\mbox{R}_{8u},\{\mbox{R}_{1u},\mbox{R}_{4u}\}
\}\,,
\\
\mbox{Con}(\mbox{R}_{9u})
&=\{
\mbox{R}_{9u},\{\mbox{R}_{3u},\mbox{R}_{7u}\}
\}\,,
\\
\mbox{Con}(\mbox{R}_{11})
&=\{
\mbox{R}_{11},\{\mbox{R}_{10u},\mbox{R}_{10d}\}
\}\,,
\\
\mbox{Self}
&=\{
\mbox{R}_{2u},\mbox{R}_{5ud},\mbox{R}_{6ud}
\}\,.
\end{align}
We also introduce also the set as
$\{\mbox{Con}(\mbox{R}_{i})\}
=\{
\mbox{Con}(\mbox{R}_{8u}),
\mbox{Con}(\mbox{R}_{9u}),
\mbox{Con}(\mbox{R}_{11})
\}$.
Referring to the explicit expressions
for $\hat{\sigma}_{subt}(\mbox{R}_{i})$
in Appendix\,\ref{ap_B},
we can prove the cancellations
of $\sigma(\mbox{Con}(\mbox{R}_{i}))$
in Eq.\,(\ref{sigcon})
for $\mbox{R}_{i}=\mbox{R}_{8u}$,
$\mbox{R}_{9u}$,
and $\mbox{R}_{11}$ as
\begin{equation}
\sigma(\mbox{Con}(\mbox{R}_{i}))=0\,.
\end{equation}
Among them, the cancellation
$\sigma(\mbox{Con}(\mbox{R}_{8u}))=0$
is concretely shown in
Sec.\,\ref{s_2_7}.
Then we obtain the cancellation
of the summation
$\sum \sigma_{subt}(\mbox{R}_{i})$
as
\begin{equation}
\sum_{i=1}^{11} 
\sigma_{subt}(\mbox{R}_{i})
=
\sum_{{\tiny \{\mbox{Con}(\mbox{R}_{i})\}}}
\sigma(\mbox{Con}(\mbox{R}_{i}))
+
\sum_{{\tiny \mbox{Self}\,\supset\,\mbox{R}_{j}}}
\sigma_{subt}(\mbox{R}_{j})=0\,.
\end{equation}
Thus the execution of the PRA for the dijet process is
completed.







\section{$n$ jets\,: $pp \to n\,jets + X$ \label{sec_5}}
In this section, we deal with the three real processes
among all those that contribute to the collider process
$pp \to n\,jets + X$ as
\begin{align}
\mbox{R}_{1} &= u\bar{u} \to (n+1)\mbox{-}g\,, \nonumber\\
\mbox{R}_{2} &= u\bar{u} \to u\bar{u}+(n-1)\mbox{-}g\,,  \nonumber\\
\mbox{R}_{3} &= u\bar{u} \to d\bar{d}+(n-1)\mbox{-}g\,,
\end{align}
and  prove the cancellation of
the cross section as 
$\sigma(\mbox{Con}(\mbox{R}_{1}))=0$
in {\bfseries Step\,6} of the PRA.
We first apply the DSA to the three processes.
Since the expressions of the created
subtraction terms are too long,
we do not show the expressions explicitly. 
Then we apply the PRA to the subtraction terms
created by the DSA.
The results of the PRA are collected
in Appendix\,\ref{ap_C}.
The results of {\bfseries Step\,1}
for $\mbox{R}_{1}$, $\mbox{R}_{2}$, and 
$\mbox{R}_{3}$, are shown in
Tables \ref{ap_C_tab1}, \ref{ap_C_tab2}, 
and \ref{ap_C_tab3}, respectively.
The results of
{\bfseries Steps\,2}--{\bfseries 5}
for $\mbox{R}_{1}$, $\mbox{R}_{2}$, and 
$\mbox{R}_{3}$ are shown after
Tables \ref{ap_C_tab1}, \ref{ap_C_tab2}, 
and \ref{ap_C_tab3}, respectively.
We would like to clarify two points
to notice in obtaining the results of the PRA.
Both of the points concern
the expressions of the integrated dipole term
$\hat{\sigma}_{\mbox{{\tiny D}}}
(\mbox{R}_{i})$.
We explain them in the following two 
paragraphs, respectively.

The first point is about the 
expressions of the integrated dipole terms
converted in {\bfseries Step\,1}.
When the reduced Born process includes
identical fields,
the field mapping, which is determined
for each dipole term, has the freedom 
to choose the emitter-spectator pair
$(y_{emi},y_{spe})$
among the identical fields
in set $\{y\}$.
Using the freedom of the field mapping,
we can transform the integrated dipole terms
into an identical expression.
We call this operation the {\em unification}
of the integrated dipole terms.
We show one example in the process
$\mbox{R}_{1}=u(x_{a})\bar{u}(x_{b}) 
\to g(x_{1})g(x_{2}) ... g(x_{n+1})$.
The creation of the dipole terms
starts from splitting (2)-1,
where the reduced Born process\
is fixed with set $\{y\}$ as
\begin{equation}
\mbox{B}1(\mbox{R}_{1})=
u(y_{a})\bar{u}(y_{b}) 
\to g(y_{1})g(y_{2}) ... g(y_{n})\,.
\label{s5b1}
\end{equation}
Then we can choose the three legs
$(x_{i}\,x_{j},x_{k})$ in the final state
in set $\{x\}$ as
\begin{align}
&(x_{1}\,x_{2},\,x_{3}), \ \ \ \ (x_{1}\,x_{2},\,x_{4}),\,
..., \ \ \ (x_{1}\,x_{2},\,x_{n+1}), 
\nonumber\\
&(x_{1}\,x_{3},\,x_{2}), \ \ \ \ (x_{1}\,x_{3},\,x_{4}),\,
..., \ \ \ (x_{1}\,x_{3},\,x_{n+1}), 
\nonumber\\
& \hspace{7mm} \vdots 
\hspace{25mm} \vdots
\hspace{33mm} \vdots
\nonumber\\
&(x_{n-1}\,x_{n},\,x_{1}),\,(x_{n-1}\,x_{n},\,x_{2}),\,
..., (x_{n-1}\,x_{n},\,x_{n-2})\,.
\label{s5xijk}
\end{align}
The total number of pairs is
${}_{n+1}C_{2}\cdot (n-1)$.
One field mapping is fixed for each pair
$(x_{i}\,x_{j},\,x_{k})$,
and specifies
the two legs of the emitter and the spectator
in set $\{y\}$
as $(y_{emi},\,y_{spe})$.
All the specified pairs
$(y_{emi},\,y_{spe})$
for the pairs
$(x_{i}\,x_{j},\,x_{k})$
in Eq.\,(\ref{s5xijk})
are written in the expression
$(y_{\alpha},\,y_{\beta})$,
where the indices 
$\alpha$ and $\beta$
take any value among
$1,...,n$,
with the condition
$\alpha \not= \beta$.
Using the freedom of the field mapping
we can reconstruct all the field mappings
in such a way that
they have an identical pair
$(y_{emi},\,y_{spe})$;
for instance,
$(y_{1},\,y_{2})$.
We call this operation {\em unification}.
After the operation of unification,
the summation of the integrated dipole terms
in splitting (2)-1 is rewritten as
\begin{equation}
\hat{\sigma}_{\mbox{{\tiny D}}}(\mbox{R}_{1},
(2)\mbox{-}1) 
=
-\frac{A_{d}}{S_{\mbox{{\tiny R}}_{1}}} 
\cdot
\frac{1}{\mbox{C}_{\mbox{{\tiny A}}}}
{\cal V}_{gg}(\ep) 
\cdot
\Phi(\mbox{B}1)_{d} \,
[ 1, 2 ]\, \times n_{deg}\,,
\label{s5d21}
\end{equation}
where the degeneracy factor,
$n_{deg}$,
is determined as
$n_{deg}={}_{n+1}C_{2}\cdot (n-1)$.
We can apply the operation of unification
to all the other integrated dipole terms as well.
The unification of all the dipole terms
may be described as follows\,:
``\,The results of {\bfseries Step\,1}
are transformed into the unification expression.''
Actually, in Tables \ref{ap_C_tab1}, 
\ref{ap_C_tab2}, and \ref{ap_C_tab3}, 
the results are represented
in the unification 
expression.
Compared to the format of
Tables \ref{ap_B_1_tab1}--\ref{ap_B_1_tab11}
in the Appendix \ref{ap_B},
the entry $(y_{a},y_{b}:y_{1},...,y_{n})$
is replaced with the entry
$(y_{emi},\,y_{spe})$,
and the new entry for the
degeneracy factor 
$n_{deg}$ is added.
The dipole term
$\hat{\sigma}_{\mbox{{\tiny D}}}(\mbox{R}_{1},
(2)\mbox{-}1)$
in Eq.\,(\ref{s5d21})
is represented by the first one,
$1.\,(y_{1},y_{2})$,
with the degeneracy factor
$n_{deg}$
in Table \ref{ap_C_tab1}.
Referring to Table \ref{ap_C_tab1},
for instance, we can read out
the integrated dipole term
$\hat{\sigma}_{\mbox{{\tiny D}}}
(\,
\mbox{R}_{1},\,
\mbox{I}
\,)$ used in {\bfseries Step\,2}
as
\begin{align}
\hat{\sigma}_{\mbox{{\tiny D}}}
(\,
\mbox{R}_{1},\,
\mbox{I}
\,)
&=
-\frac{A_{d}}{S_{\mbox{{\tiny R}}_{1}}}\,
\Phi(\mbox{B}1)_{d}\,
\biggl[
\frac{{\cal V}_{gg}(\ep)}{\mbox{C}_{\mbox{{\tiny A}}}}
\Bigl[
[1,2] \cdot {}_{n+1}C_{2}\,(n-1)
+
\bigl(\,[1,a]+[1,b]\,\bigr)\cdot {}_{n+1}C_{2}
\Bigr]
\nonumber\\
&\hspace{8mm}
+
\frac{{\cal V}_{fg} (\ep) }{\mbox{C}_{\mbox{{\tiny F}}}}
\Bigl[
\bigl(\,[a,1]+[b,1]\,\bigr)\cdot (n+1)n
+
\bigl(\,[a,b]+[b,a]\,\bigr)\cdot (n+1)
\Bigr]
\biggr]\,.
\label{s5r1di}
\end{align}
The advantage of the table format
in the unification expression
is that the length of the table
is shorter than the original format.
The disadvantage is that
the one-to-one correspondences
between the original dipole terms
and the integrated dipole terms
are lost.
The format of
Tables \ref{ap_C_tab1}, 
\ref{ap_C_tab2}, and \ref{ap_C_tab3}, 
can be a template to show the results 
in the unification expression.

The second point is about the symmetric
expression of the integrated dipole terms
used for {\bfseries Steps\,2}, {\bfseries 3},
and {\bfseries 4}.
In order that the integrated dipole terms
$\hat{\sigma}_{\mbox{{\tiny D}}}
(\mbox{R}_{i},
\mbox{I}/\mbox{P}/\mbox{K})$
cancel the I, P, and K terms
$\hat{\sigma}_{\mbox{{\tiny I/P/K}}}
(\mbox{R}_{i})$
at the level of the squared amplitude
$[y_{emi},\,y_{spe}]$
on each phase space point,
the summation of the color-correlated
Born squared amplitudes
$[y_{emi},\,y_{spe}]$
in the integrated dipole terms
$\hat{\sigma}_{\mbox{{\tiny D}}}
(\mbox{R}_{i},
\mbox{I}/\mbox{P}/\mbox{K})$
must be symmetrized
over the legs of the identical fields.
We call this operation
{\em symmetrization}
of the integrated dipole terms.
To demonstrate the symmetrization,
we take the same process $\mbox{R}_{1}$.
The reduced Born process 
$\mbox{B}1(\mbox{R}_{1})$
in Eq.\,(\ref{s5b1})
has $n$ identical fields at
in the legs $y_{1}, ...,y_{n}$.
Then the color-correlated
Born squared amplitude
$[1,2]$, for instance,
in Eq.\,(\ref{s5r1di})
can be symmetrized
over the legs
$y_{1}, ...,y_{n}$ as
\begin{equation}
[1,2]=\frac{1}{n(n-1)}\, \sum_{i,k=1}^{n}\,[i,k]\,,
\end{equation}
with the condition $i \not= k$.
We call this operation {\em symmetrization}.
The symmetrization is allowed
by the freedom of the field mapping
over the identical fields.
The freedom is the same one by which
the operation of unification
is allowed, as explained above.
When the symmetrization is applied to 
all the terms in the integrated dipole term
$\hat{\sigma}_{\mbox{{\tiny D}}}
(\mbox{R}_{1},\,
\mbox{I})$
in Eq.\,(\ref{s5r1di}),
the expression is transformed as
\begin{align}
\hat{\sigma}_{\mbox{{\tiny D}}}
(\,
\mbox{R}_{1},\,
\mbox{I}
\,)
&=
-\frac{A_{d}}{S_{\mbox{{\tiny B}}_{1}}}\,
\Phi(\mbox{B}1)_{d}\,
\Biggl[
\frac{1}{\mbox{C}_{\mbox{{\tiny A}}}}
\cdot
\half {\cal V}_{gg}(\ep)
\,
\biggl[ \
\sum_{i,k=1}^{n}\,[i,k]
+
\sum_{i=1}^{n}\,
\bigl(\,[i,a]+[i,b]\,\bigr)
\,\biggr]
\nonumber\\
&\hspace{28mm}
+
\frac{1}{\mbox{C}_{\mbox{{\tiny F}}}}{\cal V}_{fg} (\ep)
\,
\biggl[\
\sum_{k=1}^{n}\,
\bigl(\,[a,k]+[b,k]\,\bigr)
+
[a,b]+[b,a]
\,\biggr]
\Biggr]\,,
\label{s5r1disym}
\end{align}
where we use the relation of the symmetric factors
$S_{\mbox{{\tiny R}}_{1}}=
(n+1)\cdot S_{\mbox{{\tiny B}}_{1}}
=(n+1)\,!$\,.
The I term
$\hat{\sigma}_{\mbox{{\tiny I}}}
(\mbox{R}_{1})$
is created by the DSA
in such an expression that
the factor ${\cal V}_{gg}(\ep)/2$
in Eq.\,(\ref{s5r1disym})
is replaced with the factor
${\cal V}_{gg}(\ep)/2 + N_{f}
{\cal V}_{f\bar{f}}(\ep)$.
Then the integrated dipole term
$\hat{\sigma}_{\mbox{{\tiny D}}}
(\mbox{R}_{1},\,
\mbox{I})$
in Eq.\,(\ref{s5r1disym})
is able to cancel the
I term
$\hat{\sigma}_{\mbox{{\tiny I}}}
(\mbox{R}_{1})$
at the integrand level,
which means that,
on each phase-space point
of $\Phi(\mbox{B}1)_{d}$,
the color-correlated Born squared
amplitudes
$[y_{emi},\,y_{spe}]$
in the two terms
$\hat{\sigma}_{\mbox{{\tiny D}}}
(
\mbox{R}_{1},\,
\mbox{I})$
and
$\hat{\sigma}_{\mbox{{\tiny I}}}
(\mbox{R}_{1})$
cancel each other.
In this way, we prove the relation
in {\bfseries Step\,2} as
\begin{equation}
\hat{\sigma}_{\mbox{{\tiny D}}}
(\,
\mbox{R}_{1},\,
\mbox{I}
\,)
-
\hat{\sigma}_{\mbox{{\tiny I}}}
(\mbox{R}_{1})=-\,
\hat{\sigma}_{\mbox{{\tiny I}}}
(\mbox{R}_{1},(2)\mbox{-}1/2,N_{f}{\cal V}_{f\bar{f}})\,,
\end{equation}
where the term 
$\hat{\sigma}_{\mbox{{\tiny I}}}
(\mbox{R}_{1},(2)\mbox{-}1/2,N_{f}{\cal V}_{f\bar{f}})$
is written in Eq.\,(\ref{c_r1st2nf}).
In a similar way in {\bfseries Steps\,3}
and {\bfseries 4},
the integrated dipole terms
$\hat{\sigma}_{\mbox{{\tiny D}}}
(\mbox{R}_{1},\,
\mbox{P/K})$,
are symmetrized and
can be canceled against
the P/K terms
$\hat{\sigma}_{\mbox{{\tiny P/K}}}
(\mbox{R}_{1})$,
as shown in 
Eqs.\,(\ref{appc_r1st3})/(\ref{appc_r1st4}),
respectively.

Finally, we show the relation of
cancellation in {\bfseries Step\,6} as
\begin{equation}
\sigma(\mbox{Con}(\mbox{R}_{1}))=0\,.
\label{s5st6r1}
\end{equation}
The set $\mbox{Con}(\mbox{R}_{1})$
is denoted as 
$\mbox{Con}(\mbox{R}_{1})=\{\mbox{R}_{1},
\{\mbox{R}_{2}, \mbox{R}_{3}\}\}$\,.
After the execution of 
{\bfseries Steps\,1}--{\bfseries 5}
for $\mbox{R}_{1}$,\,$\mbox{R}_{2}$, and 
$\mbox{R}_{3}$,
the cross section
$\sigma(\mbox{Con}(\mbox{R}_{1}))$
is written down as
\begin{align}
\sigma(\mbox{Con}(\mbox{R}_{1}))
=
-\hat{\sigma}_{\mbox{{\tiny I}}}
(\mbox{R}_{1},(2)\mbox{-}1/2,N_{f}{\cal V}_{f\bar{f}})
-\hat{\sigma}_{\mbox{{\tiny K}}}
(\mbox{R}_{1},\,{\tt dip}1,
(3)\mbox{-}1,N_{f}h)
+
\hat{\sigma}_{\mbox{{\tiny D}}}(\mbox{R}_{2},\,{\tt dip}2)
\cdot N_{f}\,,
\label{s5st6conr1}
\end{align}
where the three terms on the right-hand side
are defined in Eqs.\,(\ref{c_r1st2nf}),
(\ref{appc_r1st4kh}), and (\ref{r2st5dip2}),
respectively. 
The integrated dipole term
$\hat{\sigma}_{\mbox{{\tiny D}}}(\mbox{R}_{2},\,{\tt dip}2)$
is separated into two terms,
$\hat{\sigma}_{\mbox{{\tiny D}}}
(\mbox{R}_{2},\,
{\tt dip}2,$
$(5)\mbox{-}1/2,
\,
{\cal V}_{f\bar{f}})$
and
$\hat{\sigma}_{\mbox{{\tiny D}}}
(\mbox{R}_{2},\,
{\tt dip}2,(5)\mbox{-}2,
\,h)$,
which are defined in
Eqs.\,(\ref{r2st5dip2a}) and 
(\ref{r2st5last}),
respectively.
After the symmetrization 
of the integrated dipole term
$\hat{\sigma}_{\mbox{{\tiny D}}}(\mbox{R}_{2},
{\tt dip}2)$,
the three terms on the right-hand side
of Eq.\,(\ref{s5st6conr1})
cancel each other in the way
shown in Eqs.\,(\ref{can1})
and (\ref{can2}).
Then the relation in 
Eq.\,(\ref{s5st6r1})
is proved.
In the present section,
we only prove the cancellation
$\sigma(\mbox{Con}(\mbox{R}_{1}))=0$,
and do not prove the whole 
relation of the cancellation 
in {\bfseries Step\,6} as
$\sum_{i=1}^{n_{\mbox{{\tiny real}}}} 
\sigma_{subt}(\mbox{R}_{i})
=0$.
The execution of the DSA and the PRA
for all the processes 
$\{\mbox{R}_{i}\}$ contributing
to the process $pp \to n\,jets$
can be regarded as the execution for
an almost general process
with an arbitrary large number $n$.
The dipole terms created by the DSA
have very long and general
expressions.
The proofs of the relations
in the PRA
also become long and general.
For these reasons, we do not 
present the execution
of the DSA and PRA for all the 
processes $\{\mbox{R}_{i}\}$
in the present article.







\section{Summary \label{sec_6}}
In the dipole subtraction procedure,
we create the subtraction terms
and write down the expressions
for the phase-space integration.
While creating the subtraction terms
and writing down the expressions,
we sometimes have the chance to 
make mistakes.
The main reason for this is that
many subtraction terms exist
for the multiparton processes
and each term is not so simple.
Among the subtraction terms
the singular parts of the dipole
and I terms are confirmed by
cancellation against the
real and virtual corrections
during the calculation
of the NLO corrections.
The P and K terms are finite and
confirmation by the cancellation is impossible.
The summation of all the 
subtraction terms created must vanish as
\begin{equation}
\sigma_{\mbox{{\tiny subt}}} 
= \sigma_{\mbox{{\tiny D}}} 
+ \sigma_{\mbox{{\tiny C}}}
- \sigma_{\mbox{{\tiny I}}} 
- \sigma_{\mbox{{\tiny P}}} - 
\sigma_{\mbox{{\tiny K}}}=0\,.
\label{subt}
\end{equation}
We call this relation the consistency
relation of the subtraction terms.
The proof of the consistency relation
provides one confirmation of
the P and K terms as well as
all the other subtraction terms.
The cancellations in the consistency
relation are realized between 
subtraction terms with
the same initial states
and the same reduced Born processes.
In Ref.\cite{Hasegawa:2014oya},
we presented the dipole splitting 
algorithm\,(DSA), which is an algorithm
to create the subtraction terms.
In the DSA, the subtraction terms are 
classified by the real processes 
$\{\mbox{R}_{i}\}$
and the kinds of splittings.
The classification can be translated
as the classification by the initial states
and the reduced Born processes.
Thanks to such a classification in the DSA,
we can construct a straightforward
algorithm to prove the consistency
relation.
In this article, we have
presented the proof algorithm\,(PRA)
with the necessary formulae
and demonstrated the PRA in the example processes.
The PRA is defined in Sec.\,\ref{sec_2}
and all the formulae are collected
in Appendix\,\ref{ap_A}.
The PRA is demonstrated in the Drell--Yan,
dijet, and $n$ jets processes in
Sec.\,\ref{sec_3}, \ref{sec_4},
and \ref{sec_5}, respectively.
The results of the PRA for
the dijet and the $n$ jets
are summarized in
Appendixes \ref{ap_B} and \ref{ap_C},
respectively.
In {\bfseries Step\,1}
of the PRA, the dipole terms are converted
to the integrated dipole terms.
We showed two templates for the tables
representing the integrated dipole
terms in Appendixes \ref{ap_B} and \ref{ap_C}.
In the tables in Appendix \ref{ap_B},
one integrated dipole term
corresponds to one original dipole term.
In the tables in Appendix \ref{ap_C},
the integrated dipole terms 
are transformed to as many identical
expressions as possible.
The transformed expression
is called the unification expression.

The executions of the PRA in
the Drell--Yan and the dijet processes
can be easily completed by hand manipulation.
For more complicated multiparton
processes, execution by hand manipulation
will be too long and
may not be realistic.
Thus, automation of the PRA
as a computer code
is desirable and may be realized in the future.
The extension of the DSA and PRA to include
massive quark cases is also left for
future work.
The extension should be straightforward
because the algorithm structure of the
dipole subtraction procedure with
massive quarks in Ref.\cite{Catani:2002hc}
is almost identical to the massless case in
Ref.\cite{Catani:1996vz}.
Beginners who start to use dipole subtraction
may have some difficulties
creating and confirming
the subtraction terms.
Thus, the DSA in Ref.\cite{Hasegawa:2014oya}
and the PRA in the present
article may serve as
supplementary materials
by which beginners can 
learn how to create and confirm
the subtraction terms.
The reason for this is that
the DSA and PRA are well defined
with all the formulae
in the documents and
what the user needs to do is just to follow
the steps in the 
straightforward algorithm.
We hope that the DSA and PRA
will help users to obtain
reliable predictions at QCD NLO accuracy.
\vspace{10mm}

\subsection*{Acknowledgments}
We are grateful to V. Ravindran
for helpful discussion.
The present article was completed during
a stay at the Harish-Chandra Research Institute.
We are grateful for the warm support
of the institute.

%

\clearpage
\appendix




\section{Formulae for PRA \label{ap_A}}
\subsection{Cross sections:
$\hat{\sigma}(\mbox{R}_{i})$ 
\label{ap_A_1}}
The partonic cross sections in Eq.\,(\ref{master})
are defined as
\begin{align}
\hat{\sigma}_{\mbox{{\tiny R}}}(\mbox{R}_{i}) 
&= \frac{1}{S_{\mbox{{\tiny R}}_{i}}} \ \Phi(\mbox{R}_{i})_{d} 
\cdot |\mbox{M}(\mbox{R}_{i})|_{d}^{2}\,,
 \label{hatr} \\
\hat{\sigma}_{\mbox{{\tiny D}}}(\mbox{R}_{i}) 
&= \frac{1}{S_{\mbox{{\tiny R}}_{i}}} \ \Phi(\mbox{R}_{i})_{d} 
\cdot \frac{1}{n_{s}(a) n_{s}(b)} \, \mbox{D}(\mbox{R}_{i})\,,
\label{hatd} \\
\hat{\sigma}_{\mbox{{\tiny V}}}(\mbox{B}1) 
&= \frac{1}{S_{\mbox{{\tiny B1}} }} \ \Phi( \mbox{B}1  )_{d} 
\cdot |\mbox{M}_{virt}(\mbox{B}1)|_{d}^{2}\,,  
\label{hatv} \\
\hat{\sigma}_{\mbox{{\tiny I}}}(\mbox{R}_{i}) 
&= \frac{1}{S_{\mbox{{\tiny B1}} }} \ \Phi(\mbox{B}1)_{d} 
\cdot \mbox{I}(\mbox{R}_{i})\,, 
\label{hati} \\
\hat{\sigma}_{\mbox{{\tiny P}}}(\mbox{R}_{i}) 
&= \int_{0}^{1}dx \sum_{\mbox{{\tiny B}}_{j}} 
\frac{1}{S_{\mbox{{\tiny B}}_{j} }}
\Phi_{a}(\mbox{R}_{i}:\mbox{B}_{j},x)_{4} \cdot 
\mbox{P}(\mbox{R}_{i},x_{a}:\mbox{B}_{j},xp_{a}) \ + \ 
(a \leftrightarrow b)\,, 
 \label{hatp} \\
\hat{\sigma}_{\mbox{{\tiny K}}}(\mbox{R}_{i}) 
&= \int_{0}^{1}dx \sum_{\mbox{{\tiny B}}_{j}} 
\frac{1}{S_{\mbox{{\tiny B}}_{j} }}
\Phi_{a}(\mbox{R}_{i}:\mbox{B}_{j},x)_{4} \cdot 
\mbox{K}(\mbox{R}_{i},x_{a}:\mbox{B}_{j},xp_{a})  \ + \ 
(a \leftrightarrow b)\,. 
\label{hatk}
\end{align}
The phase spaces 
including the flux factors
are defined as
\begin{align}
\Phi(\mbox{R}_{i})_{d} 
&= \frac{1}{{\cal F}(p_{a},p_{b})} 
\prod_{i=1}^{n+1} \int \frac{d^{d-1}p_{i}}{(2\pi)^{d-1}} 
\frac{1}{2E_{i}} \cdot (2\pi)^{d} \delta^{(d)} 
\Bigl(p_{a}+p_{b}- \sum_{i=1}^{n+1} p_{i} \Bigr),
\\
\Phi(\mbox{B}1)_{d} 
&= \frac{1}{{\cal F}(p_{a},p_{b})} 
\prod_{i=1}^{n} \int \frac{d^{d-1}p_{i}}{(2\pi)^{d-1}} 
\frac{1}{2E_{i}} \cdot (2\pi)^{d} \delta^{(d)} 
\Bigl(p_{a}+p_{b}- \sum_{i=1}^{n} p_{i} \Bigr),
\\
\Phi_{a}(\mbox{R}_{i}:\mbox{B}_{j},x)_{4} 
&= 
\frac{1}{{\cal F}(xp_{a},p_{b})} \prod_{i=1}^{n}
\int \frac{d^{3}p_{i}}{(2\pi)^{3}} \frac{1}{2E_{i}}
\cdot (2\pi)^{4} \delta^{(4)} 
\Bigl( xp_{a}+p_{b}- \sum_{i=1}^{n} p_{i} \Bigr).
\end{align}
The exact definitions of the factors and the symbols
are given in the DSA in 
Ref.\cite{Hasegawa:2014oya}.
The jet functions
$F_J^{(n/n+1)}(p_{1}, ...,p_{n/n+1})$
must be multiplied by the partonic 
cross sections
in Eqs.\,(\ref{hatr})--(\ref{hatk}).
For the real correction in Eq.\,(\ref{hatr}),
the jet function with $(n+1)$ fields 
is multiplied as
\begin{equation}
\hat{\sigma}_{\mbox{{\tiny R}}}(\mbox{R}_{i}) 
= \frac{1}{S_{\mbox{{\tiny R}}_{i}}} \ \Phi(\mbox{R}_{i})_{d} 
\cdot |\mbox{M}(\mbox{R}_{i})|_{d}^{2}
\cdot F_J^{(n+1)}(p_{1}, ...,p_{n+1})
\,.
\end{equation}
For the cross sections 
in Eqs.\,(\ref{hatv})--(\ref{hatk}),
the jet function with $n$ fields, 
$F_J^{(n)}(p_{1}, ...,p_{n})$,
is multiplied.
For the dipole term in Eq.\,(\ref{hatd}),
the jet function $F_J^{(n)}$
is multiplied and $n$ momenta
of the arguments are identified 
with the $n$ reduced momenta
$(\mbox{P}(y_{1}), ...,\mbox{P}(y_{n}))$.
The use of the jet functions
in the dipole subtraction 
is explained in Ref.\cite{Catani:1996vz}.
For compact notation, we do not show
the jet functions explicitly 
in the present article.

We summarize the PRA as follows.
We clarify the subsections
where the various cross sections
appearing in the PRA are defined
in Appendix \ref{ap_A}.
The cross section $\sigma_{subt}(\mbox{R}_{i})$
is defined as
\begin{equation}
\sigma_{subt}(\mbox{R}_{i})
=
\sigma_{\mbox{{\tiny D}}}(\mbox{R}_{i})
+
\sigma_{\mbox{{\tiny C}}}(\mbox{R}_{i})
-
\sigma_{\mbox{{\tiny I}}}(\mbox{R}_{i})
-
\sigma_{\mbox{{\tiny P}}}(\mbox{R}_{i})
-
\sigma_{\mbox{{\tiny K}}}(\mbox{R}_{i})
\,.
\end{equation}
The term 
$\sigma_{\mbox{{\tiny D}}}(\mbox{R}_{i})$
is the integrated dipole term
that is converted in {\bfseries Step 1},
and the formulae are collected
in Appendix \ref{ap_A_2}.
The integrated dipole term
$\sigma_{\mbox{{\tiny D}}}(\mbox{R}_{i})$
is separated into four terms as
\begin{equation}
\sigma_{\mbox{{\tiny D}}}(\mbox{R}_{i})=
\sigma_{\mbox{{\tiny D}}}(\mbox{R}_{i},\,\mbox{I})
+
\sigma_{\mbox{{\tiny D}}}(\mbox{R}_{i},\,\mbox{P})
+
\sigma_{\mbox{{\tiny D}}}(\mbox{R}_{i},\,\mbox{K})
+
\sigma_{\mbox{{\tiny D}}}(\mbox{R}_{i},\,{\tt dip}2)\,.
\end{equation}
The formulae for the four terms,
$\sigma_{\mbox{{\tiny D}}}(\mbox{R}_{i},
\mbox{I/P/K})$
and
$\sigma_{\mbox{{\tiny D}}}(\mbox{R}_{i},
{\tt dip}2)$,
are collected in 
Appendixes \ref{ap_A_3}, \ref{ap_A_4},
\ref{ap_A_5}, and \ref{ap_A_6},
respectively.
Then $\sigma_{subt}(\mbox{R}_{i})$
is reconstructed as
\begin{align}
\sigma_{subt}(\mbox{R}_{i})
&=
[\,\sigma_{\mbox{{\tiny D}}}(\mbox{R}_{i},\,\mbox{I})
-
\sigma_{\mbox{{\tiny I}}}(\mbox{R}_{i})
\,]
\,+\,
[\,
\sigma_{\mbox{{\tiny D}}}(\mbox{R}_{i},\,\mbox{P})
+
\sigma_{\mbox{{\tiny C}}}(\mbox{R}_{i})
-
\sigma_{\mbox{{\tiny P}}}(\mbox{R}_{i})
\,]
\nonumber \\
&\hspace{5mm}+
[\,
\sigma_{\mbox{{\tiny D}}}(\mbox{R}_{i},\,\mbox{K})
-
\sigma_{\mbox{{\tiny K}}}(\mbox{R}_{i})
\,]
\,+\,
\sigma_{\mbox{{\tiny D}}}(\mbox{R}_{i},\,{\tt dip}2)\,.
\label{apmaster}
\end{align}
The formulae for the terms
$\sigma_{\mbox{{\tiny C}}}(\mbox{R}_{i})$
and
$\sigma_{\mbox{{\tiny I/P/K}}}(\mbox{R}_{i})$
are collected in
Appendix \ref{ap_A_7}, \ref{ap_A_8},
\ref{ap_A_9}, and \ref{ap_A_10},
respectively.
In {\bfseries Steps 2}, {\bfseries 3}, 
and {\bfseries 4},
the following relations are proved\,:
\begin{align}
\mbox{{\bf Step 2.}}  & \ \ 
\sigma_{\mbox{{\tiny D}}}(\mbox{R}_{i},\,\mbox{I})
-
\sigma_{\mbox{{\tiny I}}}(\mbox{R}_{i})
=
-\sigma_{\mbox{{\tiny I}}}(\mbox{R}_{i},(2)\,
\mbox{-}1/2,N_{f}{\cal V}_{f\bar{f}})
\,, \nonumber\\
\mbox{{\bf 3.}} & \ \
\sigma_{\mbox{{\tiny D}}}(\mbox{R}_{i},\,\mbox{P})
+
\sigma_{\mbox{{\tiny C}}}(\mbox{R}_{i})
-
\sigma_{\mbox{{\tiny P}}}(\mbox{R}_{i})=0
\,, \nonumber\\ 
\mbox{{\bf 4.}} & \ \
\sigma_{\mbox{{\tiny D}}}(\mbox{R}_{i},\,\mbox{K})
-
\sigma_{\mbox{{\tiny K}}}(\mbox{R}_{i})
=
-\sigma_{\mbox{{\tiny K}}}(\mbox{R}_{i}, {\tt dip}1,
(3)/(4)\,\mbox{-}1,N_{f}h)\,,
\label{apst234}
\end{align}
where the terms
$\sigma_{\mbox{{\tiny I}}}(\mbox{R}_{i},(2)\,
\mbox{-}1/2,N_{f}{\cal V}_{f\bar{f}})$
and
$\sigma_{\mbox{{\tiny K}}}(\mbox{R}_{i}, {\tt dip}1,
(3)/(4)\,\mbox{-}1,N_{f}h)$
are defined in
Appendixes \ref{ap_A_8} and \ref{ap_A_10},
respectively.
We substitute the first three terms in
square brackets
in Eq.\,(\ref{apmaster})
by the three relations 
in Eq.\,(\ref{apst234}),
and obtain $\sigma_{subt}(\mbox{R}_{i})$
in the expression
\begin{align}
\mbox{{\bf Step 5.}} \ \
\sigma_{subt}(\mbox{R}_{i})
&=
-\sigma_{\mbox{{\tiny I}}}(\mbox{R}_{i},(2)\,\mbox{-}1/2,
N_{f}{\cal V}_{f\bar{f}})
-\sigma_{\mbox{{\tiny K}}}(\mbox{R}_{i},{\tt dip}1,
(3)/(4)\,\mbox{-}1,N_{f}h)
\nonumber\\
& \ \ \ \ +
\sigma_{\mbox{{\tiny D}}}(\mbox{R}_{i},\,{\tt dip}2)
\,.
\end{align}
In the last step,
we prove that
the summation of all the terms
$\sigma_{subt}(\mbox{R}_{i})$
vanishes as
\begin{equation}
\mbox{{\bf Step 6.}} \ \
\sum_{i=1}^{n_{\mbox{{\tiny real}}}} 
\sigma_{subt}(\mbox{R}_{i})
=0\,.
\end{equation}
%
%
%
\subsection{Integrated dipole term:
$\sigma_{\mbox{{\tiny D}}}$ 
\label{ap_A_2}}
\vspace{3mm}
\begin{figure}[h]
\begin{center}
\includegraphics[width=8cm]{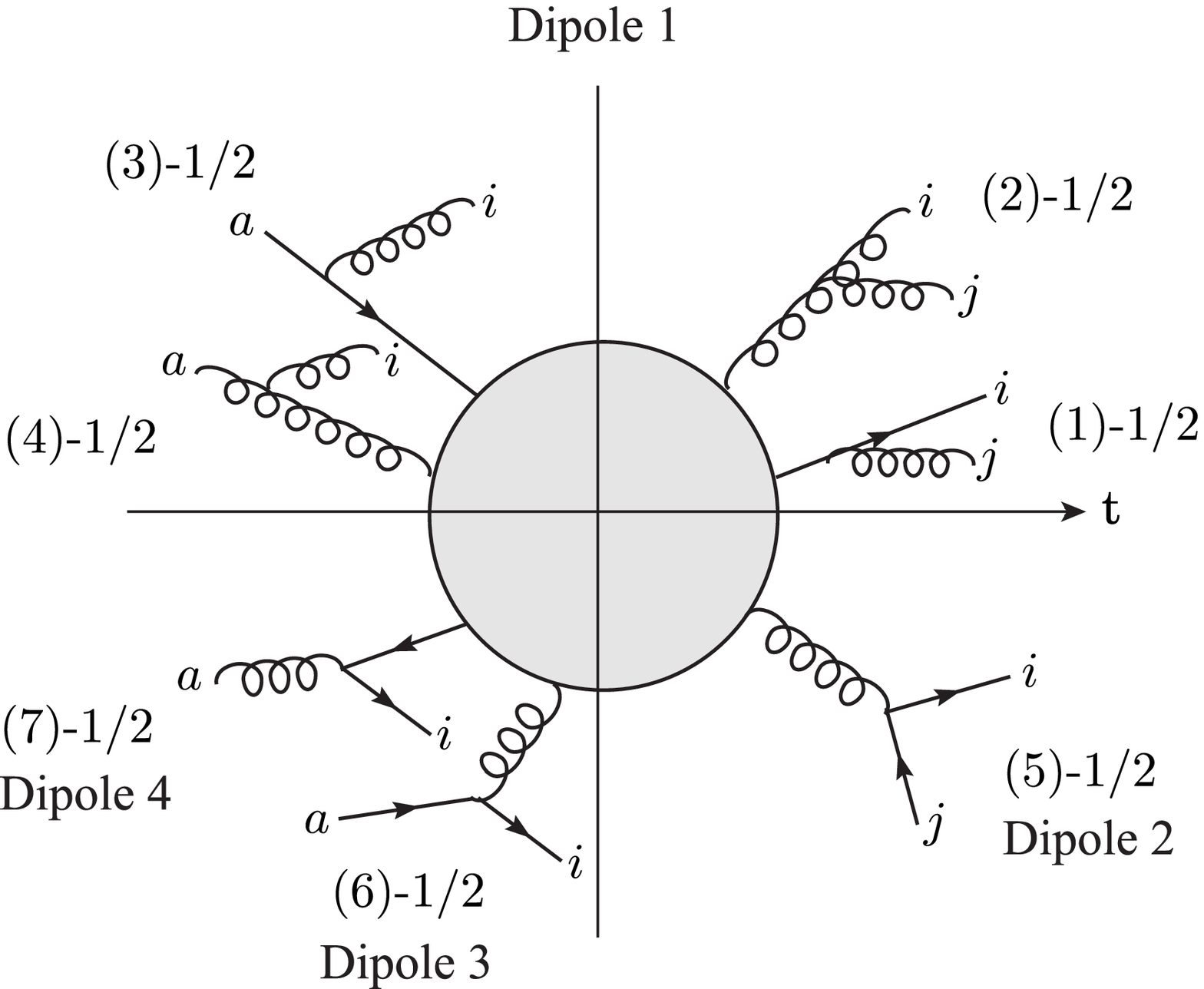}
\end{center} 
\caption{ The classification of the dipole term
$\sigma_{\mbox{{\tiny D}}}$
\label{fig_A2_int_D}}
\end{figure}
The integrated dipole term is universally written as
\begin{equation}
\hat{\sigma}_{\mbox{{\tiny D}}}(\mbox{R}_{i}) 
=
-\frac{A_{d}}{S_{\mbox{{\tiny R}}_{i}}} 
\cdot
\int_{0}^{1}dx \,
\frac{1}{\mbox{T}_{\mbox{{\tiny F}}(y_{emi})}^{2}}\, 
{\cal V}(x;\ep) 
\cdot
\Phi_{a}(\mbox{B}_{j},x)_{d} \,
[ \, y_{emi}, \,y_{spe} ]\,,
\end{equation}
where the overall factor $A_{d}$ is defined as
\begin{equation}
A_{d}=\frac{\al}{2\pi} \frac{(4\pi \mu^{2})^{\ep}}{\Gamma(1-\ep)}\,.
\label{apaad}
\end{equation}
The integrated dipole term is classified
into four types as
\begin{equation}
\hat{\sigma}_{\mbox{{\tiny D}}}(\mbox{R}_{i},\,{\tt dip}j\,)
\
\supset 
\
\hat{\sigma}_{\mbox{{\tiny D}}}(\mbox{R}_{i},\,{\tt dip}j,\,
\mbox{FF} / \mbox{FI} / \mbox{IF} / \mbox{II} \ )\,,
\end{equation}
which are defined as follows.
\vspace{12mm}

%
%
\leftline{\fbox{$\mbox{D}_{ij,k}$: Final-Final dipole}} 
\begin{verbatim}
Dipole 1 (1)-1, (2)-1,  
Dipole 2 (5)-1:
\end{verbatim}
\begin{equation}
\hat{\sigma}_{\mbox{{\tiny D}}}(\mbox{R}_{i},\,{\tt dip}j,\, 
\mbox{FF}\,) 
=
-\frac{A_{d}}{S_{\mbox{{\tiny R}}_{i}}}
\cdot
\frac{1}{\mbox{T}_{\mbox{{\tiny F}}(y_{emi})}^{2}}
{\cal V}_{\mbox{{\tiny F}}(x_{i}) \mbox{{\tiny F}}(x_{j}) } (\ep) 
\cdot 
\Phi(\mbox{B}j)_{d} \,
[ \, y_{emi}, \,y_{spe} ]\,.
\label{ffmast}
\end{equation}
\vspace{5mm}

\leftline{\underline{Definition of the symbols}}
\vspace{2mm}
\noindent
Universal singular functions\,:\,
\begin{equation}
\frac{1}{\mbox{T}_{\mbox{{\tiny F}}(y_{emi})}^{2}}
{\cal V}_{\mbox{{\tiny F}}(x_{i}) \mbox{{\tiny F}}(x_{j}) }(\ep)=
\left\{
\begin{array}{ll}
\frac{1}{\mbox{C}_{\mbox{{\tiny F}}}} 
{\cal V}_{fg}(\ep)
& : \mbox{Dipole} \ 1 \ (1)\mbox{-}1\,,
\\
\frac{1}{\mbox{C}_{\mbox{{\tiny A}}}} 
{\cal V}_{gg}(\ep) 
& : \mbox{Dipole} \ 1 \ (2)\mbox{-}1\,,
\\
\frac{1}{\mbox{C}_{\mbox{{\tiny A}}}} 
{\cal V}_{f\bar{f}}(\ep) 
& : \mbox{Dipole} \ 2 \ (5)\mbox{-}1\,.
\end{array}
\right.
\label{singff}
\end{equation}
\begin{align}
{\cal V}_{fg} (\ep) 
&= \mbox{C}_{\mbox{{\tiny F}}} \biggl[\frac{1}{\ep^2} + \frac{3}{2\ep} +5 
-\frac{\pi^2}{2}   \biggr]\,, 
\label{isfg} \\
{\cal V}_{gg} (\ep) 
&= 2 \mbox{C}_{\mbox{{\tiny A}}} \biggl[\frac{1}{\ep^2} + \frac{11}{6\ep} +
\frac{50}{9} -\frac{\pi^2}{2} \biggr]\,, 
\label{isgg} \\
{\cal V}_{f\bar{f}} (\ep) 
&= \mbox{T}_{\mbox{{\tiny R}}} \biggl[- \frac{2}{3\ep} - \frac{16}{9} 
\biggr]\,.
\label{isff} 
\end{align}
\vspace{2mm}
\noindent
Phase space\,:\,
\begin{equation}
\Phi(\mbox{B}j)_{d} = \frac{1}{{\cal F}(p_{a},p_{b})} 
\prod_{i=1}^{n} \int \frac{d^{d-1}p_{i}}{(2\pi)^{d-1}} 
\frac{1}{2E_{i}} \cdot (2\pi)^{d} \delta^{(d)} 
\Bigl(p_{a}+p_{b}- \sum_{i=1}^{n} p_{i} \Bigr).
\end{equation}
Color-correlated Born squared amplitude\,:\,
\begin{equation}
[ \, y_{emi}, \,y_{spe} ]
=(s_{y_{emi}, \,y_{spe}})^{-\ep} 
\cdot
\langle \mbox{B}j \ | 
\mbox{T}_{y_{emi}} \cdot 
\mbox{T}_{y_{spe}}
 | \ \mbox{B}j \rangle_{d}\,. 
\label{sqbk1}
\end{equation}
Lorentz scalar\,:\,
\begin{equation}
s_{y_{emi}, y_{spe}}=2 \, \mbox{P}(y_{emi}) \cdot \mbox{P}(y_{spe})\,.
\label{lsff}
\end{equation}
\vspace{5mm}

%
%
\leftline{\fbox{$\mbox{D}_{ij,a}$: Final-Initial dipole}} 
\begin{verbatim}
Dipole 1 (1)-2, (2)-2,
Dipole 2 (5)-2:
\end{verbatim}
\begin{equation}
\hat{\sigma}_{\mbox{{\tiny D}}}(\mbox{R}_{i},\,{\tt dip}j,\, 
\mbox{FI},\,x_{a/b}) 
=
-\frac{A_{d}}{S_{\mbox{{\tiny R}}_{i}}} 
\int_{0}^{1}dx \,
\frac{1}{\mbox{T}_{\mbox{{\tiny F}}(y_{emi})}^{2}} 
{\cal V}_{\mbox{{\tiny F}}(x_{i}) \mbox{{\tiny F}}(x_{j}) } (x;\ep) 
\cdot
\Phi_{a/b}(\mbox{B}_{j},x)_{d} \,
[ \, y_{emi}, \,y_{spe} ]\,.
\label{fimast}
\end{equation}
\vspace{3mm}

\leftline{\underline{Definition of the symbols}}
\vspace{2mm}
\noindent
Universal singular functions\,:\,
\begin{equation}
\frac{1}{\mbox{T}_{\mbox{{\tiny F}}(y_{emi})}^{2}}
{\cal V}_{\mbox{{\tiny F}}(x_{i}) \mbox{{\tiny F}}(x_{j}) } (x;\ep) 
=
\left\{
\begin{array}{ll}
\frac{1}{\mbox{C}_{\mbox{{\tiny F}}}} 
{\cal V}_{fg}(x\,;\ep)
& : \mbox{Dipole} \ 1 \ (1)\mbox{-}2\,,
\\
\frac{1}{\mbox{C}_{\mbox{{\tiny A}}}} 
{\cal V}_{gg}(x\,;\ep) 
& : \mbox{Dipole} \ 1 \ (2)\mbox{-}2\,,
\\
\frac{1}{\mbox{C}_{\mbox{{\tiny A}}}} 
{\cal V}_{f\bar{f}}(x\,;\ep) 
& : \mbox{Dipole} \ 2 \ (5)\mbox{-}2\,.
\end{array}
\right.
\label{singfi}
\end{equation}
\begin{align}
{\cal V}_{fg}(x\,;\ep) 
&= \delta(1-x) \,
{\cal V}_{fg}(\ep) 
+\mbox{C}_{\mbox{{\tiny F}}} \left[ 
\, g(x)
- \frac{3}{2} h(x)
\, \right],
\label{nuxepfg}
\\
{\cal V}_{gg}(x\,;\ep) 
&= 
\delta(1-x) \,
{\cal V}_{gg}(\ep)
+\mbox{C}_{\mbox{{\tiny A}}} 
\left[ 
\, 2 g(x) - \frac{11}{3} h(x)
\, \right], 
\label{nuxepgg}
\\
{\cal V}_{f \bar{f}}(x\,;\ep) 
&= 
\delta(1-x) \,
{\cal V}_{f\bar{f}}(\ep)
+\mbox{T}_{\mbox{{\tiny R}}}
\,\frac{2}{3}h(x).
\label{nuxepffb}
\end{align}
\begin{align}
g(x) &= \left( \frac{2}{1-x} \ln\frac{1}{1-x}
\right)_{+} + \frac{2}{1-x} \ln (2-x)\,, \\
h(x) &= \biggl(\frac{1}{1-x} \biggr)_{+} 
+ \delta(1-x)\,.
\label{hxd}
\end{align}
Phase spaces\,:\,
\begin{align}
\Phi_{a}(\mbox{B}_{j},x)_{d} 
&= 
\frac{1}{{\cal F}(xp_{a},p_{b})} \prod_{i=1}^{n}
\int \frac{d^{d-1}p_{i}}{(2\pi)^{d-1}} \frac{1}{2E_{i}}
\cdot (2\pi)^{d} \delta^{(d)} 
\Bigl( xp_{a}+p_{b}- \sum_{i=1}^{n} p_{i} \Bigr)\,,
\label{pkps2a}
\\
\Phi_{b}(\mbox{B}_{j},x)_{d} 
&= 
\frac{1}{{\cal F}(p_{a},xp_{b})} \prod_{i=1}^{n}
\int \frac{d^{d-1}p_{i}}{(2\pi)^{d-1}} \frac{1}{2E_{i}}
\cdot (2\pi)^{d} \delta^{(d)} 
\Bigl(p_{a}+xp_{b}- \sum_{i=1}^{n} p_{i} \Bigr)\,.
\label{pkps2b}
\end{align}
Color-correlated Born squared amplitude\,:\,
\begin{equation}
[ \, y_{emi}, \,y_{spe} ]
=(s_{y_{emi}, \,x_{a/b}})^{-\ep} 
\cdot
\langle \mbox{B}j \ | 
\mbox{T}_{y_{emi}} \cdot 
\mbox{T}_{y_{spe}}
 | \ \mbox{B}j \rangle_{d}\,. 
\label{sqbk2}
\end{equation}
Lorentz scalar\,:\,
\begin{equation}
s_{y_{emi},\,x_{a/b}}=2 \, \mbox{P}(y_{emi}) \cdot p_{a/b}\,.
\label{lsfi}
\end{equation}
\vspace{5mm}

%
%
\leftline{\fbox{$\mbox{D}_{ai,k}$: Initial-Final dipole}} 
\begin{verbatim}
Dipole 1 (3)-1, (4)-1, 
Dipole 3 (6)-1, 
Dipole 4 (7)-1:
\end{verbatim}
\begin{equation}
\hat{\sigma}_{\mbox{{\tiny D}}}(\mbox{R}_{i},\,{\tt dip}j,\, 
\mbox{IF},\,x_{a/b}) 
=
-\frac{A_{d}}{S_{\mbox{{\tiny R}}_{i}}} 
\int_{0}^{1}dx \,
\frac{1}{\mbox{T}_{\mbox{{\tiny F}}(y_{emi})}^{2}} 
{\cal V}^{\mbox{{\tiny F}}(x_{a/b}), \, \mbox{{\tiny F}}(y_{emi}) } (x;\ep)
\cdot 
\Phi_{a/b}(\mbox{B}_{j},x)_{d} 
\, [ \, y_{emi}, \,y_{spe} ]\,.
\label{ifmast}
\end{equation}
\vspace{3mm}

\leftline{\underline{Definition of the symbols}}
\vspace{2mm}
\noindent
Universal singular functions\,:\,
\begin{equation}
\frac{1}{\mbox{T}_{\mbox{{\tiny F}}(y_{emi})}^{2}}
{\cal V}^{\mbox{{\tiny F}}(x_{a/b}), \,  \mbox{{\tiny F}}(y_{emi}) } (x;\ep)
=
\left\{
\begin{array}{ll}
\frac{1}{\mbox{C}_{\mbox{{\tiny F}}}} 
{\cal V}^{f,f} (x;\ep)
& : \mbox{Dipole} \ 1 \ (3)\mbox{-}1\,,
\\
\frac{1}{\mbox{C}_{\mbox{{\tiny A}}}} 
{\cal V}^{g,g} (x;\ep)
& : \mbox{Dipole} \ 1 \ (4)\mbox{-}1\,,
\\
\frac{1}{\mbox{C}_{\mbox{{\tiny A}}}} 
{\cal V}^{f,g} (x;\ep)
& : \mbox{Dipole} \ 3 \ (6)\mbox{-}1\,,
\\
\frac{1}{\mbox{C}_{\mbox{{\tiny F}}}} 
{\cal V}^{g,f} (x;\ep)
& : \mbox{Dipole} \ 4 \ (7)\mbox{-}1\,.
\end{array}
\right.
\label{singif}
\end{equation}
\begin{align}
{\cal V}^{f,f}(x\,;\ep) 
&= 
\delta(1-x) \,
{\cal V}_{fg}(\ep) 
+ \biggl(-\frac{1}{\ep} + \ln x \biggr)
\, P^{ff}(x) 
+ {\cal V}_{other}^{f,f}(x\,;\ep)\,,
\label{sinifff}
\\
{\cal V}^{g,g}(x\,;\ep) 
&= 
\delta(1-x) {\cal V}_{g}(\ep)
+ \biggl(-\frac{1}{\ep} + \ln x \biggr)
\, P^{gg}(x) 
+ {\cal V}_{other}^{g,g}(x\,;\ep)\,,
\label{sinifgg}
\\
&{\cal V}^{f,g}(x\,;\ep) 
= 
\biggl(-\frac{1}{\ep} + \ln x \biggr)
\, P^{fg}(x) 
+ {\cal V}_{other}^{f,g}(x\,;\ep)\,,
\label{siniffg}
\\
&{\cal V}^{g,f}(x\,;\ep) 
= 
\biggl(-\frac{1}{\ep} + \ln x \biggr)
\, P^{gf}(x) 
+ {\cal V}_{other}^{g,f}(x\,;\ep)\,.
\label{sinifgf}
\end{align}
\begin{equation}
{\cal V}_{g}(\ep) = \half {\cal V}_{gg} (\ep) + 
N_{f}{\cal V}_{f\bar{f}}(\ep)\,.
\label{singgg}
\end{equation}
\begin{align}
P^{ff}(x)
&=\mbox{C}_{\mbox{{\tiny F}}} 
\biggl(\frac{1+x^2}{1-x}\biggr)_{+}
=\mbox{C}_{\mbox{{\tiny F}}} 
\biggl[
\frac{3}{2}\delta(1-x)
+
\frac{1+x^2}{(1-x)_{+}}
\biggr]\,,
\label{alpff} 
\\
P^{gg}(x)
&=
\delta(1-x)
\Bigl(\frac{11}{6}\mbox{C}_{\mbox{{\tiny A}}}
 - \frac{2}{3}N_{f}\mbox{T}_{\mbox{{\tiny R}}}
\Bigr) 
\nonumber \\
& \hspace{10mm} +2 \mbox{C}_{\mbox{{\tiny A}}}
\biggl[ \biggl(\frac{1}{1-x}\biggr)_{+}
+\frac{1-x}{x} -1 +x(1-x) \biggr]\,,  
\label{alpgg} 
\\
P^{fg}(x)
&= \mbox{C}_{\mbox{{\tiny F}}}
\frac{1+(1-x)^2}{x}\,,
\label{alpfg} 
\\
P^{gf}(x)
&=\mbox{T}_{\mbox{{\tiny R}}}
[x^2 + (1-x)^{2}]\,.
\label{alpgf}
\end{align}
\begin{align}
{\cal V}_{other}^{f,f}(x\,;\ep) 
&= 
\delta(1-x) 
\,\mbox{C}_{\mbox{{\tiny F}}}
\left( \frac{2}{3} \pi^2 - 5 \right)
- \ln x \cdot P^{ff}(x) 
\nonumber \\
+ \mbox{C}_{\mbox{{\tiny F}}} 
& \left[ - \left(\frac{4}{1-x} \ln \frac{1}{1-x} \right)_+ 
-\frac{2}{1-x} \ln(2-x) 
 +1-x - (1+x) \ln(1-x) \right],
\\
{\cal V}_{other}^{g,g}(x\,;\ep) 
&= 
\delta(1-x)\left[
\mbox{C}_{\mbox{{\tiny A}}}
\left( \frac{2}{3} \pi^2 - \frac{50}{9} \right)
+ \frac{16}{9} N_f \mbox{T}_{\mbox{{\tiny R}}} \right]
- \ln x \cdot P^{gg}(x) 
\nonumber \\
& \ \ + \mbox{C}_{\mbox{{\tiny A}}}\biggl[
- \left(\frac{4}{1-x} \ln \frac{1}{1-x} \right)_+ 
-\frac{2}{1-x} \ln(2-x)  
\nonumber \\
& \hspace{40mm}+  2 \left( -1 +x(1-x) + \frac{1-x}{x} \right)
\ln(1-x) \biggr]\,,
\\
{\cal V}_{other}^{f,g}(x\,;\ep) 
&= \ln \frac{1-x}{x} P^{fg}(x) 
+ \mbox{C}_{\mbox{{\tiny F}}}  \;x\,,
\\
{\cal V}_{other}^{g,f}(x\,;\ep) 
&= \ln \frac{1-x}{x} P^{gf}(x) 
+ \mbox{T}_{\mbox{{\tiny R}}} \;2x(1-x)\,.
\end{align}
Color correlated Born squared amplitude\,:\,
\begin{equation}
[ \, y_{emi}, \,y_{spe} ]
=(s_{x_{a/b},\,y_{spe}})^{-\ep} 
\cdot
\langle \mbox{B}j \ | 
\mbox{T}_{y_{emi}} \cdot 
\mbox{T}_{y_{spe}}
 | \ \mbox{B}j \rangle_{d}\,. 
\label{clbsaif}
\end{equation}
Lorentz scalar\,:\,
\begin{equation}
s_{x_{a/b},\,y_{spe}}=2\,p_{a/b} \cdot \mbox{P}(y_{spe})\,.
\label{lsif}
\end{equation}
\vspace{5mm}

%
%
\leftline{\fbox{$\mbox{D}_{ai,b}$: Initial-Initial dipole}} 
\begin{verbatim}
Dipole 1 (3)-2,  (4)-2, 
Dipole 3 (6)-2,  
Dipole 4 (7)-2:
\end{verbatim}
\begin{equation}
\hat{\sigma}_{\mbox{{\tiny D}}}(\mbox{R}_{i},\,{\tt dip}j,\, 
\mbox{II},\,x_{a/b}) 
=
-\frac{A_{d}}{S_{\mbox{{\tiny R}}_{i}}} 
\int_{0}^{1}dx \,
\frac{1}{\mbox{T}_{\mbox{{\tiny F}}(y_{emi})}^{2}} 
{\widetilde {\cal V}}^{\mbox{{\tiny F}}(x_{a/b}),\, 
\mbox{{\tiny F}}(y_{emi})}(x;\ep)
\cdot
\Phi_{a/b}(\mbox{B}_{j},x)_{d} 
\,[ \, y_{emi}, \,y_{spe} ]\,.
\label{iimast}
\end{equation}
\vspace{7mm}

\leftline{\underline{Definition of the symbols}}
\vspace{2mm}
\noindent
Universal singular functions\,:\,
\begin{equation}
\frac{1}{\mbox{T}_{\mbox{{\tiny F}}(y_{emi})}^{2}}
{\widetilde {\cal V}}^{\mbox{{\tiny F}}(x_{a/b}),\, 
\mbox{{\tiny F}}(y_{emi})}(x;\ep)
=
\left\{
\begin{array}{ll}
\frac{1}{\mbox{C}_{\mbox{{\tiny F}}}} 
{\widetilde {\cal V}}^{f,f} (x;\ep)
& : \mbox{Dipole} \ 1 \ (3)\mbox{-}2\,,
\\
\frac{1}{\mbox{C}_{\mbox{{\tiny A}}}} 
{\widetilde {\cal V}}^{g,g} (x;\ep)
& : \mbox{Dipole} \ 1 \ (4)\mbox{-}2\,,
\\
\frac{1}{\mbox{C}_{\mbox{{\tiny A}}}} 
{\widetilde {\cal V}}^{f,g} (x;\ep)
& : \mbox{Dipole} \ 3 \ (6)\mbox{-}2\,,
\\
\frac{1}{\mbox{C}_{\mbox{{\tiny F}}}} 
{\widetilde {\cal V}}^{g,f} (x;\ep)
& : \mbox{Dipole} \ 4 \ (7)\mbox{-}2\,.
\end{array}
\right.
\label{singii}
\end{equation}
\begin{equation}
{\widetilde {\cal V}}^{a,b}(x;\ep) = {\cal V}^{a,b}(x;\ep) +
\delta^{ab} \, 	T_a^2 \, g(x)
+ {\widetilde K}^{ab}(x)\,,
\end{equation}
\begin{align}
{\widetilde {\cal V}}^{f,f}(x;\ep) 
&= {\cal V}^{f,f}(x;\ep) +
\mbox{C}_{\mbox{{\tiny F}}} \, g(x)
+ {\widetilde K}^{ff}(x)\,,
\label{siniiff}
\\
{\widetilde {\cal V}}^{g,g}(x;\ep) 
&= {\cal V}^{g,g}(x;\ep) +
\mbox{C}_{\mbox{{\tiny A}}} \, g(x)
+ {\widetilde K}^{gg}(x)\,,
\label{siniigg}
\\
{\widetilde {\cal V}}^{f,g}(x;\ep) 
&= {\cal V}^{f,g}(x;\ep) 
\hspace{18mm}+ {\widetilde K}^{fg}(x)\,, 
\label{siniifg}
\\
{\widetilde {\cal V}}^{g,f}(x;\ep) 
&= {\cal V}^{g,f}(x;\ep) 
\hspace{18mm}+ {\widetilde K}^{gf}(x)\,.
\label{siniigf}
\end{align}
\begin{equation}
{\widetilde K}^{ab}(x) = P^{ab}_{{\rm reg}}(x) \;\ln(1-x) 
+ \;\delta^{ab} \, T_a^2 \left[ \left( \frac{2}{1-x} \ln (1-x)
\right)_+ - \frac{\pi^2}{3} \delta(1-x) \right]\,.
\end{equation}
\begin{align}
{\widetilde K}^{ff}(x) 
&= P^{ff}_{reg}(x) \;\ln(1-x) 
+ \mbox{C}_{\mbox{{\tiny F}}} \left[ \left( \frac{2}{1-x} \ln (1-x)
\right)_+ - \frac{\pi^2}{3} \delta(1-x) \right]\,,
\label{ktilff}\\
{\widetilde K}^{gg}(x)
&= P^{gg}_{reg}(x) \;\ln(1-x) 
+ \mbox{C}_{\mbox{{\tiny A}}} \left[ \left( \frac{2}{1-x} \ln (1-x)
\right)_+ - \frac{\pi^2}{3} \delta(1-x) \right]\,,
\label{ktilgg}\\
{\widetilde K}^{fg}(x) 
&= P^{fg}(x) \;\ln(1-x)\,,
\label{ktilfg}\\
{\widetilde K}^{gf}(x) 
&= P^{gf}(x) \;\ln(1-x)\,.
\label{ktilgf}
\end{align}
\begin{align}
P^{ff}_{reg}(x)
&= -\mbox{C}_{\mbox{{\tiny F}}}
(1+x)\,,  \\
P^{gg}_{reg}(x)
&= 2 \mbox{C}_{\mbox{{\tiny A}}}
\biggl[\frac{1-x}{x} -1 + x(1-x) \biggr]\,.
\end{align}
Color correlated Born squared amplitude\,:\,
\begin{equation}
[ \, y_{emi}, \,y_{spe} ]
=(s_{x_{a/b},\,y_{spe}})^{-\ep}
\cdot
\langle \mbox{B}j \ | 
\mbox{T}_{y_{emi}} \cdot 
\mbox{T}_{y_{spe}}
 | \ \mbox{B}j \rangle_{d}\,. 
\label{clbsaii}
\end{equation}
Lorentz scalar\,:\,
\begin{equation}
s_{x_{a/b},\,y_{spe}}=2\,p_{a} \cdot p_{b}\,.
\label{lsii}
\end{equation}
\newpage
\subsection{Integrated dipole term:
$\sigma_{\mbox{{\tiny D}}}(\mbox{I})$ 
\label{ap_A_3}}
{\tt Dipole\,1 (1)-1/2, (2)-1/2, (3)-1/2, (4)-1/2:}
\begin{figure}[h]
\begin{center}
\includegraphics[width=8cm]{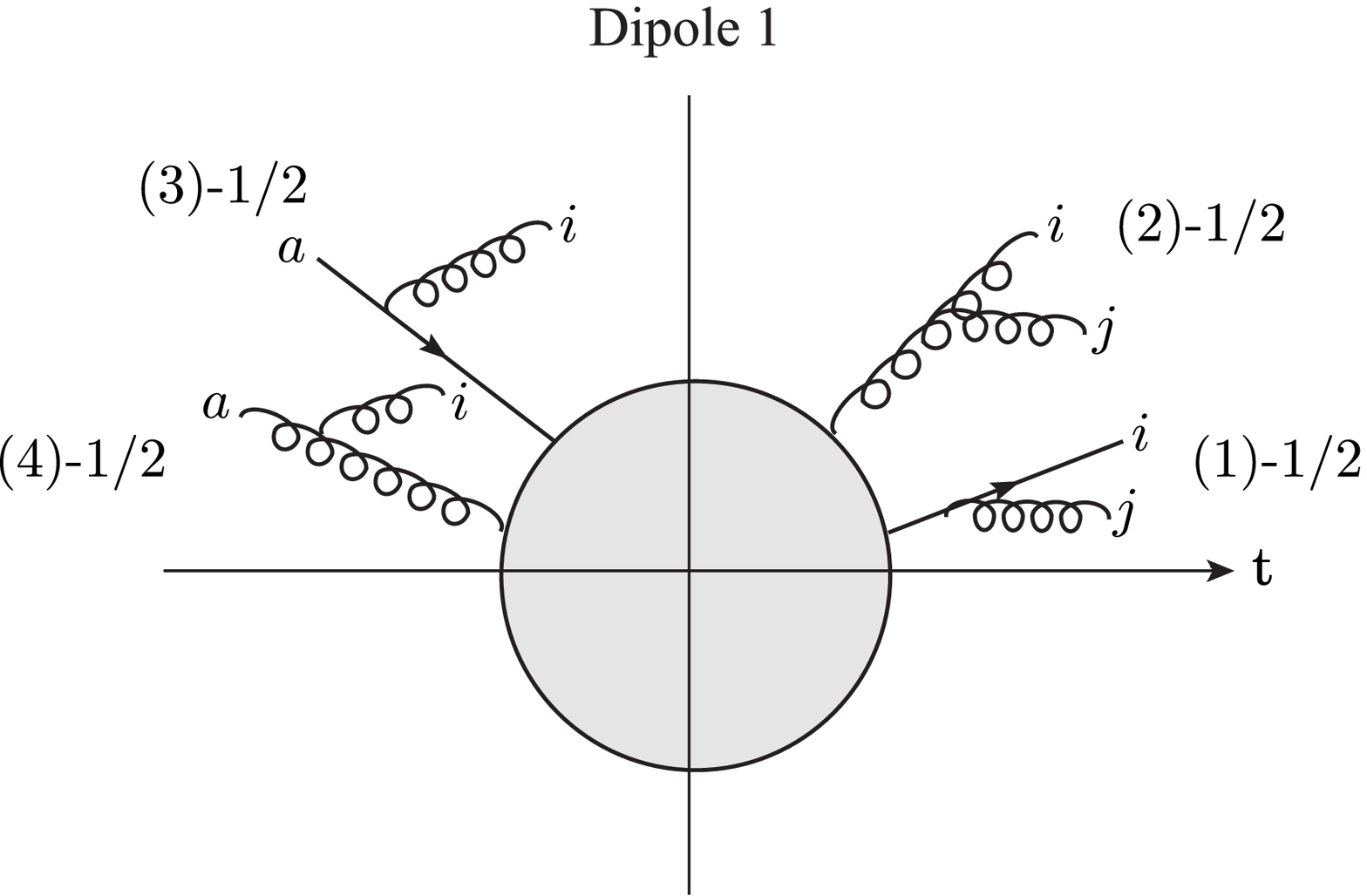}
\end{center} 
\caption{ The classification of the 
integrated dipole term
$\sigma_{\mbox{{\tiny D}}}(\mbox{I})$.
\label{fig_A3_int_I}}
\end{figure}
\begin{align}
(1)-1/2:&
\hspace{5mm}
\hat{\sigma}_{\mbox{{\tiny D}}}
(\,
\mbox{R}_{i},\,
\mbox{I},(1)\mbox{-}1/2 
\,) 
=
-\frac{A_{d}}{S_{\mbox{{\tiny R}}_{i}}} 
\cdot
\frac{1}{\mbox{C}_{\mbox{{\tiny F}}}} 
{\cal V}_{f}(\ep)
\cdot 
\Phi(\mbox{B}1)_{d} \,
[ \, y_{emi}, \,y_{spe} ]\,,
\nonumber \\
(2)-1/2:&
\hspace{5mm}
\hat{\sigma}_{\mbox{{\tiny D}}}
(\,
\mbox{R}_{i},\,
\mbox{I},(2)\mbox{-}1/2 
\,) 
=
-\frac{A_{d}}{S_{\mbox{{\tiny R}}_{i}}} 
\cdot
\frac{1}{\mbox{C}_{\mbox{{\tiny A}}}} 
{\cal V}_{gg}(\ep)
\cdot 
\Phi(\mbox{B}1)_{d} \,
[ \, y_{emi}, \,y_{spe} ]\,,
\nonumber \\
(3)-1/2:&
\hspace{5mm}
\hat{\sigma}_{\mbox{{\tiny D}}}
(\,
\mbox{R}_{i},\,
\mbox{I},(3)\mbox{-}1/2 
\,) 
=
-\frac{A_{d}}{S_{\mbox{{\tiny R}}_{i}}} 
\cdot
\frac{1}{\mbox{C}_{\mbox{{\tiny F}}}} 
{\cal V}_{f}(\ep)
\cdot 
\Phi(\mbox{B}1)_{d} \,
[ \, y_{emi}, \,y_{spe} ]\,,
\nonumber \\
(4)-1/2:&
\hspace{5mm}
\hat{\sigma}_{\mbox{{\tiny D}}}
(\,
\mbox{R}_{i},\,
\mbox{I},(4)\mbox{-}1/2 
\,) 
=
-\frac{A_{d}}{S_{\mbox{{\tiny R}}_{i}}} 
\cdot
\frac{1}{\mbox{C}_{\mbox{{\tiny A}}}} 
{\cal V}_{g}(\ep)
\cdot 
\Phi(\mbox{B}1)_{d} \,
[ \, y_{emi}, \,y_{spe} ]\,.
\end{align}
\vspace{7mm}

\leftline{\underline{Definition of the symbols}}
\vspace{2mm}
\noindent
Universal singular functions\,:\,
\begin{align}
{\cal V}_{f}(\ep)&={\cal V}_{fg}(\ep)\,,
\label{nufeq} \\
{\cal V}_{g}(\ep)&= \half {\cal V}_{gg} (\ep) + 
N_{f}{\cal V}_{f\bar{f}}(\ep)\,.
\end{align}
Color correlated Born squared amplitude\,:\,
\begin{equation}
[ \, y_{emi}, \,y_{spe} ]
=(s_{y_{emi}, \,y_{spe}})^{-\ep} 
\cdot
\langle \mbox{B}1 \ | 
\mbox{T}_{y_{emi}} \cdot 
\mbox{T}_{y_{spe}}
 | \ \mbox{B}1 \rangle_{d}\,. 
\end{equation}
Lorentz scalar\,:\,
$s_{y_{emi}, y_{spe}}=2 \, \mbox{P}(y_{emi}) 
\cdot \mbox{P}(y_{spe})$\,.

\newpage
\subsection{Integrated dipole term:
$\sigma_{\mbox{{\tiny D}}}(\mbox{P})$ 
\label{ap_A_4}}
\begin{verbatim}
Dipole 1 (3)-1/2,  (4)-1/2, 
Dipole 3 (6)-1/2,  
Dipole 4 (7)-1/2:
\end{verbatim}
\begin{figure}[h!]
\begin{center}
\includegraphics[width=7cm]{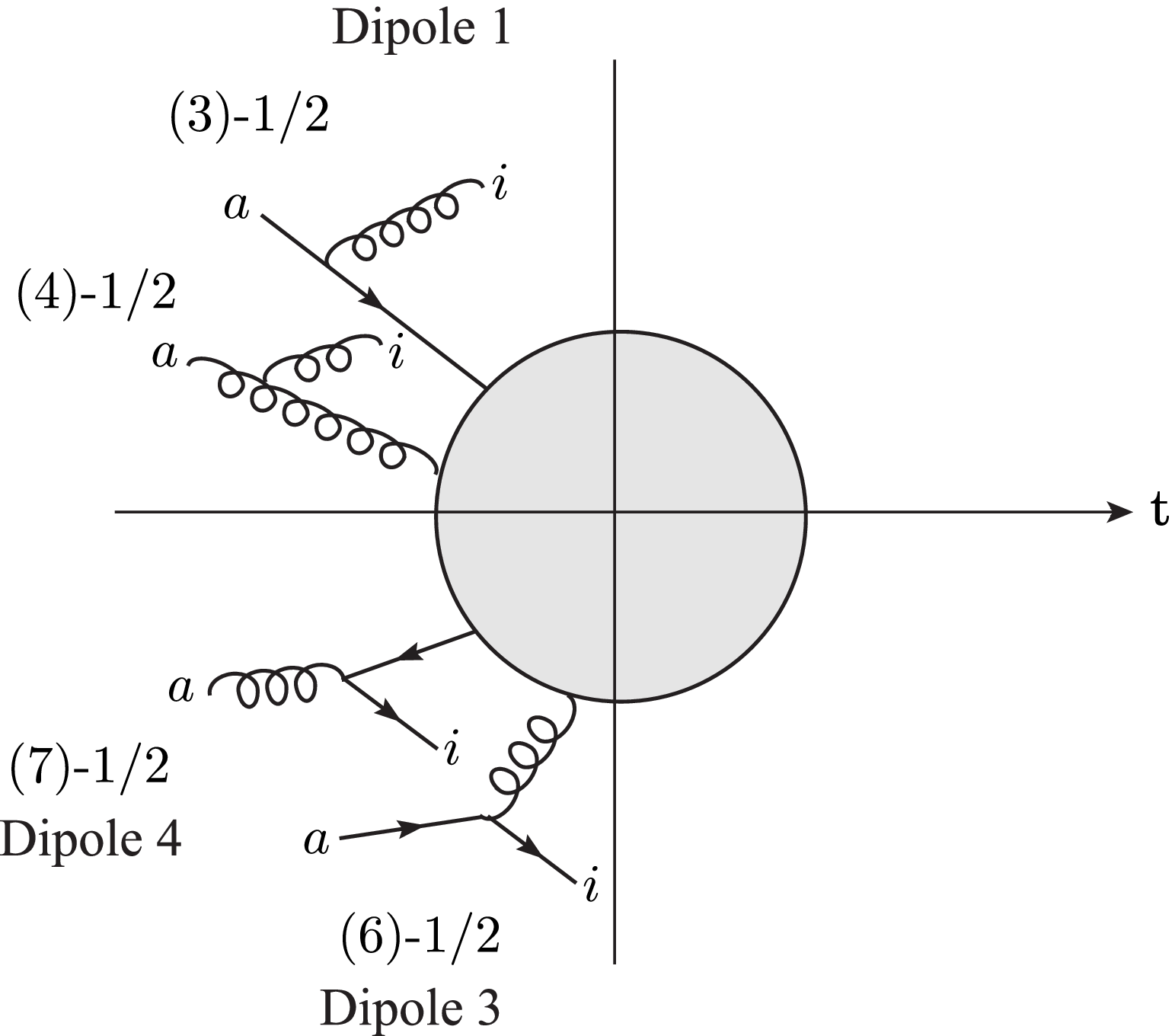}
\end{center} 
\caption{ The classification of the 
integrated dipole term
$\sigma_{\mbox{{\tiny D}}}(\mbox{P})$
\label{fig_A4_int_P}}
\end{figure}
\begin{align}
\hat{\sigma}_{\mbox{{\tiny D}}}
(\,
\mbox{R}_{i},\,
\mbox{P},\,{\tt dip}j,
\,
x_{a/b}
) 
&=
\frac{A_{d}}{S_{\mbox{{\tiny R}}_{i}}} 
\int_{0}^{1}dx \,
\biggl(\frac{1}{\ep} - \ln x \biggr)
\frac{1}{\mbox{T}_{\mbox{{\tiny F}}(y_{emi})}^{2}} 
P^{\mbox{{\tiny F}}(x_{a/b}) \, \mbox{{\tiny F}}(y_{emi}) } (x)
\nonumber \\
& \hspace{50mm}
\cdot 
\Phi_{a/b}(\mbox{B}_{j},x)_{d} 
\, [ \, y_{emi}, \,y_{spe} ]\,. 
\end{align}
\vspace{5mm}

\leftline{\underline{Definition of the symbols}}
\vspace{2mm}
\noindent
Splitting functions\,:\,
\begin{equation}
\frac{1}{\mbox{T}_{\mbox{{\tiny F}}(y_{emi})}^{2}}
P^{\mbox{{\tiny F}}(x_{a/b}) \, \mbox{{\tiny F}}(y_{emi}) } (x)
=
\left\{
\begin{array}{ll}
\frac{1}{\mbox{C}_{\mbox{{\tiny F}}}} 
P^{ff} (x)
& : \mbox{Dipole} \ 1 \ (3)\mbox{-}1/2\,,
\\
\frac{1}{\mbox{C}_{\mbox{{\tiny A}}}} 
P^{gg} (x)
& : \mbox{Dipole} \ 1 \ (4)\mbox{-}1/2\,,
\\
\frac{1}{\mbox{C}_{\mbox{{\tiny A}}}} 
P^{fg} (x)
& : \mbox{Dipole} \ 3 \ (6)\mbox{-}1/2\,,
\\
\frac{1}{\mbox{C}_{\mbox{{\tiny F}}}} 
P^{gf} (x)
& : \mbox{Dipole} \ 4 \ (7)\mbox{-}1/2\,.
\end{array}
\right.
\label{sigdpp}
\end{equation}
Color correlated Born squared amplitude\,:\,
\begin{equation}
[ \, y_{emi}, \,y_{spe} ]
=(s_{x_{a/b},\,y_{spe}})^{-\ep} 
\cdot
\langle \mbox{B}j \ | 
\mbox{T}_{y_{emi}} \cdot 
\mbox{T}_{y_{spe}}
 | \ \mbox{B}j \rangle_{d}\,. 
\label{sqbk3}
\end{equation}
Lorentz scalars\,:\,
\begin{equation}
s_{x_{a/b},\,y_{spe}}
=
\left\{
\begin{array}{ll}
2\,p_{a/b} \cdot \mbox{P}(y_{spe})
& : (3),(4),(6),(7)\mbox{-}1\,,\mbox{Initial-Final}\,,
\\
2\,p_{a} \cdot p_{b}
& : (3),(4),(6),(7)\mbox{-}2\,,\mbox{Initial-Initial}\,.
\end{array}
\right.
\label{sigdplo}
\end{equation}

\newpage
\subsection{Integrated dipole term:
$\sigma_{\mbox{{\tiny D}}}(\mbox{K})$ 
\label{ap_A_5}}
\begin{verbatim}
Dipole 1 (1)-2,   (2)-2, 
         (3)-1/2, (4)-1/2, 
Dipole 3 (6)-1/2,  
Dipole 4 (7)-1/2:
\end{verbatim}
\begin{figure}[h]
\begin{center}
\includegraphics[width=7cm]{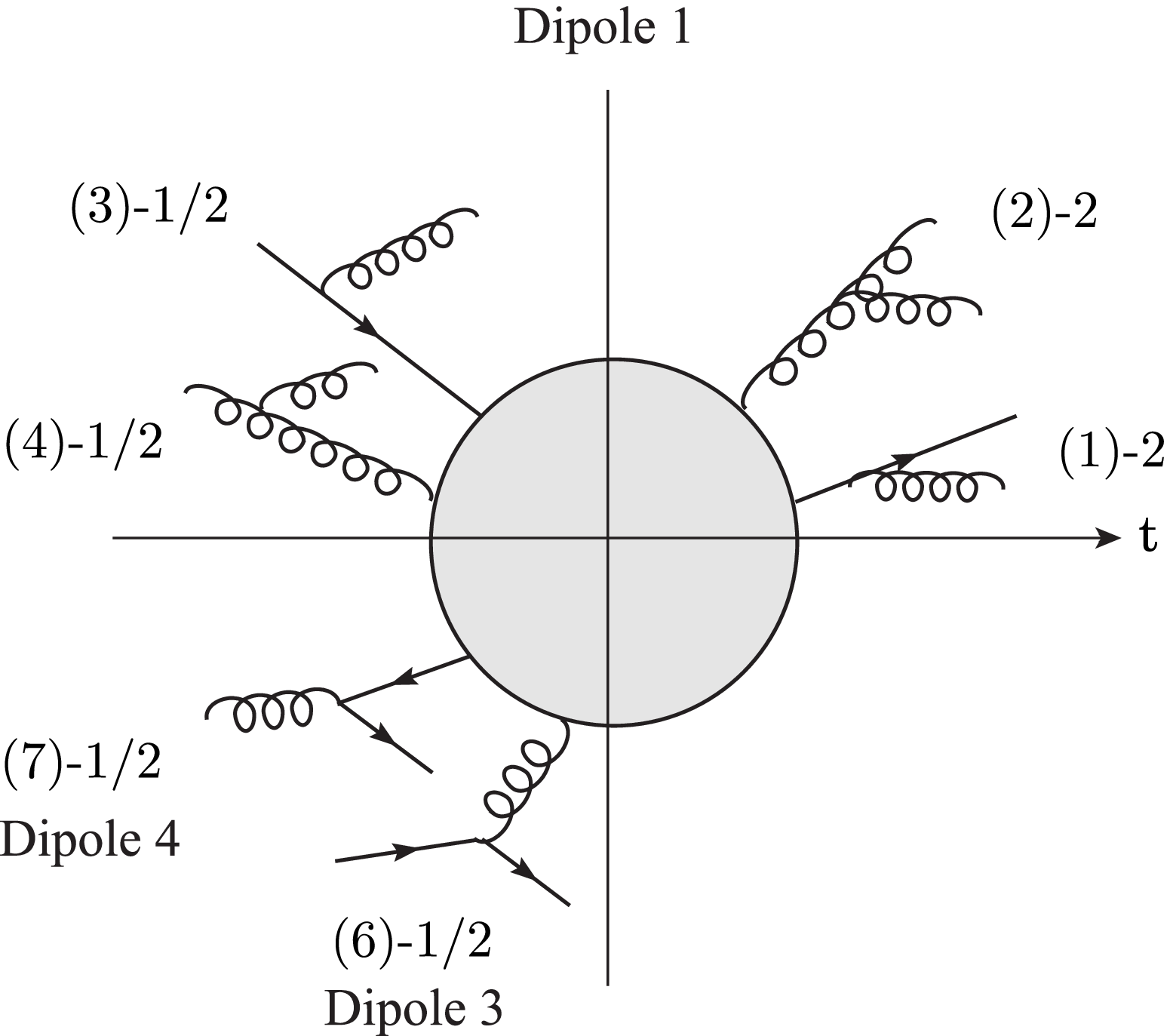}
\end{center} 
\caption{The classification of the 
integrated dipole term
$\sigma_{\mbox{{\tiny D}}}(\mbox{K})$
\label{fig_A5_int_K}}
\end{figure}
\leftline{\underline{{\tt Dipole\,1}}}
\begin{align}
(1)\mbox{-}2:&
\hspace{3mm}
\hat{\sigma}_{\mbox{{\tiny D}}}(
\mbox{R}_{i},\,
\mbox{K}, {\tt dip}1,
(1)\mbox{-}2,
x_{a}) 
=
-\frac{A_{4}}{S_{\mbox{{\tiny R}}_{i}}} 
\int_{0}^{1}dx \,
\biggl(
g(x)-\frac{3}{2}h(x)
\biggr)
\cdot 
\Phi_{a}(\mbox{B}1,x)_{4} 
\, \langle  \, y_{emi}, \,y_{spe} \,\rangle \,,
\nonumber \\
(2)\mbox{-}2:&
\hspace{3mm}
\hat{\sigma}_{\mbox{{\tiny D}}}(
\mbox{R}_{i},\,
\mbox{K}, {\tt dip}1,
(2)\mbox{-}2,
x_{a}) 
=
-\frac{A_{4}}{S_{\mbox{{\tiny R}}_{i}}} 
\int_{0}^{1}dx \,
\biggl(
2g(x)-\frac{11}{3}h(x)
\biggr)
\cdot 
\Phi_{a}(\mbox{B}1,x)_{4} 
\, \langle  \, y_{emi}, \,y_{spe} \,\rangle \,,
\label{a5eq1}
\end{align}
\begin{align}
(3)\mbox{-}1:&
\hspace{3mm}
\hat{\sigma}_{\mbox{{\tiny D}}}(
\mbox{R}_{i},\,
\mbox{K}, {\tt dip}1,
(3)\mbox{-}1,
x_{a}
) 
=
-\frac{A_{4}}{S_{\mbox{{\tiny R}}_{i}}} 
\int_{0}^{1}dx \,
\frac{1}{\mbox{C}_{\mbox{{\tiny F}}}}
{\cal V}_{other}^{f,f}(x\,;\ep)
\cdot 
\Phi_{a}(\mbox{B}1,x)_{4} 
\, \langle  \, y_{emi}, \,y_{spe} \,\rangle \,,
\nonumber \\
(4)\mbox{-}1:&
\hspace{3mm}
\hat{\sigma}_{\mbox{{\tiny D}}}(
\mbox{R}_{i},\,
\mbox{K}, {\tt dip}1,
(4)\mbox{-}1,
x_{a}) 
=
-\frac{A_{4}}{S_{\mbox{{\tiny R}}_{i}}} 
\int_{0}^{1}dx \,
\frac{1}{\mbox{C}_{\mbox{{\tiny A}}}}
{\cal V}_{other}^{g,g}(x\,;\ep)
\cdot 
\Phi_{a}(\mbox{B}1,x)_{4} 
\, \langle  \, y_{emi}, \,y_{spe} \,\rangle \,,
\label{a5eq2}
\end{align}
\begin{align}
(3)\mbox{-}2:&
\hspace{3mm}
\hat{\sigma}_{\mbox{{\tiny D}}}(
\mbox{R}_{i},\,
\mbox{K}, {\tt dip}1,
(3)\mbox{-}2,
x_{a}) 
=
-\frac{A_{4}}{S_{\mbox{{\tiny R}}_{i}}} 
\int_{0}^{1}dx \,
\frac{1}{\mbox{C}_{\mbox{{\tiny F}}}}
\biggl(
{\cal V}_{other}^{f,f}(x\,;\ep)
+
\mbox{C}_{\mbox{{\tiny F}}} \, g(x)
+ {\widetilde K}^{ff}(x)
\biggr)
\nonumber \\
& 
\hspace{70mm}
\cdot 
\Phi_{a}(\mbox{B}1,x)_{4} 
\, \langle  \, y_{emi}, \,y_{spe} \,\rangle \,,
\nonumber \\
(4)\mbox{-}2:&
\hspace{3mm}
\hat{\sigma}_{\mbox{{\tiny D}}}(
\mbox{R}_{i},\,
\mbox{K}, {\tt dip}1,
(4)\mbox{-}2,
x_{a}) 
=
-\frac{A_{4}}{S_{\mbox{{\tiny R}}_{i}}} 
\int_{0}^{1}dx \,
\frac{1}{\mbox{C}_{\mbox{{\tiny A}}}}
\biggl(
{\cal V}_{other}^{g,g}(x\,;\ep)
+
\mbox{C}_{\mbox{{\tiny A}}} \, g(x)
+ {\widetilde K}^{gg}(x)
\biggr)
\nonumber \\
& 
\hspace{70mm}
\cdot 
\Phi_{a}(\mbox{B}1,x)_{4} 
\, \langle  \, y_{emi}, \,y_{spe} \,\rangle \,.
\label{a5eq3}
\end{align}
%
%
\vspace{2mm}
\leftline{\underline{{\tt Dipole\,3/4}}}
\begin{align}
(6)\mbox{-}1:&
\hspace{3mm}
\hat{\sigma}_{\mbox{{\tiny D}}}(
\mbox{R}_{i},\,
\mbox{K}, {\tt dip}3,
(6)\mbox{-}1,
x_{a}) 
=
-\frac{A_{4}}{S_{\mbox{{\tiny R}}_{i}}} 
\int_{0}^{1}dx \,
\frac{1}{\mbox{C}_{\mbox{{\tiny A}}}}
{\cal V}_{other}^{f,g}(x\,;\ep)
\cdot 
\Phi_{a}(\mbox{B}3,x)_{4} 
\, \langle  \, y_{emi}, \,y_{spe} \,\rangle \,,
\nonumber \\
(7)\mbox{-}1:&
\hspace{3mm}
\hat{\sigma}_{\mbox{{\tiny D}}}(
\mbox{R}_{i},\,
\mbox{K}, {\tt dip}4,
(7)\mbox{-}1,
x_{a}) 
=
-\frac{A_{4}}{S_{\mbox{{\tiny R}}_{i}}} 
\int_{0}^{1}dx \,
\frac{1}{\mbox{C}_{\mbox{{\tiny F}}}}
{\cal V}_{other}^{g,f}(x\,;\ep)
\cdot 
\Phi_{a}(\mbox{B}4,x)_{4} 
\, \langle  \, y_{emi}, \,y_{spe} \,\rangle \,,
\label{a5eq4}
\end{align}
\begin{align}
(6)\mbox{-}2:&
\hspace{3mm}
\hat{\sigma}_{\mbox{{\tiny D}}}(
\mbox{R}_{i},\,
\mbox{K}, {\tt dip}3,
(6)\mbox{-}2,
x_{a}) 
=
-\frac{A_{4}}{S_{\mbox{{\tiny R}}_{i}}} 
\int_{0}^{1}dx \,
\frac{1}{\mbox{C}_{\mbox{{\tiny A}}}}
\biggl(
{\cal V}_{other}^{f,g}(x\,;\ep)
+ {\widetilde K}^{fg}(x)
\biggr)
\nonumber \\
& 
\hspace{70mm}
\cdot 
\Phi_{a}(\mbox{B}3,x)_{4} 
\, \langle  \, y_{emi}, \,y_{spe} \,\rangle \,,
\nonumber \\
(7)\mbox{-}2:&
\hspace{3mm}
\hat{\sigma}_{\mbox{{\tiny D}}}(
\mbox{R}_{i},\,
\mbox{K}, {\tt dip}4,
(7)\mbox{-}2,
x_{a}) 
=
-\frac{A_{4}}{S_{\mbox{{\tiny R}}_{i}}} 
\int_{0}^{1}dx \,
\frac{1}{\mbox{C}_{\mbox{{\tiny F}}}}
\biggl(
{\cal V}_{other}^{g,f}(x\,;\ep)
+ {\widetilde K}^{gf}(x)
\biggr)
\nonumber \\
& 
\hspace{70mm}
\cdot 
\Phi_{a}(\mbox{B}4,x)_{4} 
\, \langle  \, y_{emi}, \,y_{spe} \,\rangle \,.
\label{a5eq5}
\end{align}
\vspace{6mm}

\leftline{\underline{Reconstruction of Dipole\,1}}
\begin{align}
\sigma_{\mbox{{\tiny D}}}(\mbox{R}_{i},\,\mbox{K},
{\tt dip}1)
&=
\hat{\sigma}_{\mbox{{\tiny D}}}(\mbox{K},{\tt dip}1,
(1)\mbox{--}(4)\mbox{-}2,g
)^{\circlearrowright}
+
\hat{\sigma}_{\mbox{{\tiny D}}}(\mbox{K},{\tt dip}1,
(3)/(4)\mbox{-}1/2,
{\cal V}_{other}^{a,a}
)^{\circlearrowright}
\nonumber\\
& \ \
+
\hat{\sigma}_{\mbox{{\tiny D}}}(\mbox{K},{\tt dip}1,
(1)/(2)\mbox{-}2,h)
+
\hat{\sigma}_{\mbox{{\tiny D}}}(\mbox{K},{\tt dip}1,
(3)/(4)\mbox{-}2,\widetilde{\mbox{K}}^{aa})\,,
\end{align}
\begin{align}
\hat{\sigma}_{\mbox{{\tiny D}}}(\mbox{K},{\tt dip}1,
(1)\mbox{--}(4)\mbox{-}2,g
)
&=
-\frac{A_{4}}{S_{\mbox{{\tiny R}}_{i}}} 
\int_{0}^{1}dx \,
\Phi_{a}(\mbox{B}1,x)_{4} 
\, \langle  \, y_{emi}, \,y_{spe} \,\rangle
\, g(x) \, \times
\nonumber\\
& \ \ 
\left\{
\begin{array}{ll}
1: (1)\mbox{-},(3)\mbox{-},(4)\mbox{-}2\,,
\\
2: (2)\mbox{-}2\,.
\end{array}
\right.
\label{dkd1g}
\end{align}
\begin{align}
\hat{\sigma}_{\mbox{{\tiny D}}}(\mbox{K},{\tt dip}1,
(3)\mbox{-}1/2,
{\cal V}_{other}^{f,f})
&=-\frac{A_{4}}{S_{\mbox{{\tiny R}}_{i}}} 
\int_{0}^{1}dx \,
\frac{1}{\mbox{C}_{\mbox{{\tiny F}}}}
{\cal V}_{other}^{f,f}(x\,;\ep)
\cdot 
\Phi_{a}(\mbox{B}1,x)_{4} 
\, \langle  \, y_{emi}, \,y_{spe} \,\rangle \,,
\label{dkd1oth1}
\\
\hat{\sigma}_{\mbox{{\tiny D}}}(\mbox{K},{\tt dip}1,
(4)\mbox{-}1/2,
{\cal V}_{other}^{g,g}
)
&=-\frac{A_{4}}{S_{\mbox{{\tiny R}}_{i}}} 
\int_{0}^{1}dx \,
\frac{1}{\mbox{C}_{\mbox{{\tiny A}}}}
{\cal V}_{other}^{g,g}(x\,;\ep)
\cdot 
\Phi_{a}(\mbox{B}1,x)_{4} 
\, \langle  \, y_{emi}, \,y_{spe} \,\rangle \,,
\label{dkd1oth2}
\end{align}
\begin{align}
\hat{\sigma}_{\mbox{{\tiny D}}}(\mbox{K},{\tt dip}1,
(1)/(2)\mbox{-}2,h
)
&=
-\frac{A_{4}}{S_{\mbox{{\tiny R}}_{i}}} 
\int_{0}^{1}dx \,
\Phi_{a}(\mbox{B}1,x)_{4} 
\, \langle  \, y_{emi}, \,y_{spe} \,\rangle
\,h(x)
\,\times
\nonumber\\
& \ \ 
\left\{
\begin{array}{ll}
-3/2
\ : (1)\mbox{-}2\,,
\\
-11/3
: (2)\mbox{-}2\,,
\end{array}
\right.
\label{dkd1h}
\end{align}
\begin{align}
\hat{\sigma}_{\mbox{{\tiny D}}}(\mbox{K},{\tt dip}1,
(3)/(4)\mbox{-}2,\widetilde{\mbox{K}}^{ff/gg})
=
-\frac{A_{4}}{S_{\mbox{{\tiny R}}_{i}}} 
\int_{0}^{1}dx \,
\frac{1}{\mbox{C}_{\mbox{{\tiny F/A}}}}
{\widetilde K}^{ff/gg}(x)
\cdot
\Phi_{a}(\mbox{B}1,x)_{4} 
\, \langle  \, y_{emi}, \,y_{spe} \,\rangle \,.
\label{dkd1ktil}
\end{align}
\vspace{2mm}

\leftline{\underline{Reconstruction of Dipole\,3/4}}
\begin{align}
\sigma_{\mbox{{\tiny D}}}(\mbox{R}_{i},\,\mbox{K},
{\tt dip}3/4)
=
\hat{\sigma}_{\mbox{{\tiny D}}}(\mbox{K},{\tt dip}3/4,
(6)/(7)\mbox{-}1/2,
{\cal V}_{other}^{a,b}
)
+
\hat{\sigma}_{\mbox{{\tiny D}}}(\mbox{K},{\tt dip}3/4,
(6)/(7)\mbox{-}2,\widetilde{\mbox{K}}^{ab})\,,
\end{align}
\begin{align}
\hat{\sigma}_{\mbox{{\tiny D}}}(\mbox{K},{\tt dip}3,
(6)\mbox{-}1/2,
{\cal V}_{other}^{f,g}
)
&=-\frac{A_{4}}{S_{\mbox{{\tiny R}}_{i}}} 
\int_{0}^{1}dx \,
\frac{1}{\mbox{C}_{\mbox{{\tiny A}}}}
{\cal V}_{other}^{f,g}(x\,;\ep)
\cdot 
\Phi_{a}(\mbox{B}3,x)_{4} 
\, \langle  \, y_{emi}, \,y_{spe} \,\rangle \,,
\label{dkd34oth1}
\\
\hat{\sigma}_{\mbox{{\tiny D}}}(\mbox{K},{\tt dip}4,
(7)\mbox{-}1/2,
{\cal V}_{other}^{g,f}
)
&=-\frac{A_{4}}{S_{\mbox{{\tiny R}}_{i}}} 
\int_{0}^{1}dx \,
\frac{1}{\mbox{C}_{\mbox{{\tiny F}}}}
{\cal V}_{other}^{g,f}(x\,;\ep)
\cdot 
\Phi_{a}(\mbox{B}4,x)_{4} 
\, \langle  \, y_{emi}, \,y_{spe} \,\rangle \,,
\label{dkd34oth2}
\end{align}
\begin{align}
\hat{\sigma}_{\mbox{{\tiny D}}}(\mbox{K},{\tt dip}3/4,
(6)/(7)\mbox{-}2,\widetilde{\mbox{K}}^{fg/gf})
=
-\frac{A_{4}}{S_{\mbox{{\tiny R}}_{i}}} 
\int_{0}^{1}dx \,
\frac{1}{\mbox{C}_{\mbox{{\tiny A/F}}}}
{\widetilde K}^{fg/gf}(x)
\cdot
\Phi_{a}(\mbox{B}3/4,x)_{4} 
\, \langle  \, y_{emi}, \,y_{spe} \,\rangle \,.
\label{dkd34ktil}
\end{align}
The cross sections with leg-b\,($x_{b}$),
$\hat{\sigma}_{\mbox{{\tiny D}}}(\,
\mbox{R}_{i},\,
\mbox{K}, {\tt dip}j,
\,x_{b}
\,)$,
are obtained by the replacements
$\Phi_{a}(\mbox{B}j,x)_{4}$ 
$\to$ 
$\Phi_{b}(\mbox{B}j,x)_{4}$
in the above formulae for leg-a\,($x_{a}$).
\vspace{5mm}

\leftline{\underline{Definition of the symbols}}
\vspace{2mm}
\noindent
Color-correlated Born squared amplitude\,:\,
\begin{equation}
\langle \, y_{emi}, \,y_{spe} \rangle\,
=
\langle \mbox{B}j \ | 
\mbox{T}_{y_{emi}} \cdot 
\mbox{T}_{y_{spe}}
 | \ \mbox{B}j \rangle_{4}\,.
\end{equation}

\newpage
\subsection{Integrated dipole term:
$\sigma_{\mbox{{\tiny D}}}(\mbox{dip}2)$ 
\label{ap_A_6}}
\begin{verbatim}
Dipole 2 (5)-1/2:
\end{verbatim}
\begin{figure}[h]
\begin{center}
\includegraphics[width=6cm]{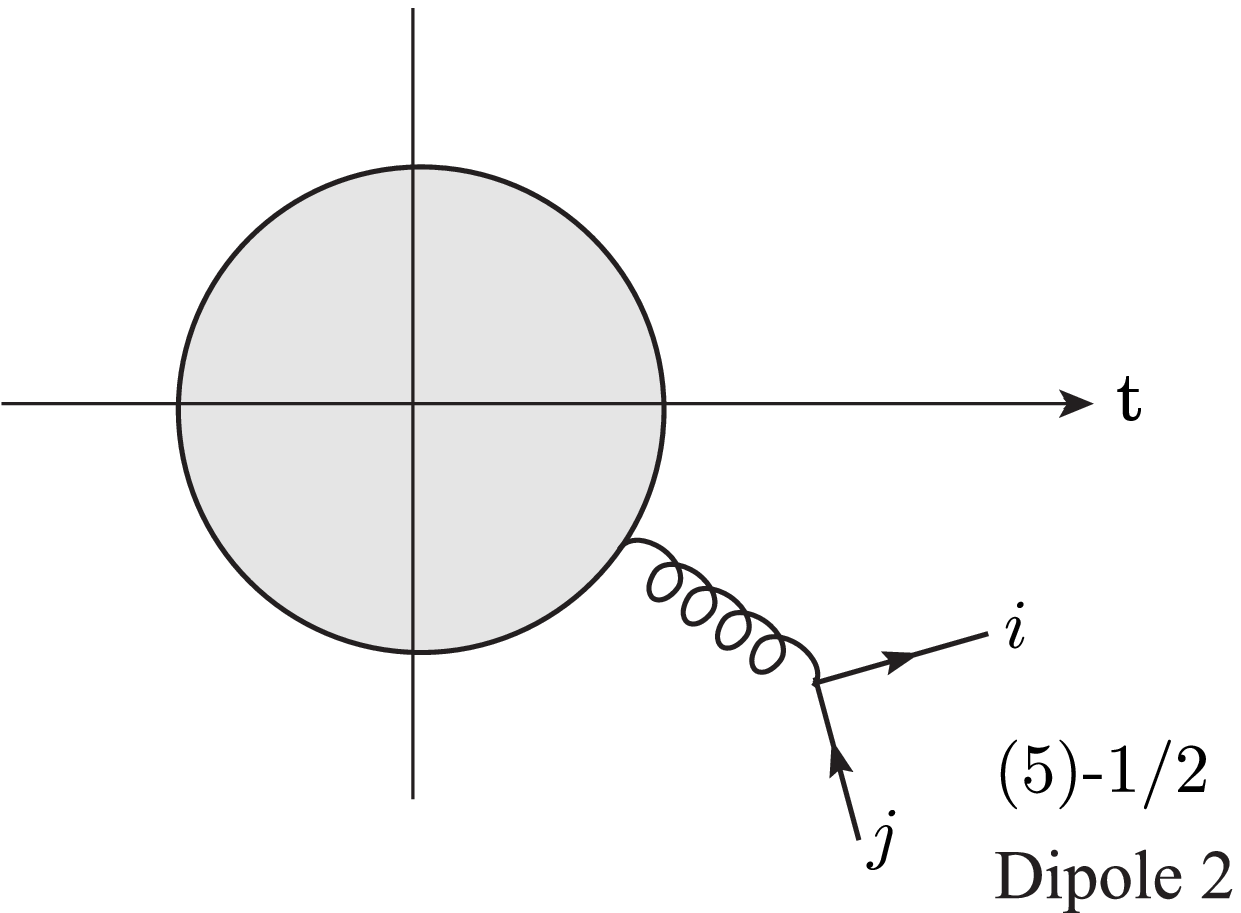}
\end{center} 
\caption{The classification of the 
integrated dipole term
 $\sigma_{\mbox{{\tiny D}}}({\tt dip}2)$
\label{fig_A6_int_dip2}}
\end{figure}
\leftline{\underline{{\tt Dipole\,2}}}
\begin{align}
&(5)\mbox{-}1:
\hspace{1mm}
\hat{\sigma}_{\mbox{{\tiny D}}}
(\,
\mbox{R}_{i},\,
{\tt dip}2,(5)\mbox{-}1) 
=
-\frac{A_{d}}{S_{\mbox{{\tiny R}}_{i}}} 
\cdot
\frac{1}{\mbox{C}_{\mbox{{\tiny A}}}} 
{\cal V}_{f\bar{f}} (\ep) 
\cdot 
\Phi(\mbox{B}2)_{d} \,
[ \, y_{emi}, \,y_{spe} ]\,,
\label{d251}
\\
&(5)\mbox{-}2:
\hspace{1mm}
\hat{\sigma}_{\mbox{{\tiny D}}}
(\,
\mbox{R}_{i},\,
{\tt dip}2,(5)\mbox{-}2,x_{a/b}) 
=
-\frac{A_{d}}{S_{\mbox{{\tiny R}}_{i}}} 
\int_{0}^{1}dx \,
\frac{1}{\mbox{C}_{\mbox{{\tiny A}}}}
{\cal V}_{f\bar{f}}(x\,;\ep)
\cdot 
\Phi_{a/b}(\mbox{B}2,x)_{d}\,
[ \, y_{emi}, \,y_{spe} ]\,.
\label{d252}
\end{align}
\vspace{3mm}

\leftline{\underline{Reconstruction of Dipole\,2}}
\begin{align}
&\hat{\sigma}_{\mbox{{\tiny D}}}
(\,
\mbox{R}_{i},\,
{\tt dip}2,(5)\mbox{-}1/2,
\,
{\cal V}_{f\bar{f}}
\,) 
=
-\frac{A_{d}}{S_{\mbox{{\tiny R}}_{i}}} 
\cdot
\frac{1}{\mbox{C}_{\mbox{{\tiny A}}}} 
{\cal V}_{f\bar{f}} (\ep) 
\cdot 
\Phi(\mbox{B}2)_{d} \,
[ \, y_{emi}, \,y_{spe} ]\,,
\label{d25ffb}
\\
&
\hat{\sigma}_{\mbox{{\tiny D}}}
(\,
\mbox{R}_{i},\,
{\tt dip}2,(5)\mbox{-}2,
\,h,x_{a/b}
\,) 
=
-\frac{A_{4}}{S_{\mbox{{\tiny R}}_{i}}} 
\int_{0}^{1}dx \,
\frac{\mbox{T}_{\mbox{{\tiny R}}}}
{\mbox{C}_{\mbox{{\tiny A}}}} 
\,\frac{2}{3}h(x)
\cdot 
\Phi_{a/b}(\mbox{B}2,x)_{4}\,
\langle \, y_{emi}, \,y_{spe} \rangle_{4}\,.
\label{d2h}
\end{align}
\vspace{7mm}

\leftline{\underline{Definition of the symbols}}
\vspace{2mm}
\noindent
Color-correlated Born squared amplitude\,:\,
\begin{align}
[ \, y_{emi}, \,y_{spe} ]
&=
(s_{y_{emi}, \,y_{spe}})^{-\ep} 
\cdot
\langle \, y_{emi}, \,y_{spe} \rangle_{d}\,\,,
\\
\langle \, y_{emi}, \,y_{spe} \rangle_{d/4}\,
&=
\langle \mbox{B}2 \ | 
\mbox{T}_{y_{emi}} \cdot 
\mbox{T}_{y_{spe}}
 | \ \mbox{B}2 \rangle_{d/4}\,.
\end{align}
Lorentz scalar\,:\,
$s_{y_{emi},\,y_{spe}} =2 \, \mbox{P}(y_{emi}) \cdot 
\mbox{P}(y_{spe})$\,.

\newpage
\subsection{Collinear subtraction term: $\sigma_{\mbox{{\tiny C}}}$ 
\label{ap_A_7}}
{\tt Input:$\mbox{R}_{i}$}
\begin{verbatim}
Dipole 1 (3), (4), 
Dipole 3 (6),  
Dipole 4 (7):
\end{verbatim}
\begin{figure}[h]
\begin{center}
\includegraphics[width=7cm]{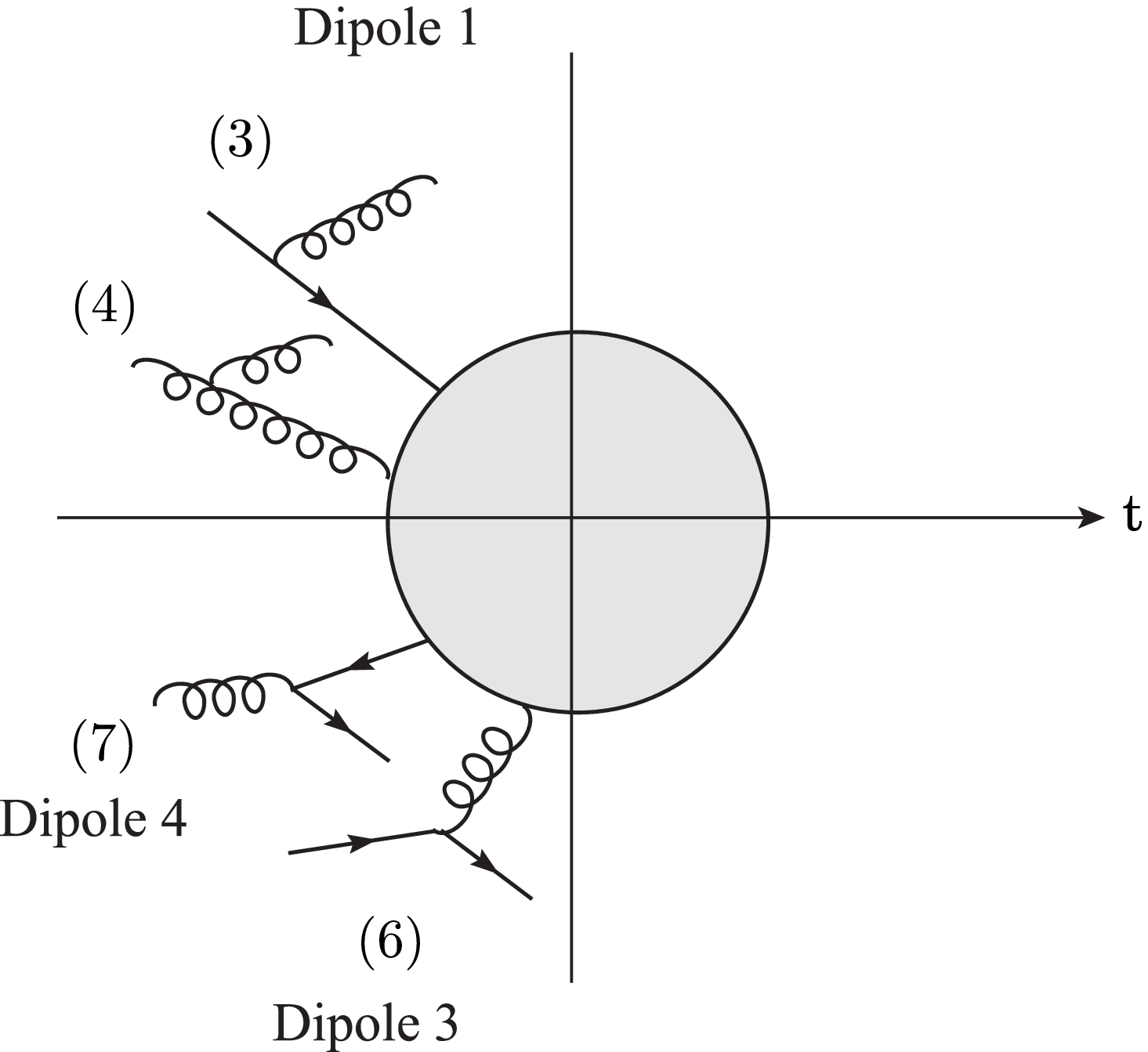}
\end{center} 
\caption{The classification of the 
collinear subtraction term
$\sigma_{\mbox{{\tiny C}}}$
\label{fig_A7_coll}}
\end{figure}
\begin{equation}
\hat{\sigma}_{\mbox{{\tiny C}}}(\mbox{R}_{i}) 
= \sum_{\mbox{{\tiny dip}}j} 
\hat{\sigma}_{\mbox{{\tiny C}}}(\mbox{R}_{i},{\tt dip}j)\,.
\end{equation}
\begin{equation}
\hat{\sigma}_{\mbox{{\tiny C}}}(\mbox{R}_{i},{\tt dip}j) 
=\hat{\sigma}_{\mbox{{\tiny C}}}(\mbox{R}_{i},{\tt dip}j,\,x_{a}) 
+
\hat{\sigma}_{\mbox{{\tiny C}}}(\mbox{R}_{i},{\tt dip}j,\,x_{b})\,.
\end{equation}
\begin{align}
\hat{\sigma}_{\mbox{{\tiny C}}}(\mbox{R}_{i},{\tt dip}j,\,x_{a/b}) 
&=\frac{\al}{2\pi}\frac{1}{\Gamma(1-\ep)}
\int_{0}^{1}dx 
\biggl[
\frac{1}{\ep}
\biggl(
\frac{4\pi \mu^{2}}{\mu_{F}^{2}}
\biggr)^{\ep}
P^{
\mbox{{\tiny F}}(x_{a/b})\mbox{{\tiny F}}(y_{emi})
}(x) 
\biggr]
\nonumber\\
&\hspace{60mm} \times\frac{1}{S_{\mbox{{\tiny B}}j}}
\Phi_{a/b}(\mbox{B}_{j},x)_{d}\,
\langle \mbox{B}j \rangle\,, \\
&= 
\frac{A_{d}}{S_{\mbox{{\tiny B}}j}}
\int_{0}^{1}dx\,
\Bigl(\,
\frac{1}{\ep}
-
\ln \mu_{F}^{2}
\Bigr)
P^{
\mbox{{\tiny F}}(x_{a/b})\mbox{{\tiny F}}(y_{emi})
}(x) 
\cdot
\Phi_{a/b}(\mbox{B}_{j},x)_{d}\,
\langle \mbox{B}j \rangle\,.
\end{align}
\vspace{5mm}

\leftline{\underline{Definition of the symbols}}
\vspace{2mm}
\noindent
Born squared amplitude\,:\,
$\langle \mbox{B}j \rangle=
\langle \mbox{B}j \,|  \mbox{B}j \rangle_{d}$\,.
\vspace{3mm}

\noindent
Splitting functions\,:\,
\begin{equation}
P^{
\mbox{{\tiny F}}(x_{a/b})\mbox{{\tiny F}}(y_{emi})
}(x)
=
\left\{
\begin{array}{ll}
P^{ff}(x)
& : \mbox{Dipole}\,1 \ (3)\,,
\\
P^{gg}(x)
& : \mbox{Dipole}\,1 \ (4)\,,
\\
P^{fg}(x)
& : \mbox{Dipole}\,3 \ (6)\,,
\\
P^{gf}(x)
& : \mbox{Dipole}\,4 \ (7)\,.
\end{array}
\right.
\label{collap}
\end{equation}

\newpage
\subsection{I term: $\sigma_{\mbox{{\tiny I}}}$ \label{ap_A_8}}
{\tt Input:\,$\mbox{B}1(\mbox{R}_{i})$}
\begin{verbatim}
Legs: (1)-1/2, (2)-1/2, (3)-1/2, (4)-1/2.
\end{verbatim}
\begin{figure}[h]
\begin{center}
\includegraphics[width=7cm]{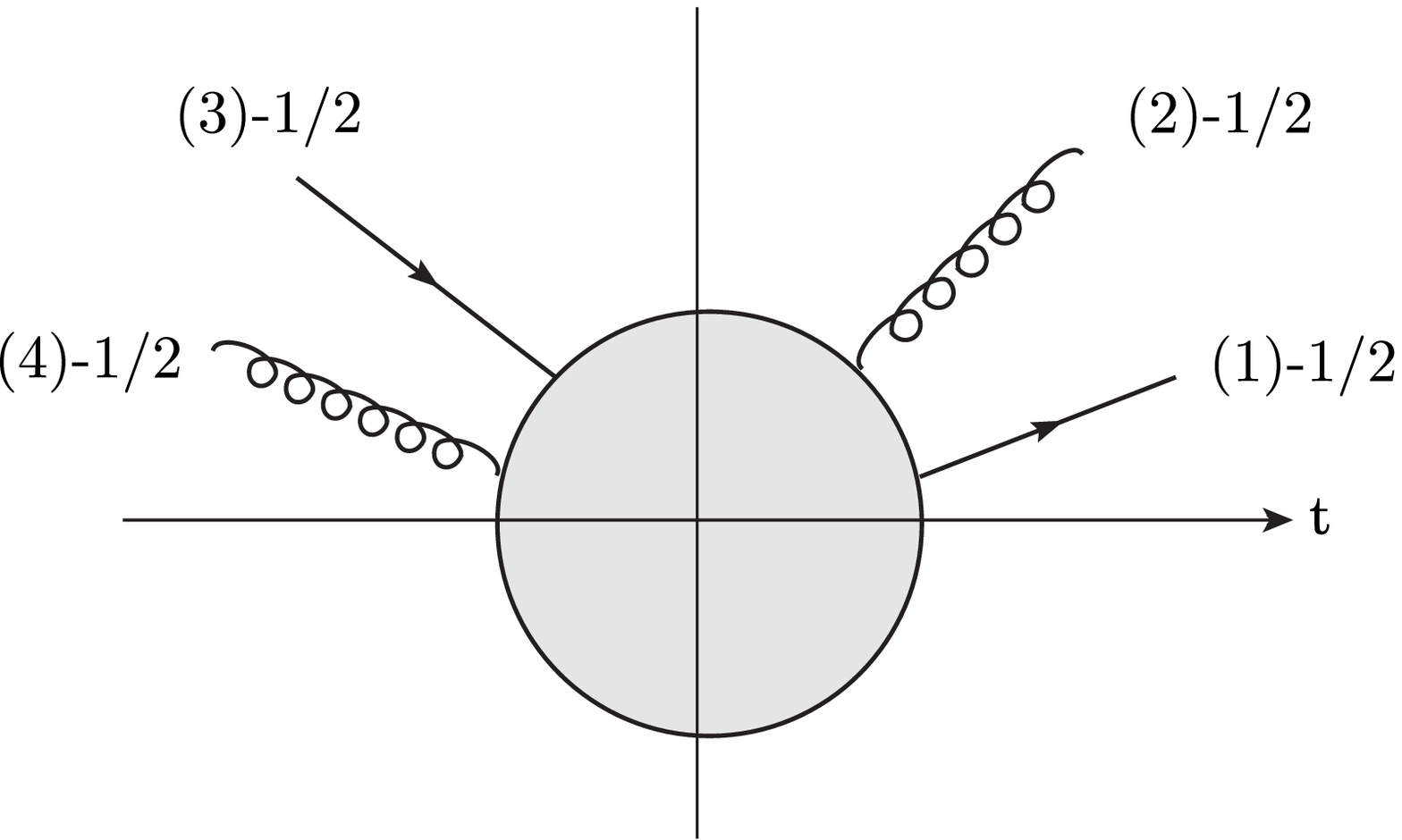}
\end{center} 
\caption{The classification of the 
I term
 $\sigma_{\mbox{{\tiny I}}}$
\label{fig_A8_I}}
\end{figure}
\begin{equation}
\hat{\sigma}_{\mbox{{\tiny I}}}(\mbox{R}_{i})_{IK}
= -\frac{A_{d}}{S_{\mbox{{\tiny B1}} }} \cdot
\frac{1}{\mbox{T}_{\mbox{{\tiny F}}(I)}^{2}}
{\cal V}_{\mbox{{\tiny F}}(I)}
\cdot
\ \Phi(\mbox{B}1)_{d} \,
[\,I,K \,] \,.
\end{equation}
\vspace{7mm}

\leftline{\underline{Definition of the symbols}}
\vspace{2mm}
\noindent
Common factor\,:\,$A_{d}$ is the same as in Eq.\,(\ref{ad}).

\noindent
Universal singular functions\,:\,
\begin{equation}
\frac{1}{\mbox{T}_{\mbox{{\tiny F}}(I)}^{2}}
{\cal V}_{\mbox{{\tiny F}}(I)}
=
\left\{
\begin{array}{ll}
\frac{1}{\mbox{C}_{\mbox{{\tiny F}}}}
{\cal V}_{f}(\ep)
=
\frac{1}{\mbox{C}_{\mbox{{\tiny F}}}}
{\cal V}_{fg}(\ep)
& : (1),(3)-1/2\,,
\\
\frac{1}{\mbox{C}_{\mbox{{\tiny A}}}} 
{\cal V}_{g}(\ep)
=
\frac{1}{\mbox{C}_{\mbox{{\tiny A}}}}
\Bigl(
\half {\cal V}_{gg} (\ep) + 
N_{f}{\cal V}_{f\bar{f}}(\ep)
\Bigr)
& : (2),(4)-1/2\,.
\end{array}
\right.
\end{equation}
Color-correlated Born squared amplitude\,:\,
\begin{equation}
[ \, I, \, K \,]
=s_{IK}^{-\ep} 
\,.
\langle \mbox{B}1 \ | 
\mbox{T}_{I} \cdot 
\mbox{T}_{K}
 | \ \mbox{B}1 \rangle_{d}\,. 
\label{sqbk4}
\end{equation}
\vspace{4mm}

\leftline{\underline{$\hat{\sigma}_{\mbox{{\tiny I}}}(
\mbox{R}_{i},(2)\,
\mbox{-}1/2,N_{f}{\cal V}_{f\bar{f}}
)_{IK}$}}
\begin{equation}
\hat{\sigma}_{\mbox{{\tiny I}}}(
\mbox{R}_{i},(2)\,
\mbox{-}1/2,N_{f}{\cal V}_{f\bar{f}}
)_{IK} 
= -\frac{A_{d}}{S_{\mbox{{\tiny B1}}}} \cdot
\frac{N_{f}}{\mbox{C}_{\mbox{{\tiny A}}}}
{\cal V}_{f\bar{f}}(\ep)
\cdot
\ \Phi(\mbox{B}1)_{d} \,
[\,I,K \,] \,.
\end{equation}
\newpage
\subsection{P term: $\sigma_{\mbox{{\tiny P}}}$ \label{ap_A_9}}
{\tt Input:$\mbox{R}_{i}$}
\begin{verbatim}
Dipole 1 (3)-1/2, (4)-1/2, 
Dipole 3 (6)-1/2,  
Dipole 4 (7)-1/2:
\end{verbatim}
\begin{figure}[h]
\begin{center}
\includegraphics[width=7cm]{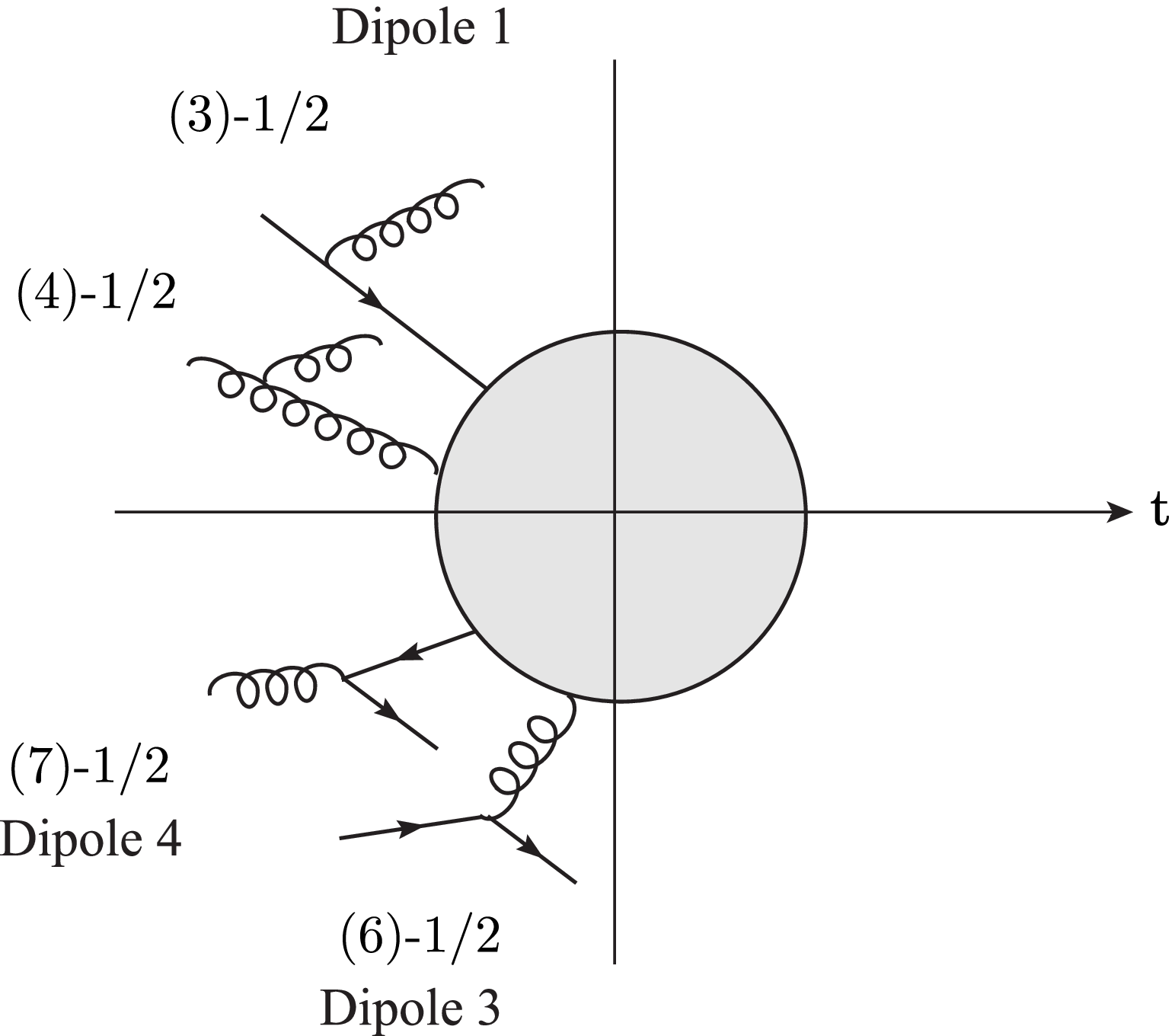}
\end{center} 
\caption{The classification of the 
P term
$\sigma_{\mbox{{\tiny P}}}$.
\label{fig_A9_P}}
\end{figure}
%
\begin{align}
\hat{\sigma}_{\mbox{{\tiny P}}}(\mbox{R}_{i},\, {\tt dip}j,
x_{a/b}) 
&= 
\frac{A_{4}}{S_{\mbox{{\tiny B}}_{j} }}
\int_{0}^{1}dx \,
\frac{1}{\mbox{T}_{\mbox{{\tiny F}}(y_{emi})}^{2}} \,
P^{
\mbox{{\tiny F}}(x_{a/b})\mbox{{\tiny F}}(y_{emi})
}(x) \cdot
\ln \frac{\mu_{F}^{2}}{x\,s_{x_{a/b},y_{spe}}}
\nonumber \\
& \hspace{25mm}
\cdot \Phi_{a/b}(\mbox{B}_{j},x)_{4} \,
\langle \, y_{emi}, \,y_{spe} \rangle\,.
\end{align}
\vspace{7mm}

\leftline{\underline{Definition of the symbols}}
\vspace{2mm}
\noindent
Common factor\,:\,
\begin{equation}
A_{4}=\frac{\al}{2\pi}\,.
\end{equation}
Splitting functions\,:\,
$P^{
\mbox{{\tiny F}}(x_{a/b})\,\mbox{{\tiny F}}(y_{emi})
}(x)/
\mbox{T}_{\mbox{{\tiny F}}(y_{emi})}^{2}
$ is the same as in Eq.\,(\ref{sigdpp}).
\\
Lorentz scalar\,:\,
$s_{x_{a/b},\,y_{spe}}$ is the same as in Eq.\,(\ref{sigdplo}).
\\
Color-correlated Born squared amplitude\,:\,
\begin{equation}
\langle \, y_{emi}, \,y_{spe} \rangle\,
=
\langle \mbox{B}j \ | 
\mbox{T}_{y_{emi}} \cdot 
\mbox{T}_{y_{spe}}
 | \ \mbox{B}j \rangle_{4}\,.
\label{ccbsa4}
\end{equation}

\newpage
\subsection{K term: $\sigma_{\mbox{{\tiny K}}}$ \label{ap_A_10}}
{\tt Input:$\mbox{R}_{i}$}
\begin{verbatim}
Dipole 1 (3)-0/1/2, (4)-0/1/2, 
Dipole 3 (6)-0/2,  
Dipole 4 (7)-0/2:
\end{verbatim}
\begin{figure}[h]
\begin{center}
\includegraphics[width=7cm]{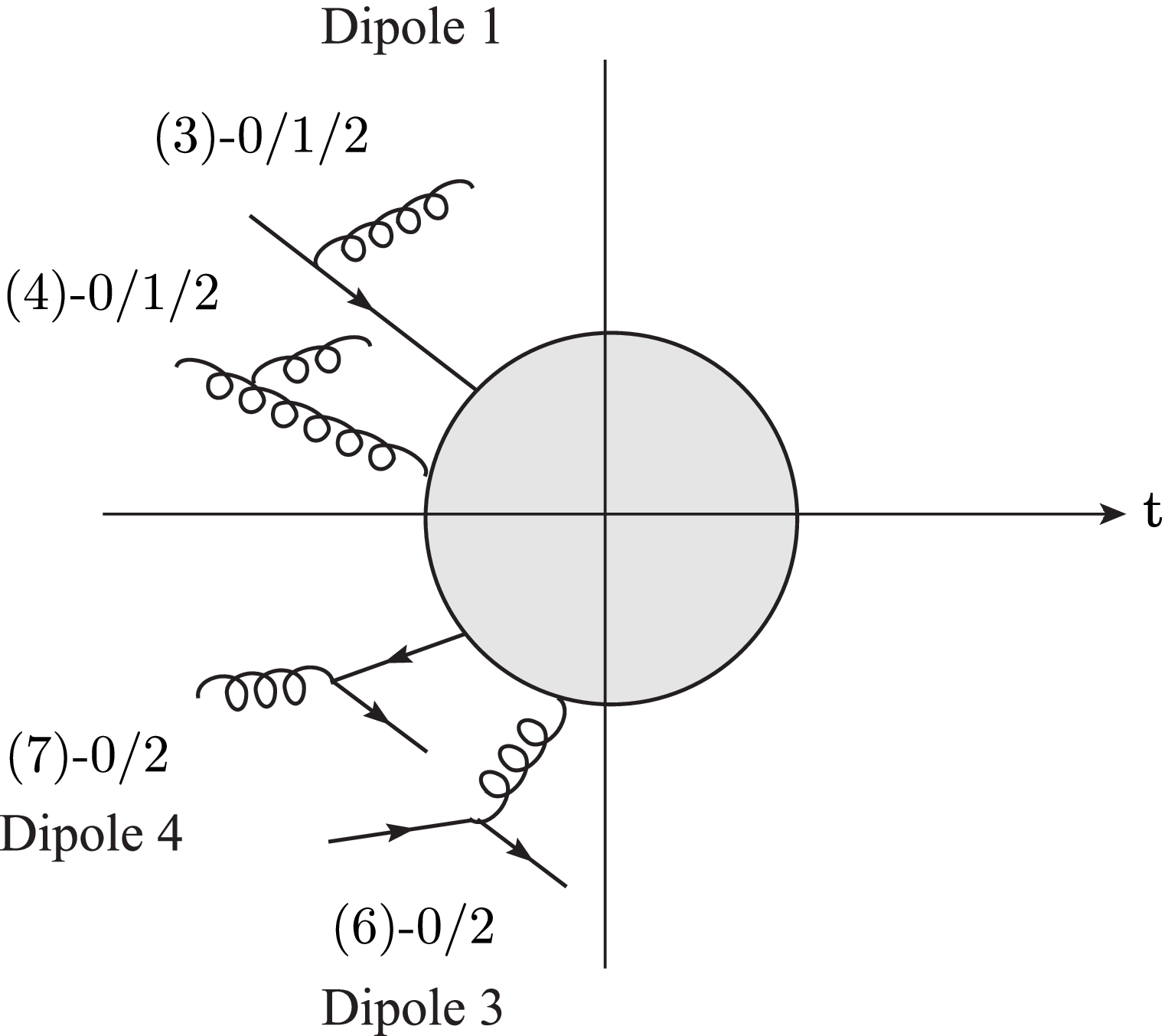}
\end{center} 
\caption{The classification of the 
K term
$\sigma_{\mbox{{\tiny K}}}$
\label{fig_A10_K}}
\end{figure}
\leftline{\underline{ {\tt Dipole\,1} (3)/(4)}} 
\begin{align}
\hat{\sigma}_{\mbox{{\tiny K}}}(\mbox{R}_{i},{\tt dip}1,\,
(3)/(4)\mbox{-}0,x_{a}) 
&= 
\frac{A_{4}}{S_{\mbox{{\tiny B}}_{1}}}
\int_{0}^{1}dx \
\overline{\mbox{K}}^{ff/gg}(x)
\cdot
\Phi_{a}(\mbox{B}_{1},x)_{4} \,
\langle \mbox{B}1 \rangle\,, 
\label{a10eq1}
\\
\hat{\sigma}_{\mbox{{\tiny K}}}(\mbox{R}_{i},{\tt dip}1,\,
(3)/(4)\mbox{-}1,x_{a}) 
&= 
\frac{A_{4}}{S_{\mbox{{\tiny B}}_{1}}}
\int_{0}^{1}dx \
\frac{\gamma_{\mbox{{\tiny F}}(y_{spe})}}
{\mbox{T}_{\mbox{{\tiny F}}(y_{spe})}^{2}} \, h(x)
\cdot
\Phi_{a}(\mbox{B}_{1},x)_{4} \,
\langle \, y_{emi}, \,y_{spe} \rangle\,,
\label{a10eq2}
\\
\hat{\sigma}_{\mbox{{\tiny K}}}(\mbox{R}_{i},{\tt dip}1,\,
(3)/(4)\mbox{-}2,x_{a}
) 
&= 
\frac{A_{4}}{S_{\mbox{{\tiny B}}_{1}}}
\int_{0}^{1}dx \
\frac{-1}{\mbox{C}_{\mbox{{\tiny F}}/\mbox{{\tiny A}}}}
\widetilde{\mbox{K}}^{ff/gg}(x)
\cdot
\Phi_{a}(\mbox{B}_{1},x)_{4} \,
\langle \, y_{emi}, \,y_{spe} \rangle\,.
\label{a10eq3}
\end{align}

\leftline{\underline{ {\tt Dipole\,3/4} (6)/(7)}} 
\begin{align}
\hat{\sigma}_{\mbox{{\tiny K}}}(\mbox{R}_{i},{\tt dip}3/4,\,
(6)/(7)\mbox{-}0,x_{a}) 
&= 
\frac{A_{4}}{S_{\mbox{{\tiny B}}_{3/4}}}
\int_{0}^{1}dx \
\overline{\mbox{K}}^{fg/gf}(x)
\cdot
\Phi_{a}(\mbox{B}_{3/4},x)_{4} \,
\langle \mbox{B}3/4 \rangle\,, 
\label{a10eq5}
\\
\hat{\sigma}_{\mbox{{\tiny K}}}(\mbox{R}_{i},{\tt dip}3/4,\,
(6)/(7)\mbox{-}2,x_{a}) 
&= 
\frac{A_{4}}{S_{\mbox{{\tiny B}}_{3/4}}}
\int_{0}^{1}dx \
\frac{-1}{\mbox{C}_{\mbox{{\tiny A}}/\mbox{{\tiny F}}}}
\widetilde{\mbox{K}}^{fg/gf}(x) 
\cdot
\Phi_{a}(\mbox{B}_{3/4},x)_{4} \,
\langle \, y_{emi}, \,y_{spe} \rangle\,.
\label{a10eq7}
\end{align}
The cross sections with leg-b\,($x_{b}$),
$\hat{\sigma}_{\mbox{{\tiny K}}}(\mbox{R}_{i},{\tt dip}j,
x_{b})$,
are obtained by the replacements
$\Phi_{a}(\mbox{B}j,x)_{4}$ 
$\to$
$\Phi_{b}(\mbox{B}j,x)_{4}$
in the formulae for leg-a\,($x_{a}$).
\vspace{7mm}

\leftline{\underline{Definition of the symbols}}
\vspace{2mm}
\noindent
Color-correlated Born squared amplitude\,:\,
$\langle \, y_{emi}, \,y_{spe} \rangle$
is the same as in Eq.\,(\ref{ccbsa4}).
\vspace{3mm}

\noindent
Relations of the functions,
$\overline{\mbox{K}}^{ff/gg}(x)$,
${\cal V}_{other}^{f,f/g,g}(x\,;\ep)$,
and
$g(x)$\,:\,
\begin{align}
\overline{\mbox{K}}^{ff/gg}(x)
&={\cal V}_{other}^{f,f/g,g}(x\,;\ep) 
+ \mbox{C}_{\mbox{{\tiny F}}/\mbox{{\tiny A}}} \, g(x)\,,
\label{kbaroth1}
\\
\overline{\mbox{K}}^{fg/gf}(x) 
&={\cal V}_{other}^{f,g/g,f}(x\,;\ep)\,.
\label{kbaroth2}
\end{align}
The factor 
$\mbox{T}_{\mbox{{\tiny F}}(y_{spe})}^{2}/
\gamma_{\mbox{{\tiny F}}(y_{spe})}$\,:\,
\begin{equation}
\frac{\gamma_{\mbox{{\tiny F}}(y_{spe})}}
{\mbox{T}_{\mbox{{\tiny F}}(y_{spe})}^{2}}
=
\left\{
\begin{array}{ll}
\frac{3}{2}
& : \mbox{F}(y_{spe})=\mbox{quark} \,,
\\
\frac{11}{6}
- \frac{2}{3}
\frac{\mbox{T}_{\mbox{{\tiny R}}} \mbox{$N$}_{f}}
{\mbox{C}_{\mbox{{\tiny A}}}}
& : \mbox{F}(y_{spe})=\mbox{gluon} \,.
\end{array}
\right.
\label{gamts}
\end{equation}
\vspace{7mm}

\leftline{\underline{
$\sigma_{\mbox{{\tiny K}}}(\mbox{R}_{i}, {\tt dip}1,
(3)/(4)\,\mbox{-}1,N_{f}h,x_{a/b})$}}
\begin{equation}
\sigma_{\mbox{{\tiny K}}}(\mbox{R}_{i}, {\tt dip}1,
(3)/(4)\,\mbox{-}1,N_{f}h,x_{a/b})
= 
\frac{A_{4}}{S_{\mbox{{\tiny B}}_{1}}}
\int_{0}^{1}dx \
\biggl(-\frac{2}{3}
\frac{\mbox{T}_{\mbox{{\tiny R}}} \mbox{$N$}_{f}}
{\mbox{C}_{\mbox{{\tiny A}}}}
\biggr)
\, h(x)
\cdot
\Phi_{a/b}(\mbox{B}_{1},x)_{4} \,
\langle \, y_{emi}, \,y_{spe} \rangle\,.
\label{a10eqla}
\end{equation}


\clearpage




\section{Summary for the dijet process \label{ap_B}}
\subsection{$\hat{\sigma}_{\mbox{{\tiny subt}}}(\mbox{R}_{1u})$  
\label{ap_B_1}}
%
%
\begin{table}[h!]
  \centering
\begin{align}
& \fbox{Step1} \ \ \ 
\hat{\sigma}_{\mbox{{\tiny D}}}(\mbox{R}_{i}) 
=
-\frac{A_{d}}{S_{\mbox{{\tiny R}}_{i}}} 
\cdot
\bigl(\mbox{Factor 1} \bigr)
\cdot
\Phi_{a}(\mbox{R}_{i}:\mbox{B}_{j},x)_{d} \,
[ \, y_{emi}, \,y_{spe} ]\,.
\nonumber\\
&\hat{\sigma}_{\mbox{{\tiny D}}}
(\mbox{R}_{1u} = u\bar{u} \to u\bar{u}g): 
\ \ S_{\mbox{{\tiny R}}_{1}}=1,
\nonumber\\
&
\begin{array}{|c|c|c|c|c|c|} \hline
{\tt Dip}\,j
& \mbox{B}j 
& \mbox{\small Splitting} 
& (y_{a},y_{b};y_{1},y_{2}) 
& \mbox{Factor 1}
& \Phi(\mbox{B}_{j})
\,[y_{emi}, y_{spe}]
\\[4pt] \hline
& & & & &  \\[-12pt]
{\tt Dip\,1}
& u\bar{u} \to u\bar{u}
& (1)-1   
& 1.\,(a,b\,;\widetilde{13},\widetilde{2} ) 
& {\cal V}_{fg}/\mbox{C}_{\mbox{{\tiny F}}} 
& \Phi(\mbox{B}1) \, [1,2]
\\[4pt]  
&  
&   
& 2.\,(a,b\,;\widetilde{1},\widetilde{23}) 
&
& \Phi(\mbox{B}1) \, [2,1]
\\[4pt]  
&  
& (1)-2  
& 3.\,(\widetilde{a},b\,;\widetilde{13},2) 
& \int dx 
{\cal V}_{fg}(x)/\mbox{C}_{\mbox{{\tiny F}}} 
& \Phi_{a}(\mbox{B}1)  \, [1,a]
\\[4pt]  
&  
&   
& 4.\,(a,\widetilde{b}\,;\widetilde{13},2) 
&
& \Phi_{b}(\mbox{B}1)  \, [1,b]
\\[4pt]  
&  
&  
& 5.\,(\widetilde{a},b\,;1,\widetilde{23}) 
&
& \Phi_{a}(\mbox{B}1)  \, [2,a]
\\[4pt]  
&  
&   
& 6.\,(a,\widetilde{b}\,;1,\widetilde{23}) 
&
& \Phi_{b}(\mbox{B}1)  \, [2,b]
\\[4pt]  
&  
& (3)-1  
& 7.\,(\widetilde{a3},b\,;\widetilde{1},2) 
& \int dx  
{\cal V}^{f,f} (x)/\mbox{C}_{\mbox{{\tiny F}}}
& \Phi_{a}(\mbox{B}1)  \, [a,1]
\\[4pt]  
&  
&   
& 8.\,(\widetilde{a3},b\,;1,\widetilde{2}) 
&
& \Phi_{a}(\mbox{B}1)  \, [a,2]
\\[4pt]  
&  
&   
& 9.\,(a,\widetilde{b3}\,;\widetilde{1},2) 
&
& \Phi_{b}(\mbox{B}1)  \, [b,1]
\\[4pt]  
&  
&   
& 10.\,(a,\widetilde{b3}\,;1,\widetilde{2})
&
& \Phi_{b}(\mbox{B}1)  \, [b,2]
\\[4pt]  
&  
& (3)-2  
& 11.\,(\widetilde{a3},\widetilde{b}\,;1,2) 
& \int dx
{\widetilde {\cal V}}^{f,f} (x)
/\mbox{C}_{\mbox{{\tiny F}}}
& \Phi_{a}(\mbox{B}1)  \, [a,b]
\\[4pt]  
&  
&   
& 12.\,(\widetilde{a},\widetilde{b3}\,;1,2) 
&
& \Phi_{b}(\mbox{B}1)  \, [b,a]
\\[4pt]  \hline
& & & & &  \\[-12pt]
{\tt Dip\,2u}  
& u\bar{u} \to gg
& (5)-1  
& 13.\,(a,b\,;\widetilde{12},\widetilde{3}) 
& {\cal V}_{f\bar{f}}/ \mbox{C}_{\mbox{{\tiny A}}}
& \Phi(\mbox{B}2u) \, [1,2]
\\[4pt]  
&  
& (5)-2    
& 14.\,(\widetilde{a},b\,;\widetilde{12},3) 
& \int dx {\cal V}_{f\bar{f}}(x)/ \mbox{C}_{\mbox{{\tiny A}}}
& \Phi_{a}(\mbox{B}2u)  \, [1,a]
\\[4pt]  
&  
&   
& 15.\,(a,\widetilde{b}\,;\widetilde{12},3) 
&
& \Phi_{b}(\mbox{B}2u)  \, [1,b]
\\[4pt]  \hline
& & & & &  \\[-12pt]
{\tt Dip\,3u}  
&  g\bar{u} \to \bar{u}g
& (6)-1    
& 16.\,(\widetilde{a1},b\,;\widetilde{2},3) 
& \int dx {\cal V}^{f,g} (x)/\mbox{C}_{\mbox{{\tiny A}}}
& \Phi_{a}(\mbox{B}3u)  \, [a,1]
\\[4pt]  
&  
&   
& 17.\,(\widetilde{a1},b\,;2,\widetilde{3}) 
& 
& \Phi_{a}(\mbox{B}3u)  \, [a,2]
\\[4pt]  
&  
& (6)-2  
& 18.\,(\widetilde{a1},\widetilde{b}\,;2,3)
& \int dx {\widetilde {\cal V}}^{f,g} (x)/
\mbox{C}_{\mbox{{\tiny A}}}
& \Phi_{a}(\mbox{B}3u)  \, [a,b]
\\[4pt]  \hline
& & & & &  \\[-12pt]
{\tt Dip\,3\bar{u}}  
& ug \to ug
& (6)-1   
& 19.\,(a,\widetilde{b2}\,;\widetilde{1},3)  
& \int dx {\cal V}^{f,g} (x)/\mbox{C}_{\mbox{{\tiny A}}}
& \Phi_{b}(\mbox{B}3\bar{u})  \, [b,1]
\\[4pt]  
&  
&   
& 20.\,(a,\widetilde{b2}\,;1,\widetilde{3})  
&
& \Phi_{b}(\mbox{B}3\bar{u})  \, [b,2]
\\[4pt]  
&  
& (6)-2   
& 21.\,(\widetilde{a},\widetilde{b2}\,;1,3)  
& \int dx {\widetilde {\cal V}}^{f,g} (x)/
\mbox{C}_{\mbox{{\tiny A}}}
& \Phi_{b}(\mbox{B}3\bar{u})  \, [b,a]
\\[4pt] \hline
\end{array} \nonumber
\end{align}
\caption{
Summary table of
$\hat{\sigma}_{\mbox{{\tiny D}}}(\mbox{R}_{1u})$.
The universal singular functions are 
abbreviated as
${\cal V}_{ij}(\ep)={\cal V}_{ij}$,
${\cal V}_{ij}(x\,;\ep)={\cal V}_{ij}(x)$,
${\cal V}^{a,a'} (x;\ep)={\cal V}^{a,a'}(x)$, and
${\widetilde {\cal V}}^{a,a'} (x;\ep)=
{\widetilde {\cal V}}^{a,a'}(x)$\,.
\label{ap_B_1_tab1}}
\end{table}
\newpage
\rightline{
\underline{
$\hat{\sigma}_{\mbox{{\tiny subt}}}(\mbox{R}_{1u}=u\bar{u} \to u\bar{u}g)$  
}
}
\begin{enumerate}
\item[\fbox{Step2}] 
\begin{equation}
\hat{\sigma}_{\mbox{{\tiny D}}}
(\,
\mbox{R}_{1},\,
\mbox{I}
\,)
-
\hat{\sigma}_{\mbox{{\tiny I}}}
(\mbox{R}_{1})=0\,.
\label{r1st2}
\end{equation}
\item[\fbox{Step3}]
\begin{equation}
\hat{\sigma}_{\mbox{{\tiny D}}}
(\,
\mbox{R}_{1},\,
\mbox{P}
\,)
+
\hat{\sigma}_{\mbox{{\tiny C}}}(\mbox{R}_{1}) 
-
\hat{\sigma}_{\mbox{{\tiny P}}}
(\mbox{R}_{1})=0\,,
\label{r1st3}
\end{equation}
which is separated into three relations 
for Dipoles 1, 3$u$, and 3$\bar{u}$ as
\begin{equation}
\hat{\sigma}_{\mbox{{\tiny D}}}
(\,
\mbox{R}_{1},\,
\mbox{P},\,{\tt dip}j
\,)
+
\hat{\sigma}_{\mbox{{\tiny C}}}(\mbox{R}_{1},\,
{\tt dip}j
) 
-
\hat{\sigma}_{\mbox{{\tiny P}}}
(\mbox{R}_{1},
\,{\tt dip}j)=0\,.
\label{r1st3ind}
\end{equation}
\item[\fbox{Step4}]
\begin{equation}
\hat{\sigma}_{\mbox{{\tiny D}}}
(\,
\mbox{R}_{1},\,
\mbox{K}
\,)
-
\hat{\sigma}_{\mbox{{\tiny K}}}
(\mbox{R}_{1})=0\,,
\label{r1st4}
\end{equation}
which is separated into three relations 
for Dipoles 1, 3$u$, and 3$\bar{u}$ as
\begin{equation}
\hat{\sigma}_{\mbox{{\tiny D}}}
(\,
\mbox{R}_{1},\,
\mbox{K},\,{\tt dip}j
\,)
-
\hat{\sigma}_{\mbox{{\tiny K}}}
(\mbox{R}_{1},
\,{\tt dip}j)=0\,.
\label{r1st41}
\end{equation}
\item[\fbox{Step5}]
\begin{equation}
\hat{\sigma}_{subt}(\mbox{R}_{1})=
\hat{\sigma}_{\mbox{{\tiny D}}}(\mbox{R}_{1},\,{\tt dip}2)\,.
\label{r1st51}
\end{equation}
\begin{equation}
\hat{\sigma}_{\mbox{{\tiny D}}}(\mbox{R}_{1},\,{\tt dip}2)
=
\hat{\sigma}_{\mbox{{\tiny D}}}
(\,
\mbox{R}_{1},\,
{\tt dip}2,(5)\mbox{-}1/2,
\,
{\cal V}_{f\bar{f}}
\,)
+
\hat{\sigma}_{\mbox{{\tiny D}}}
(\,
\mbox{R}_{1},\,
{\tt dip}2,(5)\mbox{-}2,
\,
h)\,.
\end{equation}
\begin{align}
\hat{\sigma}_{\mbox{{\tiny D}}}
(
\mbox{R}_{1},\,
{\tt dip}2,(5)\mbox{-}1/2,
\,
{\cal V}_{f\bar{f}}
\,)
&=-\frac{A_{d}}{S_{\mbox{{\tiny R}}_{1}}} 
\cdot
\frac{1}{\mbox{C}_{\mbox{{\tiny A}}}} 
{\cal V}_{f\bar{f}} (\ep) 
\cdot 
\Phi(\mbox{B}2)_{d} \,
\bigl(
\,
[1,2]+
[1,a]+
[1,b]
\,
\bigr)\,.
\\
\hat{\sigma}_{\mbox{{\tiny D}}}
(
\mbox{R}_{1},\,
{\tt dip}2,(5)\mbox{-}2,
\,
h)
&=
-\frac{A_{4}}{S_{\mbox{{\tiny R}}_{1}}} 
\int_{0}^{1}dx \,
\frac{\mbox{T}_{\mbox{{\tiny R}}}}
{\mbox{C}_{\mbox{{\tiny A}}}} 
\,\frac{2}{3}h(x)
\ \times
\nonumber\\
& \hspace{20mm}
\Bigl(
\Phi_{a}(\mbox{B}2,x)_{4}\,
\langle 1,a \rangle
+
\Phi_{b}(\mbox{B}2,x)_{4}\,
\langle 1,b \rangle
\Bigr)\,.
\label{r1st5last}
\end{align}
\end{enumerate}
%
\clearpage
\subsection{$\hat{\sigma}_{\mbox{{\tiny subt}}}(\mbox{R}_{2u})$  
\label{ap_B_2}}
%
%
\begin{table}[h!]
  \centering
\begin{align}
&\fbox{Step1} \ \ \
\hat{\sigma}_{\mbox{{\tiny D}}}
(\mbox{R}_{2u} = uu \to uug\,): 
\ \ S_{\mbox{{\tiny R}}_{2}}=2,
\nonumber\\
&
\begin{array}{|c|c|c|c|c|c|} \hline
{\tt Dip}\,j
& \mbox{B}j 
& \mbox{\small Splitting} 
& (y_{a},y_{b}:y_{1},y_{2}) 
& \mbox{Factor 1}
&\Phi(\mbox{B}_{j})
\,[y_{emi}, y_{spe}]
\\[4pt] \hline
& & & & & \\[-12pt]
{\tt Dip\,1}  
& uu \to uu
& (1)-1   
& 1.\,(a,b\,;\widetilde{13},\widetilde{2} ) 
& {\cal V}_{fg}/\mbox{C}_{\mbox{{\tiny F}}}
& \Phi(\mbox{B}1) \, [1,2]
\\[4pt]  
&  
&   
& 2.\,(a,b\,;\widetilde{1},\widetilde{23}) 
&
& \Phi(\mbox{B}1) \, [2,1]
\\[4pt]  
&  
& (1)-2  
& 3.\,(\widetilde{a},b\,;\widetilde{13},2) 
& \int dx 
{\cal V}_{fg}(x)/\mbox{C}_{\mbox{{\tiny F}}}
& \Phi_{a}(\mbox{B}1)  \, [1,a]
\\[4pt]  
&  
&   
& 4.\,(a,\widetilde{b}\,;\widetilde{13},2) 
& 
& \Phi_{b}(\mbox{B}1)  \, [1,b]
\\[4pt]  
&  
&  
& 5.\,(\widetilde{a},b\,;1,\widetilde{23}) 
&
& \Phi_{a}(\mbox{B}1)  \, [2,a]
\\[4pt]  
&  
&   
& 6.\,(a,\widetilde{b}\,;1,\widetilde{23}) 
&
& \Phi_{b}(\mbox{B}1)  \, [2,b]
\\[4pt]  
&  
& (3)-1  
& 7.\,(\widetilde{a3},b\,;\widetilde{1},2) 
& \int dx  
{\cal V}^{f,f} (x)/\mbox{C}_{\mbox{{\tiny F}}}
& \Phi_{a}(\mbox{B}1)  \, [a,1]
\\[4pt]  
&  
&   
& 8.\,(\widetilde{a3},b\,;1,\widetilde{2}) 
&
& \Phi_{a}(\mbox{B}1)  \, [a,2]
\\[4pt]  
&  
&   
& 9.\,(a,\widetilde{b3}\,;\widetilde{1},2) 
&
& \Phi_{b}(\mbox{B}1)  \, [b,1]
\\[4pt]  
&  
&   
& 10.\,(a,\widetilde{b3}\,;1,\widetilde{2})
&
& \Phi_{b}(\mbox{B}1)  \, [b,2]
\\[4pt]  
&  
& (3)-2  
& 11.\,(\widetilde{a3},\widetilde{b}\,;1,2) 
& \int dx
{\widetilde {\cal V}}^{f,f} (x)
/\mbox{C}_{\mbox{{\tiny F}}}
& \Phi_{a}(\mbox{B}1)  \, [a,b]
\\[4pt]  
&  
&   
& 12.\,(\widetilde{a},\widetilde{b3}\,;1,2) 
&
& \Phi_{b}(\mbox{B}1)  \, [b,a]
\\[4pt]  \hline
& & & & &  \\[-12pt]
{\tt Dip\,3u}  
&  gu \to ug
& (6)-1    
& 13.\,(\widetilde{a1},b\,;\widetilde{2},3) 
& \int dx {\cal V}^{f,g} (x)/\mbox{C}_{\mbox{{\tiny A}}}
& \Phi_{a}(\mbox{B}3u)  \, [a,1]
\\[4pt]  
&  
&   
& 14.\,(\widetilde{a1},b\,;2,\widetilde{3}) 
&
& \Phi_{a}(\mbox{B}3u)  \, [a,2]
\\[4pt]  
&  
&   
& 15.\,(\widetilde{a2},b\,;\widetilde{1},3) 
&
& \Phi_{a}(\mbox{B}3u)  \, [a,1]
\\[4pt]  
&  
&   
& 16.\,(\widetilde{a2},b\,;1,\widetilde{3}) 
& 
& \Phi_{a}(\mbox{B}3u)  \, [a,2]
\\[4pt]  
&  
&   
& 17.\,(\widetilde{b1},a\,;\widetilde{2},3) 
&
& \Phi_{b}(\mbox{B}3u)  \, [a,1]
\\[4pt]  
&  
&   
& 18.\,(\widetilde{b1},a\,;2,\widetilde{3}) 
&
& \Phi_{b}(\mbox{B}3u)  \, [a,2]
\\[4pt]  
&  
&   
& 19.\,(\widetilde{b2},a\,;\widetilde{1},3) 
&
& \Phi_{b}(\mbox{B}3u)  \, [a,1]
\\[4pt]  
&  
&   
& 20.\,(\widetilde{b2},a\,;1,\widetilde{3}) 
&
& \Phi_{b}(\mbox{B}3u)  \, [a,2]
\\[4pt]  
&  
& (6)-2  
& 21.\,(\widetilde{a1},\widetilde{b}\,;2,3) 
& \int dx {\widetilde {\cal V}}^{f,g} (x)/
\mbox{C}_{\mbox{{\tiny A}}}
& \Phi_{a}(\mbox{B}3u)  \, [a,b]
\\[4pt]  
&  
& 
& 22.\,(\widetilde{a2},\widetilde{b}\,;1,3) 
&
& \Phi_{a}(\mbox{B}3u)  \, [a,b]
\\[4pt]  
&  
& 
& 23.\,(\widetilde{b1},\widetilde{a}\,;2,3) 
&
& \Phi_{b}(\mbox{B}3u)  \, [a,b]
\\[4pt]  
&  
& 
& 24.\,(\widetilde{b2},\widetilde{a}\,;1,3) 
&
& \Phi_{b}(\mbox{B}3u)  \, [a,b]
\\[4pt]  \hline
\end{array} \nonumber
\end{align}
\caption{
Summary table of 
$\hat{\sigma}_{\mbox{{\tiny D}}}(\mbox{R}_{2u})$.
\label{ap_B_1_tab2}}
\end{table}
\newpage
\rightline{
\underline{
$\hat{\sigma}_{\mbox{{\tiny subt}}}
(\mbox{R}_{2u} = uu \to uug\,)$
}
}
\begin{enumerate}
\item[\fbox{Step2}]
\begin{equation}
\hat{\sigma}_{\mbox{{\tiny D}}}
(\,
\mbox{R}_{2u},\,
\mbox{I}
\,)
-
\hat{\sigma}_{\mbox{{\tiny I}}}
(\mbox{R}_{2u})=0\,.
\end{equation}
\item[\fbox{Step3}]
\ \ 
\begin{equation}
\hat{\sigma}_{\mbox{{\tiny D}}}
(\,
\mbox{R}_{2u},\,
\mbox{P}
\,)
+
\hat{\sigma}_{\mbox{{\tiny C}}}(\mbox{R}_{2u}) 
-
\hat{\sigma}_{\mbox{{\tiny P}}}
(\mbox{R}_{2u})=0\,,
\end{equation}
which is separated into two relations 
for Dipoles\,1 and 3$u$ as
\begin{align}
\hat{\sigma}_{\mbox{{\tiny D}}}
(\,
\mbox{R}_{2u},\,
\mbox{P},\,{\tt dip}j
\,)
+
\hat{\sigma}_{\mbox{{\tiny C}}}(\mbox{R}_{2u},\,
{\tt dip}j
) 
-
\hat{\sigma}_{\mbox{{\tiny P}}}
(\mbox{R}_{2u},
\,{\tt dip}j)=0\,.
\end{align}
\item[\fbox{Step4}]
\begin{equation}
\hat{\sigma}_{\mbox{{\tiny D}}}
(\,
\mbox{R}_{2u},\,
\mbox{K}
\,)
-
\hat{\sigma}_{\mbox{{\tiny K}}}
(\mbox{R}_{2u})=0\,,
\end{equation}
which is separated into two relations 
for Dipoles\,1 and 3$u$ as
\begin{align}
\hat{\sigma}_{\mbox{{\tiny D}}}
(\,
\mbox{R}_{2u},\,
\mbox{K},\,{\tt dip}j
\,)
-
\hat{\sigma}_{\mbox{{\tiny K}}}
(\mbox{R}_{2u},
\,{\tt dip}j)=0\,.
\end{align}
\item[\fbox{Step5}]
\begin{equation}
\hat{\sigma}_{subt}(\mbox{R}_{2u})=0\,.
\end{equation}
\end{enumerate}
%
\clearpage
\subsection{$\hat{\sigma}_{\mbox{{\tiny subt}}}(\mbox{R}_{3u})$  
\label{ap_B_3}}
%
%
\begin{table}[h!]
  \centering
\begin{align}
&\fbox{Step1} \ \ \ 
\hat{\sigma}_{\mbox{{\tiny D}}}
(\mbox{R}_{3u}= ug \to uu\bar{u}): 
\ \ S_{\mbox{{\tiny R}}_{3}}=2,
\nonumber\\
&
\begin{array}{|c|c|c|c|c|c|} \hline
{\tt Dip}\,j
& \mbox{B}j 
& \mbox{\small Splitting} 
& (y_{a},y_{b}:y_{1},y_{2}) 
& \mbox{Factor 1}
& \Phi(\mbox{B}_{j})
\,[y_{emi}, y_{spe}]
\\[4pt] \hline
& & & & & \\[-12pt]
{\tt Dip\,2u}  
& ug \to gu
& (5)-1  
& 1.\,(a,b\,;\widetilde{13},\widetilde{2}) 
& {\cal V}_{f\bar{f}}/ \mbox{C}_{\mbox{{\tiny A}}}
& \Phi(\mbox{B}2u) \, [1,2]
\\[4pt] 
& 
&   
& 2.\,(a,b\,;\widetilde{23},\widetilde{1}) 
&
& \Phi(\mbox{B}2u) \, [1,2]
\\[4pt] 
&  
& (5)-2    
& 3.\,(\widetilde{a},b\,;\widetilde{13},2) 
& \int dx {\cal V}_{f\bar{f}}(x)/ \mbox{C}_{\mbox{{\tiny A}}}
& \Phi_{a}(\mbox{B}2u)  \, [1,a]
\\[4pt]  
&  
&   
& 4.\,(a,\widetilde{b}\,;\widetilde{13},2) 
&
& \Phi_{b}(\mbox{B}2u)  \, [1,b]
\\[4pt]  
&  
&   
& 5.\,(\widetilde{a},b\,;\widetilde{23},1) 
&
& \Phi_{a}(\mbox{B}2u)  \, [1,a]
\\[4pt] 
&  
&   
& 6.\,(a,\widetilde{b}\,;\widetilde{23},1) 
&
& \Phi_{b}(\mbox{B}2u)  \, [1,b]
\\[4pt]  \hline
& & & & & \\[-12pt]
{\tt Dip\,3u}  
& gg \to u\bar{u}
& (6)-1    
& 7.\,(\widetilde{a1},b\,;\widetilde{2},3) 
& \int dx {\cal V}^{f,g} (x)/\mbox{C}_{\mbox{{\tiny A}}}
& \Phi_{a}(\mbox{B}3u)  \, [a,1]
\\[4pt]  
&  
&   
& 8.\,(\widetilde{a1},b\,;2,\widetilde{3}) 
&
& \Phi_{a}(\mbox{B}3u)  \, [a,2]
\\[4pt] 
&  
&   
& 9.\,(\widetilde{a2},b\,;\widetilde{1},3) 
&
& \Phi_{a}(\mbox{B}3u)  \, [a,1]
\\[4pt]  
&  
&   
& 10.\,(\widetilde{a2},b\,;1,\widetilde{3}) 
&
& \Phi_{a}(\mbox{B}3u)  \, [a,2]
\\[4pt]  
&  
& (6)-2  
& 11.\,(\widetilde{a1},\widetilde{b}\,;2,3) 
& \int dx {\widetilde {\cal V}}^{f,g} (x)/
\mbox{C}_{\mbox{{\tiny A}}}
& \Phi_{a}(\mbox{B}3u)  \, [a,b]
\\[4pt]
&  
&  
& 12.\,(\widetilde{a2},\widetilde{b}\,;1,3) 
&
& \Phi_{a}(\mbox{B}3u)  \, [a,b]
\\[4pt]  \hline
& & & & & \\[-12pt]
{\tt Dip\,4u}  
& u\bar{u} \to u\bar{u}
& (7)-1   
& 13.\,(a,\widetilde{b1}\,;\widetilde{2},3)  
& \int dx {\cal V}^{g,f} (x)/\mbox{C}_{\mbox{{\tiny F}}}
& \Phi_{b}(\mbox{B}4u)  \, [b,1]
\\[4pt]  
&  
&   
& 14.\,(a,\widetilde{b1}\,;2,\widetilde{3})  
&
& \Phi_{b}(\mbox{B}4u)  \, [b,2]
\\[4pt]  
&  
&   
& 15.\,(a,\widetilde{b2}\,;\widetilde{1},3)  
&
& \Phi_{b}(\mbox{B}4u)  \, [b,1]
\\[4pt]  
&  
&   
& 16.\,(a,\widetilde{b2}\,;1,\widetilde{3})  
&
& \Phi_{b}(\mbox{B}4u)  \, [b,2] 
\\[4pt]  
&  
& (7)-2   
& 17.\,(\widetilde{a},\widetilde{b1}\,;2,3)  
& \int dx {\widetilde {\cal V}}^{g,f} (x)/
\mbox{C}_{\mbox{{\tiny F}}}
& \Phi_{b}(\mbox{B}4u)  \, [b,a] 
\\[4pt] 
&  
&   
& 18.\,(\widetilde{a},\widetilde{b2}\,;1,3)  
&
& \Phi_{b}(\mbox{B}4u)  \, [b,a] 
\\[4pt] \hline
& & & & & \\[-12pt]
{\tt Dip\,4\bar{u}}  
& uu \to uu
& (7)-1   
& 19.\,(a,\widetilde{b3}\,;\widetilde{1},2)  
& \int dx {\cal V}^{g,f} (x)/\mbox{C}_{\mbox{{\tiny F}}}
& \Phi_{b}(\mbox{B}4\bar{u})  \, [b,1]
\\[4pt]  
&  
&   
& 20.\,(a,\widetilde{b3}\,;1,\widetilde{2})  
&
& \Phi_{b}(\mbox{B}4\bar{u})  \, [b,2]
\\[4pt]  
&  
& (7)-2   
& 21.\,(\widetilde{a},\widetilde{b3}\,;1,2)  
& \int dx {\widetilde {\cal V}}^{g,f} (x)/
\mbox{C}_{\mbox{{\tiny F}}}
& \Phi_{b}(\mbox{B}4\bar{u})  \, [b,a]
\\[4pt] \hline
\end{array} \nonumber
\end{align}
\caption{
Summary table of 
$\hat{\sigma}_{\mbox{{\tiny D}}}(\mbox{R}_{3u})$.
\label{ap_B_1_tab3}}
\end{table}
\newpage
\rightline{
\underline{
$\hat{\sigma}_{\mbox{{\tiny subt}}}
(\mbox{R}_{3u}= ug \to uu\bar{u})$
}
}
\begin{enumerate}
\item[\fbox{Step2}]
\begin{equation}
\hat{\sigma}_{\mbox{{\tiny D}}}
(\,
\mbox{R}_{3u},\,
\mbox{I}
\,)=0 
\ \ \ \mbox{and} \ \ \
\hat{\sigma}_{\mbox{{\tiny I}}}
(\mbox{R}_{3u})=0\,.
\end{equation}
\item[\fbox{Step3}]
\begin{equation}
\hat{\sigma}_{\mbox{{\tiny D}}}
(\,
\mbox{R}_{3u},\,
\mbox{P}
\,)
+
\hat{\sigma}_{\mbox{{\tiny C}}}(\mbox{R}_{3u}) 
-
\hat{\sigma}_{\mbox{{\tiny P}}}
(\mbox{R}_{3u})=0\,,
\end{equation}
which is separated into three relations 
for Dipoles\,3$u$, 4$u$, and 4$\bar{u}$ as
\begin{align}
\hat{\sigma}_{\mbox{{\tiny D}}}
(\,
\mbox{R}_{3u},\,
\mbox{P},\,{\tt dip}j
\,)
+
\hat{\sigma}_{\mbox{{\tiny C}}}(\mbox{R}_{3u},\,
{\tt dip}j
) 
-
\hat{\sigma}_{\mbox{{\tiny P}}}
(\mbox{R}_{3u},
\,{\tt dip}j)=0\,.
\end{align}
\item[\fbox{Step4}]
\begin{equation}
\hat{\sigma}_{\mbox{{\tiny D}}}
(\,
\mbox{R}_{3u},\,
\mbox{K}
\,)
-
\hat{\sigma}_{\mbox{{\tiny K}}}
(\mbox{R}_{3u})=0\,,
\end{equation}
which is separated into three relations 
for Dipoles 3$u$, 4$u$, and 4$\bar{u}$ as
\begin{align}
\hat{\sigma}_{\mbox{{\tiny D}}}
(\,
\mbox{R}_{3u},\,
\mbox{K},\,{\tt dip}j
\,)
-
\hat{\sigma}_{\mbox{{\tiny K}}}
(\mbox{R}_{3u},
\,{\tt dip}j)=0\,.
\end{align}
\item[\fbox{Step5}]
\begin{equation}
\hat{\sigma}_{subt}(\mbox{R}_{3u})=
\hat{\sigma}_{\mbox{{\tiny D}}}(\mbox{R}_{3u},\,{\tt dip}2)\,.
\end{equation}
\begin{equation}
\hat{\sigma}_{\mbox{{\tiny D}}}(\mbox{R}_{3u},\,{\tt dip}2)
=
\hat{\sigma}_{\mbox{{\tiny D}}}
(\,
\mbox{R}_{3u},\,
{\tt dip}2,(5)\mbox{-}1/2,
\,
{\cal V}_{f\bar{f}}
\,)
+
\hat{\sigma}_{\mbox{{\tiny D}}}
(\,
\mbox{R}_{3u},\,
{\tt dip}2,(5)\mbox{-}2,
\,
h)\,.
\end{equation}
\begin{align}
\hat{\sigma}_{\mbox{{\tiny D}}}
(\mbox{R}_{3u},\,
{\tt dip}2,(5)\mbox{-}1/2,
\,
{\cal V}_{f\bar{f}}
\,)
&=-\frac{A_{d}}{S_{\mbox{{\tiny R}}_{3u}}} 
\cdot
\frac{1}{\mbox{C}_{\mbox{{\tiny A}}}} 
{\cal V}_{f\bar{f}} (\ep) 
\cdot 
\Phi(\mbox{B}2)_{d} \,
\bigl(
\,
[1,2]+
[1,a]+
[1,b]
\,
\bigr)\cdot 2\,.
\\
\hat{\sigma}_{\mbox{{\tiny D}}}
(\mbox{R}_{3u},\,
{\tt dip}2,(5)\mbox{-}2,
\,
h)
&=
-\frac{A_{4}}{S_{\mbox{{\tiny R}}_{3u}}} 
\int_{0}^{1}dx \,
\frac{\mbox{T}_{\mbox{{\tiny R}}}}
{\mbox{C}_{\mbox{{\tiny A}}}} 
\,\frac{2}{3}h(x)
\ \times
\nonumber\\
& \hspace{10mm}
\Bigl(
\Phi_{a}(\mbox{B}2,x)_{4}\,
\langle 1,a \rangle
+
\Phi_{b}(\mbox{B}2,x)_{4}\,
\langle 1,b \rangle
\Bigr)\cdot 2\,.
\end{align}
\end{enumerate}
%
\clearpage
\subsection{$\hat{\sigma}_{\mbox{{\tiny subt}}}(\mbox{R}_{4u})$  
\label{ap_B_4}}
%
%
\begin{table}[h!]
  \centering
\begin{align}
&\fbox{Step1}  \ \ \
\hat{\sigma}_{\mbox{{\tiny D}}}
(\mbox{R}_{4u} = u\bar{u} \to d\bar{d}g): 
\ \ S_{\mbox{{\tiny R}}_{4}}=1,
\nonumber\\
&
\begin{array}{|c|c|c|c|c|c|} \hline
{\tt Dip}\,j
& \mbox{B}j 
& \mbox{\small Splitting} 
& (y_{a},y_{b}:y_{1},y_{2}) 
& \mbox{Factor 1}
& \Phi(\mbox{B}_{j})
\,[y_{emi}, y_{spe}]
\\[4pt] \hline
& & & & & \\[-12pt]
{\tt Dip\,1}  
& u\bar{u} \to d\bar{d}
& (1)-1   
& 1.\,(a,b\,;\widetilde{13},\widetilde{2} ) 
& {\cal V}_{fg}/\mbox{C}_{\mbox{{\tiny F}}}
& \Phi(\mbox{B}1) \, [1,2]
\\[4pt]  
&  
&   
& 2.\,(a,b\,;\widetilde{1},\widetilde{23}) 
&
& \Phi(\mbox{B}1) \, [2,1]
\\[4pt]  
&  
& (1)-2  
& 3.\,(\widetilde{a},b\,;\widetilde{13},2) 
& \int dx 
{\cal V}_{fg}(x)/\mbox{C}_{\mbox{{\tiny F}}} 
& \Phi_{a}(\mbox{B}1)  \, [1,a]
\\[4pt]  
&  
&   
& 4.\,(a,\widetilde{b}\,;\widetilde{13},2) 
&
& \Phi_{b}(\mbox{B}1)  \, [1,b]
\\[4pt]  
&  
&  
& 5.\,(\widetilde{a},b\,;1,\widetilde{23}) 
&
& \Phi_{a}(\mbox{B}1)  \, [2,a]
\\[4pt]  
&  
&   
& 6.\,(a,\widetilde{b}\,;1,\widetilde{23}) 
&
& \Phi_{b}(\mbox{B}1)  \, [2,b]
\\[4pt]  
&  
& (3)-1  
& 7.\,(\widetilde{a3},b\,;\widetilde{1},2) 
& \int dx  
{\cal V}^{f,f} (x)/\mbox{C}_{\mbox{{\tiny F}}}
& \Phi_{a}(\mbox{B}1)  \, [a,1]
\\[4pt]  
&  
&   
& 8.\,(\widetilde{a3},b\,;1,\widetilde{2}) 
&
& \Phi_{a}(\mbox{B}1)  \, [a,2]
\\[4pt]  
&  
&   
& 9.\,(a,\widetilde{b3}\,;\widetilde{1},2) 
&
& \Phi_{b}(\mbox{B}1)  \, [b,1]
\\[4pt]  
&  
&   
& 10.\,(a,\widetilde{b3}\,;1,\widetilde{2})
&
& \Phi_{b}(\mbox{B}1)  \, [b,2]
\\[4pt]  
&  
& (3)-2  
& 11.\,(\widetilde{a3},\widetilde{b}\,;1,2) 
& \int dx
{\widetilde {\cal V}}^{f,f} (x)
/\mbox{C}_{\mbox{{\tiny F}}}
& \Phi_{a}(\mbox{B}1)  \, [a,b]
\\[4pt]  
&  
&   
& 12.\,(\widetilde{a},\widetilde{b3}\,;1,2) 
&
& \Phi_{b}(\mbox{B}1)  \, [b,a]
\\[4pt]  \hline
& & & & & \\[-12pt]
{\tt Dip\,2d}  
& u\bar{u} \to gg
& (5)-1  
& 13.\,(a,b\,;\widetilde{12},\widetilde{3}) 
& {\cal V}_{f\bar{f}}/ \mbox{C}_{\mbox{{\tiny A}}}
& \Phi(\mbox{B}2d)  \, [1,2]
\\[4pt]  
&  
& (5)-2    
& 14.\,(\widetilde{a},b\,;\widetilde{12},3) 
& \int dx {\cal V}_{f\bar{f}}(x)/ \mbox{C}_{\mbox{{\tiny A}}}
& \Phi_{a}(\mbox{B}2d)  \, [1,a]
\\[4pt]  
&  
&   
& 15.\,(a,\widetilde{b}\,;\widetilde{12},3) 
&
& \Phi_{b}(\mbox{B}2d)  \, [1,b]
\\[4pt]  \hline
\end{array} \nonumber
\end{align}
\caption{
Summary table of 
$\hat{\sigma}_{\mbox{{\tiny D}}}(\mbox{R}_{4u})$.
\label{ap_B_1_tab4}}
\end{table}
\newpage
\rightline{
\underline{
$\hat{\sigma}_{\mbox{{\tiny subt}}}
(\mbox{R}_{4u} = u\bar{u} \to d\bar{d}g)$  
}
}
\begin{enumerate}
\item[\fbox{Step2}]
\begin{equation}
\hat{\sigma}_{\mbox{{\tiny D}}}
(\,
\mbox{R}_{4u},\,
\mbox{I}
\,)
-
\hat{\sigma}_{\mbox{{\tiny I}}}
(\mbox{R}_{4u})=0\,.
\end{equation}
\item[\fbox{Step3}]
\begin{equation}
\hat{\sigma}_{\mbox{{\tiny D}}}
(\,
\mbox{R}_{4u},\,
\mbox{P}
\,)
+
\hat{\sigma}_{\mbox{{\tiny C}}}(\mbox{R}_{4u}) 
-
\hat{\sigma}_{\mbox{{\tiny P}}}
(\mbox{R}_{4u})=0\,,
\end{equation}
which includes only Dipole\,1.
\item[\fbox{Step4}]
\begin{equation}
\hat{\sigma}_{\mbox{{\tiny D}}}
(\,
\mbox{R}_{4u},\,
\mbox{K}
\,)
-
\hat{\sigma}_{\mbox{{\tiny K}}}
(\mbox{R}_{4u})=0\,,
\end{equation}
which includes only Dipole\,1.
\item[\fbox{Step5}]
\begin{equation}
\hat{\sigma}_{subt}(\mbox{R}_{4u})=
\hat{\sigma}_{\mbox{{\tiny D}}}(\mbox{R}_{4u},\,{\tt dip}2d)\,.
\end{equation}
\begin{equation}
\hat{\sigma}_{\mbox{{\tiny D}}}(\mbox{R}_{4u},\,{\tt dip}2d)
=
\hat{\sigma}_{\mbox{{\tiny D}}}
(
\mbox{R}_{4u},\,
{\tt dip}2d,(5)\mbox{-}1/2,
\,
{\cal V}_{f\bar{f}}
\,)
+
\hat{\sigma}_{\mbox{{\tiny D}}}
(
\mbox{R}_{4u},\,
{\tt dip}2d,(5)\mbox{-}2,
\,
h)\,.
\end{equation}
\begin{align}
\hat{\sigma}_{\mbox{{\tiny D}}}
(\mbox{R}_{4u},\,
{\tt dip}2d,(5)\mbox{-}1/2,
\,
{\cal V}_{f\bar{f}}
\,)
&=-\frac{A_{d}}{S_{\mbox{{\tiny R}}_{2d}}} 
\cdot
\frac{1}{\mbox{C}_{\mbox{{\tiny A}}}} 
{\cal V}_{f\bar{f}} (\ep) 
\cdot 
\Phi(\mbox{B}2)_{d} \,
\bigl(
\,
[1,2]+
[1,a]+
[1,b]
\,
\bigr)\,.
\\
\hat{\sigma}_{\mbox{{\tiny D}}}
(\mbox{R}_{4u},\,
{\tt dip}2d,(5)\mbox{-}2,
\,
h)
&=
-\frac{A_{4}}{S_{\mbox{{\tiny R}}_{2d}}} 
\int_{0}^{1}dx \,
\frac{\mbox{T}_{\mbox{{\tiny R}}}}
{\mbox{C}_{\mbox{{\tiny A}}}} 
\,\frac{2}{3}h(x)
\ \times
\nonumber\\
& \hspace{10mm}
\Bigl(
\Phi_{a}(\mbox{B}2,x)_{4}\,
\langle 1,a \rangle
+
\Phi_{b}(\mbox{B}2,x)_{4}\,
\langle 1,b \rangle
\Bigr)\,.
\end{align}
\end{enumerate}
%
\clearpage
\subsection{$\hat{\sigma}_{\mbox{{\tiny subt}}}(\mbox{R}_{5ud})$  
\label{ap_B_5}}
%
%
\begin{table}[h!]
  \centering
\begin{align}
&\fbox{Step1}  \ \ \
\hat{\sigma}_{\mbox{{\tiny D}}}
(\mbox{R}_{5ud} = ud \to udg): 
\ \ S_{\mbox{{\tiny R}}_{5}}=1,
\nonumber\\
&
\begin{array}{|c|c|c|c|c|c|} \hline
{\tt Dip}\,j
& \mbox{B}j 
& \mbox{\small Splitting} 
& (y_{a},y_{b}:y_{1},y_{2}) 
& \mbox{Factor 1}
& \Phi(\mbox{B}_{j})
\,[y_{emi}, y_{spe}]
\\[4pt] \hline
& & & & & \\[-12pt]
{\tt Dip\,1}  
& ud \to ud
& (1)-1   
& 1.\,(a,b\,;\widetilde{13},\widetilde{2} ) 
& {\cal V}_{fg}/\mbox{C}_{\mbox{{\tiny F}}}
& \Phi(\mbox{B}1) \, [1,2]
\\[4pt]  
&  
&   
& 2.\,(a,b\,;\widetilde{1},\widetilde{23}) 
&
& \Phi(\mbox{B}1) \, [2,1]
\\[4pt]  
&  
& (1)-2  
& 3.\,(\widetilde{a},b\,;\widetilde{13},2) 
& \int dx {\cal V}_{fg}(x)/\mbox{C}_{\mbox{{\tiny F}}}
& \Phi_{a}(\mbox{B}1)  \, [1,a]
\\[4pt]  
&  
&   
& 4.\,(a,\widetilde{b}\,;\widetilde{13},2) 
&
& \Phi_{b}(\mbox{B}1)  \, [1,b]
\\[4pt]  
&  
&  
& 5.\,(\widetilde{a},b\,;1,\widetilde{23}) 
& 
& \Phi_{a}(\mbox{B}1)  \, [2,a]
\\[4pt]  
&  
&   
& 6.\,(a,\widetilde{b}\,;1,\widetilde{23}) 
&
& \Phi_{b}(\mbox{B}1)  \, [2,b]
\\[4pt]  
&  
& (3)-1  
& 7.\,(\widetilde{a3},b\,;\widetilde{1},2) 
& \int dx  
{\cal V}^{f,f} (x)/\mbox{C}_{\mbox{{\tiny F}}}
& \Phi_{a}(\mbox{B}1)  \, [a,1]
\\[4pt]  
&  
&   
& 8.\,(\widetilde{a3},b\,;1,\widetilde{2}) 
&
& \Phi_{a}(\mbox{B}1)  \, [a,2]
\\[4pt]  
&  
&   
& 9.\,(a,\widetilde{b3}\,;\widetilde{1},2) 
&
& \Phi_{b}(\mbox{B}1)  \, [b,1]
\\[4pt]  
&  
&   
& 10.\,(a,\widetilde{b3}\,;1,\widetilde{2})
&
& \Phi_{b}(\mbox{B}1)  \, [b,2]
\\[4pt]  
&  
& (3)-2  
& 11.\,(\widetilde{a3},\widetilde{b}\,;1,2) 
& \int dx
{\widetilde {\cal V}}^{f,f} (x)
/\mbox{C}_{\mbox{{\tiny F}}}
& \Phi_{a}(\mbox{B}1)  \, [a,b]
\\[4pt]  
&  
&   
& 12.\,(\widetilde{a},\widetilde{b3}\,;1,2) 
&
& \Phi_{b}(\mbox{B}1)  \, [b,a]
\\[4pt]  \hline
& & & & & \\[-12pt]
{\tt Dip\,3u}  
& gd \to dg
& (6)-1    
& 13.\,(\widetilde{a1},b\,;\widetilde{2},3) 
& \int dx {\cal V}^{f,g} (x)/\mbox{C}_{\mbox{{\tiny A}}}
& \Phi_{a}(\mbox{B}3u)  \, [a,1]
\\[4pt]  
&  
&   
& 14.\,(\widetilde{a1},b\,;2,\widetilde{3}) 
&
& \Phi_{a}(\mbox{B}3u)  \, [a,2]
\\[4pt]  
&  
& (6)-2  
& 15.\,(\widetilde{a1},\widetilde{b}\,;2,3) 
& \int dx {\widetilde {\cal V}}^{f,g} (x)/
\mbox{C}_{\mbox{{\tiny A}}}
& \Phi_{a}(\mbox{B}3u)  \, [a,b]
\\[4pt]  \hline
& & & & & \\[-12pt]
{\tt Dip\,3d}  
& ug \to ug
& (6)-1   
& 16.\,(a,\widetilde{b2}\,;\widetilde{1},3)  
& \int dx {\cal V}^{f,g} (x)/\mbox{C}_{\mbox{{\tiny A}}}
& \Phi_{b}(\mbox{B}3d)  \, [b,1]
\\[4pt]  
&  
&   
& 17.\,(a,\widetilde{b2}\,;1,\widetilde{3})  
&
& \Phi_{b}(\mbox{B}3d)  \, [b,2]
\\[4pt]  
&  
& (6)-2   
& 18.\,(\widetilde{a},\widetilde{b2}\,;1,3)  
& \int dx {\widetilde {\cal V}}^{f,g} (x)/
\mbox{C}_{\mbox{{\tiny A}}}
& \Phi_{b}(\mbox{B}3d)  \, [b,a]
\\[4pt] \hline
\end{array} \nonumber
\end{align}
\caption{
Summary table of
$\hat{\sigma}_{\mbox{{\tiny D}}}(\mbox{R}_{5ud})$. 
\label{ap_B_1_tab5}}
\end{table}
\newpage
\rightline{
\underline{
$\hat{\sigma}_{\mbox{{\tiny subt}}}
(\mbox{R}_{5ud} = ud \to udg)$  
}
}
\begin{enumerate}
\item[\fbox{Step2}]
\begin{equation}
\hat{\sigma}_{\mbox{{\tiny D}}}
(\,
\mbox{R}_{5ud},\,
\mbox{I}
\,)
-
\hat{\sigma}_{\mbox{{\tiny I}}}
(\mbox{R}_{5ud})=0\,.
\end{equation}
\item[\fbox{Step3}]
\begin{equation}
\hat{\sigma}_{\mbox{{\tiny D}}}
(\,
\mbox{R}_{5ud},\,
\mbox{P}
\,)
+
\hat{\sigma}_{\mbox{{\tiny C}}}(\mbox{R}_{5ud}) 
-
\hat{\sigma}_{\mbox{{\tiny P}}}
(\mbox{R}_{5ud})=0\,,
\end{equation}
which is separated into three relations 
for Dipoles 1, 3$u$, and 3$d$ as
\begin{align}
\hat{\sigma}_{\mbox{{\tiny D}}}
(\,
\mbox{R}_{5ud},\,
\mbox{P},\,{\tt dip}j
\,)
+
\hat{\sigma}_{\mbox{{\tiny C}}}(\mbox{R}_{5ud},\,
{\tt dip}j
) 
-
\hat{\sigma}_{\mbox{{\tiny P}}}
(\mbox{R}_{5ud},
\,{\tt dip}j)=0\,.
\end{align}
\item[\fbox{Step4}]
\begin{equation}
\hat{\sigma}_{\mbox{{\tiny D}}}
(\,
\mbox{R}_{5ud},\,
\mbox{K}
\,)
-
\hat{\sigma}_{\mbox{{\tiny K}}}
(\mbox{R}_{5ud})=0\,,
\end{equation}
which is separated into three relations 
for Dipole 1, 3$u$, and 3$d$ as
\begin{align}
\hat{\sigma}_{\mbox{{\tiny D}}}
(\,
\mbox{R}_{5ud},\,
\mbox{K},\,{\tt dip}j
\,)
-
\hat{\sigma}_{\mbox{{\tiny K}}}
(\mbox{R}_{5ud},
\,{\tt dip}j)=0\,.
\end{align}
\item[\fbox{Step5}]
\begin{equation}
\hat{\sigma}_{subt}(\mbox{R}_{5ud})=0\,.
\end{equation}
\end{enumerate}
%
\clearpage
\subsection{$\hat{\sigma}_{\mbox{{\tiny subt}}}(\mbox{R}_{6u\bar{d}})$  
\label{ap_B_6}}
%
%
\begin{table}[h!]
  \centering
\begin{align}
&\fbox{Step1} \ \ \
\hat{\sigma}_{\mbox{{\tiny D}}}
(\mbox{R}_{6u\bar{d}}=u\bar{d} \to u\bar{d}g): 
\ \ S_{\mbox{{\tiny R}}_{6}}=1,
\nonumber\\
&
\begin{array}{|c|c|c|c|c|c|} \hline
{\tt Dip}\,j
& \mbox{B}j 
& \mbox{\small Splitting} 
& (y_{a},y_{b}:y_{1},y_{2}) 
& \mbox{Factor 1}
& \Phi(\mbox{B}_{j})
\,[y_{emi}, y_{spe}]
\\[4pt] \hline
& & & & & \\[-12pt]
{\tt Dip\,1}  
& u\bar{d} \to u\bar{d}
& (1)-1   
& 1.\,(a,b\,;\widetilde{13},\widetilde{2} ) 
& {\cal V}_{fg}/\mbox{C}_{\mbox{{\tiny F}}}
& \Phi(\mbox{B}1) \, [1,2]
\\[4pt]  
&  
&   
& 2.\,(a,b\,;\widetilde{1},\widetilde{23}) 
&
& \Phi(\mbox{B}1) \, [2,1]
\\[4pt]  
&  
& (1)-2  
& 3.\,(\widetilde{a},b\,;\widetilde{13},2) 
& \int dx {\cal V}_{fg}(x)/\mbox{C}_{\mbox{{\tiny F}}}
& \Phi_{a}(\mbox{B}1)  \, [1,a]
\\[4pt]  
&  
&   
& 4.\,(a,\widetilde{b}\,;\widetilde{13},2) 
&
& \Phi_{b}(\mbox{B}1)  \, [1,b]
\\[4pt]  
&  
&  
& 5.\,(\widetilde{a},b\,;1,\widetilde{23}) 
&
& \Phi_{a}(\mbox{B}1)  \, [2,a]
\\[4pt]  
&  
&   
& 6.\,(a,\widetilde{b}\,;1,\widetilde{23}) 
&
& \Phi_{b}(\mbox{B}1)  \, [2,b]
\\[4pt]  
&  
& (3)-1  
& 7.\,(\widetilde{a3},b\,;\widetilde{1},2) 
& \int dx  
{\cal V}^{f,f} (x)/\mbox{C}_{\mbox{{\tiny F}}}
& \Phi_{a}(\mbox{B}1)  \, [a,1]
\\[4pt]  
&  
&   
& 8.\,(\widetilde{a3},b\,;1,\widetilde{2}) 
&
& \Phi_{a}(\mbox{B}1)  \, [a,2]
\\[4pt]  
&  
&   
& 9.\,(a,\widetilde{b3}\,;\widetilde{1},2) 
&
& \Phi_{b}(\mbox{B}1)  \, [b,1]
\\[4pt]  
&  
&   
& 10.\,(a,\widetilde{b3}\,;1,\widetilde{2})
&
& \Phi_{b}(\mbox{B}1)  \, [b,2]
\\[4pt]  
&  
& (3)-2  
& 11.\,(\widetilde{a3},\widetilde{b}\,;1,2) 
& \int dx
{\widetilde {\cal V}}^{f,f} (x)
/\mbox{C}_{\mbox{{\tiny F}}}
& \Phi_{a}(\mbox{B}1)  \, [a,b]
\\[4pt]  
&  
&   
& 12.\,(\widetilde{a},\widetilde{b3}\,;1,2) 
&
& \Phi_{b}(\mbox{B}1)  \, [b,a]
\\[4pt]  \hline
& & & & & \\[-12pt]
{\tt Dip\,3u}  
& g\bar{d} \to \bar{d}g
& (6)-1    
& 13.\,(\widetilde{a1},b\,;\widetilde{2},3) 
& \int dx {\cal V}^{f,g} (x)/\mbox{C}_{\mbox{{\tiny A}}}
& \Phi_{a}(\mbox{B}3u)  \, [a,1]
\\[4pt]  
&  
&   
& 14.\,(\widetilde{a1},b\,;2,\widetilde{3}) 
&
& \Phi_{a}(\mbox{B}3u)  \, [a,2]
\\[4pt]  
&  
& (6)-2  
& 15.\,(\widetilde{a1},\widetilde{b}\,;2,3) 
& \int dx {\widetilde {\cal V}}^{f,g} (x)/
\mbox{C}_{\mbox{{\tiny A}}}
& \Phi_{a}(\mbox{B}3u)  \, [a,b]
\\[4pt]  \hline
& & & & & \\[-12pt]
{\tt Dip\,3\bar{d}}  
& ug \to ug
& (6)-1   
& 16.\,(a,\widetilde{b2}\,;\widetilde{1},3)  
& \int dx {\cal V}^{f,g} (x)/\mbox{C}_{\mbox{{\tiny A}}}
& \Phi_{b}(\mbox{B}3\bar{d})  \, [b,1]
\\[4pt]  
&  
&   
& 17.\,(a,\widetilde{b2}\,;1,\widetilde{3})  
&
& \Phi_{b}(\mbox{B}3\bar{d})  \, [b,2]
\\[4pt]  
&  
& (6)-2   
& 18.\,(\widetilde{a},\widetilde{b2}\,;1,3)  
& \int dx {\widetilde {\cal V}}^{f,g} (x)/
\mbox{C}_{\mbox{{\tiny A}}}
& \Phi_{b}(\mbox{B}3\bar{d})  \, [b,a]
\\[4pt] \hline
\end{array} \nonumber
\end{align}
\caption{
Summary table of 
$\hat{\sigma}_{\mbox{{\tiny D}}}(\mbox{R}_{6u\bar{d}})$.
\label{ap_B_1_tab6}}
\end{table}
\newpage
\rightline{
\underline{
$\hat{\sigma}_{\mbox{{\tiny subt}}}
(\mbox{R}_{6u\bar{d}}=u\bar{d} \to u\bar{d}g)$  
}
}
\begin{enumerate}
\item[\fbox{Step2}]
\begin{equation}
\hat{\sigma}_{\mbox{{\tiny D}}}
(\,
\mbox{R}_{6u\bar{d}},\,
\mbox{I}
\,)
-
\hat{\sigma}_{\mbox{{\tiny I}}}
(\mbox{R}_{6u\bar{d}})=0\,.
\end{equation}
\item[\fbox{Step3}]
\begin{equation}
\hat{\sigma}_{\mbox{{\tiny D}}}
(\,
\mbox{R}_{6u\bar{d}},\,
\mbox{P}
\,)
+
\hat{\sigma}_{\mbox{{\tiny C}}}(\mbox{R}_{6u\bar{d}}) 
-
\hat{\sigma}_{\mbox{{\tiny P}}}
(\mbox{R}_{6u\bar{d}})=0\,,
\end{equation}
which is separated into three relations 
for Dipoles 1, 3$u$, and 3$\bar{d}$ as
\begin{align}
\hat{\sigma}_{\mbox{{\tiny D}}}
(\,
\mbox{R}_{6u\bar{d}},\,
\mbox{P},\,{\tt dip}j
\,)
+
\hat{\sigma}_{\mbox{{\tiny C}}}(\mbox{R}_{6u\bar{d}},\,
{\tt dip}j
) 
-
\hat{\sigma}_{\mbox{{\tiny P}}}
(\mbox{R}_{6u\bar{d}},
\,{\tt dip}j)=0\,.
\end{align}
\item[\fbox{Step4}]
\begin{equation}
\hat{\sigma}_{\mbox{{\tiny D}}}
(\,
\mbox{R}_{6u\bar{d}},\,
\mbox{K}
\,)
-
\hat{\sigma}_{\mbox{{\tiny K}}}
(\mbox{R}_{6u\bar{d}})=0\,,
\end{equation}
which is separated into three relations 
for Dipoles 1, 3$u$, and 3$\bar{d}$ as
\begin{align}
\hat{\sigma}_{\mbox{{\tiny D}}}
(\,
\mbox{R}_{6u\bar{d}},\,
\mbox{K},\,{\tt dip}j
\,)
-
\hat{\sigma}_{\mbox{{\tiny K}}}
(\mbox{R}_{6u\bar{d}},
\,{\tt dip}j)=0\,.
\end{align}
\item[\fbox{Step5}]
\begin{equation}
\hat{\sigma}_{subt}(\mbox{R}_{6u\bar{d}})=0\,.
\end{equation}
\end{enumerate}
%
\clearpage
\subsection{$\hat{\sigma}_{\mbox{{\tiny subt}}}(\mbox{R}_{7u})$  
\label{ap_B_7}}
%
%
\begin{table}[h!]
  \centering
\begin{align}
&\fbox{Step1}  \ \ \
\hat{\sigma}_{\mbox{{\tiny D}}}
(\mbox{R}_{7u} = ug \to ud\bar{d}): 
\ \ S_{\mbox{{\tiny R}}_{7}}=1,
\nonumber\\
&
\begin{array}{|c|c|c|c|c|c|} \hline
{\tt Dip}\,j
& \mbox{B}j 
& \mbox{\small Splitting} 
& (y_{a},y_{b}:y_{1},y_{2}) 
& \mbox{Factor 1}
& \Phi(\mbox{B}_{j})
\,[y_{emi}, y_{spe}]
\\[4pt] \hline
& & & & & \\[-12pt]
{\tt Dip\,2u}  
& ug \to ug
& (5)-1  
& 1.\,(a,b\,;\widetilde{1},\widetilde{23}) 
& {\cal V}_{f\bar{f}}/ \mbox{C}_{\mbox{{\tiny A}}}
& \Phi(\mbox{B}2u) \, [2,1]
\\[4pt] 
&  
& (5)-2    
& 2.\,(\widetilde{a},b\,;1,\widetilde{23}) 
& \int dx {\cal V}_{f\bar{f}}(x)/ \mbox{C}_{\mbox{{\tiny A}}}
& \Phi_{a}(\mbox{B}2u)  \, [2,a]
\\[4pt]  
&  
&   
& 3.\,(a,\widetilde{b}\,;1,\widetilde{23}) 
&
& \Phi_{b}(\mbox{B}2u)  \, [2,b]
\\[4pt]  \hline
& & & & & \\[-12pt]
{\tt Dip\,3u}  
& gg \to d\bar{d}
& (6)-1    
& 4.\,(\widetilde{a1},b\,;\widetilde{2},3) 
& \int dx {\cal V}^{f,g} (x)/\mbox{C}_{\mbox{{\tiny A}}}
& \Phi_{a}(\mbox{B}3u)  \, [a,1]
\\[4pt]  
&  
&   
& 5.\,(\widetilde{a1},b\,;2,\widetilde{3}) 
&
& \Phi_{a}(\mbox{B}3u)  \, [a,2]
\\[4pt] 
&  
& (6)-2  
& 6.\,(\widetilde{a1},\widetilde{b}\,;2,3) 
& \int dx {\widetilde {\cal V}}^{f,g} (x)/
\mbox{C}_{\mbox{{\tiny A}}}
& \Phi_{a}(\mbox{B}3u)  \, [a,b]
\\[4pt] \hline
& & & & & \\[-12pt]
{\tt Dip\,4u}  
& u\bar{u} \to d\bar{d}
& (7)-1   
& 7.\,(a,\widetilde{b1}\,;\widetilde{2},3)  
& \int dx {\cal V}^{g,f} (x)/\mbox{C}_{\mbox{{\tiny F}}}
& \Phi_{b}(\mbox{B}4u)  \, [b,1]
\\[4pt]  
&  
&   
& 8.\,(a,\widetilde{b1}\,;2,\widetilde{3})  
&
& \Phi_{b}(\mbox{B}4u)  \, [b,2]
\\[4pt] 
&  
& (7)-2 
& 9.\,(\widetilde{a},\widetilde{b1}\,;2,3)  
& \int dx {\widetilde {\cal V}}^{g,f} (x)/
\mbox{C}_{\mbox{{\tiny F}}}
& \Phi_{b}(\mbox{B}4u)  \, [b,a]
\\[4pt] \hline
& & & & & \\[-12pt]
{\tt Dip\,4d}  
& u\bar{d} \to u\bar{d}
& (7)-1   
& 10.\,(a,\widetilde{b2}\,;\widetilde{1},3)  
& \int dx {\cal V}^{g,f} (x)/\mbox{C}_{\mbox{{\tiny F}}}
& \Phi_{b}(\mbox{B}4d)  \, [b,1]
\\[4pt]  
&  
&   
& 11.\,(a,\widetilde{b2}\,;1,\widetilde{3})  
&
& \Phi_{b}(\mbox{B}4d)  \, [b,2]
\\[4pt]  
&  
& (7)-2   
& 12.\,(\widetilde{a},\widetilde{b2}\,;1,3)  
& \int dx {\widetilde {\cal V}}^{g,f} (x)/
\mbox{C}_{\mbox{{\tiny F}}}
& \Phi_{b}(\mbox{B}4d)  \, [b,a]
\\[4pt] \hline
& & & & & \\[-12pt]
{\tt Dip\,4\bar{d}}  
& ud \to ud
& (7)-1   
& 13.\,(a,\widetilde{b3}\,;\widetilde{1},2)  
& \int dx {\cal V}^{g,f} (x)/\mbox{C}_{\mbox{{\tiny F}}}
& \Phi_{b}(\mbox{B}4\bar{d})  \, [b,1]
\\[4pt]  
&  
&   
& 14.\,(a,\widetilde{b3}\,;1,\widetilde{2})  
&
& \Phi_{b}(\mbox{B}4\bar{d})  \, [b,2]
\\[4pt]  
&  
& (7)-2   
& 15.\,(\widetilde{a},\widetilde{b3}\,;1,2)  
& \int dx {\widetilde {\cal V}}^{g,f} (x)/
\mbox{C}_{\mbox{{\tiny F}}}
& \Phi_{b}(\mbox{B}4\bar{d})  \, [b,a]
\\[4pt] \hline
\end{array} \nonumber
\end{align}
\caption{
Summary table of
$\hat{\sigma}_{\mbox{{\tiny D}}}(\mbox{R}_{7u})$.
\label{ap_B_1_tab7}}
\end{table}
\newpage
\rightline{
\underline{
$\hat{\sigma}_{\mbox{{\tiny subt}}}
(\mbox{R}_{7u} = ug \to ud\bar{d})$  
}
}
\begin{enumerate}
\item[\fbox{Step2}]
\begin{equation}
\hat{\sigma}_{\mbox{{\tiny D}}}
(\,
\mbox{R}_{7u},\,
\mbox{I}
\,)=0
\ \ \ \mbox{and} \ \ \
\hat{\sigma}_{\mbox{{\tiny I}}}
(\mbox{R}_{7u})=0\,.
\end{equation}
\item[\fbox{Step3}]
\begin{equation}
\hat{\sigma}_{\mbox{{\tiny D}}}
(\,
\mbox{R}_{7u},\,
\mbox{P}
\,)
+
\hat{\sigma}_{\mbox{{\tiny C}}}(\mbox{R}_{7u}) 
-
\hat{\sigma}_{\mbox{{\tiny P}}}
(\mbox{R}_{7u})=0\,,
\end{equation}
which is separated into four relations 
for Dipoles 3$u$, 4$u$, 4$d$, and 4$\bar{d}$ as
\begin{align}
\hat{\sigma}_{\mbox{{\tiny D}}}
(\,
\mbox{R}_{7u},\,
\mbox{P},\,{\tt dip}j
\,)
+
\hat{\sigma}_{\mbox{{\tiny C}}}(\mbox{R}_{7u},\,
{\tt dip}j
) 
-
\hat{\sigma}_{\mbox{{\tiny P}}}
(\mbox{R}_{7u},
\,{\tt dip}j)=0\,.
\end{align}
\item[\fbox{Step4}]
\begin{equation}
\hat{\sigma}_{\mbox{{\tiny D}}}
(\,
\mbox{R}_{7u},\,
\mbox{K}
\,)
-
\hat{\sigma}_{\mbox{{\tiny K}}}
(\mbox{R}_{7u})=0\,,
\end{equation}
which is separated into four relations 
for Dipoles 3$u$, 4$u$, 4$d$, and 4$\bar{d}$ as
\begin{align}
\hat{\sigma}_{\mbox{{\tiny D}}}
(\,
\mbox{R}_{7u},\,
\mbox{K},\,{\tt dip}j
\,)
-
\hat{\sigma}_{\mbox{{\tiny K}}}
(\mbox{R}_{7u},
\,{\tt dip}j)=0\,.
\end{align}
\item[\fbox{Step5}]
\begin{equation}
\hat{\sigma}_{subt}(\mbox{R}_{7u})=
\hat{\sigma}_{\mbox{{\tiny D}}}(\mbox{R}_{7u},\,{\tt dip}2)\,.
\end{equation}
\begin{equation}
\hat{\sigma}_{\mbox{{\tiny D}}}(\mbox{R}_{7u},\,{\tt dip}2))
=
\hat{\sigma}_{\mbox{{\tiny D}}}
(
\mbox{R}_{7u},\,
{\tt dip}2,(5)\mbox{-}1/2,
\,
{\cal V}_{f\bar{f}}
\,)
+
\hat{\sigma}_{\mbox{{\tiny D}}}
(
\mbox{R}_{7u},\,
{\tt dip}2,(5)\mbox{-}2,
\,
h)\,.
\end{equation}
\begin{align}
\hat{\sigma}_{\mbox{{\tiny D}}}
(
\mbox{R}_{7u},\,
{\tt dip}2,(5)\mbox{-}1/2,
\,
{\cal V}_{f\bar{f}}
\,)
&=-\frac{A_{d}}{S_{\mbox{{\tiny R}}_{7u}}} 
\cdot
\frac{1}{\mbox{C}_{\mbox{{\tiny A}}}} 
{\cal V}_{f\bar{f}} (\ep) 
\cdot 
\Phi(\mbox{B}2)_{d} \,
\bigl(
\,
[2,1]+
[2,a]+
[2,b]
\,
\bigr)\,.
\\
\hat{\sigma}_{\mbox{{\tiny D}}}
(\,
\mbox{R}_{7u},\,
{\tt dip}2,(5)\mbox{-}2,
\,
h)
&=
-\frac{A_{4}}{S_{\mbox{{\tiny R}}_{7u}}} 
\int_{0}^{1}dx \,
\frac{\mbox{T}_{\mbox{{\tiny R}}}}
{\mbox{C}_{\mbox{{\tiny A}}}} 
\,\frac{2}{3}h(x)
\ \times
\nonumber\\
& \hspace{15mm}
\Bigl(
\Phi_{a}(\mbox{B}2,x)_{4}\,
\langle 2,a \rangle
+
\Phi_{b}(\mbox{B}2,x)_{4}\,
\langle 2,b \rangle
\Bigr)\,.
\end{align}
\end{enumerate}
%
\clearpage
\subsection{$\hat{\sigma}_{\mbox{{\tiny subt}}}(\mbox{R}_{8u})$  
\label{ap_B_8}}
%
%
\begin{table}[h!]
  \centering
\begin{align}
&\fbox{Step1} \ \ \ 
\hat{\sigma}_{\mbox{{\tiny D}}}
(\mbox{R}_{8u} = u\bar{u} \to ggg): 
\ \ S_{\mbox{{\tiny R}}_{8}}=6,
\nonumber\\
&
\begin{array}{|c|c|c|c|c|c|} \hline
{\tt Dip}\,j
& \mbox{B}j 
& \mbox{\small Splitting} 
& (y_{a},y_{b}:y_{1},y_{2}) 
& \mbox{Factor 1}
& \Phi(\mbox{B}_{j})
\,[y_{emi}, y_{spe}]
\\[4pt] \hline
& & & & & \\[-12pt]
{\tt Dip\,1}  
& u\bar{u} \to gg
& (2)-1   
& 1.\,(a,b\,;\widetilde{12},\widetilde{3} ) 
& {\cal V}_{gg}/\mbox{C}_{\mbox{{\tiny A}}}
& \Phi(\mbox{B}1) \, [1,2]
\\[4pt]  
&  
&   
& 2.\,(a,b\,;\widetilde{13},\widetilde{2}) 
&
& \Phi(\mbox{B}1) \, [1,2]
\\[4pt]  
&  
&   
& 3.\,(a,b\,;\widetilde{23},\widetilde{1}) 
&
& \Phi(\mbox{B}1) \, [1,2]
\\[4pt]  
&  
& (2)-2  
& 4.\,(\widetilde{a},b\,;\widetilde{12},3) 
& \int dx 
{\cal V}_{gg}(x)/\mbox{C}_{\mbox{{\tiny A}}}
& \Phi_{a}(\mbox{B}1)  \, [1,a]
\\[4pt]  
&  
&   
& 5.\,(a,\widetilde{b}\,;\widetilde{12},3) 
& 
& \Phi_{b}(\mbox{B}1)  \, [1,b]
\\[4pt]  
&  
&  
& 6.\,(\widetilde{a},b\,;\widetilde{13},2)
&
& \Phi_{a}(\mbox{B}1)  \, [1,a]
\\[4pt]  
&  
&   
& 7.\,(a,\widetilde{b}\,;\widetilde{13},2) 
&
& \Phi_{b}(\mbox{B}1)  \, [1,b]
\\[4pt] 
&  
&   
& 8.\,(\widetilde{a},b\,;\widetilde{23},1)
&
& \Phi_{a}(\mbox{B}1)  \, [1,a]
\\[4pt] 
&  
&   
& 9.\,(a,\widetilde{b}\,;\widetilde{23},1) 
&
& \Phi_{b}(\mbox{B}1)  \, [1,b]
\\[4pt] 
&  
& (3)-1  
& 10.\,(\widetilde{a1},b\,;\widetilde{2},3) 
& \int dx  
{\cal V}^{f,f} (x)/\mbox{C}_{\mbox{{\tiny F}}}
& \Phi_{a}(\mbox{B}1)  \, [a,1]
\\[4pt]  
&  
&   
& 11.\,(\widetilde{a1},b\,;2,\widetilde{3}) 
& 
& \Phi_{a}(\mbox{B}1)  \, [a,2]
\\[4pt]  
&  
&   
& 12.\,(\widetilde{a2},b\,;\widetilde{1},3)  
&
& \Phi_{a}(\mbox{B}1)  \, [a,1]
\\[4pt]  
&  
&   
& 13.\,(\widetilde{a2},b\,;1,\widetilde{3}) 
&
& \Phi_{a}(\mbox{B}1)  \, [a,2]
\\[4pt]  
&  
&   
& 14.\,(\widetilde{a3},b\,;\widetilde{1},2)
&
& \Phi_{a}(\mbox{B}1)  \, [a,1]
\\[4pt]  
&  
&   
& 15.\,(\widetilde{a3},b\,;1,\widetilde{2})
&
& \Phi_{a}(\mbox{B}1)  \, [a,2]
\\[4pt]   
&  
&   
& 16.\,(a,\widetilde{b1}\,;\widetilde{2},3)
&
& \Phi_{b}(\mbox{B}1)  \, [b,1]
\\[4pt]  
&  
&   
& 17.\,(a,\widetilde{b1}\,;2,\widetilde{3})
&
& \Phi_{b}(\mbox{B}1)  \, [b,2]
\\[4pt]  
&  
&   
& 18.\,(a,\widetilde{b2}\,;\widetilde{1},3)
&
& \Phi_{b}(\mbox{B}1)  \, [b,1]
\\[4pt]  
&  
&   
& 19.\,(a,\widetilde{b2}\,;1,\widetilde{3})
&
& \Phi_{b}(\mbox{B}1)  \, [b,2]
\\[4pt]  
&  
&   
& 20.\,(a,\widetilde{b3}\,;\widetilde{1},2)
&
& \Phi_{b}(\mbox{B}1)  \, [b,1]
\\[4pt]   
&  
&   
& 21.\,(a,\widetilde{b3}\,;1,\widetilde{2})
&
& \Phi_{b}(\mbox{B}1)  \, [b,2]
\\[4pt]  
&  
& (3)-2  
& 22.\,(\widetilde{a1},\widetilde{b}\,;2,3) 
& \int dx
{\widetilde {\cal V}}^{f,f} (x)
/\mbox{C}_{\mbox{{\tiny F}}}
& \Phi_{a}(\mbox{B}1)  \, [a,b]
\\[4pt]  
&  
& 
& 23.\,(\widetilde{a2},\widetilde{b}\,;1,3) 
&
& \Phi_{a}(\mbox{B}1)  \, [a,b]
\\[4pt] 
&  
& 
& 24.\,(\widetilde{a3},\widetilde{b}\,;1,2) 
&
& \Phi_{a}(\mbox{B}1)  \, [a,b]
\\[4pt] 
&  
& 
& 25.\,(\widetilde{a},\widetilde{b1}\,;2,3) 
&
& \Phi_{b}(\mbox{B}1)  \, [b,a]
\\[4pt] 
&  
& 
& 26.\,(\widetilde{a},\widetilde{b2}\,;1,3) 
&
& \Phi_{b}(\mbox{B}1)  \, [b,a]
\\[4pt] 
&  
&   
& 27.\,(\widetilde{a},\widetilde{b3}\,;1,2) 
&
& \Phi_{b}(\mbox{B}1)  \, [b,a]
\\[4pt]  \hline
\end{array} \nonumber
\end{align}
\caption{
Summary table of 
$\hat{\sigma}_{\mbox{{\tiny D}}}(\mbox{R}_{8u})$.
\label{ap_B_1_tab8}}
\end{table}
\newpage
\rightline{
\underline{
$\hat{\sigma}_{\mbox{{\tiny subt}}}
(\mbox{R}_{8u} = u\bar{u} \to ggg)$  
}
}
\begin{enumerate}
\item[\fbox{Step2}]
\begin{equation}
\hat{\sigma}_{\mbox{{\tiny D}}}
(\,
\mbox{R}_{8u},\,
\mbox{I}
\,)
-
\hat{\sigma}_{\mbox{{\tiny I}}}
(\mbox{R}_{8u})
=
-\,
\hat{\sigma}_{\mbox{{\tiny I}}}
(\mbox{R}_{8u},(2)\mbox{-}1/2,N_{f}{\cal V}_{f\bar{f}})\,.
\end{equation}
\begin{align}
\hat{\sigma}_{\mbox{{\tiny I}}}
(\mbox{R}_{8u},(2)\mbox{-}1/2,N_{f}{\cal V}_{f\bar{f}})
&=
-\frac{A_{d}}{S_{\mbox{{\tiny B}}_{1}}} 
\cdot
\frac{N_{f}}{\mbox{C}_{\mbox{{\tiny A}}}} 
{\cal V}_{f\bar{f}} (\ep) 
\Phi(\mbox{B}1)_{d} \,
\cdot
\nonumber\\
&\ \ \ 
\bigl(
\,
[1,2]+
[2,1]+
[1,a]+
[1,b]+
[2,a]+
[2,b]
\,
\bigr)\,.
\label{r8st2nf}
\end{align}
\item[\fbox{Step3}]
\begin{equation}
\hat{\sigma}_{\mbox{{\tiny D}}}
(\,
\mbox{R}_{8u},\,
\mbox{P}
\,)
+
\hat{\sigma}_{\mbox{{\tiny C}}}(\mbox{R}_{8u}) 
-
\hat{\sigma}_{\mbox{{\tiny P}}}
(\mbox{R}_{8u})=0\,,
\end{equation}
which includes only Dipole\,1\,.
\item[\fbox{Step4}]
\begin{equation}
\hat{\sigma}_{\mbox{{\tiny D}}}
(\,
\mbox{R}_{8u},\,
\mbox{K}
\,)
-
\hat{\sigma}_{\mbox{{\tiny K}}}
(\mbox{R}_{8u})=
-\,
\hat{\sigma}_{\mbox{{\tiny K}}}
(\mbox{R}_{8u},\,{\tt dip}1,
(3)\mbox{-}1,N_{f}h)\,,
\end{equation}
which includes only Dipole\,1\,.
\begin{align}
\hat{\sigma}_{\mbox{{\tiny K}}}
(\mbox{R}_{8u},\,{\tt dip}1,
(3)\mbox{-}1,N_{f}h)
&=
-\frac{A_{4}}{S_{\mbox{{\tiny B}}_{1}}} 
\int_{0}^{1}dx \,
\frac{\mbox{T}_{\mbox{{\tiny R}}}N_{f}}
{\mbox{C}_{\mbox{{\tiny A}}}} 
\,\frac{2}{3}h(x)
\ \times
\nonumber\\
\Bigl[
\Phi_{a}(\mbox{B}1,x)_{4}\,
&\bigl(
\langle a,1 \rangle
+
\langle a,2 \rangle
\bigr)
+
\Phi_{b}(\mbox{B}1,x)_{4}\,
\bigl(
\langle b,1 \rangle
+
\langle b,2 \rangle
\bigr)
\Bigr]\,.
\label{r8st4nf}
\end{align}
\item[\fbox{Step5}]
\begin{equation}
\hat{\sigma}_{subt}(\mbox{R}_{8u})=
-\,
\hat{\sigma}_{\mbox{{\tiny I}}}
(\mbox{R}_{8u},(2)\mbox{-}1/2,N_{f}{\cal V}_{f\bar{f}})
-\,
\hat{\sigma}_{\mbox{{\tiny K}}}
(\mbox{R}_{8u},\,{\tt dip}1,
(3)\mbox{-}1,N_{f}h)\,.
\label{r8st5}
\end{equation}
\end{enumerate}
%
\clearpage
\subsection{$\hat{\sigma}_{\mbox{{\tiny subt}}}(\mbox{R}_{9u})$  
\label{ap_B_9}}
%
%
\begin{table}[h!]
  \centering
\begin{align}
&\fbox{Step1}  \ \ \
\hat{\sigma}_{\mbox{{\tiny D}}}
(\mbox{R}_{9u} = ug \to ugg): 
\ \ S_{\mbox{{\tiny R}}_{9}}=2,
\nonumber\\
&
\begin{array}{|c|c|c|c|c|c|} \hline
{\tt Dip}\,j
& \mbox{B}j 
& \mbox{\small Splitting} 
& (y_{a},y_{b}:y_{1},y_{2}) 
& \mbox{Factor 1}
& \Phi(\mbox{B}_{j})
\,[y_{emi}, y_{spe}]
\\[4pt] \hline
& & & & & \\[-12pt]
{\tt Dip\,1}  
& ug \to ug
& (1)-1   
& 1.\,(a,b\,;\widetilde{12},\widetilde{3} ) 
& {\cal V}_{fg}/\mbox{C}_{\mbox{{\tiny F}}}
& \Phi(\mbox{B}1) \, [1,2]
\\[4pt]  
&  
&   
& 2.\,(a,b\,;\widetilde{13},\widetilde{2}) 
&
& \Phi(\mbox{B}1) \, [1,2]
\\[4pt]  
&  
& (1)-2  
& 3.\,(\widetilde{a},b\,;\widetilde{12},3) 
& \int dx 
{\cal V}_{fg}(x)/\mbox{C}_{\mbox{{\tiny F}}}
& \Phi_{a}(\mbox{B}1)  \, [1,a]
\\[4pt]  
&  
&   
& 4.\,(a,\widetilde{b}\,;\widetilde{12},3) 
&
& \Phi_{b}(\mbox{B}1)  \, [1,b]
\\[4pt]  
&  
&  
& 5.\,(\widetilde{a},b\,;\widetilde{13},2) 
&
& \Phi_{a}(\mbox{B}1)  \, [1,a]
\\[4pt]  
&  
&   
& 6.\,(a,\widetilde{b}\,;\widetilde{13},2) 
&
& \Phi_{b}(\mbox{B}1)  \, [1,b]
\\[4pt]  
&  
& (2)-1  
& 7.\,(a,b\,;\widetilde{1},\widetilde{23}) 
& {\cal V}_{gg}/\mbox{C}_{\mbox{{\tiny A}}}
& \Phi(\mbox{B}1) \, [2,1]
\\[4pt]  
&  
& (2)-2
& 8.\,(\widetilde{a},b\,;1,\widetilde{23}) 
& \int dx 
{\cal V}_{gg}(x)/\mbox{C}_{\mbox{{\tiny A}}}
& \Phi_{a}(\mbox{B}1)  \, [2,a]
\\[4pt]  
&  
&   
& 9.\,(a,\widetilde{b}\,;1,\widetilde{23}) 
&
& \Phi_{b}(\mbox{B}1)  \, [2,b]
\\[4pt]  
&  
& (3)-1  
& 10.\,(\widetilde{a2},b\,;\widetilde{1},3)
& \int dx  
{\cal V}^{f,f} (x)/\mbox{C}_{\mbox{{\tiny F}}}
& \Phi_{a}(\mbox{B}1)  \, [a,1]
\\[4pt]  
&  
&
& 11.\,(\widetilde{a2},b\,;1,\widetilde{3}) 
&  
& \Phi_{a}(\mbox{B}1)  \, [a,2]
\\[4pt]  
&  
&   
& 12.\,(\widetilde{a3},b\,;\widetilde{1},2) 
&
& \Phi_{a}(\mbox{B}1)  \, [a,1]
\\[4pt]
&  
& 
& 13.\,(\widetilde{a3},b\,;1,\widetilde{2}) 
&
& \Phi_{a}(\mbox{B}1)  \, [a,2]
\\[4pt]
&  
& (3)-2
& 14.\,(\widetilde{a2},\widetilde{b}\,;1,3) 
& \int dx
{\widetilde {\cal V}}^{f,f} (x)
/\mbox{C}_{\mbox{{\tiny F}}}
& \Phi_{a}(\mbox{B}1)  \, [a,b]
\\[4pt]
&  
&   
& 15.\,(\widetilde{a3},\widetilde{b}\,;1,2) 
&
& \Phi_{a}(\mbox{B}1)  \, [a,b]
\\[4pt]
&  
& (4)-1  
& 16.\,(a,\widetilde{b2}\,;\widetilde{1},3)
& \int dx  
{\cal V}^{g,g} (x)/\mbox{C}_{\mbox{{\tiny A}}}
& \Phi_{b}(\mbox{B}1)  \, [b,1]
\\[4pt]  
&  
&
& 17.\,(a,\widetilde{b2}\,;1,\widetilde{3}) 
&
& \Phi_{b}(\mbox{B}1)  \, [b,2]
\\[4pt]  
&  
&   
& 18.\,(a,\widetilde{b3}\,;\widetilde{1},2) 
&
& \Phi_{b}(\mbox{B}1)  \, [b,1]
\\[4pt]
&  
& 
& 19.\,(a,\widetilde{b3}\,;1,\widetilde{2}) 
&
& \Phi_{b}(\mbox{B}1)  \, [b,2]
\\[4pt]
&  
& (4)-2
& 20.\,(\widetilde{a},\widetilde{b2}\,;1,3) 
& \int dx
{\widetilde {\cal V}}^{g,g} (x)
/\mbox{C}_{\mbox{{\tiny A}}}
& \Phi_{b}(\mbox{B}1)  \, [b,a]
\\[4pt]
&  
&   
& 21.\,(\widetilde{a},\widetilde{b3}\,;1,2) 
&
& \Phi_{b}(\mbox{B}1)  \, [b,a]
\\[4pt]  \hline
& & & & & \\[-12pt]
{\tt Dip\,3u}  
& gg \to gg
& (6)-1    
& 22.\,(\widetilde{a1},b\,;\widetilde{2},3) 
& \int dx {\cal V}^{f,g} (x)/\mbox{C}_{\mbox{{\tiny A}}}
& \Phi_{a}(\mbox{B}3u)  \, [a,1]
\\[4pt]  
&  
&   
& 23.\,(\widetilde{a1},b\,;2,\widetilde{3}) 
&
& \Phi_{a}(\mbox{B}3u)  \, [a,2]
\\[4pt]  
&  
& (6)-2  
& 24.\,(\widetilde{a1},\widetilde{b}\,;2,3) 
& \int dx {\widetilde {\cal V}}^{f,g} (x)/
\mbox{C}_{\mbox{{\tiny A}}}
& \Phi_{a}(\mbox{B}3u)  \, [a,b]
\\[4pt]  \hline
& & & & & \\[-12pt]
{\tt Dip\,4u}  
& u\bar{u} \to gg
& (7)-1   
& 25.\,(a,\widetilde{b1}\,;\widetilde{2},3)  
& \int dx {\cal V}^{g,f} (x)/\mbox{C}_{\mbox{{\tiny F}}}
& \Phi_{b}(\mbox{B}4u)  \, [b,1]
\\[4pt]  
&  
&   
& 26.\,(a,\widetilde{b1}\,;2,\widetilde{3})  
&
& \Phi_{b}(\mbox{B}4u)  \, [b,2]
\\[4pt]  
&  
& (7)-2   
& 27.\,(\widetilde{a},\widetilde{b1}\,;2,3)  
& \int dx {\widetilde {\cal V}}^{g,f} (x)/
\mbox{C}_{\mbox{{\tiny F}}}
& \Phi_{b}(\mbox{B}4u)  \, [b,a]
\\[4pt] \hline
\end{array} \nonumber
\end{align}
\caption{
Summary table of 
$\hat{\sigma}_{\mbox{{\tiny D}}}(\mbox{R}_{9u})$.
\label{ap_B_1_tab9}}
\end{table}
\newpage
\rightline{
\underline{
$\hat{\sigma}_{\mbox{{\tiny subt}}}
(\mbox{R}_{9u} = ug \to ugg)$  
}
}
\begin{enumerate}
\item[\fbox{Step2}]
\begin{equation}
\hat{\sigma}_{\mbox{{\tiny D}}}
(\,
\mbox{R}_{9u},\,
\mbox{I}
\,)
-
\hat{\sigma}_{\mbox{{\tiny I}}}
(\mbox{R}_{9u})
=
-\,
\hat{\sigma}_{\mbox{{\tiny I}}}
(\mbox{R}_{9u},(2)\mbox{-}1/2,N_{f}{\cal V}_{f\bar{f}})\,.
\end{equation}
\begin{align}
\hat{\sigma}_{\mbox{{\tiny I}}}
(\mbox{R}_{9u},(2)\mbox{-}1/2,N_{f}{\cal V}_{f\bar{f}})
&=
-\frac{A_{d}}{S_{\mbox{{\tiny B}}_{1}}} 
\cdot
\frac{N_{f}}{\mbox{C}_{\mbox{{\tiny A}}}} 
{\cal V}_{f\bar{f}} (\ep) 
\,\Phi(\mbox{B}1)_{d} \,
\cdot
\bigl(
\,
[2,1]+
[2,a]+
[2,b]
\,
\bigr)\,.
\end{align}
\item[\fbox{Step3}]
\begin{equation}
\hat{\sigma}_{\mbox{{\tiny D}}}
(\,
\mbox{R}_{9u},\,
\mbox{P}
\,)
+
\hat{\sigma}_{\mbox{{\tiny C}}}(\mbox{R}_{9u}) 
-
\hat{\sigma}_{\mbox{{\tiny P}}}
(\mbox{R}_{9u})=0\,,
\end{equation}
which is separated into three relations 
for Dipoles 1, 3$u$, and 4$u$ as
\begin{align}
\hat{\sigma}_{\mbox{{\tiny D}}}
(\,
\mbox{R}_{9u},\,
\mbox{P},\,{\tt dip}j
\,)
+
\hat{\sigma}_{\mbox{{\tiny C}}}(\mbox{R}_{9u},\,
{\tt dip}j
) 
-
\hat{\sigma}_{\mbox{{\tiny P}}}
(\mbox{R}_{9u},
\,{\tt dip}j)=0\,.
\end{align}
\item[\fbox{Step4}]
\begin{equation}
\hat{\sigma}_{\mbox{{\tiny D}}}
(\,
\mbox{R}_{9u},\,
\mbox{K}
\,)
-
\hat{\sigma}_{\mbox{{\tiny K}}}
(\mbox{R}_{9u})=
-\,
\hat{\sigma}_{\mbox{{\tiny K}}}
(\mbox{R}_{9u},\,{\tt dip}1,
(3)/(4)\mbox{-}1,N_{f}h)\,,
\end{equation}
which is separated into three relations 
for Dipoles 1, 3$u$, and 4$u$ as
\begin{align}
\hat{\sigma}_{\mbox{{\tiny D}}}
(\,
\mbox{R}_{9u},\,
\mbox{K},\,{\tt dip}1
\,)
-
\hat{\sigma}_{\mbox{{\tiny K}}}
(\mbox{R}_{9u},
\,{\tt dip}1)
&=
-\,
\hat{\sigma}_{\mbox{{\tiny K}}}
(\mbox{R}_{9u},\,{\tt dip}1,
(3)/(4)\mbox{-}1,N_{f}h)\,,
\\
\hat{\sigma}_{\mbox{{\tiny D}}}
(\,
\mbox{R}_{9u},\,
\mbox{K},\,{\tt dip}3u
\,)
-
\hat{\sigma}_{\mbox{{\tiny K}}}
(\mbox{R}_{9u},
\,{\tt dip}3u)
&=0\,,
\\
\hat{\sigma}_{\mbox{{\tiny D}}}
(\,
\mbox{R}_{9u},\,
\mbox{K},\,{\tt dip}4u
\,)
-
\hat{\sigma}_{\mbox{{\tiny K}}}
(\mbox{R}_{9u},
\,{\tt dip}4u)
&=0\,.
\end{align}
\begin{align}
\hat{\sigma}_{\mbox{{\tiny K}}}
(\mbox{R}_{9u},\,{\tt dip}1,
(3)/(4)\mbox{-}1,N_{f}h)
&=
-\frac{A_{4}}{S_{\mbox{{\tiny B}}_{1}}} 
\int_{0}^{1}dx \,
\frac{\mbox{T}_{\mbox{{\tiny R}}}N_{f}}
{\mbox{C}_{\mbox{{\tiny A}}}} 
\,\frac{2}{3}h(x)
\ \times
\nonumber\\
&
\Bigl[
\Phi_{a}(\mbox{B}1,x)_{4}\,
\langle a,2 \rangle
+
\Phi_{b}(\mbox{B}1,x)_{4}\,
\langle b,2 \rangle
\Bigr]\,.
\end{align}
\item[\fbox{Step5}]
\begin{equation}
\hat{\sigma}_{subt}(\mbox{R}_{9u})=
-\,
\hat{\sigma}_{\mbox{{\tiny I}}}
(\mbox{R}_{9u},(2)\mbox{-}1/2,N_{f}{\cal V}_{f\bar{f}})
-\,
\hat{\sigma}_{\mbox{{\tiny K}}}
(\mbox{R}_{9u},\,{\tt dip}1,
(3)/(4)\mbox{-}1,N_{f}h)\,.
\end{equation}
\end{enumerate}
%
\clearpage
\subsection{$\hat{\sigma}_{\mbox{{\tiny subt}}}(\mbox{R}_{10u})$  
\label{ap_B_10}}
%
%
\begin{table}[h!]
  \centering
\begin{align}
&\fbox{Step1} \ \ \ 
\hat{\sigma}_{\mbox{{\tiny D}}}
(\mbox{R}_{10u}=gg \to u\bar{u}g): 
\ \ S_{\mbox{{\tiny R}}_{10}}=1,
\nonumber\\
&
\begin{array}{|c|c|c|c|c|c|} \hline
{\tt Dip}\,j
& \mbox{B}j 
& \mbox{\small Splitting} 
& (y_{a},y_{b}:y_{1},y_{2}) 
& \mbox{Factor 1}
& \Phi(\mbox{B}_{j})
\,[y_{emi}, y_{spe}]
\\[4pt] \hline
& & & & & \\[-12pt]
{\tt Dip\,1}  
& gg \to u\bar{u}
& (1)-1   
& 1.\,(a,b\,;\widetilde{13},\widetilde{2} ) 
& {\cal V}_{fg}/\mbox{C}_{\mbox{{\tiny F}}} 
& \Phi(\mbox{B}1) \, [1,2]
\\[4pt]  
&  
&   
& 2.\,(a,b\,;\widetilde{1},\widetilde{23}) 
&
& \Phi(\mbox{B}1) \, [2,1]
\\[4pt]  
&  
& (1)-2  
& 3.\,(\widetilde{a},b\,;\widetilde{13},2) 
& \int dx 
{\cal V}_{fg}(x)/\mbox{C}_{\mbox{{\tiny F}}}
& \Phi_{a}(\mbox{B}1)  \, [1,a]
\\[4pt]  
&  
&   
& 4.\,(a,\widetilde{b}\,;\widetilde{13},2) 
&
& \Phi_{b}(\mbox{B}1)  \, [1,b]
\\[4pt]  
&  
&  
& 5.\,(\widetilde{a},b\,;1,\widetilde{23}) 
&
& \Phi_{a}(\mbox{B}1)  \, [2,a]
\\[4pt]  
&  
&   
& 6.\,(a,\widetilde{b}\,;1,\widetilde{23}) 
&
& \Phi_{b}(\mbox{B}1)  \, [2,b]
\\[4pt]  
&  
& (4)-1  
& 7.\,(\widetilde{a3},b\,;\widetilde{1},2) 
& \int dx  
{\cal V}^{g,g} (x)/\mbox{C}_{\mbox{{\tiny A}}}
& \Phi_{a}(\mbox{B}1)  \, [a,1]
\\[4pt]  
&  
&   
& 8.\,(\widetilde{a3},b\,;1,\widetilde{2}) 
&
& \Phi_{a}(\mbox{B}1)  \, [a,2]
\\[4pt]  
&  
&   
& 9.\,(a,\widetilde{b3}\,;\widetilde{1},2) 
&
& \Phi_{b}(\mbox{B}1)  \, [b,1]
\\[4pt]  
&  
&   
& 10.\,(a,\widetilde{b3}\,;1,\widetilde{2})
&
& \Phi_{b}(\mbox{B}1)  \, [b,2]
\\[4pt]  
&  
& (4)-2  
& 11.\,(\widetilde{a3},\widetilde{b}\,;1,2) 
& \int dx
{\widetilde {\cal V}}^{g,g} (x)
/\mbox{C}_{\mbox{{\tiny A}}}
& \Phi_{a}(\mbox{B}1)  \, [a,b]
\\[4pt]  
&  
&   
& 12.\,(\widetilde{a},\widetilde{b3}\,;1,2) 
&
& \Phi_{b}(\mbox{B}1)  \, [b,a]
\\[4pt]  \hline
& & & & & \\[-12pt]
{\tt Dip\,2u}  
& gg \to gg
& (5)-1  
& 13.\,(a,b\,;\widetilde{12},\widetilde{3}) 
& {\cal V}_{f\bar{f}}/ \mbox{C}_{\mbox{{\tiny A}}}
& \Phi(\mbox{B}2u) \, [1,2]
\\[4pt]  
&  
& (5)-2    
& 14.\,(\widetilde{a},b\,;\widetilde{12},3) 
& \int dx {\cal V}_{f\bar{f}}(x)/ \mbox{C}_{\mbox{{\tiny A}}}
& \Phi_{a}(\mbox{B}2u)  \, [1,a]
\\[4pt]  
&  
&   
& 15.\,(a,\widetilde{b}\,;\widetilde{12},3) 
&
& \Phi_{b}(\mbox{B}2u)  \, [1,b]
\\[4pt]  \hline
& & & & & \\[-12pt]
{\tt Dip\,4u}  
& \bar{u}g \to \bar{u}g
& (7)-1    
& 16.\,(\widetilde{a1},b\,;\widetilde{2},3) 
& \int dx {\cal V}^{g,f} (x)/\mbox{C}_{\mbox{{\tiny F}}}
& \Phi_{a}(\mbox{B}4u)  \, [a,1]
\\[4pt]  
&  
&   
& 17.\,(\widetilde{a1},b\,;2,\widetilde{3}) 
&
& \Phi_{a}(\mbox{B}4u)  \, [a,2]
\\[4pt] 
&  
&   
& 18.\,(\widetilde{b1},a\,;\widetilde{2},3) 
&
& \Phi_{b}(\mbox{B}4u)  \, [a,1]
\\[4pt] 
&  
&   
& 19.\,(\widetilde{b1},a\,;2,\widetilde{3}) 
&
& \Phi_{b}(\mbox{B}4u)  \, [a,2]
\\[4pt] 
&  
& (7)-2  
& 20.\,(\widetilde{a1},\widetilde{b}\,;2,3) 
& \int dx {\widetilde {\cal V}}^{g,f} (x)/
\mbox{C}_{\mbox{{\tiny F}}}
& \Phi_{a}(\mbox{B}4u)  \, [a,b]
\\[4pt] 
&  
&  
& 21.\,(\widetilde{b1},\widetilde{a}\,;2,3) 
&
& \Phi_{b}(\mbox{B}4u)  \, [a,b]
\\[4pt]  \hline
& & & & & \\[-12pt]
{\tt Dip\,4\bar{u}}  
& ug \to ug
& (7)-1   
& 22.\,(\widetilde{a2},b\,;\widetilde{1},3)  
& \int dx {\cal V}^{g,f} (x)/\mbox{C}_{\mbox{{\tiny F}}}
& \Phi_{a}(\mbox{B}4\bar{u})  \, [a,1]
\\[4pt]  
&  
&   
& 23.\,(\widetilde{a2},b\,;1,\widetilde{3})  
&
& \Phi_{a}(\mbox{B}4\bar{u})  \, [a,2]
\\[4pt] 
&  
&   
& 24.\,(\widetilde{b2},a\,;\widetilde{1},3)  
&
& \Phi_{b}(\mbox{B}4\bar{u})  \, [a,1]
\\[4pt] 
&  
&   
& 25.\,(\widetilde{b2},a\,;1,\widetilde{3})  
&
& \Phi_{b}(\mbox{B}4\bar{u})  \, [a,2]
\\[4pt] 
&  
& (7)-2   
& 26.\,(\widetilde{a2},\widetilde{b}\,;1,3)  
& \int dx {\widetilde {\cal V}}^{g,f} (x)/
\mbox{C}_{\mbox{{\tiny F}}}
& \Phi_{a}(\mbox{B}4\bar{u})  \, [a,b]
\\[4pt]
&  
&   
& 27.\,(\widetilde{b2},\widetilde{a}\,;1,3)  
&
& \Phi_{b}(\mbox{B}4\bar{u})  \, [a,b]
\\[4pt] \hline
\end{array} \nonumber
\end{align}
\caption{
Summary table of
$\hat{\sigma}_{\mbox{{\tiny D}}}(\mbox{R}_{10u})$. 
\label{ap_B_1_tab10}}
\end{table}
\newpage
\rightline{
\underline{
$
\hat{\sigma}_{\mbox{{\tiny subt}}}
(\mbox{R}_{10u}=gg \to u\bar{u}g)
$  
}
}
\begin{enumerate}
\item[\fbox{Step2}]
\begin{equation}
\hat{\sigma}_{\mbox{{\tiny D}}}
(\,
\mbox{R}_{10u},\,
\mbox{I}
\,)
-
\hat{\sigma}_{\mbox{{\tiny I}}}
(\mbox{R}_{10u})=0\,.
\end{equation}
\item[\fbox{Step3}]
\begin{equation}
\hat{\sigma}_{\mbox{{\tiny D}}}
(\,
\mbox{R}_{10u},\,
\mbox{P}
\,)
+
\hat{\sigma}_{\mbox{{\tiny C}}}(\mbox{R}_{10u}) 
-
\hat{\sigma}_{\mbox{{\tiny P}}}
(\mbox{R}_{10u})=0\,,
\end{equation}
which is separated into three relations 
for Dipoles 1, 4$u$, and 4$\bar{u}$ as
\begin{align}
\hat{\sigma}_{\mbox{{\tiny D}}}
(\,
\mbox{R}_{10u},\,
\mbox{P},\,{\tt dip}j
\,)
+
\hat{\sigma}_{\mbox{{\tiny C}}}(\mbox{R}_{10u},\,
{\tt dip}j
) 
-
\hat{\sigma}_{\mbox{{\tiny P}}}
(\mbox{R}_{10u},
\,{\tt dip}j)=0\,.
\end{align}
\item[\fbox{Step4}]
\begin{equation}
\hat{\sigma}_{\mbox{{\tiny D}}}
(\,
\mbox{R}_{10u},\,
\mbox{K}
\,)
-
\hat{\sigma}_{\mbox{{\tiny K}}}
(\mbox{R}_{10u})=0\,,
\end{equation}
which is separated into three relations 
for Dipoles 1, 4$u$, and 4$\bar{u}$ as
\begin{align}
\hat{\sigma}_{\mbox{{\tiny D}}}
(\,
\mbox{R}_{10u},\,
\mbox{K},\,{\tt dip}j
\,)
-
\hat{\sigma}_{\mbox{{\tiny K}}}
(\mbox{R}_{10u},
\,{\tt dip}j)=0\,.
\end{align}
\item[\fbox{Step5}]
\begin{equation}
\hat{\sigma}_{subt}(\mbox{R}_{10u})=
\hat{\sigma}_{\mbox{{\tiny D}}}(\mbox{R}_{10u},\,{\tt dip}2)\,.
\end{equation}
\begin{equation}
\hat{\sigma}_{\mbox{{\tiny D}}}(\mbox{R}_{10u},\,{\tt dip}2)
=
\hat{\sigma}_{\mbox{{\tiny D}}}
(
\mbox{R}_{10u},\,
{\tt dip}2,(5)\mbox{-}1/2,
\,
{\cal V}_{f\bar{f}}
\,)
+
\hat{\sigma}_{\mbox{{\tiny D}}}
(
\mbox{R}_{10u},\,
{\tt dip}2,(5)\mbox{-}2,
\,
h)\,.
\end{equation}
\begin{align}
\hat{\sigma}_{\mbox{{\tiny D}}}
(
\mbox{R}_{10u},\,
{\tt dip}2,(5)\mbox{-}1/2,
\,
{\cal V}_{f\bar{f}}
\,)
&=-\frac{A_{d}}{S_{\mbox{{\tiny R}}_{10u}}} 
\cdot
\frac{1}{\mbox{C}_{\mbox{{\tiny A}}}} 
{\cal V}_{f\bar{f}} (\ep) 
\cdot 
\Phi(\mbox{B}2)_{d} \,
\bigl(
\,
[1,2]+
[1,a]+
[1,b]
\,
\bigr)\,.
\\
\hat{\sigma}_{\mbox{{\tiny D}}}
(
\mbox{R}_{10u},\,
{\tt dip}2,(5)\mbox{-}2,
\,
h)
&=
-\frac{A_{4}}{S_{\mbox{{\tiny R}}_{10u}}} 
\int_{0}^{1}dx \,
\frac{\mbox{T}_{\mbox{{\tiny R}}}}
{\mbox{C}_{\mbox{{\tiny A}}}} 
\,\frac{2}{3}h(x)
\ \times
\nonumber\\
& \hspace{15mm}
\Bigl(
\Phi_{a}(\mbox{B}2,x)_{4}\,
\langle 1,a \rangle
+
\Phi_{b}(\mbox{B}2,x)_{4}\,
\langle 1,b \rangle
\Bigr)\,.
\end{align}
\end{enumerate}
%
\clearpage
\subsection{$\hat{\sigma}_{\mbox{{\tiny subt}}}(\mbox{R}_{11})$  
\label{ap_B_11}}
%
%
\begin{table}[h!]
  \centering
\begin{align}
&\fbox{Step1}  \ \ \
\hat{\sigma}_{\mbox{{\tiny D}}}
(\mbox{R}_{11}=gg \to ggg): 
\ \ S_{\mbox{{\tiny R}}_{11}}=6,
\nonumber\\
&
\begin{array}{|c|c|c|c|c|c|} \hline
{\tt Dip}\,j
& \mbox{B}j 
& \mbox{\small Splitting} 
& (y_{a},y_{b}:y_{1},y_{2}) 
& \mbox{Factor 1}
& \Phi(\mbox{B}_{j})
\,[y_{emi}, y_{spe}]
\\[4pt] \hline
& & & & & \\[-12pt]
{\tt Dip\,1}  
& gg \to gg
& (2)-1   
& 1.\,(a,b\,;\widetilde{12},\widetilde{3})
& {\cal V}_{gg}/\mbox{C}_{\mbox{{\tiny A}}}
& \Phi(\mbox{B}1) \, [1,2]
\\[4pt]  
&  
&   
& 2.\,(a,b\,;\widetilde{13},\widetilde{2}) 
&
& \Phi(\mbox{B}1) \, [1,2]
\\[4pt]  
&  
&   
& 3.\,(a,b\,;\widetilde{23},\widetilde{1}) 
&
& \Phi(\mbox{B}1) \, [1,2]
\\[4pt]  
&  
& (2)-2  
& 4.\,(\widetilde{a},b\,;\widetilde{12},3) 
& \int dx 
{\cal V}_{gg}(x)/\mbox{C}_{\mbox{{\tiny A}}}
& \Phi_{a}(\mbox{B}1)  \, [1,a]
\\[4pt]  
&  
&   
& 5.\,(a,\widetilde{b}\,;\widetilde{12},3) 
&
& \Phi_{b}(\mbox{B}1)  \, [1,b]
\\[4pt]  
&  
&  
& 6.\,(\widetilde{a},b\,;\widetilde{13},2) 
& 
& \Phi_{a}(\mbox{B}1)  \, [1,a]
\\[4pt]  
&  
&   
& 7.\,(a,\widetilde{b}\,;\widetilde{13},2) 
&
& \Phi_{b}(\mbox{B}1)  \, [1,b]
\\[4pt]  
&  
&   
& 8.\,(\widetilde{a},b\,;\widetilde{23},1) 
&
& \Phi_{a}(\mbox{B}1)  \, [1,a]
\\[4pt]  
&  
&   
& 9.\,(a,\widetilde{b}\,;\widetilde{23},1) 
&
& \Phi_{b}(\mbox{B}1)  \, [1,b]
\\[4pt]  
&  
& (4)-1  
& 10.\,(\widetilde{a1},b\,;\widetilde{2},3) 
& \int dx  
{\cal V}^{g,g} (x)/\mbox{C}_{\mbox{{\tiny A}}}
& \Phi_{a}(\mbox{B}1)  \, [a,1]
\\[4pt]  
&  
&   
& 11.\,(\widetilde{a1},b\,;2,\widetilde{3}) 
&
& \Phi_{a}(\mbox{B}1)  \, [a,2]
\\[4pt]  
&  
&   
& 12.\,(\widetilde{a2},b\,;\widetilde{1},3) 
&
& \Phi_{a}(\mbox{B}1)  \, [a,1]
\\[4pt]  
&  
&   
& 13.\,(\widetilde{a2},b\,;1,\widetilde{3})
&
& \Phi_{a}(\mbox{B}1)  \, [a,2]
\\[4pt]  
&  
&   
& 14.\,(\widetilde{a3},b\,;\widetilde{1},2)
&
& \Phi_{a}(\mbox{B}1)  \, [a,1]
\\[4pt]  
&  
&   
& 15.\,(\widetilde{a3},b\,;1,\widetilde{2})
&
& \Phi_{a}(\mbox{B}1)  \, [a,2]
\\[4pt]  
&  
&   
& 16.\,(a,\widetilde{b1}\,;\widetilde{2},3)
&
& \Phi_{b}(\mbox{B}1)  \, [b,1]
\\[4pt]  
&  
&   
& 17.\,(a,\widetilde{b1}\,;2,\widetilde{3})
&
& \Phi_{b}(\mbox{B}1)  \, [b,2]
\\[4pt]  
&  
&   
& 18.\,(a,\widetilde{b2}\,;\widetilde{1},3)
&
& \Phi_{b}(\mbox{B}1)  \, [b,1]
\\[4pt]  
&  
&   
& 19.\,(a,\widetilde{b2}\,;1,\widetilde{3})
&
& \Phi_{b}(\mbox{B}1)  \, [b,2]
\\[4pt]  
&  
&   
& 20.\,(a,\widetilde{b3}\,;\widetilde{1},2)
&
& \Phi_{b}(\mbox{B}1)  \, [b,1]
\\[4pt]  
&  
&   
& 21.\,(a,\widetilde{b3}\,;1,\widetilde{2})
&
& \Phi_{b}(\mbox{B}1)  \, [b,2]
\\[4pt]  
&  
& (4)-2  
& 22.\,(\widetilde{a1},\widetilde{b}\,;2,3) 
& \int dx
{\widetilde {\cal V}}^{g,g} (x)
/\mbox{C}_{\mbox{{\tiny A}}}
& \Phi_{a}(\mbox{B}1)  \, [a,b]
\\[4pt]  
&  
& 
& 23.\,(\widetilde{a2},\widetilde{b}\,;1,3) 
&
& \Phi_{a}(\mbox{B}1)  \, [a,b]
\\[4pt]  
&  
& 
& 24.\,(\widetilde{a3},\widetilde{b}\,;1,2) 
&
& \Phi_{a}(\mbox{B}1)  \, [a,b]
\\[4pt]  
&  
& 
& 25.\,(\widetilde{a},\widetilde{b1}\,;2,3) 
&
& \Phi_{b}(\mbox{B}1)  \, [b,a]
\\[4pt]  
&  
& 
& 26.\,(\widetilde{a},\widetilde{b2}\,;1,3) 
&
& \Phi_{b}(\mbox{B}1)  \, [b,a]
\\[4pt]  
&  
&   
& 27.\,(\widetilde{a},\widetilde{b3}\,;1,2) 
&
& \Phi_{b}(\mbox{B}1)  \, [b,a]
\\[4pt]  \hline
\end{array} \nonumber
\end{align}
\caption{
Summary table of 
$\hat{\sigma}_{\mbox{{\tiny D}}}(\mbox{R}_{11})$.
\label{ap_B_1_tab11}}
\end{table}
\newpage
\rightline{
\underline{
$\hat{\sigma}_{\mbox{{\tiny subt}}}
(\mbox{R}_{11}=gg \to ggg)$  
}
}
\begin{enumerate}
\item[\fbox{Step2}]
\begin{equation}
\hat{\sigma}_{\mbox{{\tiny D}}}
(\,
\mbox{R}_{11},\,
\mbox{I}
\,)
-
\hat{\sigma}_{\mbox{{\tiny I}}}
(\mbox{R}_{11})
=
-\,
\hat{\sigma}_{\mbox{{\tiny I}}}
(\mbox{R}_{11},(2)\mbox{-}1/2,N_{f}{\cal V}_{f\bar{f}})\,.
\end{equation}
\begin{align}
\hat{\sigma}_{\mbox{{\tiny I}}}
(\mbox{R}_{11},(2)\mbox{-}1/2,N_{f}{\cal V}_{f\bar{f}})
&=
-\frac{A_{d}}{S_{\mbox{{\tiny B}}_{1}}} 
\cdot
\frac{N_{f}}{\mbox{C}_{\mbox{{\tiny A}}}} 
{\cal V}_{f\bar{f}} (\ep) 
\Phi(\mbox{B}1)_{d} \,
\cdot
\nonumber\\
&\ \ \ 
\bigl(
\,
[1,2]+
[2,1]+
[1,a]+
[1,b]+
[2,a]+
[2,b]
\,
\bigr)\,.
\end{align}
\item[\fbox{Step3}]
\begin{equation}
\hat{\sigma}_{\mbox{{\tiny D}}}
(\,
\mbox{R}_{11},\,
\mbox{P}
\,)
+
\hat{\sigma}_{\mbox{{\tiny C}}}(\mbox{R}_{11}) 
-
\hat{\sigma}_{\mbox{{\tiny P}}}
(\mbox{R}_{11})=0\,,
\end{equation}
which includes only Dipole\,1\,.
\item[\fbox{Step4}]
\begin{equation}
\hat{\sigma}_{\mbox{{\tiny D}}}
(\,
\mbox{R}_{11},\,
\mbox{K}
\,)
-
\hat{\sigma}_{\mbox{{\tiny K}}}
(\mbox{R}_{11})=
-\,
\hat{\sigma}_{\mbox{{\tiny K}}}
(\mbox{R}_{11},\,{\tt dip}1,
(4)\mbox{-}1,N_{f}h)\,,
\end{equation}
which includes only Dipole\,1\,.
\begin{align}
\hat{\sigma}_{\mbox{{\tiny K}}}
(\mbox{R}_{11},\,{\tt dip}1,
(4)\mbox{-}1,N_{f}h)
&=
-\frac{A_{4}}{S_{\mbox{{\tiny B}}_{1}}} 
\int_{0}^{1}dx \,
\frac{\mbox{T}_{\mbox{{\tiny R}}}N_{f}}
{\mbox{C}_{\mbox{{\tiny A}}}} 
\,\frac{2}{3}h(x)
\ \times
\nonumber\\
\Bigl[
\Phi_{a}(\mbox{B}1,x)_{4}\,
&\bigl(
\langle a,1 \rangle
+
\langle a,2 \rangle
\bigr)
+
\Phi_{b}(\mbox{B}1,x)_{4}\,
\bigl(
\langle b,1 \rangle
+
\langle b,2 \rangle
\bigr)
\Bigr]\,.
\end{align}
\item[\fbox{Step5}]
\begin{equation}
\hat{\sigma}_{subt}(\mbox{R}_{11})=
-\,
\hat{\sigma}_{\mbox{{\tiny I}}}
(\mbox{R}_{11},(2)\mbox{-}1/2,N_{f}{\cal V}_{f\bar{f}})
-\,
\hat{\sigma}_{\mbox{{\tiny K}}}
(\mbox{R}_{11},\,{\tt dip}1,
(4)\mbox{-}1,N_{f}h)\,.
\end{equation}
\end{enumerate}


\clearpage




\section{Summary for the $n$ jets process \label{ap_C}}
\subsection{$\hat{\sigma}_{\mbox{{\tiny subt}}}(\mbox{R}_{1})$  
\label{ap_C_1}}
%
%
\begin{table}[h!]
  \centering
\begin{align}
& \fbox{Step1} \ \ \ 
\hat{\sigma}_{\mbox{{\tiny D}}}(\mbox{R}_{i}) 
=
-\frac{A_{d}}{S_{\mbox{{\tiny R}}_{i}}} 
\cdot
\bigl(\mbox{Factor 1} \bigr)
\cdot
\Phi_{a/b}(\mbox{R}_{i}:\mbox{B}_{j},x)_{d} \,
[ \, y_{emi}, \,y_{spe} ]\, \times n_{deg}\,.
\nonumber\\
&\hat{\sigma}_{\mbox{{\tiny D}}}
(\mbox{R}_{1} = u\bar{u} \to (n+1)\mbox{-}g): 
\ \ S_{\mbox{{\tiny R}}_{1}}=(n+1)\,!\,.
\nonumber\\
&
\begin{array}{|c|c|c|c|c|c|c|} \hline
{\tt Dip}\,j
& \mbox{B}j 
& \mbox{\small Splitting} 
& (y_{emi},y_{spe}) 
& \mbox{Factor 1}
& \Phi(\mbox{B}_{j})
\,[y_{emi}, y_{spe}]
& n_{deg}
\\[4pt] \hline
& & & & & &  \\[-12pt]
{\tt Dip\,1}
& u\bar{u} \to n\mbox{-}g
& (2)-1   
& 1.\,(y_{1},y_{2}) 
& {\cal V}_{gg}/\mbox{C}_{\mbox{{\tiny A}}} 
& \Phi(\mbox{B}1) \, [1,2]
& {\scriptscriptstyle {}_{n+1}C_{2} \cdot (n-1)}
\\[4pt]
&  
& (2)-2  
& 2.\,(y_{1},y_{a})
& \int dx 
{\cal V}_{gg}(x)/\mbox{C}_{\mbox{{\tiny A}}} 
& \Phi_{a}(\mbox{B}1)  \, [1,a]
& {\scriptscriptstyle {}_{n+1}C_{2} \cdot 1}
\\[4pt]  
&  
&   
& 3.\,(y_{1},y_{b}) 
&
& \Phi_{b}(\mbox{B}1)  \, [1,b]
& {\scriptscriptstyle {}_{n+1}C_{2} \cdot 1}
\\[4pt]  
&  
& (3)-1  
& 4.\,(y_{a},y_{1}) 
& \int dx  
{\cal V}^{f,f} (x)/\mbox{C}_{\mbox{{\tiny F}}}
& \Phi_{a}(\mbox{B}1)  \, [a,1]
& {\scriptscriptstyle (n+1)\cdot n}
\\[4pt]  
&  
&   
& 5.\,(y_{b},y_{1}) 
&
& \Phi_{b}(\mbox{B}1)  \, [b,1]
& {\scriptscriptstyle (n+1)\cdot n}
\\[4pt]  
&  
& (3)-2
& 6.\,(y_{a},y_{b}) 
& \int dx
{\widetilde {\cal V}}^{f,f} (x)
/\mbox{C}_{\mbox{{\tiny F}}}
& \Phi_{a}(\mbox{B}1)  \, [a,b]
&{\scriptscriptstyle (n+1)\cdot 1}
\\[4pt]  
&  
&   
& 7.\,(y_{b},y_{a}) 
&
& \Phi_{b}(\mbox{B}1)  \, [b,a]
& {\scriptscriptstyle (n+1)\cdot 1}
\\[4pt] \hline
\end{array} \nonumber
\end{align}
\caption{
Summary table of
$\hat{\sigma}_{\mbox{{\tiny D}}}(\mbox{R}_{1})$.
\label{ap_C_tab1}}
\end{table}
\newpage
\rightline{
\underline{
$\hat{\sigma}_{\mbox{{\tiny subt}}}
(\mbox{R}_{1} = u\bar{u} \to (n+1)\mbox{-}g
)$  
}
}
\begin{enumerate}
\item[\fbox{Step2}] 
\begin{equation}
\hat{\sigma}_{\mbox{{\tiny D}}}
(\,
\mbox{R}_{1},\,
\mbox{I}
\,)
-
\hat{\sigma}_{\mbox{{\tiny I}}}
(\mbox{R}_{1})=-\,
\hat{\sigma}_{\mbox{{\tiny I}}}
(\mbox{R}_{1},(2)\mbox{-}1/2,N_{f}{\cal V}_{f\bar{f}})\,.
\label{c_r1st2}
\end{equation}
\begin{align}
\hat{\sigma}_{\mbox{{\tiny I}}}
(\mbox{R}_{1},(2)\mbox{-}1/2,N_{f}{\cal V}_{f\bar{f}})
&=
-\frac{A_{d}}{S_{\mbox{{\tiny B}}_{1}}}\,
\cdot
\frac{N_{f}}{\mbox{C}_{\mbox{{\tiny A}}}}\,
{\cal V}_{f\bar{f}} (\ep) 
\cdot
\Phi(\mbox{B}1)_{d}\,
\sum_{i,\,K}\,
[\,i,K\,]
\,,
\label{c_r1st2nf}
\end{align}
where the indices of the summation take the
values as
$i=1, ...,n$ and $K=1, ..., n, a, b$,
with the condition $i \not= K$.
\item[\fbox{Step3}]
\begin{equation}
\hat{\sigma}_{\mbox{{\tiny D}}}
(\,
\mbox{R}_{1},\,
\mbox{P}
\,)
+
\hat{\sigma}_{\mbox{{\tiny C}}}(\mbox{R}_{1}) 
-
\hat{\sigma}_{\mbox{{\tiny P}}}
(\mbox{R}_{1})=0\,,
\label{appc_r1st3}
\end{equation}
which includes only Dipole\,1\,.
\item[\fbox{Step4}]
\begin{equation}
\hat{\sigma}_{\mbox{{\tiny D}}}
(\,
\mbox{R}_{1},\,
\mbox{K}
\,)
-
\hat{\sigma}_{\mbox{{\tiny K}}}
(\mbox{R}_{1})=
-\,
\hat{\sigma}_{\mbox{{\tiny K}}}
(\mbox{R}_{1},\,{\tt dip}1,
(3)\mbox{-}1,N_{f}h)\,,
\label{appc_r1st4}
\end{equation}
which includes only Dipole\,1\,.
\begin{align}
\hat{\sigma}_{\mbox{{\tiny K}}}
(\mbox{R}_{1},\,{\tt dip}1,
(3)\mbox{-}1,N_{f}h)
&=-\frac{A_{4}}{S_{\mbox{{\tiny B}}_{1}}} 
\int_{0}^{1}dx \,
\biggl(\frac{2}{3}
\frac{\mbox{T}_{\mbox{{\tiny R}}}N_{f}}
{\mbox{C}_{\mbox{{\tiny A}}}}
\biggr)
\,h(x)
\ \times
\nonumber\\
&
\Bigl[
\Phi_{a}(\mbox{B}1,x)_{4}\,
\sum_{k=1}^{n}
\langle a,k \rangle
+
\Phi_{b}(\mbox{B}1,x)_{4}\,
\sum_{k=1}^{n}
\langle b,k \rangle
\Bigr]\,.
\label{appc_r1st4kh}
\end{align}
\item[\fbox{Step5}]
\begin{equation}
\hat{\sigma}_{subt}(\mbox{R}_{1})=
-\hat{\sigma}_{\mbox{{\tiny I}}}
(\mbox{R}_{1},(2)\mbox{-}1/2,N_{f}{\cal V}_{f\bar{f}})
-\hat{\sigma}_{\mbox{{\tiny K}}}
(\mbox{R}_{1},\,{\tt dip}1,
(3)\mbox{-}1,N_{f}h)\,.
\label{appc_r1st51}
\end{equation}
\end{enumerate}
%
\clearpage
\subsection{$\hat{\sigma}_{\mbox{{\tiny subt}}}(\mbox{R}_{2})$  
\label{ap_C_2}}
%
%
\begin{table}[h!]
  \centering
\begin{align}
&\fbox{Step1} \ \ \
\hat{\sigma}_{\mbox{{\tiny D}}}
(\mbox{R}_{2} = u\bar{u} \to u\bar{u}\,(n-1)\mbox{-}g \,): 
\ \ S_{\mbox{{\tiny R}}_{2}}=(n-1)\,!\,.
\nonumber\\
&
\begin{array}{|c|c|c|c|c|c|c|} \hline
{\tt Dip}\,j
& \mbox{B}j 
& \mbox{\small Splitting} 
& (y_{emi},y_{spe}) 
& \mbox{Factor 1}
& {\scriptscriptstyle \Phi(\mbox{\tiny B}_{j})
\,[y_{emi}, y_{spe}]}
& n_{deg}
\\[4pt] \hline
& & & & & & \\[-12pt]
{\tt Dip\,1}  
& u\bar{u} \to u\bar{u}
& (1)-1   
& 1.\,(y_{1},y_{2})
& {\cal V}_{fg}/\mbox{C}_{\mbox{{\tiny F}}}
& \Phi(\mbox{B}1) \, [1,2]
& {\scriptscriptstyle (n-1) \cdot 1}
\\[2pt]  
& +(n-2)\mbox{-}g
&   
& 2.\,(y_{1},y_{3})
&
& \Phi(\mbox{B}1) \, [1,3]
& {\scriptscriptstyle (n-1) \cdot (n-2)}
\\[2pt]
& 
&   
& 3.\,(y_{2},y_{1})
&
& \Phi(\mbox{B}1) \, [2,1]
& {\scriptscriptstyle (n-1) \cdot 1}
\\[2pt] 
& 
&   
& 4.\,(y_{2},y_{3})
&
& \Phi(\mbox{B}1) \, [2,3]
& {\scriptscriptstyle (n-1) \cdot (n-2)}
\\[2pt] 
&  
& (1)-2  
& 5.\,(y_{1},y_{a})
& \int dx 
{\cal V}_{fg}(x)/\mbox{C}_{\mbox{{\tiny F}}}
& \Phi_{a}(\mbox{B}1)  \, [1,a]
& {\scriptscriptstyle (n-1) \cdot 1}
\\[2pt]  
&  
&   
& 6.\,(y_{1},y_{b})
& 
& \Phi_{b}(\mbox{B}1)  \, [1,b]
& {\scriptscriptstyle (n-1) \cdot 1}
\\[2pt]
&  
&   
& 7.\,(y_{2},y_{a})
& 
& \Phi_{a}(\mbox{B}1)  \, [2,a]
& {\scriptscriptstyle (n-1) \cdot 1}
\\[2pt]
&  
&   
& 8.\,(y_{2},y_{b})
& 
& \Phi_{b}(\mbox{B}1)  \, [2,b]
& {\scriptscriptstyle (n-1) \cdot 1}
\\[2pt]
& 
& (2)-1   
& 9.\,(y_{3},y_{1})
& {\cal V}_{gg}/\mbox{C}_{\mbox{{\tiny A}}}
& \Phi(\mbox{B}1) \, [3,1]
& {\scriptscriptstyle {}_{n-1}C_{2} \cdot 1}
\\[2pt]  
&
&   
& 10.\,(y_{3},y_{2})
&
& \Phi(\mbox{B}1) \, [3,2]
& {\scriptscriptstyle {}_{n-1}C_{2} \cdot 1}
\\[2pt]
& 
&   
& 11.\,(y_{3},y_{4})
&
& \Phi(\mbox{B}1) \, [3,4]
& {\scriptscriptstyle {}_{n-1}C_{2} \cdot (n-3)}
\\[2pt] 
& 
& (2)-2
& 12.\,(y_{3},y_{a})
& \int dx 
{\cal V}_{gg}(x)/\mbox{C}_{\mbox{{\tiny A}}}
& \Phi_{a}(\mbox{B}1) \, [3,a]
& {\scriptscriptstyle {}_{n-1}C_{2} \cdot 1}
\\[2pt] 
&  
&  
& 13.\,(y_{3},y_{b})
& 
& \Phi_{b}(\mbox{B}1)  \, [3,b]
& {\scriptscriptstyle {}_{n-1}C_{2} \cdot 1}
\\[2pt]  
&  
& (3)-1  
& 14.\,(y_{a},y_{1})
& \int dx  
{\cal V}^{f,f} (x)/\mbox{C}_{\mbox{{\tiny F}}}
& \Phi_{a}(\mbox{B}1)  \, [a,1]
& {\scriptscriptstyle (n-1) \cdot 1}
\\[2pt]  
&  
&   
& 15.\,(y_{a},y_{2})
& 
& \Phi_{a}(\mbox{B}1)  \, [a,2]
& {\scriptscriptstyle (n-1) \cdot 1}
\\[2pt]
&  
&   
& 16.\,(y_{a},y_{3})
& 
& \Phi_{a}(\mbox{B}1)  \, [a,3]
& {\scriptscriptstyle (n-1) \cdot (n-2)}
\\[2pt]
&  
&
& 17.\,(y_{b},y_{1})
& 
& \Phi_{b}(\mbox{B}1)  \, [b,1]
& {\scriptscriptstyle (n-1) \cdot 1}
\\[2pt]  
&  
&   
& 18.\,(y_{b},y_{2})
&
& \Phi_{b}(\mbox{B}1)  \, [b,2]
& {\scriptscriptstyle (n-1) \cdot 1}
\\[2pt]
&  
&   
& 19.\,(y_{b},y_{3})
&
& \Phi_{b}(\mbox{B}1)  \, [b,3]
& {\scriptscriptstyle (n-1) \cdot (n-2)}
\\[2pt]
&  
& (3)-2  
& 20.\,(y_{a},y_{b})
& \int dx
{\widetilde {\cal V}}^{f,f} (x)
/\mbox{C}_{\mbox{{\tiny F}}}
& \Phi_{a}(\mbox{B}1)  \, [a,b]
& {\scriptscriptstyle (n-1) \cdot 1}
\\[2pt]  
&  
&   
& 21.\,(y_{b},y_{a})
&
& \Phi_{b}(\mbox{B}1)  \, [b,a]
& {\scriptscriptstyle (n-1) \cdot 1}
\\[2pt]  \hline
& & & & & & \\[-12pt]
{\tt Dip\,2u}  
&  u\bar{u} \to n\mbox{-}g
& (5)-1    
& 22.\,(y_{1},y_{2})
& {\cal V}_{f\bar{f}}/ \mbox{C}_{\mbox{{\tiny A}}}
& \Phi(\mbox{B}2u)  \, [1,2]
& {\scriptscriptstyle 1 \cdot (n-1)}
\\[2pt]
&  
& (5)-2  
& 23.\,(y_{1},y_{a})
& \int dx {\cal V}_{f\bar{f}}(x)/ \mbox{C}_{\mbox{{\tiny A}}}
& \Phi_{a}(\mbox{B}2u)  \, [1,a]
& {\scriptscriptstyle 1 \cdot 1}
\\[2pt]  
&  
& 
& 24.\,(y_{1},y_{b})
&
& \Phi_{b}(\mbox{B}2u)  \, [1,b]
& {\scriptscriptstyle 1 \cdot 1}
\\[2pt]  \hline
& & & & & & \\[-12pt]
{\tt Dip\,3u}  
&  g\bar{u} \to \bar{u}+
& (6)-1    
& 25.\,(y_{a},y_{1})
& \int dx {\cal V}^{f,g} (x)/\mbox{C}_{\mbox{{\tiny A}}}
& \Phi_{a}(\mbox{B}3u)  \, [a,1]
& {\scriptscriptstyle 1 \cdot 1}
\\[2pt]  
& (n-1)\mbox{-}g
&   
& 26.\,(y_{a},y_{2})
&
& \Phi_{a}(\mbox{B}3u)  \, [a,2]
& {\scriptscriptstyle 1 \cdot (n-1)}
\\[2pt]  
&  
& (6)-2  
& 27.\,(y_{a},y_{b})
& \int dx {\widetilde {\cal V}}^{f,g} (x)/
\mbox{C}_{\mbox{{\tiny A}}}
& \Phi_{a}(\mbox{B}3u)  \, [a,b]
& {\scriptscriptstyle 1 \cdot 1}
\\[2pt]  \hline
& & & & & & \\[-12pt]
{\tt Dip\,3\bar{u}}  
&  ug \to u+
& (6)-1    
& 28.\,(y_{b},y_{1})
& \int dx {\cal V}^{f,g} (x)/\mbox{C}_{\mbox{{\tiny A}}}
& \Phi_{b}(\mbox{B}3\bar{u})  \, [b,1]
& {\scriptscriptstyle 1 \cdot 1}
\\[2pt]  
& (n-1)\mbox{-}g
&   
& 29.\,(y_{b},y_{2})
&
& \Phi_{b}(\mbox{B}3\bar{u})  \, [b,2]
& {\scriptscriptstyle 1 \cdot (n-1)}
\\[2pt]  
&  
& (6)-2  
& 30.\,(y_{b},y_{a})
& \int dx {\widetilde {\cal V}}^{f,g} (x)/
\mbox{C}_{\mbox{{\tiny A}}}
& \Phi_{b}(\mbox{B}3\bar{u})  \, [b,a]
& {\scriptscriptstyle 1 \cdot 1}
\\[2pt] 
\hline
\end{array} \nonumber
\end{align}
\caption{
Summary table of 
$\hat{\sigma}_{\mbox{{\tiny D}}}(\mbox{R}_{2u})$.
\label{ap_C_tab2}}
\end{table}
\newpage
\rightline{
\underline{
$\hat{\sigma}_{\mbox{{\tiny subt}}}
(\mbox{R}_{2} = u\bar{u} \to u\bar{u}\,(n-1)\mbox{-}g 
\,)$
}
}
\begin{enumerate}
\item[\fbox{Step2}]
\begin{equation}
\hat{\sigma}_{\mbox{{\tiny D}}}
(\,
\mbox{R}_{2},\,
\mbox{I}
\,)
-
\hat{\sigma}_{\mbox{{\tiny I}}}
(\mbox{R}_{2})=-\,
\hat{\sigma}_{\mbox{{\tiny I}}}
(\mbox{R}_{2},(2)\mbox{-}1/2,N_{f}{\cal V}_{f\bar{f}})\,.
\label{c_r2st2}
\end{equation}
\begin{align}
\hat{\sigma}_{\mbox{{\tiny I}}}
(\mbox{R}_{2},(2)\mbox{-}1/2,N_{f}{\cal V}_{f\bar{f}})
&=
-\frac{A_{d}}{S_{\mbox{{\tiny B}}_{1}}}\,
\cdot
\frac{N_{f}}{\mbox{C}_{\mbox{{\tiny A}}}}\,
{\cal V}_{f\bar{f}} (\ep) 
\cdot
\Phi(\mbox{B}1)_{d}\,
\sum_{i,\,K}\,
[\,i,K\,]
\,,
\label{c_r2st2nf}
\end{align}
where the indices of the summation take the
values
$i=3, ...,n$ and $K=1, ..., n, a, b$
with the condition $i \not= K$.
\item[\fbox{Step3}]
\ \ 
\begin{equation}
\hat{\sigma}_{\mbox{{\tiny D}}}
(\,
\mbox{R}_{2},\,
\mbox{P}
\,)
+
\hat{\sigma}_{\mbox{{\tiny C}}}(\mbox{R}_{2}) 
-
\hat{\sigma}_{\mbox{{\tiny P}}}
(\mbox{R}_{2})=0\,,
\end{equation}
which is separated into three relations 
for Dipoles 1, 3$u$, and 3$\bar{u}$ as
\begin{align}
\hat{\sigma}_{\mbox{{\tiny D}}}
(\,
\mbox{R}_{2},\,
\mbox{P},\,{\tt dip}j
\,)
+
\hat{\sigma}_{\mbox{{\tiny C}}}(\mbox{R}_{2},\,
{\tt dip}j
) 
-
\hat{\sigma}_{\mbox{{\tiny P}}}
(\mbox{R}_{2},
\,{\tt dip}j)=0\,.
\end{align}
\item[\fbox{Step4}]
\begin{equation}
\hat{\sigma}_{\mbox{{\tiny D}}}
(\,
\mbox{R}_{2},\,
\mbox{K}
\,)
-
\hat{\sigma}_{\mbox{{\tiny K}}}
(\mbox{R}_{2})=
-\,
\hat{\sigma}_{\mbox{{\tiny K}}}
(\mbox{R}_{2},\,{\tt dip}1,
(3)\mbox{-}1,N_{f}h)\,,
\label{appc_r2st4}
\end{equation}
which is separated into three relations 
for Dipoles 1, 3$u$, and 3$\bar{u}$ as
\begin{align}
\hat{\sigma}_{\mbox{{\tiny D}}}
(\,
\mbox{R}_{2},\,
\mbox{K},\,{\tt dip}1
\,)
-
\hat{\sigma}_{\mbox{{\tiny K}}}
(\mbox{R}_{2},
\,{\tt dip}1)
&=
-\,
\hat{\sigma}_{\mbox{{\tiny K}}}
(\mbox{R}_{2},\,{\tt dip}1,
(3)\mbox{-}1,N_{f}h)\,,
\\
\hat{\sigma}_{\mbox{{\tiny D}}}
(\,
\mbox{R}_{2},\,
\mbox{K},\,{\tt dip}3u
\,)
-
\hat{\sigma}_{\mbox{{\tiny K}}}
(\mbox{R}_{2},
\,{\tt dip}3u)
&=0\,,
\\
\hat{\sigma}_{\mbox{{\tiny D}}}
(\,
\mbox{R}_{2},\,
\mbox{K},\,{\tt dip}3\bar{u}
\,)
-
\hat{\sigma}_{\mbox{{\tiny K}}}
(\mbox{R}_{2},
\,{\tt dip}3\bar{u})
&=0\,.
\end{align}
\begin{align}
\hat{\sigma}_{\mbox{{\tiny K}}}
(\mbox{R}_{2},\,{\tt dip}1,
(3)\mbox{-}1,N_{f}h)
&=-\frac{A_{4}}{S_{\mbox{{\tiny B}}_{1}}} 
\int_{0}^{1}dx \,
\biggl(
\frac{2}{3}
\frac{\mbox{T}_{\mbox{{\tiny R}}}N_{f}}
{\mbox{C}_{\mbox{{\tiny A}}}}
\biggr)
\,h(x)
\ \times
\nonumber\\
&
\Bigl[
\Phi_{a}(\mbox{B}1,x)_{4}\,
\sum_{k=3}^{n}
\langle a,k \rangle
+
\Phi_{b}(\mbox{B}1,x)_{4}\,
\sum_{k=3}^{n}
\langle b,k \rangle
\Bigr]\,.
\end{align}
\item[\fbox{Step5}]
\begin{equation}
\hat{\sigma}_{subt}(\mbox{R}_{2})=
-
\hat{\sigma}_{\mbox{{\tiny I}}}
(\mbox{R}_{2},(2)\mbox{-}1/2,N_{f}{\cal V}_{f\bar{f}})
-
\hat{\sigma}_{\mbox{{\tiny K}}}(\mbox{R}_{2},
{\tt dip}1,(3)\mbox{-}1,N_{f}h)
+
\hat{\sigma}_{\mbox{{\tiny D}}}(\mbox{R}_{2},\,{\tt dip}2)\,.
\label{r2st51}
\end{equation}
\begin{equation}
\hat{\sigma}_{\mbox{{\tiny D}}}(\mbox{R}_{2},\,{\tt dip}2)
=
\hat{\sigma}_{\mbox{{\tiny D}}}
(\,
\mbox{R}_{2},\,
{\tt dip}2,(5)\mbox{-}1/2,
\,
{\cal V}_{f\bar{f}}
\,)
+
\hat{\sigma}_{\mbox{{\tiny D}}}
(\,
\mbox{R}_{2},\,
{\tt dip}2,(5)\mbox{-}2,
\,h\,)\,.
\label{r2st5dip2}
\end{equation}
\begin{align}
\hat{\sigma}_{\mbox{{\tiny D}}}
(\,
\mbox{R}_{2},\,
{\tt dip}2,(5)\mbox{-}1/2,
\,
{\cal V}_{f\bar{f}}
\,)
&=-\frac{A_{d}}{S_{\mbox{{\tiny R}}_{2}}} 
\cdot
\frac{1}{\mbox{C}_{\mbox{{\tiny A}}}} 
{\cal V}_{f\bar{f}} (\ep) 
\cdot 
\Phi(\mbox{B}2)_{d}\, \times
\nonumber\\
&\hspace{20mm} \bigl(
\,
[1,2] \cdot (n-1)+
[1,a]+
[1,b]
\,
\bigr)\,.
\label{r2st5dip2a}
\\
\hat{\sigma}_{\mbox{{\tiny D}}}
(\,
\mbox{R}_{2},\,
{\tt dip}2,(5)\mbox{-}2,
\,h\,)
&=
-\frac{A_{4}}{S_{\mbox{{\tiny R}}_{2}}} 
\int_{0}^{1}dx \,
\biggl(
\frac{2}{3}
\frac{\mbox{T}_{\mbox{{\tiny R}}}}
{\mbox{C}_{\mbox{{\tiny A}}}} 
\biggr)
\,h(x)
\times
\nonumber\\
& \hspace{20mm}
\Bigl(
\Phi_{a}(\mbox{B}2,x)_{4}\,
\langle 1,a \rangle
+
\Phi_{b}(\mbox{B}2,x)_{4}\,
\langle 1,b \rangle
\Bigr)\,.
\label{r2st5last}
\end{align}
\end{enumerate}
%
\clearpage
\subsection{$\hat{\sigma}_{\mbox{{\tiny subt}}}(\mbox{R}_{3})$  
\label{ap_C_3}}
%
%
\begin{table}[h!]
  \centering
\begin{align}
&\fbox{Step1} \ \ \
\hat{\sigma}_{\mbox{{\tiny D}}}
(\mbox{R}_{3} = u\bar{u} \to d\bar{d}+(n-1)\mbox{-}g \,): 
\ \ S_{\mbox{{\tiny R}}_{3}}=(n-1)\,!\,.
\nonumber\\
&
\begin{array}{|c|c|c|c|c|c|c|} \hline
{\tt Dip}\,j
& \mbox{B}j 
& \mbox{\small Splitting} 
& (y_{emi},y_{spe}) 
& \mbox{Factor 1}
& {\scriptscriptstyle \Phi(\mbox{\tiny B}_{j})
\,[y_{emi}, y_{spe}]}
& n_{deg}
\\[4pt] \hline
& & & & & & \\[-12pt]
{\tt Dip\,1}  
& u\bar{u} \to d\bar{d}
& (1)-1   
& 1.\,(y_{1},y_{2})
& {\cal V}_{fg}/\mbox{C}_{\mbox{{\tiny F}}}
& \Phi(\mbox{B}1) \, [1,2]
& {\scriptscriptstyle (n-1) \cdot 1}
\\[4pt]  
& +(n-2)\mbox{-}g
&   
& 2.\,(y_{1},y_{3})
&
& \Phi(\mbox{B}1) \, [1,3]
& {\scriptscriptstyle (n-1) \cdot (n-2)}
\\[4pt]
& 
&   
& 3.\,(y_{2},y_{1})
&
& \Phi(\mbox{B}1) \, [2,1]
& {\scriptscriptstyle (n-1) \cdot 1}
\\[4pt] 
& 
&   
& 4.\,(y_{2},y_{3})
&
& \Phi(\mbox{B}1) \, [2,3]
& {\scriptscriptstyle (n-1) \cdot (n-2)}
\\[4pt] 
&  
& (1)-2  
& 5.\,(y_{1},y_{a})
& \int dx 
{\cal V}_{fg}(x)/\mbox{C}_{\mbox{{\tiny F}}}
& \Phi_{a}(\mbox{B}1)  \, [1,a]
& {\scriptscriptstyle (n-1) \cdot 1}
\\[4pt]  
&  
&   
& 6.\,(y_{1},y_{b})
& 
& \Phi_{b}(\mbox{B}1)  \, [1,b]
& {\scriptscriptstyle (n-1) \cdot 1}
\\[4pt]
&  
&   
& 7.\,(y_{2},y_{a})
& 
& \Phi_{a}(\mbox{B}1)  \, [2,a]
& {\scriptscriptstyle (n-1) \cdot 1}
\\[4pt]
&  
&   
& 8.\,(y_{2},y_{b})
& 
& \Phi_{b}(\mbox{B}1)  \, [2,b]
& {\scriptscriptstyle (n-1) \cdot 1}
\\[4pt]
& 
& (2)-1   
& 9.\,(y_{3},y_{1})
& {\cal V}_{gg}/\mbox{C}_{\mbox{{\tiny A}}}
& \Phi(\mbox{B}1) \, [3,1]
& {\scriptscriptstyle {}_{n-1}C_{2} \cdot 1}
\\[4pt]  
&
&   
& 10.\,(y_{3},y_{2})
&
& \Phi(\mbox{B}1) \, [3,2]
& {\scriptscriptstyle {}_{n-1}C_{2} \cdot 1}
\\[4pt]
& 
&   
& 11.\,(y_{3},y_{4})
&
& \Phi(\mbox{B}1) \, [3,4]
& {\scriptscriptstyle {}_{n-1}C_{2} \cdot (n-3)}
\\[4pt] 
& 
& (2)-2
& 12.\,(y_{3},y_{a})
& \int dx 
{\cal V}_{gg}(x)/\mbox{C}_{\mbox{{\tiny A}}}
& \Phi_{a}(\mbox{B}1) \, [3,a]
& {\scriptscriptstyle {}_{n-1}C_{2} \cdot 1}
\\[4pt] 
&  
&  
& 13.\,(y_{3},y_{b})
& 
& \Phi_{a}(\mbox{B}1)  \, [3,b]
& {\scriptscriptstyle {}_{n-1}C_{2} \cdot 1}
\\[4pt]  
&  
& (3)-1  
& 14.\,(y_{a},y_{1})
& \int dx  
{\cal V}^{f,f} (x)/\mbox{C}_{\mbox{{\tiny F}}}
& \Phi_{a}(\mbox{B}1)  \, [a,1]
& {\scriptscriptstyle (n-1) \cdot 1}
\\[4pt]  
&  
&   
& 15.\,(y_{a},y_{2})
& 
& \Phi_{a}(\mbox{B}1)  \, [a,2]
& {\scriptscriptstyle (n-1) \cdot 1}
\\[4pt]
&  
&   
& 16.\,(y_{a},y_{3})
& 
& \Phi_{a}(\mbox{B}1)  \, [a,3]
& {\scriptscriptstyle (n-1) \cdot (n-2)}
\\[4pt]
&  
&
& 17.\,(y_{b},y_{1})
& 
& \Phi_{b}(\mbox{B}1)  \, [b,1]
& {\scriptscriptstyle (n-1) \cdot 1}
\\[4pt]  
&  
&   
& 18.\,(y_{b},y_{2})
&
& \Phi_{b}(\mbox{B}1)  \, [b,2]
& {\scriptscriptstyle (n-1) \cdot 1}
\\[4pt]
&  
&   
& 19.\,(y_{b},y_{3})
&
& \Phi_{b}(\mbox{B}1)  \, [b,3]
& {\scriptscriptstyle (n-1) \cdot (n-2)}
\\[4pt]
&  
& (3)-2  
& 20.\,(y_{a},y_{b})
& \int dx
{\widetilde {\cal V}}^{f,f} (x)
/\mbox{C}_{\mbox{{\tiny F}}}
& \Phi_{a}(\mbox{B}1)  \, [a,b]
& {\scriptscriptstyle (n-1) \cdot 1}
\\[4pt]  
&  
&   
& 21.\,(y_{b},y_{a})
&
& \Phi_{b}(\mbox{B}1)  \, [b,a]
& {\scriptscriptstyle (n-1) \cdot 1}
\\[4pt]  \hline
& & & & & & \\[-12pt]
{\tt Dip\,2u}  
&  u\bar{u} \to n\mbox{-}g
& (5)-1    
& 22.\,(y_{1},y_{2})
& {\cal V}_{f\bar{f}}/ \mbox{C}_{\mbox{{\tiny A}}}
& \Phi(\mbox{B}2u)  \, [1,2]
& {\scriptscriptstyle 1 \cdot (n-1)}
\\[4pt]
&  
& (5)-2  
& 23.\,(y_{1},y_{a})
& \int dx {\cal V}_{f\bar{f}}(x)/ \mbox{C}_{\mbox{{\tiny A}}}
& \Phi_{a}(\mbox{B}2u)  \, [1,a]
& {\scriptscriptstyle 1 \cdot 1}
\\[4pt]  
&  
& 
& 24.\,(y_{1},y_{b})
&
& \Phi_{b}(\mbox{B}2u)  \, [1,b]
& {\scriptscriptstyle 1 \cdot 1}
\\[4pt] 
\hline
\end{array} \nonumber
\end{align}
\caption{
Summary table of 
$\hat{\sigma}_{\mbox{{\tiny D}}}(\mbox{R}_{3u})$.
\label{ap_C_tab3}}
\end{table}
\newpage
\rightline{
\underline{
$\hat{\sigma}_{\mbox{{\tiny subt}}}
(\,\mbox{R}_{3} = u\bar{u} \to d\bar{d}\,(n-1)\mbox{-}g\,)$
}
}
\begin{enumerate}
\item[\fbox{Step2}]
\begin{equation}
\hat{\sigma}_{\mbox{{\tiny D}}}
(\,
\mbox{R}_{3},\,
\mbox{I}
\,)
-
\hat{\sigma}_{\mbox{{\tiny I}}}
(\mbox{R}_{3})=-\,
\hat{\sigma}_{\mbox{{\tiny I}}}
(\mbox{R}_{3},(2)\mbox{-}1/2,N_{f}{\cal V}_{f\bar{f}})\,.
\label{c_r3st2}
\end{equation}
\begin{align}
\hat{\sigma}_{\mbox{{\tiny I}}}
(\mbox{R}_{3},(2)\mbox{-}1/2,N_{f}{\cal V}_{f\bar{f}})
&=
-\frac{A_{d}}{S_{\mbox{{\tiny B}}_{1}}}\,
\cdot
\frac{N_{f}}{\mbox{C}_{\mbox{{\tiny A}}}}\,
{\cal V}_{f\bar{f}} (\ep) 
\cdot
\Phi(\mbox{B}1)_{d}\,
\sum_{i,\,K}\,
[\,i,K\,]
\,,
\label{c_r3st2nf}
\end{align}
where the indices of the summation take the
values
$i=3, ...,n$ and $K=1, ..., n, a, b$
with the condition $i \not= K$.
\item[\fbox{Step3}]
\ \ 
\begin{equation}
\hat{\sigma}_{\mbox{{\tiny D}}}
(\,
\mbox{R}_{3},\,
\mbox{P}
\,)
+
\hat{\sigma}_{\mbox{{\tiny C}}}(\mbox{R}_{3}) 
-
\hat{\sigma}_{\mbox{{\tiny P}}}
(\mbox{R}_{3})=0\,,
\end{equation}
which includes Dipole\,1.
\item[\fbox{Step4}]
\begin{equation}
\hat{\sigma}_{\mbox{{\tiny D}}}
(\,
\mbox{R}_{3},\,
\mbox{K}
\,)
-
\hat{\sigma}_{\mbox{{\tiny K}}}
(\mbox{R}_{3})=
-\,
\hat{\sigma}_{\mbox{{\tiny K}}}
(\mbox{R}_{3},\,{\tt dip}1,
(3)\mbox{-}1,N_{f}h)\,,
\label{appc_r3st4}
\end{equation}
which includes Dipole\,1.
\begin{align}
\hat{\sigma}_{\mbox{{\tiny K}}}
(\mbox{R}_{3},\,{\tt dip}1,
(3)\mbox{-}1,N_{f}h)
&=-\frac{A_{4}}{S_{\mbox{{\tiny B}}_{1}}} 
\int_{0}^{1}dx \,
\biggl(
\frac{2}{3}
\frac{\mbox{T}_{\mbox{{\tiny R}}}N_{f}}
{\mbox{C}_{\mbox{{\tiny A}}}} 
\biggr)
\,h(x)
\ \times
\nonumber\\
&
\Bigl[
\Phi_{a}(\mbox{B}1,x)_{4}\,
\sum_{k=3}^{n}
\langle a,k \rangle
+
\Phi_{b}(\mbox{B}1,x)_{4}\,
\sum_{k=3}^{n}
\langle b,k \rangle
\Bigr]\,.
\end{align}
\item[\fbox{Step5}]
\begin{equation}
\hat{\sigma}_{subt}(\mbox{R}_{3})=
-
\hat{\sigma}_{\mbox{{\tiny I}}}
(\mbox{R}_{3},(2)\mbox{-}1/2,N_{f}{\cal V}_{f\bar{f}})
-
\hat{\sigma}_{\mbox{{\tiny K}}}(\mbox{R}_{3},
{\tt dip}1,(3)\mbox{-}1,N_{f}h)
+
\hat{\sigma}_{\mbox{{\tiny D}}}(\mbox{R}_{3},\,{\tt dip}2)\,.
\label{r3st51}
\end{equation}
\begin{equation}
\hat{\sigma}_{\mbox{{\tiny D}}}(\mbox{R}_{3},\,{\tt dip}2)
=
\hat{\sigma}_{\mbox{{\tiny D}}}
(\,
\mbox{R}_{3},\,
{\tt dip}2,(5)\mbox{-}1/2,
\,
{\cal V}_{f\bar{f}}
\,)
+
\hat{\sigma}_{\mbox{{\tiny D}}}
(\,
\mbox{R}_{3},\,
{\tt dip}2,(5)\mbox{-}2,
\,h\,)\,.
\end{equation}
\begin{align}
\hat{\sigma}_{\mbox{{\tiny D}}}
(\,
\mbox{R}_{3},\,
{\tt dip}2,(5)\mbox{-}1/2,
\,
{\cal V}_{f\bar{f}}
\,)
&=-\frac{A_{d}}{S_{\mbox{{\tiny R}}_{3}}} 
\cdot
\frac{1}{\mbox{C}_{\mbox{{\tiny A}}}} 
{\cal V}_{f\bar{f}} (\ep) 
\cdot 
\Phi(\mbox{B}2)_{d}\, \times
\nonumber\\
&\hspace{20mm} \bigl(
\,
[1,2] \cdot (n-1)+
[1,a]+
[1,b]
\,
\bigr)\,.
\\
\hat{\sigma}_{\mbox{{\tiny D}}}
(\,
\mbox{R}_{3},\,
{\tt dip}2,(5)\mbox{-}2,
\,h\,)
&=
-\frac{A_{4}}{S_{\mbox{{\tiny R}}_{3}}} 
\int_{0}^{1}dx \,
\biggl(
\frac{2}{3}
\frac{\mbox{T}_{\mbox{{\tiny R}}}}
{\mbox{C}_{\mbox{{\tiny A}}}}
\biggr)
\,h(x)
\times
\nonumber\\
& \hspace{20mm}
\Bigl(
\Phi_{a}(\mbox{B}2,x)_{4}\,
\langle 1,a \rangle
+
\Phi_{b}(\mbox{B}2,x)_{4}\,
\langle 1,b \rangle
\Bigr)\,.
\label{r3st5last}
\end{align}
\end{enumerate}


\clearpage
{\footnotesize


\providecommand{\href}[2]{#2}

}

\end{document}